Universiteit Antwerpen

Faculteit Wetenschappen
Departement Natuurkunde

# Optical Investigation of Electrical Spin Injection into Semiconductors

Proefschrift voorgelegd tot het behalen van de graad van doctor in de Wetenschappen aan de Universiteit Antwerpen te verdedigen door

VASYL MOTSNYI

Promotoren: Prof. Dr. E. Goovaerts
Prof. Dr. G. Borghs

Antwerpen, 2003

In samenwerking met
*IMEC* vzw,
Interuniversitair Micro-Elektronica Centrum

# Abstract

Spintronics, or spin-dependent electronics aims at combination of the intrinsic properties of charge carriers, the charge, as well as their spin. Here, the quantum mechanical concept of spin brings an amazing new functionality into mainstream of charge-based electronics. It allows engineering devices with lower power consumption and higher functionality. Moreover, the spin of electron or proton can be used to store and process information locally, on the nanoscale. Though, semiconductors offer the most of functional advantages (very long electron spin scattering time, for example), the fabrication of semiconductor-based spintronic device, where spins are manipulated, stored or processed in a semiconductor, up to now remains a challenge. The lack of efficient way to create spin-polarized charge ensemble in a semiconductor by electrical means, spin injection, leaves semiconductor spintronics on the level of many others, only nice ideas.

At the same time, the conventional ferromagnetic metals, like Co or Fe, have very large electron spin polarization, even at room temperature. Their physical properties and fabrication technology are well known. All these fundamental properties make them almost an ideal candidate for the utilization as all kinds of spin sources for spintronic applications. Unfortunately, due to the very different character of the charge transport in ferromagnetic metals and semiconductors, the transport of spin-polarized carriers through their interface is not so evident and even questionable. Fortunately, the III-V semiconductors, like GaAs, offer a unique opportunity for this type of investigation. As in these semiconductors, the radiative recombination of spin-polarized electrons with holes leads to transfer of the angular momentum of electron, the spin, into the angular momentum of light, the polarization.

This doctoral research, which was performed in the MN-group (Magnetoelectronics & Nanotechnology) of IMEC, consists of optical investigation of electrical spin injection in a III-V semiconductor heterostructure from a ferromagnetic metal. It is a result of collaboration between University of Antwerp, the ECMP-group (Experimental Condensed Matter Physics) and IMEC.

In this research, the potential of the MIS-type heterostructures (Metal/ Insulator/ Semiconductor), well-known among semiconductor device engineers, for the electrical injection of spin-polarized electrons into a semiconductor from a ferromagnetic metal is investigated in detail. It is shown how one can achieve more than 60% spin transparency of the ferromagnetic metal / semiconductor interface at low and room temperatures. In addition, a new experimental method for optical investigation of electrical spin injection has been developed. It is based on the electron spin precession in the external magnetic field, once spin-polarized carriers have been injected into a semiconductor. This allows clear separation of the spin injection from the side effects that mask the spin injection and even can be entirely responsible for the measured quantities. Finally, the efficient dynamic polarization of spins of lattice nuclei, due to the hyperfine interaction with the spin-polarized electrons, electrically injected into a semiconductor is demonstrated. At the moment it is believed that such spin-polarized nuclei will allow fabrication of a new generation of very dense memories, or even to perform a new class of very efficient and sophisticated computational algorithms, the quantum computing.

# Acknowledgements

Having this opportunity I would like to gratefully acknowledge all people I was lucky to work and communicate with during my Ph.D. years. If I forgot somebody, this does not mean the opposite in any way…

Thanks to

Prof. Dr. Etienne Goovaerts and Prof. Dr. Gustaaf Borghs, my promoters, for giving me this opportunity to perform research, for believing and encouraging me, for the countless fruitful and stimulating discussions, for being curious, and their constructive criticism.

Prof. Dr. ir. Jo De Boeck, my group leader, who practically shared the responsibility of promoter at IMEC, for creation of excellent conditions for doing research, for advices and support on a daily basis, and for being a good Group Leader.

Prof. Dr. Viacheslav Safarov, whom I was lucky to meet once in Italy on our numerous EC Project meetings. Looking back, I see that this meeting had an enormous impact on me personally, as well as on the way we have been dealing with the spin injection problem in general.

Dr. Wim Van Roy, for the backup I always had concerning magnetism, device processing or MBE growth.

Pol Van Dorpe, my successor, who succeeded not only in quick assimilation of existing knowledge, but in bringing new ideas also. And of course, as well as to

Dr. ir. Hans Bove, Dr. ir. Jo Das and Mayke Nijboer for the development of the state-of-the-art technology of the tunnel oxide spin-injectors.

Dr. Stefan Degroote, Willem van de Graaf and Dr. Stefan Nemeth for the MBE sample growth.

Dr. ir. Reiner Windisch and Cathleen Rooman for the LED hints, and frequent use of their setup for the fast LED inspections.

Dr. ir. Liesbet Lagae and Wouter Eyckmans for MOKE measurements.

Dr. Barundeb Dutta for engineering hints and numerous discussions about life in general, and world of R&D in particular.

Johan Feyaerts and Erwin Vandenplas, for the technical support.

Albert Debie for the fabrication of my electronic needs.

Benny Charliers and Karel Van Ranst for the fabrication of my mechanical ‘wonders’.

Chantal Deboes for being a secretary and a travel agent not only for the "CEOs", but for me sometimes also.

Dr. ir. Paul Heremans, Dr. ir. Chris Van Hoof, Prof. Dr. Vladimir Arkhipov, Prof. Dr. Vladimir Fomin, Prof.Dr Michail Baklanov, the gurus of device architectures and processing, physics of organics semiconductor, etc., for the very interesting communication I had.

other my colleagues and friends

Dr. Karen Attenborough, Dr. Joost Bekaert, Liu Zhiyu (Guy), Dr. Jean-Louis Primus, Roel Wirix-Speetjens, Dr.ir. Kristof Dessein, Kristof Daemen, Dr. ir. Wouter Ruythooren, Kristiaan De Greve, Iwijn De Vlaminck, Koen De Keersmaecker, Dr. Marianne Germain, Dr. Maarten Leys, Raf Vandersmissen, Dr. Kang-Hoon Choi, Stijn De Jonge, Dimitri Janssen, Stijn De Vusser, Dr. Joachim John, Dr. ir. Lihuan Song, Dr. Vesselin Vassilev, Dr. ir. Lars Zemmerman, Johan Reynaert,

and just friends of mine

Dr. Michail Abramov, Natasha and Sergej Kokorev and their kids, Lucas Irazabal, Padre Miguel Gonzalez Chandia, Leyre Castro Ruiz, Alberto Pezzutto, Gaitano Fortunato and Laura Collada, Joan Puig Vall, Jordi Moral-Cardoner, Sofie Depreitere and her family for sharing part, or all of these years together with me in Belgium.

The financial support for this work has been provided under the framework of collaboration between IMEC and Flemish universities, and as IMEC Innovation Project, which was greatly appreciated.

And the last, but not the least to

Prof. Dr. Eugenia Buzaneva, who actually gave me this crazy idea to go abroad for doing Ph.D. At the end, she was right…

and my family: my mother Maria and father Prof.Dr. Fedir Motsnyi, together with my sister Olena.

Здоров'я, добробуту, і хай Вам завжди щастить…

September 10, 2003
Leuven, Belgium

Vasyl Motsnyi

# Table of Contents

# List of Acronyms

| | |
|---|---|
| 2DEG | Two-Dimensional Electron Gas |
| 3D | Three-Dimensional |
| AMR | Anisotropic MagnetoResistance |
| BAP | Birr-Aronov-Pikus (mechanism of spin scattering) |
| CGR | Compound Growth Rate |
| CIP | Current-In-Plane |
| CPP | Current-Perpendicular-to-the Plane |
| *c* | Conduction (band) |
| ccp | Cubic Close-Packed (lattice type) |
| DMS | Diluted Magnetic Semiconductor |
| DOS | Density Of States |
| DP | D’yakonov-Perel (mechanism of spin scattering) |
| EMMI | EMission MIcroscopy |
| EY | Elliot-Yafet (mechanism of spin scattering) |
| FET | Field Effect Transistor |
| FM | Ferromagnetic Metal |
| fcc | Face-Centered Cubic (lattice type) |
| GMR | Giant MagnetoResistance |
| hcp | Hexagonal Close-Packed (lattice type) |
| *hh* | Heavy-Hole (band) |
| ITRS | International Technology Roadmap for Semiconductors |
| IMEC | Interuniversity MicroElectronics Center |
| LED | Light Emitting Diode |
| *lh* | Light-Hole (band) |
| MCD | Magnetooptical Circular Dichroism |
| MIS | Metal-Insulator-Semiconductor |
| MOKE | MagnetoOptical Kerr Effect |
| MRAM | Magnetic Random Access Memory |
| MTJ | Magnetic Tunnel Junction |
| NM | Normal Metal (Conventional paramagnetic metal like Cu) |
| QW | Qunatum Well |
| RT | Room Temperature |
| SEM | Scanning Electron Microscopy |
| SIA | Semiconductor Industry Association |
| STM | Scanning Tunneling Microscopy |
| SVT | Spin Valve Transistor |
| spin-LED | spin Light Emitting Diode |
| TB | Tunnel Barrier |
| TEM | Transmission Electron Microscopy |
| TMR | Tunnel MagnetoResistance |
| UHV | Ultra High Vaccum |

# List of Symbols

| Symbol | Description | Units |
|---|---|---|
| $B$ | Magnetic induction | T |
| $B$ | Magnetic field, B-field [1] ( $B = \mu_0 \cdot H$ ) | T |
| $B_N$ | Nuclear magnetic field | T |
| $B_L$ | Local fluctuating magnetic field, due to the dipol-dipol interactions of nuclei | T |
| $\Delta B$ | Half-width of Hanle curve | T |
| $D$ | Diffusion constant | $m^2/s$ |
| $D$ | Circular polarization of light due to Magnetooptical Circular Dichroism | % |
| $E$ | Energy | eV |
| $E_C$ | Conduction band edge | eV |
| $E_F$ | Fermi energy / Fermi level | eV |
| $E_g$ | Semiconductor band gap | eV |
| $E_V$ | Valence band edge | eV |
| $e$ | Charge of a single electron (=$1.6022\times10^{-19}$) | C |
| $G$ | Electrical conductance | S |
| $g$ | Landé factor (giromagnetic ration for an electron) | |
| $g^*$ | Effective g-factor | |

| Symbol | Description | Units |
|---|---|---|
| $H$ | Magnetic field, H-field [1] | A/m |
| $H_C$ | Coercivity | A/m |
| $h$ | Planck's constant (=6.6262×10$^{-34}$) | J·s |
| $I$ | Total magnetic moment of magnetized solid | A·m$^2$ |
| $I$ | Electrical current | A |
| $I^+$ $(I^-)$ | Intensities of right (left) circularly polarized compoents of light | W |
| $J$ | Angular quantum momentum | |
| $j$ | Current density | A/μm$^2$ |
| $k$ | Boltzmann constant (=1.38066×10$^{-23}$) | J/K |
| $L$ | Orbital quantum momentum | |
| $M$ | Magnetization | A/m |
| $M_0$ | Saturation magnetization | A/m |
| $M_R$ | Remanence | A/m |
| $m_J$ | Magnetic quantum number | |
| $m^*$ | Effective mass of an electron | kg |
| $N(E_F)$ | Number of electrons at Fermi level | |
| $N_A$ | Acceptor concentration | cm$^{-3}$ |
| $N_D$ | Donor concentration | cm$^{-3}$ |
| $n$ | Number of electrons | |
| $n$ | Electron concentration | cm$^{-3}$ |
| $P$ | Degree of circular polarization of light | % |
| $p$ | Hole concentration | cm$^{-3}$ |
| $R$ | Electrical resistance | Ω |
| $S$ | Average electron spin | J·s |
| $s$ | Spin of individual electron ( $s = \hbar/2$ ) | J·s |
| $T$ | Temperature | K |
| $T_C$ | Curie temperature | K |
| $T_S$ | Spin lifetime | s |
| $T^+$ $(T^-)$ | Transmission coeficients for the right (left) circularly polarized component of light | % |

| Symbol | Description | Units |
|---|---|---|
| $U$ | Electrical bias | V |
| $V$ | Volume of a solid | $m^3$ |
| $v_F$ | Fermi velocity | m/s |
| $\chi$ | Magnetic susceptibility | |
| $\Phi$ | Work function | eV |
| $\lambda$ | Wavelength of optical wave | m |
| $\lambda_{sf}$ | Spin-flip length | m |
| $\mu_B$ | Bohr magneton ($=9.2740\times10^{-24}$) | J/T |
| $\mu_0$ | Permeability of free space ($=4\pi\times10^{-7}$) | $Wb\cdot A^{-1}\cdot m^{-1}$ |
| $\mu$ | Electrochemical potential | eV |
| $\nu$ | Frequency of electromagnetic wave | Hz |
| $\Pi$ | Degree of spin polarization | % |
| $\pi$ | Linearly polarized light | |
| $\sigma$ | Electrical Conductance | S |
| $\sigma^+$ ($\sigma^-$) | Right (left) circularly polarized compoents of light | W |
| $\tau$ | Electron lifetime | s |
| $\tau_s$ | Spin scattering time | s |
| $\tau_{sf}$ | Spin-flip time | s |
| $\Omega$ | Larmor frequency | Hz |

# Scientific Notations

| | |
|---|---|
| $A$ | Scalar |
| $\vec{A}$ | Vector |
| $A \cdot B$ | Multiplication of $A$ and $B$ |
| $\left(\vec{A} \cdot \vec{B}\right)$ | Scalar multiplication of $\vec{A}$ and $\vec{B}$ |
| $\left[\vec{A} \times \vec{B}\right]$ | Vector multiplication of $\vec{A}$ and $\vec{B}$ |

# List of Figures

# List of Tables

# Optical Investigation of Electrical Spin Injection into Semiconductors

# 1. Introduction

Since the invention of the bipolar transistor by John Bardeen, William Shockley, and Walter Brattain in 1947, and consequent invention of the integrated circuits by Jack Kilby and Robert Noyce in 1959 [2], the conventional way to improve the functionality of the electrical circuits remains the traditional downscaling of device dimensions. Since 1959 the device downscaling is well-described by the so-called Moore's Law [3], which states that the number of components fabricated on chip doubles every 18 months. Following this law one will end up only with one electron per device around 2020. It is obvious that this cannot go on forever, the existing device architectures as well as material properties have fundamental limitations far beyond that point. It is generally accepted that in the near future device dimensions are going to approach their physical limits. Moreover, it seems that this point is going to be reached already in the very near future. Fig.1.1 shows the assigned technology nodes for the semiconductor industry development following the Semiconductor Industry Association (SIA) and International Technology Roadmap for Semiconductors (ITRS) [4]. The wall, known as 'red brick wall' indicates the point at which there are no known solutions for most technical areas and where an essential research breakthrough is needed. The wall is still there, where it was a couple of years ago.

Under such circumstances the development of new device architectures, which may enable future increase of chip functionality within existing technology, is, without any doubt, an important advance.

One of such concepts is very rapidly evolving field of spintronics. Here the quantum mechanical concept of electron spin brings an amazing new functionality into the mainstream of charge-based electronics [5, 6, 7]. It allows engineering devices with higher performance with regards to power consumption, functionality and is an enabler of new device architectures.

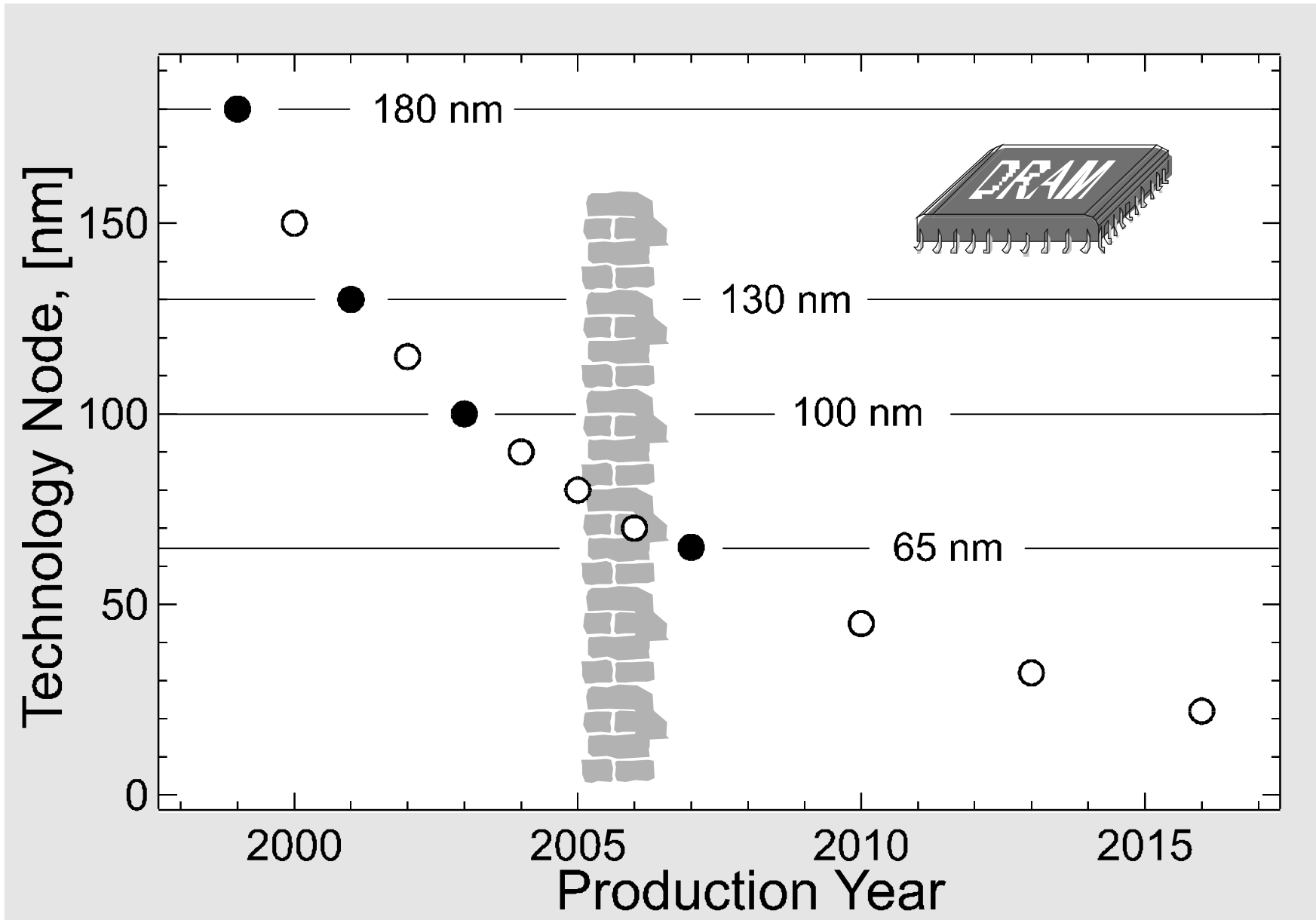

Fig.1.1. SIA Roadmap of the semiconductor industry development [4]. The wall shows the current physical limitations of the known technological solutions.

As it was shown in 1925 by Goudsmith and Uhlenbeck, that apart of its charge, an electron has an intrinsic angular momentum, which is known as spin, and connected with it magnetic momentum $M = g \cdot \mu_B \cdot s/\hbar$, where $s = \hbar/2$ is electron spin. The quantization of the spin for free electron implies that during measurements along a certain direction it can be found only in two possible states, namely spin-up and spin-down. It further turns out that in the conventional metals or semiconductors the number of spin-up charges is equal to spin-down ones, hence the electrical current does not contain any spin information. However, in the ferromagnetic solids there is imbalance of spin-up and spin-down states due to the 'exchange interactions'. It follows that application of electrical bias to such solid leads to transfer of spin and connected with it magnetic momentum. Obviously, one could try to use this phenomenon for device implementation.

Indeed the spin-dependent electron scattering in magnetic multilayers, namely Giant Magnetoresistance Junctions (GMR) has been a driving force for the increase of the hard drive capacity already for a number of years (Fig.1.2) [8, 5-7].

The effect of spin-dependent tunneling, Tunnel Magnetoresistance (TMR) in Magnetic Tunnel Junctions (MTJ) has allowed development of Magnetic Random Access Memories (MRAM), which have potential of replacing CMOS based non-volatile FLASH memories in the following couple of years [9, 10, 5-7].

Moreover, the spin-dependent architectures are promising to be successful in the downscaling run also, as device dimension limits are lying in the range of couple of nanometers. On the other hand, the GMR and MTJ device architectures are passive electrical devices as they contain only metallic multilayers. From this point of view the semiconductor-based architectures are much more attractive, as creation of active device exploiting electron spin in order to change its properties should be feasible.

At present, there is a strong belief that use of spin-depending architectures in the semiconductor-based devices is going to revolutionize modern world of computation and data storage. The relatively long spin memory times in the conventional semiconductors like Si or GaAs encourage thinking minds worldwide for exploration of unknown. Such devices could use spin itself to store and process data without any need to move charges at all, which requires much less power than conventional electronics. Moreover, while 'the mystical property of electron spin is revolutionizing the memory business", if 'it can do the same with logic, electronics will become spintronics' [10]. The quantum mechanical nature of spin combined with quantum algorithms would allow creation of completely new computational devices, quantum computers [11, 12, 13, 5]. This type of thinking rises new and new supporters around the globe as traditional world of electronics starts to show more and more the quantum character of the nature.

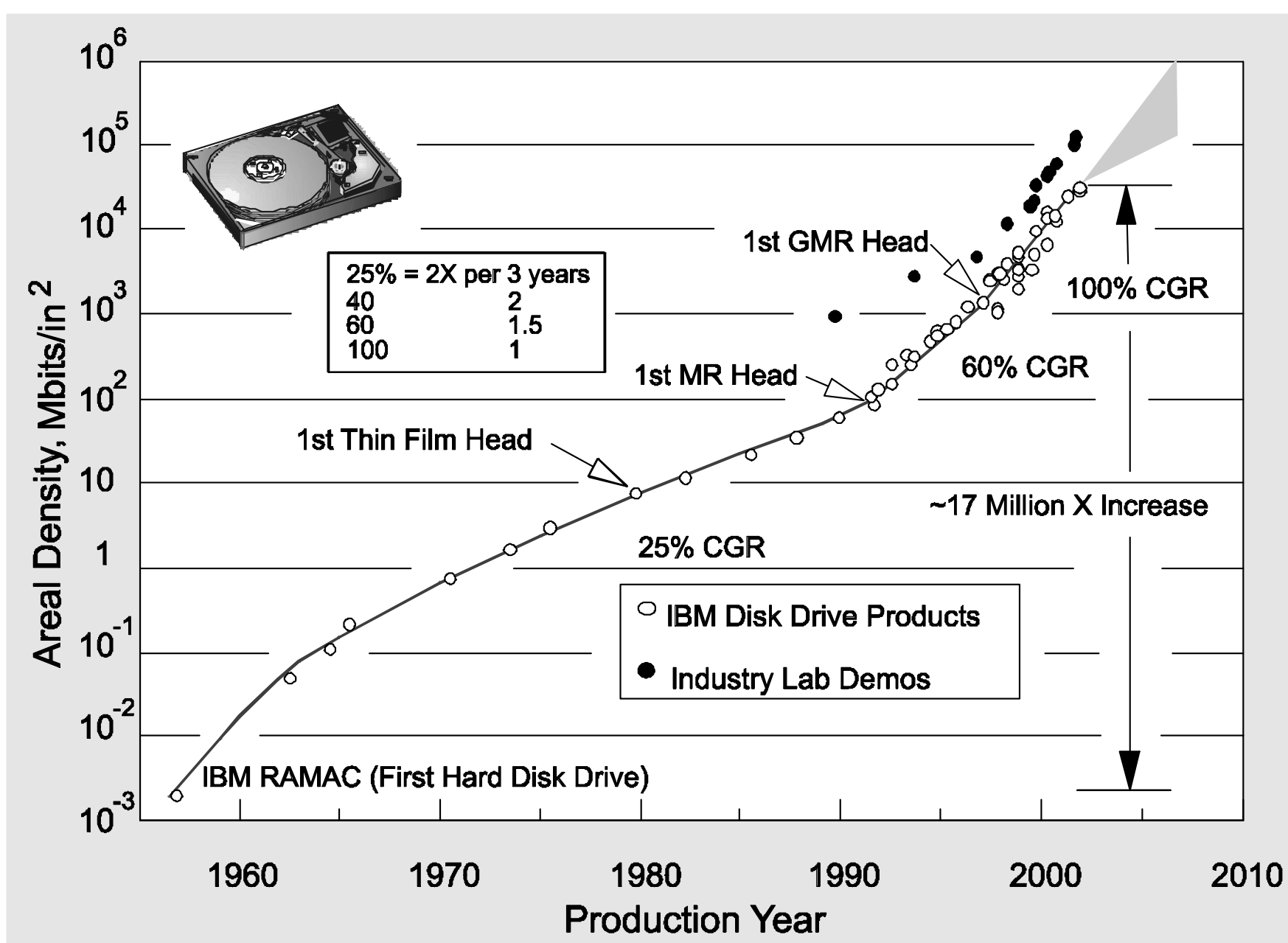

Fig.1.2. Historical outline of the areal recording density in the conventional hard drive [8]. The thin film heads utilize the change of the resistance due to the Lorentz force, like in the conventional Hall bar. MR heads utilize the effect of Anisotropic Magnetoresistance (AMR), which is caused by the interaction of spin and lattice [14]. The GMR head utilizes the spin-dependent scattering in magnetic multilayers. CGR stands for compound growth rate.

All these benefits require efficient creation of a spin-polarized charge ensemble within a semiconductor, preferably in the direct electrical contact at room temperature. Ferromagnetic metals completely satisfy these needs, as relatively high spin polarization exists in these metals even at room temperature. Moreover, their fabrication technology and physical properties are well studied. Unfortunately, the preliminary experiments have shown that electrical spin injection into a semiconductor from a ferromagnetic metal in the direct electrical contact is not a trivial task and is nearly impossible [15, 16, 17]. None of the fabricated devices, combining different three terminal geometries for electrical spin injection and detection of spin-polarized electrons on the ferromagnetic metal / semiconductor interface have been able to show clear spin-dependent effects.

GaAs and other III-V semiconductors are already known for a long time for their ability to efficiently convert angular momentum of light into electron spin and vice versa [18]. Moreover, the electron and spin relaxation processes are well studied in these semiconductors. Further, as there is a known correlation between spin polarization of injected charges and polarization of the emitted light, they provide a unique opportunity for optical investigation of electrical spin injection into semiconductors in a light emitting diode (spin-LED) type heterostructures - across a single ferromagnetic metal / semiconductor interface only.

Taking into account the frustration of the preliminary experiments and expertise existing in the Magnetoelectronic Group in IMEC on both, the fabrication of III-V based semiconductor heterostructures with different functionality and unique expertise of fabrication high quality magnetoelectronic components (GMR, TMR devices, etc.):

**The aim of this doctoral research was the realization and optical investigation of electrical spin injection into a semiconductor from a ferromagnetic metal in the direct electrical contact,**

and evaluation of side effects, which mask the spin-dependent effects and may be entirely responsible for the measured quantities.

In this thesis, Chapters 2 and 3 introduce important concepts used in spin-dependent electronics and supply the most important references containing more detailed study of specific areas.

Chapter 2 introduces the concept of spin-polarized density of states which exists in ferromagnetic solids due to exchange interactions. Here it is shown how this polarization is transferred into the spin polarization of electrical current, how one can measure the latter in the independent experiment and how one can make an extremely sensitive electrical devices with a new functionality based on the ‘new’ property of electron- its spin. Further, a brief introduction is given to the new types of ferromagnetic materials: ferromagnetic semiconductors and half-metals. The concluding section gives a comparison between the most common ferromagnetic materials used in spin-dependent electronics and intrinsic spin polarization of charge carriers.

Chapter 3 introduces the known and the near-term manufacturable semiconductor-based spintronic devices. It describes the first preliminary experiments targeting electrical spin injection and detection, the first disappointments and specifies the first challenges. It further, describes the intrinsic properties of electron spin in GaAs and presents the state-of-the-art experiments targeting electrical spin injection into semiconductors.

Chapter 4 describes the experimental restriction of a ferromagnetic metal / GaAs system, different geometries of observation of spin-dependent effects - different types of spin-LEDs, and possible experimental artifacts. Further, it introduces the experimental approach developed during work on the topic of this thesis. It allows fundamental separation of effects caused by spin injection from the side ones, which is impossible in any other geometry, reported previously. In addition, it reveals the important information on electron spin kinetics in the semiconductor simultaneously.

Chapter 5 gives practical considerations concerning electrical injection of electrons into conduction band of GaAs, which is somehow similar to the problems arising on the metal/ Si interface. Here it is shown that fundamental problem of ohmic contacts to these semiconductors significantly differentiate spin injection into GaAs or Si from the case of InAs reported earlier [16, 17]. Further, the technological aspects of fabrication of the ferromagnetic metal / $AlO_X$/ semiconductor MIS spin-LEDs, the spin-LED types fabricated during work presented in this thesis, and their preliminary characterization are given.

Chapter 6 compares the spin injection achieved in the fabricated spin-LEDs of different types at low and room temperatures and discusses different physical phenomena observed for the first time in these types of heterostructures.

# I. Electron Spin in Electronics: A Key to Understanding

# 2. Ferromagnetism, Spin Degree of Freedom, New Device Concepts

A solid, being introduced in an external magnetic field $\vec{H}$, generally obtains a magnetic moment $\vec{I}$ and connected with it magnetization $\vec{M} = \vec{I}/V$, where $V$ is the volume of a solid. As result, depending on the value $\chi = M/H$, which is called the magnetic susceptibility and which describes the response of a solid to the external magnetic field, any material can be attributed to the one of the following classes: diamagnetic ($\chi < 0$), paramagnetic ($0 < \chi < 1$) or ferromagnetic ($\chi \gg 1$). Among the solids containing only one element of the Mendeleev table, the ferromagnetic are some transition metals (Fe, Co, Ni) and more heavy elements of lanthanide group (Gd, Dy, Ho, Er). In magnetic measurements, the ferromagnetic order of a solid generally is characterized by the well-known parameters, like coercivity $H_C$, remanence $M_R$, saturation magnetization $M_0$ and Curie temperature of the transition into the ferromagnetic state $T_C$. In electrical measurements, the ferromagnetic order in a solid can be responsible for a full range of fundamentally new interesting effects. Even though, ferromagnetism nowadays has been discovered in a large amount of materials including organic molecules containing no magnetic atoms [19, 20, 21], the most attractive materials for device implementation, due to well-studied physical properties and technology available, remain traditional transition metals and their alloys.

## 2.1. Ferromagnetic Metals, Exchange Spin Splitting, Spin Polarization

The early realization that magnetic properties of ferromagnetic metals simply reflect the imbalance of spin-up and spin-down electron ensembles [22], and that the band structure of ferromagnetic metal in order to avoid some electrons having large energy has splitting for different electron spin orientations, directly leads to conclusion that there is a difference in the density of states for spin-up and spin-down electrons on the Fermi level of ferromagnetic metal as well.

Indeed such imbalance exists, Fig.2.1 shows the Density Of States (**DOS**) versus occupation energy for spin-up (↑) and spin-down (↓) electrons for a conventional noble metal, like Cu, and for conventional ferromagnetic metals, like Ni, Co and Fe, the corresponding atomic numbers and electronic configurations for the ground state neutral gaseous atom.

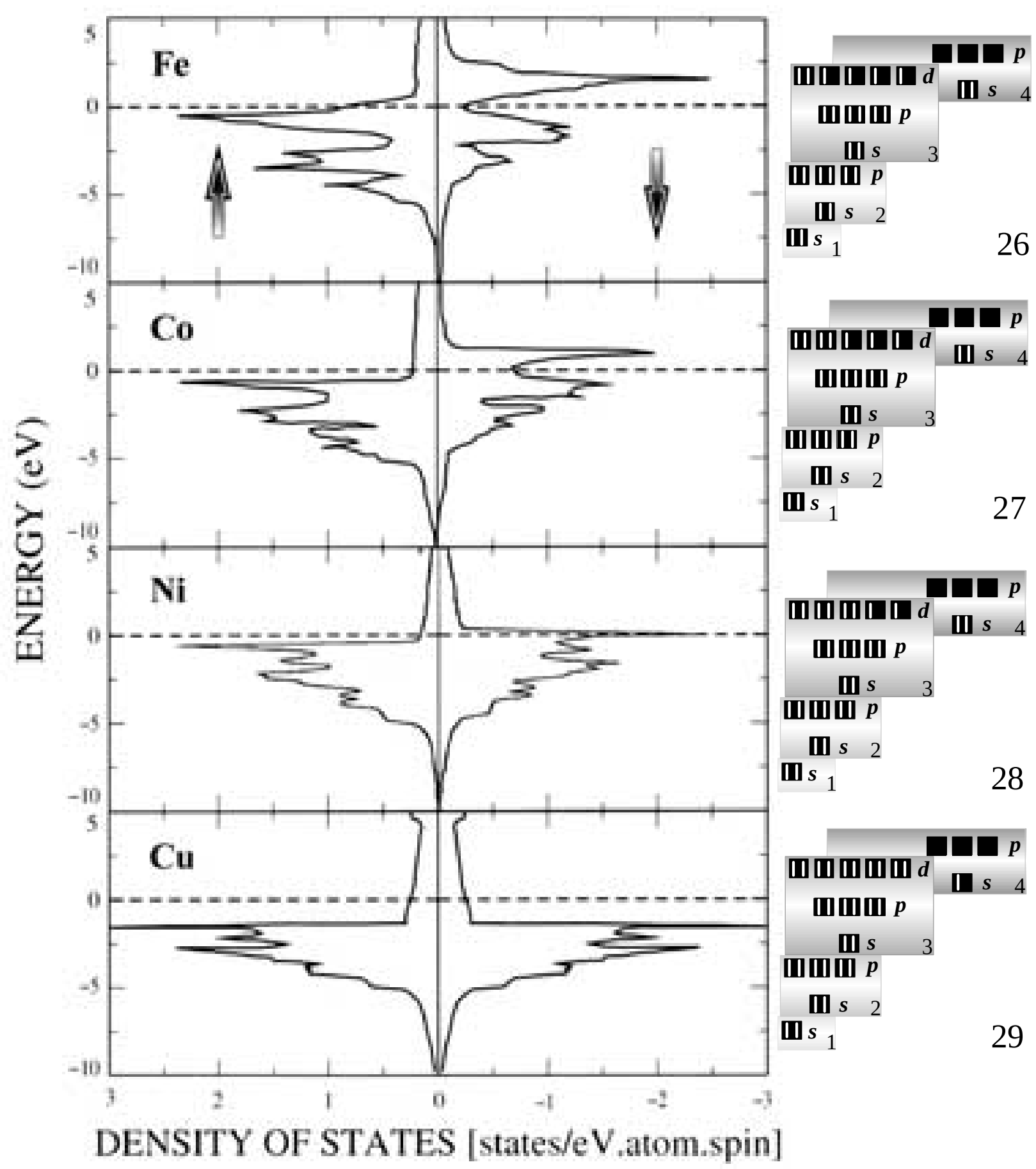

Fig.2.1. Spin resolved density of states for ferromagnetic Fe, Co and Ni, conventional paramagnetic metal Cu, the corresponding atomic numbers and electronic configurations for the ground state neutral gaseous atom.

In fact the density of states is composed of a broad sp-hybridized band with low-density of states superimposed on a narrow 3d band with high density of states. In Cu, generally having cubic close-packed (ccp) crystal structure, the d-band is completely filled and is lying below the Fermi level. As result, the density of states at the Fermi level is formed only by the sp-hybridized band and is low.

In the ferromagnetic metals the d-band is not completely filled and is spin-split (for ferromagnetic transition metals $\Delta E \simeq 1$ eV) by the so-called exchange interaction. Iron, generally having body-centered cubic (bcc) crystal structure, has the largest atomic spin moment of 2.22 Bohr magnetons ($\mu_B$) and is a 'weak' ferromagnet, as there are both $3d^{\uparrow}$ and $3d^{\downarrow}$ electrons at the Fermi level. Cobalt having hexagonal close-packed (hcp) or face-centered cubic (fcc) and nickel with fcc crystal structures have atomic spin moments of $\sim 1.7 \cdot \mu_B$ and $0.6 \cdot \mu_B$, respectively, and are 'strong' ferromagnets as $3d^{\downarrow}$ states lie entirely below the Fermi level.

The spin-splitting leads to a different density of states and thus to a different number of spin-up and spin-down electrons at the Fermi level, or spin polarization. The straightforward definition of spin polarization $\Pi$, which is in fact the most common characteristic of a ferromagnetic material in spin-dependent electronics, is

$$\Pi = \frac{N^{\uparrow}(E_F) - N^{\downarrow}(E_F)}{N^{\uparrow}(E_F) + N^{\downarrow}(E_F)} \qquad (2.1)$$

where, $N^{\uparrow}(E_F)$ and $N^{\downarrow}(E_F)$ are the number of spin-up and spin-down electrons at the Fermi level, respectively.

Further, such fundamental difference in band structure implies a difference in conductivity [23] as well. Copper as other noble metals is known to be a better conductor than transition metals contrary to its small density of states at the Fermi level, comparing to Co for example. As it follows from the band structure calculation, electrons in the d-band have more localized character with larger effective mass compared to the s-electrons [24, 25, 26, 27, 28], so that mainly itinerant s-electrons carry out the electrical current. It further follows that effect of electron scattering from impurities in these metals also differ significantly. Taking into account the Pauli exclusion principle, an electron can be scattered from an impurity only to quantum states, which are not occupied by the other electrons. At low temperatures, all the states with the energies E below the Fermi energy $E_F$ are occupied and those with $E>E_F$ are empty. Since scattering from impurities is elastic (no loss of energy during scattering occurs), electrons at the Fermi level can be scattered only to the states in the immediate vicinity of the Fermi level. As result, the scattering probability is proportional to the density of states at the Fermi level. In Cu the d-band is completely occupied and density of states at the Fermi level is low, so the electron scattering probability is also low. On contrary, the ferromagnetic metals have only partially occupied d-band with high density of states that acts as a new channel for scattering of the conduction electrons into d-band (the so-called Mott scattering), lowering the conductivity. Moreover, as density of states for spin-up and spin-down electrons in the d-band of ferromagnetic metal differ significantly, the s-electrons in the two different spin subbands experience different scattering rate, thus having different mobility. As result the current in the ferromagnetic metal is carried out by spin-polarized s-electrons.

## 2.2. What is Spin Polarization of the Current ?

As it was pointed out in the previous section, the difference in the density of states leads to imbalance of spin-up and spin-down electrons at the Fermi level. The application of electrical bias produces electrical current governed by a number of different physical phenomena that result in the spin polarization of the last one, but following the rules strictly, of opposite sign to the one defined as simple difference in the density of states (Eq.2.1).

It appears that if the transport involves direct tunneling between two solids or in the case of ballistic transport the polarization of the current is more complicated function of the spin-polarized density of states at the Fermi level [29]. It should be weighted by the Fermi velocities $v_F$ and averaged over different electron subbands:

$$\Pi_1 = \frac{\left\langle N^{\uparrow}(E_F)\cdot v_F^{\uparrow}\right\rangle - \left\langle N^{\downarrow}(E_F)\cdot v_F^{\downarrow}\right\rangle}{\left\langle N^{\uparrow}(E_F)\cdot v_F^{\uparrow}\right\rangle + \left\langle N^{\downarrow}(E_F)\cdot v_F^{\downarrow}\right\rangle} \qquad (2.2)$$

In the case of diffusive transport the spin polarization of the current can also be derived from the density of states, but it should be weighted already by square of Fermi velocities:

$$\Pi_2 = \frac{\left\langle N^{\uparrow}(E_F)\cdot\left(v_F^{\uparrow}\right)^2\right\rangle - \left\langle N^{\downarrow}(E_F)\cdot\left(v_F^{\downarrow}\right)^2\right\rangle}{\left\langle N^{\uparrow}(E_F)\cdot\left(v_F^{\uparrow}\right)^2\right\rangle + \left\langle N^{\downarrow}(E_F)\cdot\left(v_F^{\downarrow}\right)^2\right\rangle} \qquad (2.3)$$

Thus, the only case when spin polarization, defined by Eq.2.1, is a real measure for the spin polarization of the current is the case when electrons are emitted into free space as result of photo or field emission.

## 2.3. Assessing Spin Polarization

As it has been discussed in the previous section, the spin polarization of the current in the ferromagnetic metal is a quite interesting phenomenon arising just from the band structure of material. Theoretical calculations can give some inside view on these very complicated phenomena, however an experimental approach is needed in order to classify materials and test their usefulness for practical applications. In the next sections a short overview of the available techniques is given. The summary of these data including modern improvements of material fabrication and characterization is presented in the Section 2.6.

### 2.3.1. Field and Photo Emission

The idea that application of a strong electrical field to a ferromagnet should result in the emission of the spin-polarized electrons dates the beginning of the last century [30]. It took another 34 years before this effect was actually observed [31, 32, 33, 34, 35, 36, 37]. In these experiments the spin polarization of the electrons emitted within extreme vicinity $\sim 100\ \mathrm{meV}$ of the Fermi surface from the apex of the etched metal tips was examined. The measurements have revealed $\Pi_{Ni}(polycrytalline) = +13\%$, $\Pi_{Ni}(100) = -3\%$, $\Pi_{Ni}(110) = +5\%$, $\Pi_{Fe}(100) = +25\%$, $\Pi_{Fe}(110) = -5\%$, $\Pi_{Fe}(111) = +20\%$. As one can see, these values differ significantly in sign, as well as numerically for electrons emitted even from the different crystallographic orientations of the same material, indicating the extreme complexity of the phenomenum. However, the analysis of such data generally is complicated due to the influence of the high electric field on the surface electronic structure, and due to the contamination with other elements even in UHV, as electrons are emitted from the very last surface layer of the metal.

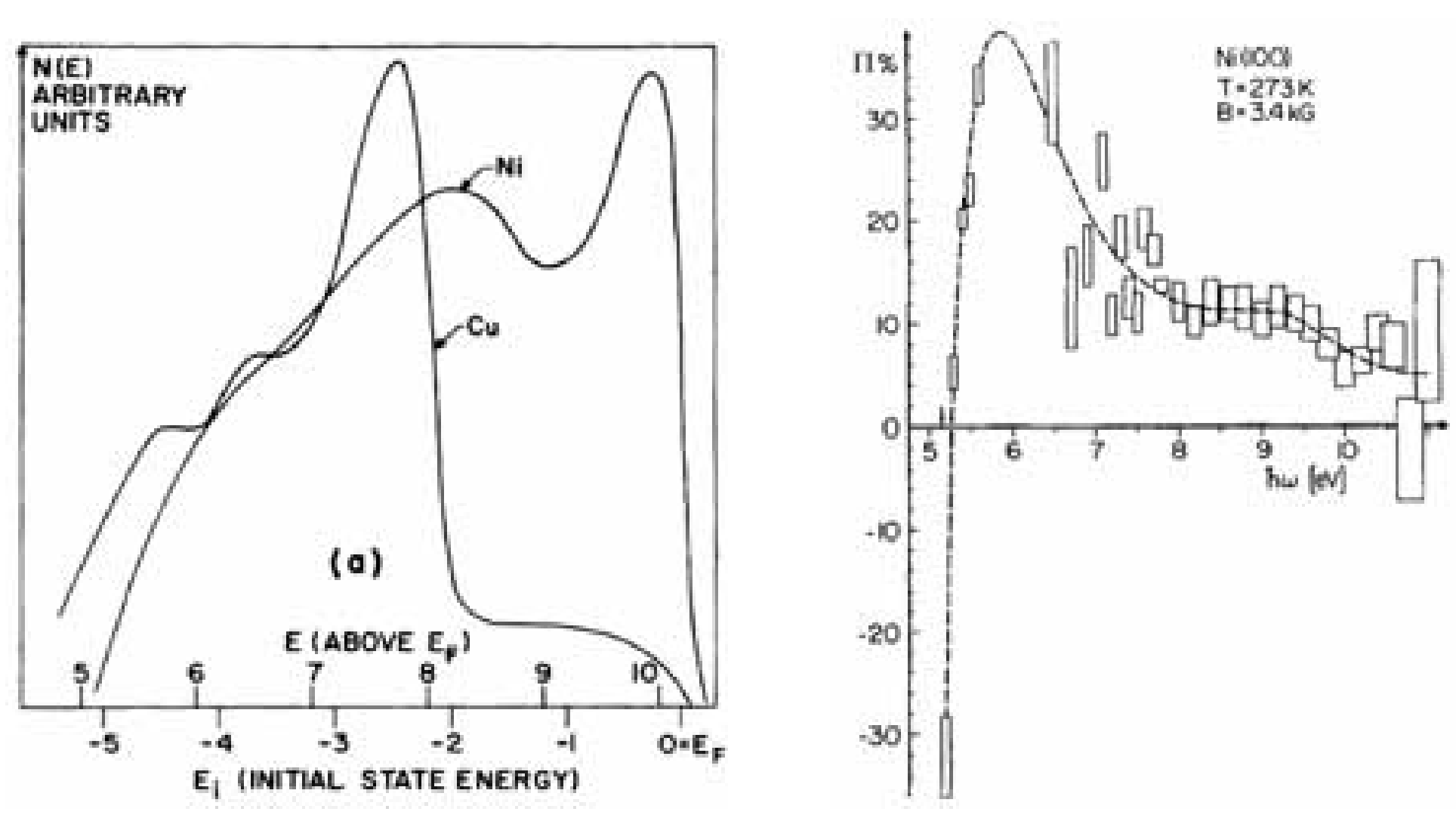

Fig.2.2. Optical density of states and spin polarization of photoemitted electrons from paramagnetic Cu and ferromagnetic Ni after Ref.[48] and Ref.[42].

In photoemission [38, 39, 40, 41, 42, 43, 44, 45, 46, 47] the illumination of the metallic sample with ultraviolet light ($h \cdot \upsilon = 4 \ldots 10\ \mathrm{eV}$) results in the emission of electrons from an extreme vicinity ~2nm of the surface, making it less dependent on the very final surface layer, comparing to the field emission for example. Moreover, the photoemission allows sampling the total (optical) density of states (Eq.2.1) and profiling of the entire band structure of the solid. Fig.2.2 (left) shows the measured optical density of states for polycrystalline Cu and Ni after Ref.[48] (to be compared with Fig.2.1). On Fig.2.2 (right) the spin resolved density of states for Ni after Ref.[42] is shown, where improved sensitivity and energy resolution allowed observation the predominance of the minority carriers at $h \cdot \nu - \Phi < 0.05$ eV, while at slightly higher energies spin polarization changes sign and becomes positive. In these experiments, the spin polarization at the Fermi level $\Pi_{Gd} = +5.5\%$, $\Pi_{Fe} = +54\%$, $\Pi_{Co} = +21\%$ and $\Pi_{Ni} = \pm 30\%$ was reported.

### 2.3.2. Zeeman Splitting of the Electron Levels in Superconductors

The idea of using the spin-split electron levels in a superconductor for measuring the spin polarization of the ferromagnetic metal was introduced by Tedrow and Meservey [49, 50, 51, 52, 53]. In their experiments, electrons tunneled through a thin nonmagnetic insulating barrier into a superconducting film that acted as a spin detector, when magnetic field $B$ was applied to the structure. The applied magnetic field defines the orientation of the magnetic moment and therefore the spin orientation in the magnetic film. It also splits the sharply peaked density of states in the superconducting film into spin-up and spin-down states separated by energy $\pm \mu_B \cdot B$ (Fig.2.3a).

In the case of electron tunneling into such spin detector from a normal metal (no spin-polarized density of states) the conduction curve Fig.2.3b shows 4 peaks related to density of states in the normal metal and spin-split density of states in the superconducting state, as indicated by spin-resolved conductances. However, if the metal under examination has a spin-polarized density of states, then each of the spin conductances is weighted by the relative density of states for that spin channel (Fig.2.3.c). As result, by careful analysis of the current $I^{\uparrow}$ and $I^{\downarrow}$ transmitted through the tunnel barrier into the superconductor spin states, as function of electrical bias and applied magnetic field, the spin polarization of the tunneling electrons can be estimated by the relative heights of the four conductance peaks:

$$\Pi_{Ef} = \frac{(\sigma_4 - \sigma_2) - (\sigma_1 - \sigma_3)}{(\sigma_4 - \sigma_2) + (\sigma_1 - \sigma_3)} \qquad (2.4)$$

These experiments have shown $\Pi_{Fe} = +44\%$, $\Pi_{Co} = +34\%$, $\Pi_{Ni} = +11\%$ and $\Pi_{Gd} = +4.3\%$.

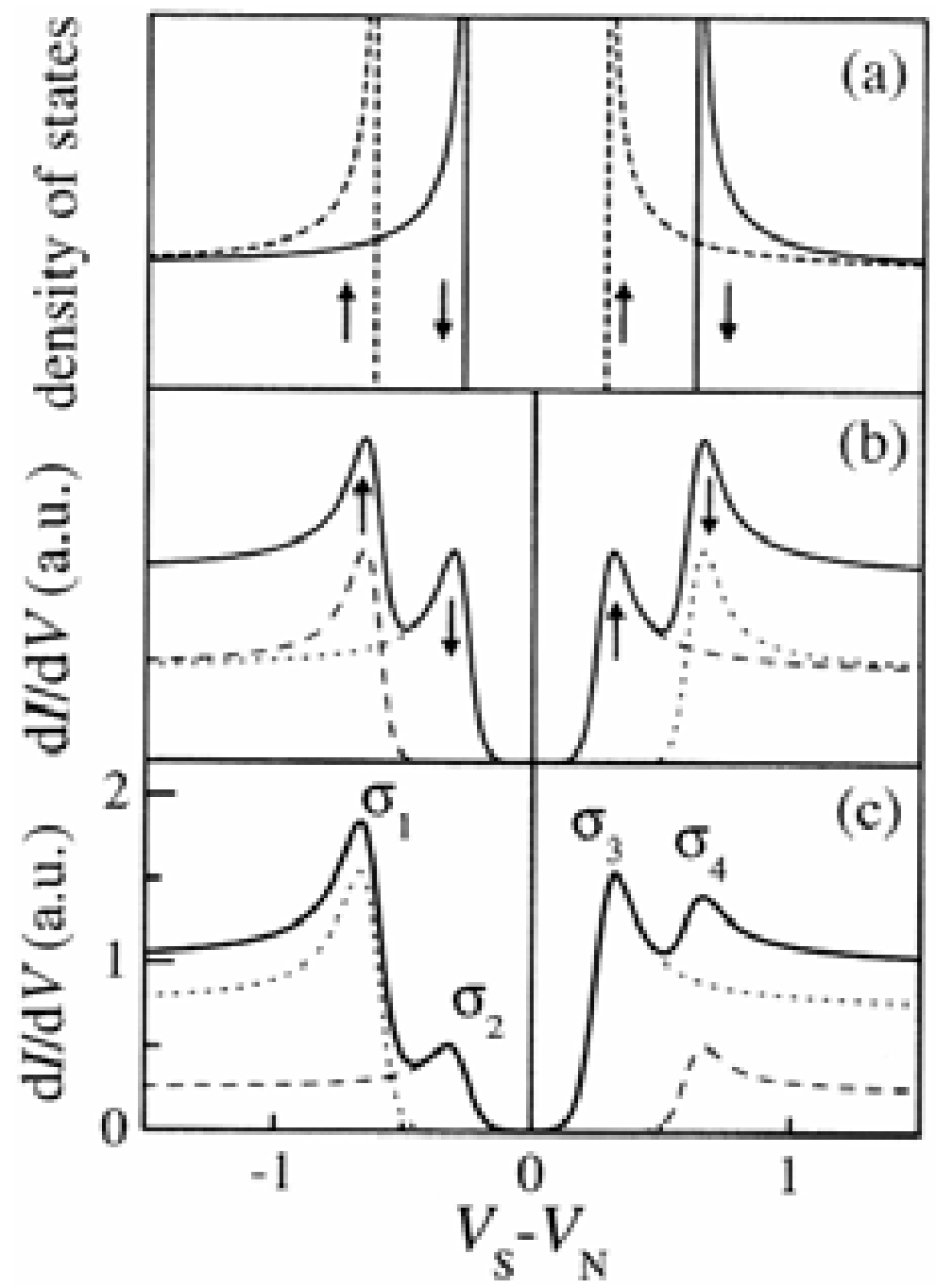

Fig.2.3. Magnetic field splitting of the quasi-particle states into spin-up and spin-down densities of states (a). Spin resolved conductances and resulting total conductance (solid line) of NM/I/Superconductor (b) and FM/I/Superconductor (c) heterostructures.

### 2.3.3. Andreev Reflection in Quantum Point Contact

The idea of using the quantum mechanical phenomena in the direct electrical contact for the assessment of the spin polarization was proposed in [54, 55]. The effect of the conversion of the normal current to the supercurrent at the border of a metal and a superconductor is called Andreev reflection [56].

Fig.2.4a shows a spin-up electron in a normal metal ($\Pi = 0$ %) propagating towards the interface with superconductor. For the electron to enter the superconducting state, it must be a member of a Cooper pair. Because a superconducting pair is composed of a spin-up and spin-down electron, an incident spin-up electron in the metal requires a spin-down electron to be removed from the metal as well for conversion to supercurrent. The removal of the spin-down electron leaves a spin-up hole that is Andreev reflected back into the metal. The Andreev reflected holes act as a parallel conduction channel to the initial electron current, doubling the normal-state conductance $G_n$ ($G = dI/dV$) of the point contact for the applied voltages $e \cdot V < \Delta$. Where $\Delta$ is the interface superconducting gap. In an I-V measurement, the supercurrent conversion appears as excess current added to the ohmic response at the interface. This effect is shown in Fig.2.4b for a superconducting niobium (Nb) point contact to a Cu foil at $T = 1.6$ K. At low voltage the normalized conductance is twice that of the normal state.

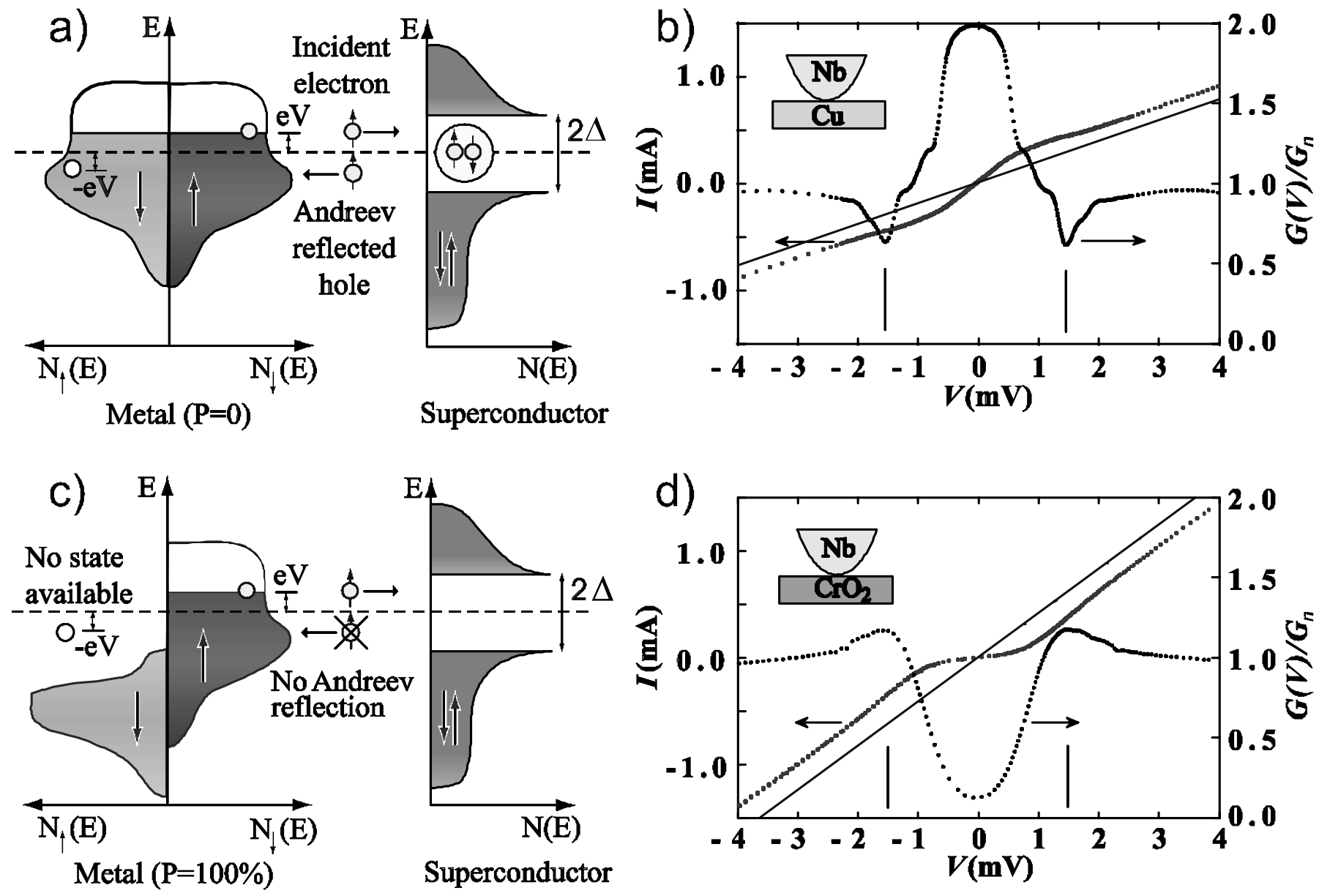

Fig.2.4. Supercurrent conversion at the metal-superconductor interface. The schematic representation of the process and experimental measurements of the I-V and dI/dV characteristics for the NM/Superconductor (a, b) and FM/Superconductor (c, d) interfaces. The solid lines correspond to the I-Vs of the junctions, when superconducting contact is in the normal state. After Ref. [54, 55].

However this is not the case for the metal with $\Pi \neq 0$ %. Fig.2.4c,d show a case of a superconducting Nb point contact to an epitaxial film of $CrO_2$, which is known to have a very large spin polarization (see Section 2.5.1). Now there are no spin-down states in the metal to provide the other member of the superconducting Cooper pair for Andreev reflection. Supercurrent conversion via Andreev reflection at the interface is blocked, allowing only single-particle excitations to contribute to the conductance. These single-particle states see the gap in the energy spectrum of the superconductor, thus suppressing the conductance $G$ for $e \cdot V < \Delta$. As one can see nearly all of the Andreev reflection has been suppressed, showing very low conductance and thus, very high spin polarization for $CrO_2$.

## 2.3.4. Other Techniques

Other techniques that can be used for measurements of the spin polarization include the Electron Capture Spectroscopy [57, 58, 59], Secondary Electron Emission [60, 61, 62, 63, 64] and Spin-Polarized Metastable-Atom De-Excitation Spectroscopy [65, 66], which are not so often used, due to difficulties with interpretation of measurements and complicated experimental approach.

## 2.4. Magnetoelectronics

In the previous section it was shown how one can measure the spin polarization in a ferromagnetic solid in an independent experiment. This section describes how one can fabricate new devices with characteristics completely relying on the spin polarization and its magnitude.

### 2.4.1. Tunnel Magnetoresistance

Further development of Tedrow and Meservey ideas was performed by Julliere in 1975 [67]. In his experiments a superconducting film was replaced by another ferromagnetic film. One can think that electrons originating from one spin state at the Fermi level of the first film would be accepted by unfilled states of the same spin at the Fermi level of the second film (Fig.2.5).

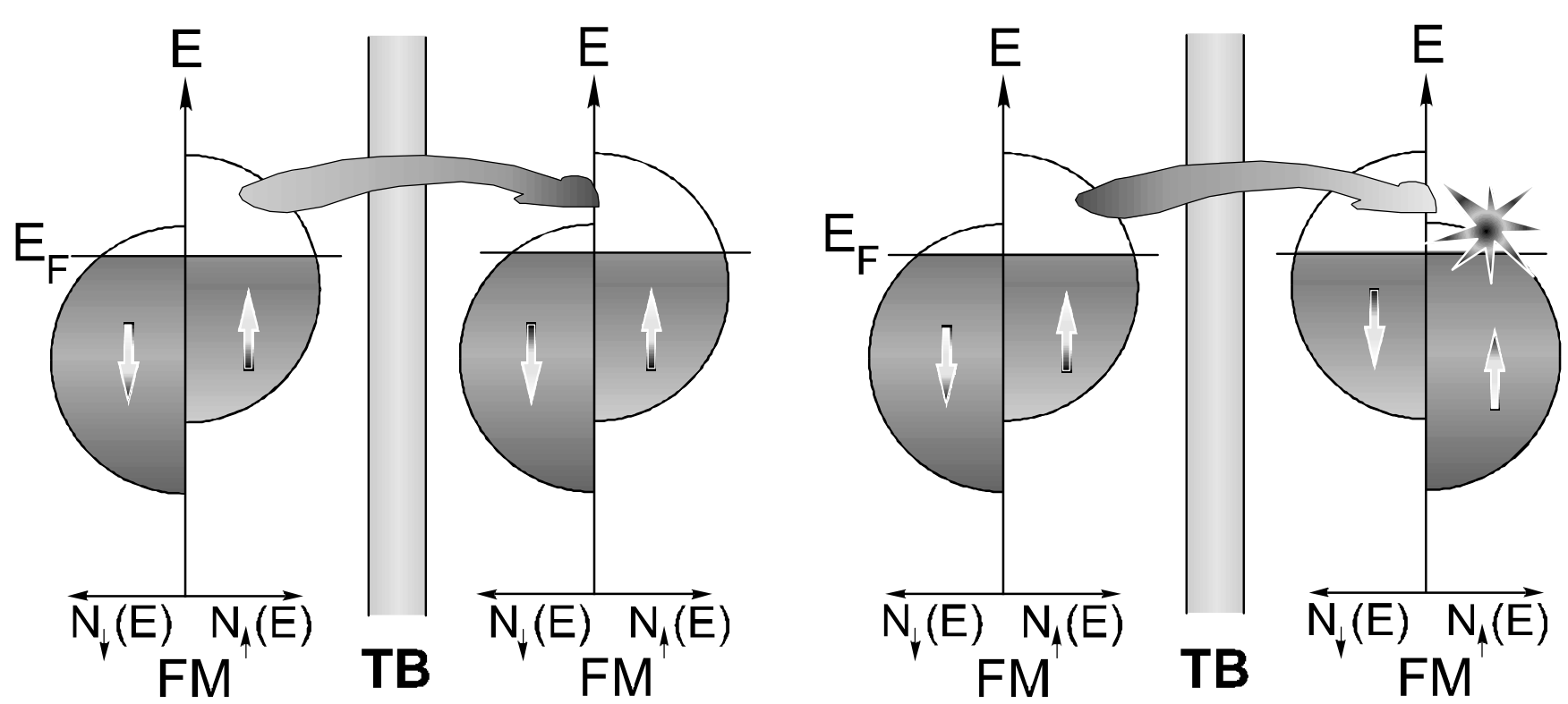

Fig.2.5. The electron tunneling between two ferromagnetic metals in a magnetic tunnel junction. The difference in the density of states results in higher conduction for parallel than for antiparallel alignments. A simplified Stoner diagram is used to represent the spin-polarized density of states in the ferromagnetic metals.

If two ferromagnetic films are magnetized parallel to each other, then minority electrons can pass into minority states and majority electrons can pass into majority states. Howver, if two films are magnetized in opposite directions, the identity of majority and minority electrons is reversed and minority electrons from the first film seek empty majority states in the second one, just as majority electrons from the first one seek minority empty states in the second. One can see that if the density of states is spin-dependent then the parallel arrangement should yield much higher conductance through the barrier than does the antiparallel arrangement. Indeed, in this experiment a 14% change in the conductance was observed for electrons tunneling between Fe and Co ferromagnetic films through a Ge barrier at low temperature.

Moreover, he realized that following the simple consideration that tunnel probability is proportional to the density of states at the Fermi level, the resistance of the junction at zero bias should be proportional to the density of states in both of electrodes and thus can be evaluated just from their spin polarization

$$\frac{\Delta G}{G^{\uparrow\uparrow}} = \frac{\Delta R}{R^{\uparrow\downarrow}} = \frac{2 \cdot \Pi_1 \cdot \Pi_2}{1 + \Pi_1 \cdot \Pi_2} \tag{2.5}$$

where $\frac{\Delta G}{G^{\uparrow\uparrow}}$ ($\frac{\Delta R}{R^{\uparrow\downarrow}}$) is the relative change of the conductivity (resistance) of the junction in parallel and antiparallel configurations, $G^{\uparrow\uparrow}$ ($R^{\uparrow\downarrow}$) is conductivity (resistance) in parallel (antiparallel) configuration, $\Pi_1$ and $\Pi_2$ are the spin polarizations in the left and right ferromagnetic metals. In fact, Eq.2.5 is representing another approach for measuring the spin polarization and is very often referred to as such.

Furthermore, if two ferromagnetic metals forming a junction have very different electronic structure, so that in the case of parallel alignment in the first FM the conductivity is carried by the majority while in the second one by the minority electrons, the conductivity of the junction is higher in the case of antiparallel alignment. Such Tunnel Magnetoresistance (TMR) junctions show a negative TMR effect [68].

As it was pointed above, the functionality of the TMR junction completely relies on such quantum mechanical phenomena like the spin-dependent electron tunneling between two ferromagnetic electrodes. This implies, that the resistance of the junction is high, limiting the overall thickness of the insulating layer to about $1 \ldots 4\,\text{nm}$ and making them extremely sensitive to the structural imperfections of the insulating layer and its interfaces [69, 70, 71, 72, 73, 74]. As result, the fabrication of the heterojunction with abrupt, perfect interfaces is required for observation of a large TMR effect. Up to date the highest reported TMR value at RT is 60% [75]. It is achieved in CoFe/$Al_2O_3$/CoFe heterojunctions with plasma oxidation of the Al layer in oxygen atmosphere.

Generally, the magnetic properties of ferromagnetic metals are considered to be non-volatile. Indeed, the magnetization state of the ferromagnetic metal can be preserved infinitely long, once the operating temperature is below the Curie temperature $T_C$ (which is above $700^\circ C$ for most common ferromagnetic metals). In addition this state can be changed almost infinite amount of times, making ferromagnetic metals a perfect candidate for all sorts of memory applications. In the case of TMR junction the magnetic state can be easily read out by electrical means, being extremely sensitive to the smallest change of the external magnetic field. Taking into account the scaling possibilities of this technology (the smallest device can have dimensions up to $\sim 8nm$) and low power consumption (due to high resistance), a new type of Magnetic Random Access Memory (MRAM) [5-7, 9] is expected to revolutionize the world of computation and data storage in the very near future.

### 2.4.2. Giant Magnetoresistance

In 1936, based on certain anomalies of electrical transport in ferromagnetic alloys, Mott [76] introduced an important concept, which was later confirmed experimentally [77], that in ferromagnetic metals the electrical current can be thought consisting of two independent components with different conductivities, one consisting of spin-up and another one of spin-down electrons. This implies that spin-flip scattering is sufficiently rare on the timescale of all other scattering processes and mixing of electrons from one channel to the other may be ignored, leading to relative independence of two channels. Making one step ahead, taking analogy with the TMR junctions described in the previous section, one can expect similar effects in the magnetic multilayers as well.

The Giant Magnetoresistance (GMR) effect was discovered at the end of $80^{th}$ [78, 79]. Investigation of the magnetoresistance in thin magnetic multilayers in the so-called Current In-Plane (CIP) geometry (Fig.2.6 left) have revealed a very large change of the resistance in the antiferromagnetically coupled Fe/Cr multilayers. This effect was much larger than the one observed in any single metallic film before. Later the same effect was observed in the other so-called Current Perpendicular-to-the-Plane (CPP) geometry [80] Fig.2.6 (right).

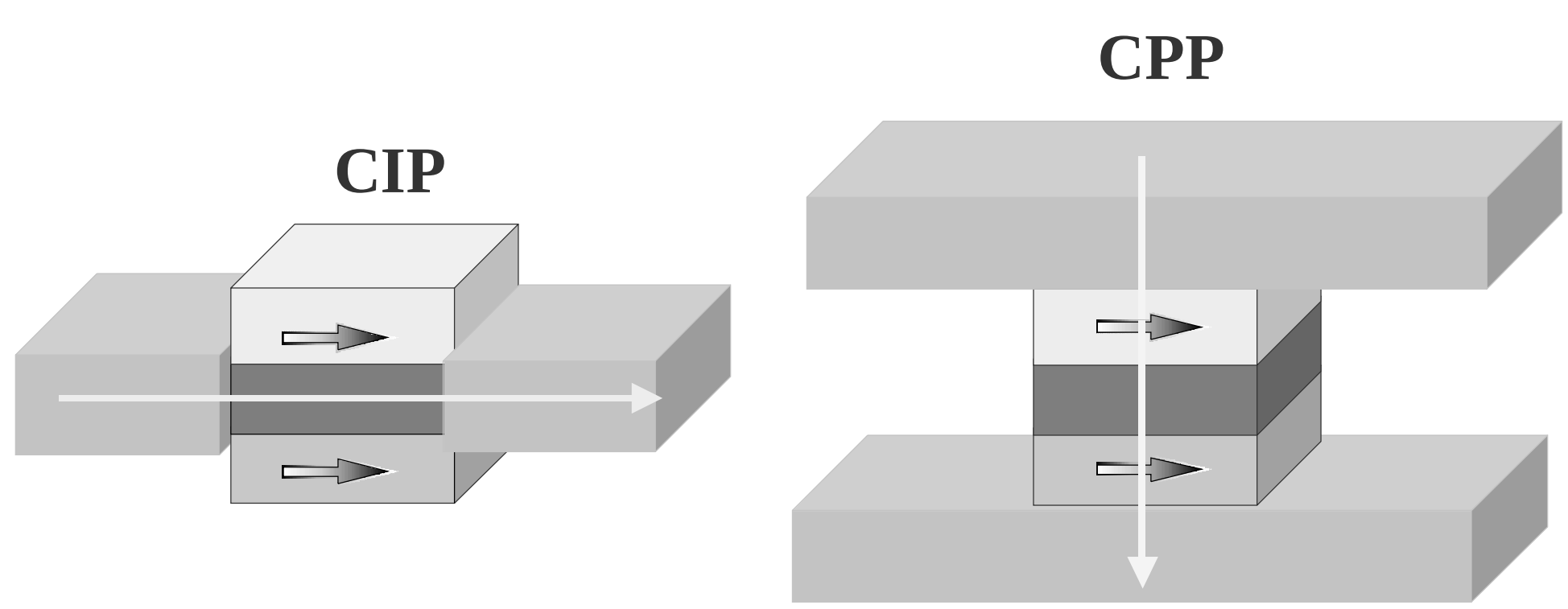

Fig.2.6. Exploration of the Giant Magnetoresistance effect in the Current-In-Plane and Current-Perpendicular-to-the-Plane geometries.

The fundamental physical phenomenon lying behind such large change of resistance is the so-called spin valve effect. Let's consider the case of CPP geometry depicted in Fig.2.7. The simplest device is a metallic multilayer consisting of two ferromagnetic layers separated by a non-magnetic conductive spacer. The spacer layer provides the ability to change the magnetic interaction between FM layers and, hence, allows changing their relative magnetization by an external magnetic field from parallel into antiparallel and vice versa. Such functionality can be realized having ferromagnetic layers with different coercivity, for example by varying relative thickness of ferromagnetic layers, using different composition, different alloys, etc. Or, as it has been demonstrated in the first experiment the variation of the thickness of the non-magnetic spacer layer can result in the antiferromagnetic coupling between ferromagnetic layers themselves.

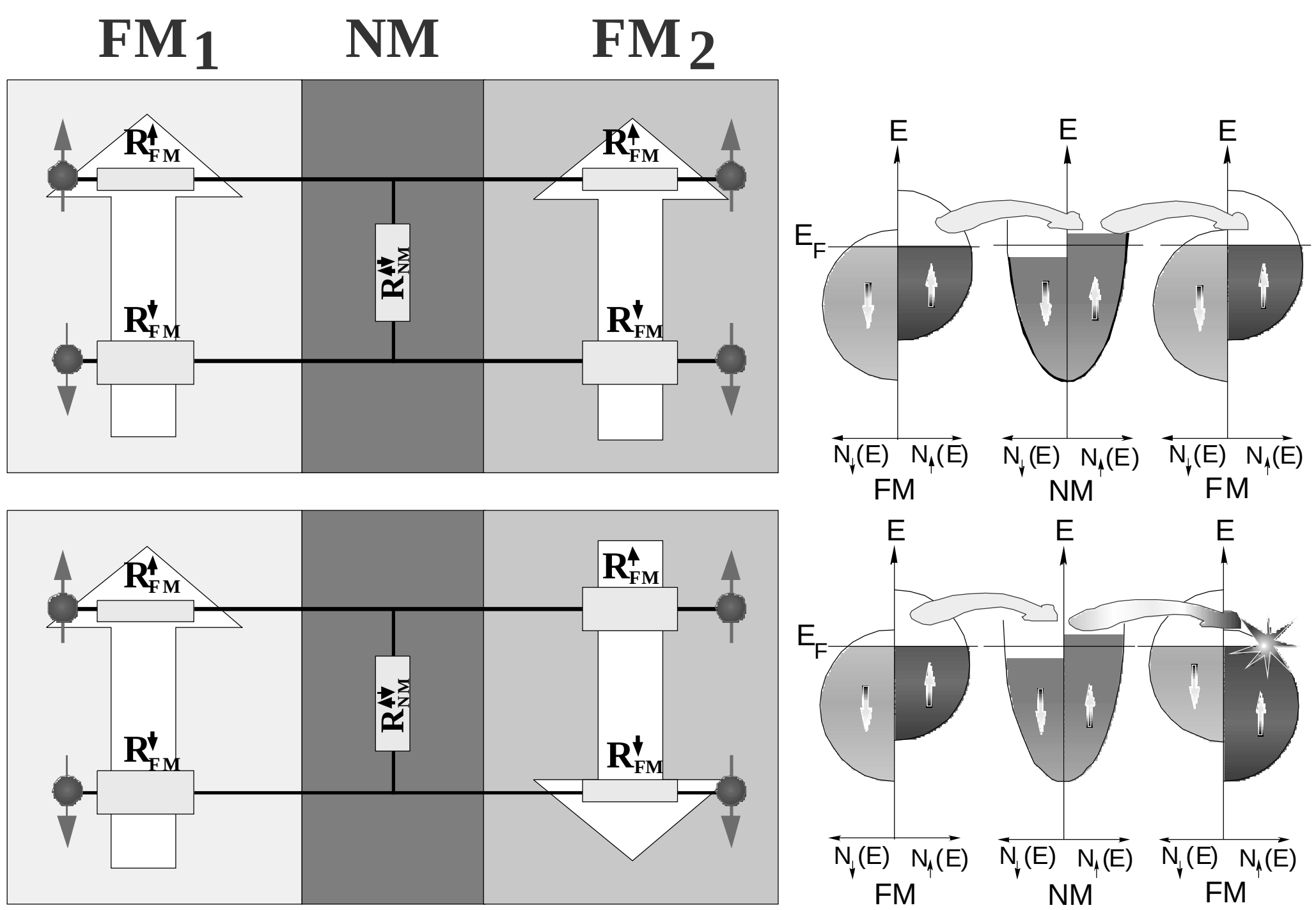

Fig.2.7. The electrical spin transport in GMR junction formed by two ferromagnetic metals separated by a non-magnetic metal. The difference in the channel resistances for the spin-up and spin-down electrons results in higher conduction for the parallel than for the antiparallel alignments. Simplified Stoner diagram is used to represent the spin-polarized density of states in the ferromagnetic metals.

As been discussed above, in such heterostructure the total electrical current can be thought consisting of two parts, spin-up and spin-down electrons. Let's consider the case of applied bias so that electrons transport occurs from left to right (Fig.2.7). In this case, in the first ferromagnetic metal all current is carried by majority spin-up electrons ( $R_{FM}^{\uparrow} < R_{FM}^{\downarrow}$ ) and thus is spin-polarized. If the $FM_1$/NM interface does not contain large number of spin scattering magnetic impurities (what is again the case for abrupt interfaces), then the spin-polarized electrons are injected into non-magnetic spacer. Now, if the thickness of nonmagnetic spacer is smaller than the electron spin-flip length $\lambda_{sf}$ , the electrons arriving at the interface with the second ferromagnetic metal interface will be spin-polarized as well, having preferred spin orientation that can be tracked backed to the first ferromagnetic metal. In the case of parallel alignment, the current in the second ferromagnetic metal also is carried by spin-up electrons, as result the resistance of the junction is small. However if the alignment is antiparallel then the current in the second ferromagnetic metal is carried by spin-down electrons, the spin-up electrons in the spacer layer seek the empty states to enter the second ferromagnetic metal and resistance of the junction is large.

As such there is no fundamental physical difference between CPP and CIP geometries. In the CPP geometry electrons are traveling across the heterojunction interfaces, as result all electrons have to pass through all interfaces. In the CIP geometry electrons are traveling along the interfaces, as result not all electrons have to pass through all interfaces. It follows that CPP geometry enables higher magnetic field sensitivity compared to CIP. However, the last one is the easiest for practical realization, since the resistance of the device in CPP geometry is too low to allow direct measurements and a numbered technological solutions, like superconducting leads [81], sub-micron micropillars [82, 83] and V-grooves [84], have to be implemented in order to overcome this limitation.

The typical material combinations in GMR devices are ferromagnetic Fe, Co, NiFe, separated by thin spacer layers of Cr, Cu, Ag, Au, Re, Ru with typical thickness of $\sim 1 \ldots 5\ \mathrm{nm}$. The magnetic sensitivity can be increased combining a large number of such magnetic multilayers ($FM_1/NM/FM_2/NM/FM_1/NM/FM_2$), so that typical device contains typically $\sim 8 \ldots 80$ of basic trilayers. These GMR junctions in the relatively week external magnetic fields show extremely large change of the resistance $\Delta R/R = 220\%$ at low [85] and $\Delta R/R = 100\%$ at RT [86], what makes them an ideal device for utilization as all sorts of magnetic sensors. Nowadays, almost every PC contains at least one of them as read head of magnetic hard drive.

## 2.5. Other Ferromagnetic Materials

As it was pointed earlier, in nature exist different ferromagnetic materials. Some of them, due to their electrical properties have a potential to replace the traditional ferromagnetic metals in the area of the spin-dependent electronics, maybe in the very near future already.

### 2.5.1. Half-Metals

The half-metallic properties were first predicted by de Groot et al in 1983 [87] based on band structure calculation for bulk crystals of NiMnSb and PtMnSb. Some of half-metallic ferromagnets are ferromagnetic oxides ($CrO_2$, $Fe_3O_4$), perovskites ($La_{0.7}Sr_{0.3}MnO_3$, $La_{0.7}Ca_{0.3}MnO_3$, $LaV_{0.5}Cu_{0.5}O_3$), Heusler alloys (NiMnSb, PtMnSb,

CoMnSb and FeMnSb) [88], semimetallic $Tl_2Mn_2O_7$ [89] and some others. Recently these materials received a huge attention due to their unusual electronic structure leading to 100% spin polarization at the Fermi surface.

In these compounds only one spin band is participating in the electrical current, while they preserve ferromagnetic order. The simplest realization of such state is the case of band structure, when for one of the spin subbands there is a gap in the spin-polarized density of states, like in a semiconductor or insulator (Fig.2.8.), while other shows a metallic behavior. The more complicated case is when two spin sub-bands are present at the Fermi level, but only one of them is actually participating in the electrical current. It happens when electrons in one of the spin subbands have more localized character than in the other one (due to different effective mass, for example) [90].

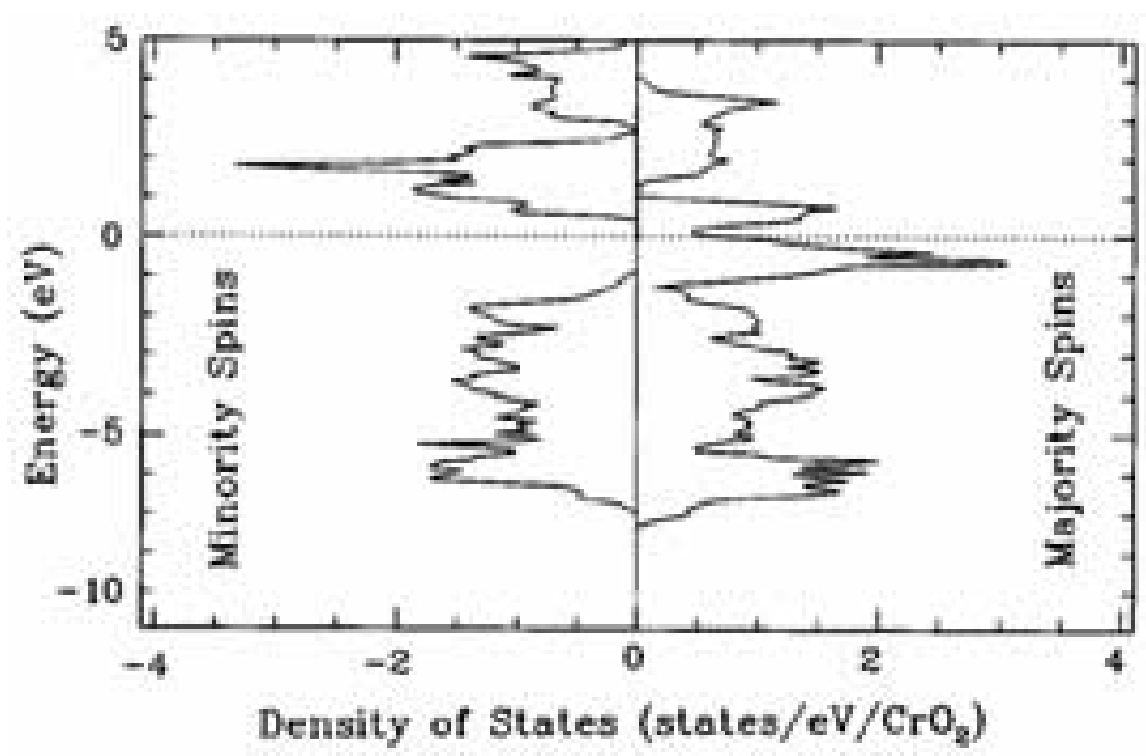

Fig.2.8. The calculated spin resolved density of states for $CrO_2$ after Ref.[91].

Taking into account that the Curie temperature in some of these materials can be far above room temperature [92], such fundamental properties make them a very promising candidate for the replacement of traditional ferromagnetic transition metals in the area of spin-dependent electronics. However, up to now nobody has succeeded in measuring 100% spin polarization for these alloys [93, 94], although the growth of the films showing much higher spin polarization than traditional ferromagnetic metals is already demonstrated [54, 55, 95, 96]. The common current limitation is, as it was shown theoretically [97], the lattice imperfections, which can never be completely suppressed in a real structure, have a dramatic impact on the half-metallic band structure and spin polarization, dramatically surpressing it at threshold below the ideal value of 100%.

### 2.5.2. Diluted Magnetic Semiconductors

Ferromagnetic semiconductors are already known for a long time [98]. Some of them are europium and chromium chalcogenides (EuS, EuO, $CdCr_2S_4$, $CdCr_2Se_4$). For these compounds the Curie temperature does not exceed $T_C \leq 100\ K$, the crystal structure is quite different from GaAs or Si and growth is extremely difficult.

Recent developments in growth techniques allowed the creation of a new type of magnetic semiconductors, the II-VI, III-V, IV-VI -based, Diluted Magnetic Semiconductors (DMS) [99, 100, 101, 102, 103]. In these materials the ferromagnetic order is achieved by incorporating a high concentration of magnetic ions, typically Mn, into lattice. The solubility limit of Mn ions is overcome by the growth at low temperatures.

In GaAs and InAs for example, below a certain limit $N_{Mn} < 10\%$, Mn atoms tend to replace Ga in the lattice, acting as acceptor. Typically one needs at least $N_{Mn} = 10^{18}\,\mathrm{cm}^{-3}$ for appearance of the magnetic order. Ferromagnetism in these compounds could be understood as a result of multiple exchange interactions. First, there is antiferromagnetic coupling between the spin of Mn core and the spin of the hole surrounding it. Second, interactions between the spins of holes result in a parallel alignment and are responsible for the ferromagnetic ordering. The direct exchange interaction between two Mn cores, which is antiferromagnetic, is negligible. This ferromagnetic ordering results in the spin-splitting of the valence and conduction bands, leading to the spin-polarized density of states (Fig.2.9). Recent TMR studies of GaMnAs magnetic tunnel junctions indicate a very large spin polarization ($\Pi \approx 78\%$) at the Fermi level in this compound [104].

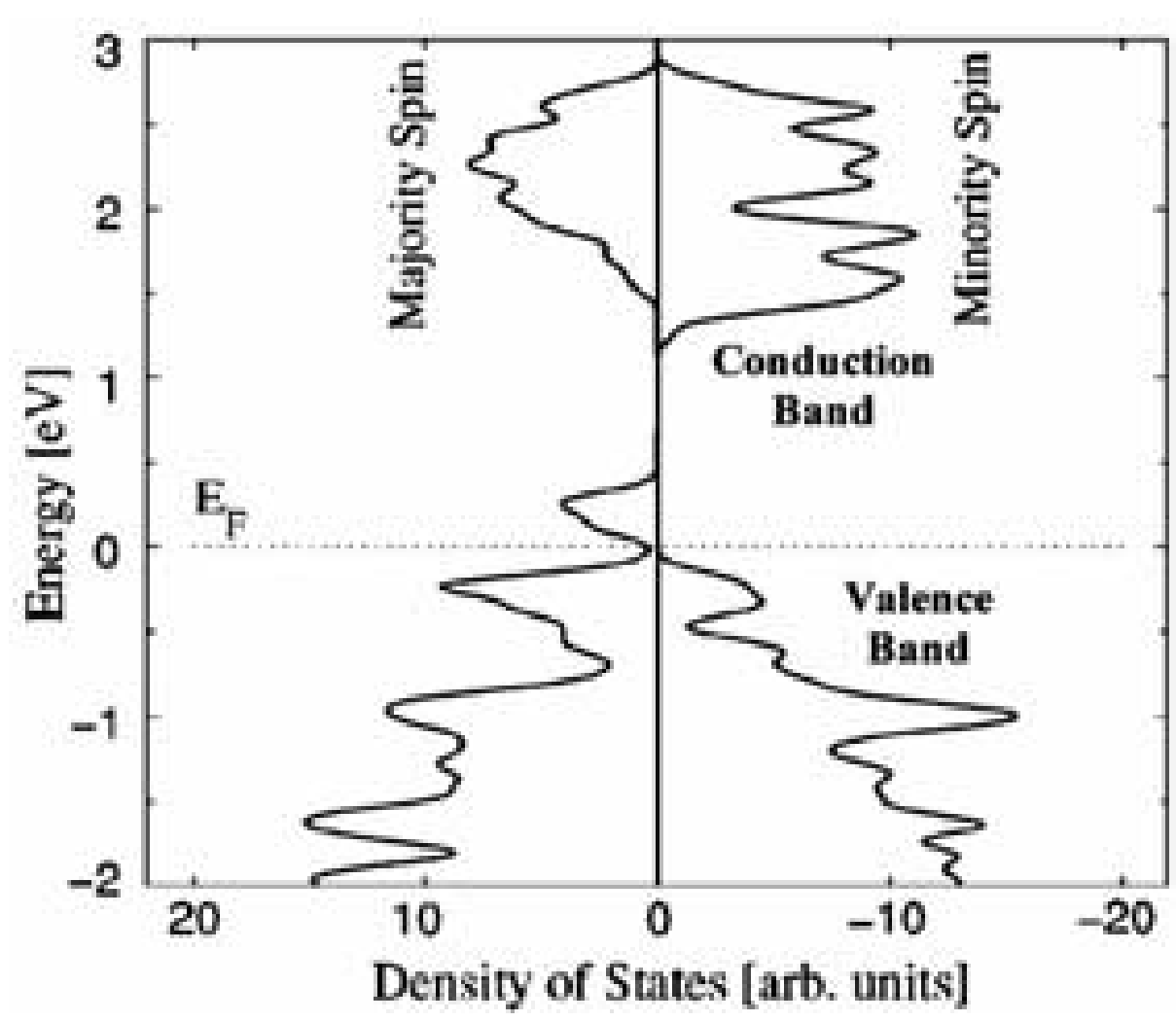

Fig.2.9. Calculated density of states vs occupation energy for $Mn_{0.063}Ga_{0.937}As$ [105]

For device implementations these materials suffer from low Curie temperature. Calculations based on the Zener model predict increase of $T_C$ in GaMnAs with increase of the Mn content and/or the free hole concentration in the alloy [106]. But at higher concentration Mn atoms be likely to create interstitial lattice imperfections, which act as a donor, and excess hole concentration tends to stabilize $N_P \approx 10^{19} \ldots 10^{20}\,\mathrm{cm}^{-3}$. Recent studies [107, 108] of GaMnAs samples doped with Be have shown that the concentration of free holes and ferromagnetically active Mn spins tends to stabilize $N_P \approx\sim 5 \cdot 10^{20}\,\text{cm}^{-3}$ and is governed by the position of the Fermi level, which controls the formation energy of the compensating interstitial Mn donors, limiting the Curie temperature. Other diluted magnetic semiconductors are expected to have a higher $T_C$, even above room temperature (Fig.2.10.), however their growth have not been demonstrated so far.

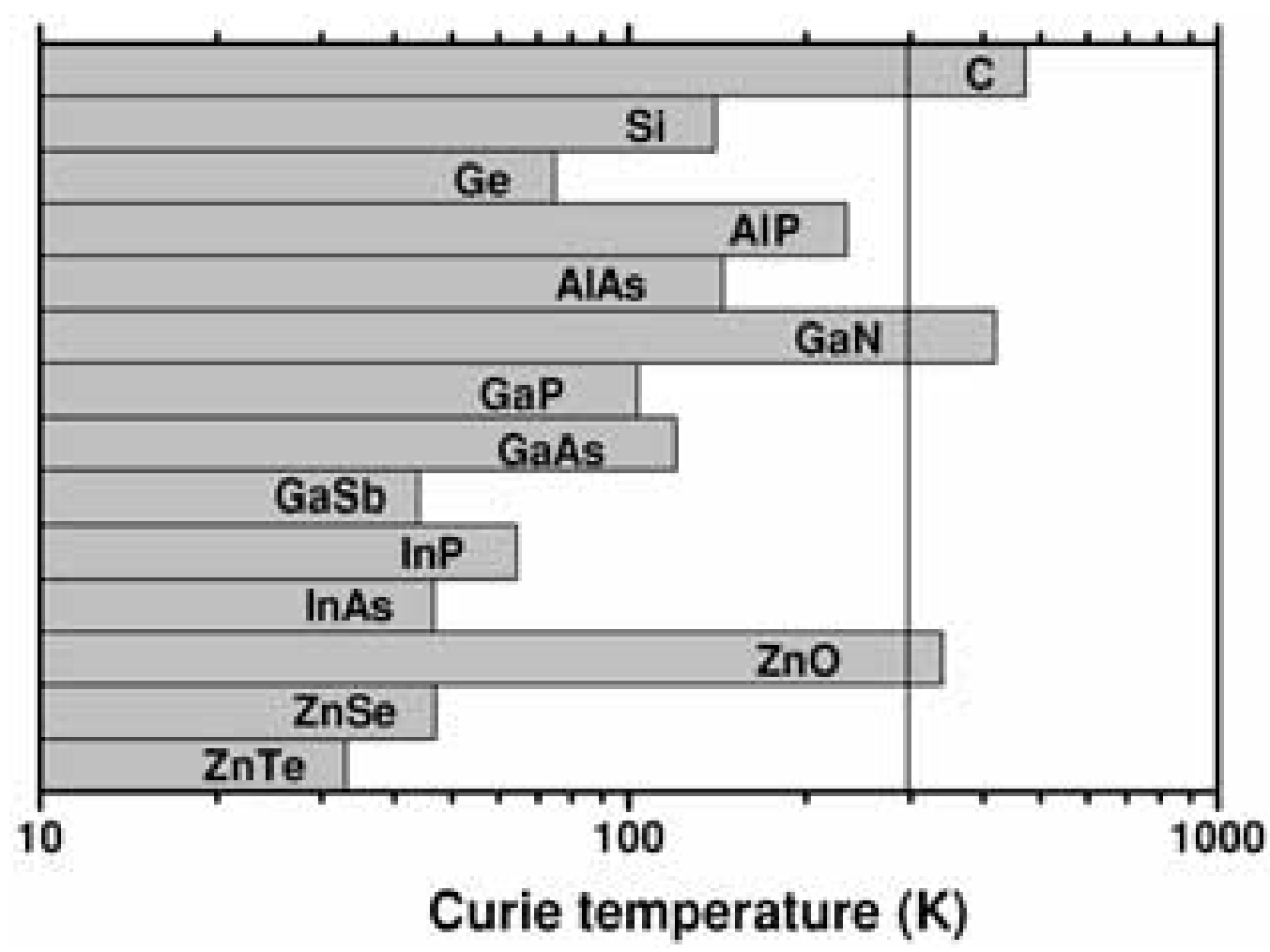

Fig.2.10. Calculated values of the Curie temperature $T_c$ for various p-type semiconductors containing 5 % of Mn and $N_P = 3.5 \cdot 10^{20}\,\text{cm}^{-3}$ [106].

At low temperatures, below $T_C$, diluted magnetic semiconductors initiate a new mean of study ferromagnetism itself, as magnetic properties in these materials could be controlled by the strain and carrier concentration. So that samples having tensile strain (grown on GaAs) show in-plane magnetic anisotropy [109], while for samples having compressive strain (grown on InGaAs) the easy axis is lying in the perpendicular direction [110]. Moreover, modulation of the carrier concentration by mean of electrical gate, like in the conventional field effect transistor, changes the hole concentration and hence the magnetic interaction between Mn ions, allowing switching of the ferromagnetic phase into paramagnetic and vice versa, performing the so-called electric field control of ferromagnetism [111].

## 2.6. Ferromagnetic Materials and Spin Polarization

As it has been argued in the previous sections the spin-dependent effects in the electronic devices rely on the spin polarization in the ferromagnetic layers. Very often a right choice of materials is needed in order to achieve an optimal performance.

Measurements of the spin polarization is not a trivial task, as different types of charge transport [29], surface contamination [31-47], chemical bonds on the interface [50-53, 25-28] have a tremendous influence on the spin polarization of the current.

Photoemission (see Section 2.3.1) allows direct probing of the spin-polarized density of states, but it has difficulties with surface contamination and a lack of energy resolution. For assessment of the spin polarization using the spin-dependent tunneling into the spin-split superconducting states (see Section 2.3.2), which is perfect method for comparison with TMR junctions (see Section 2.4.1), the fabrication of the planar heterostructure is needed. Contrary, Andreev reflection (see Section 2.3.3) can be used when fabrication of the planar junction is difficult, it has a very high energy resolution, which may be mandatory in the case of complicated electron transport like in the half-metals (see Section 2.5.1), but for reliable implementation it requires a perfect point contact in order to have a transparent interface.

Fig.2.11 shows an overview of the current state of the spin-polarization available in the traditional ferromagnetic transition metals and their alloys, like Fe, Co, Ni, CoFe and NiFe [53, 54, 112], ferromagnetic semiconductor GaMnAs (estimated from Ref.[104]) and half-metallic NiMnSb, LaSrMnO and $CrO_2$ [54, 96].

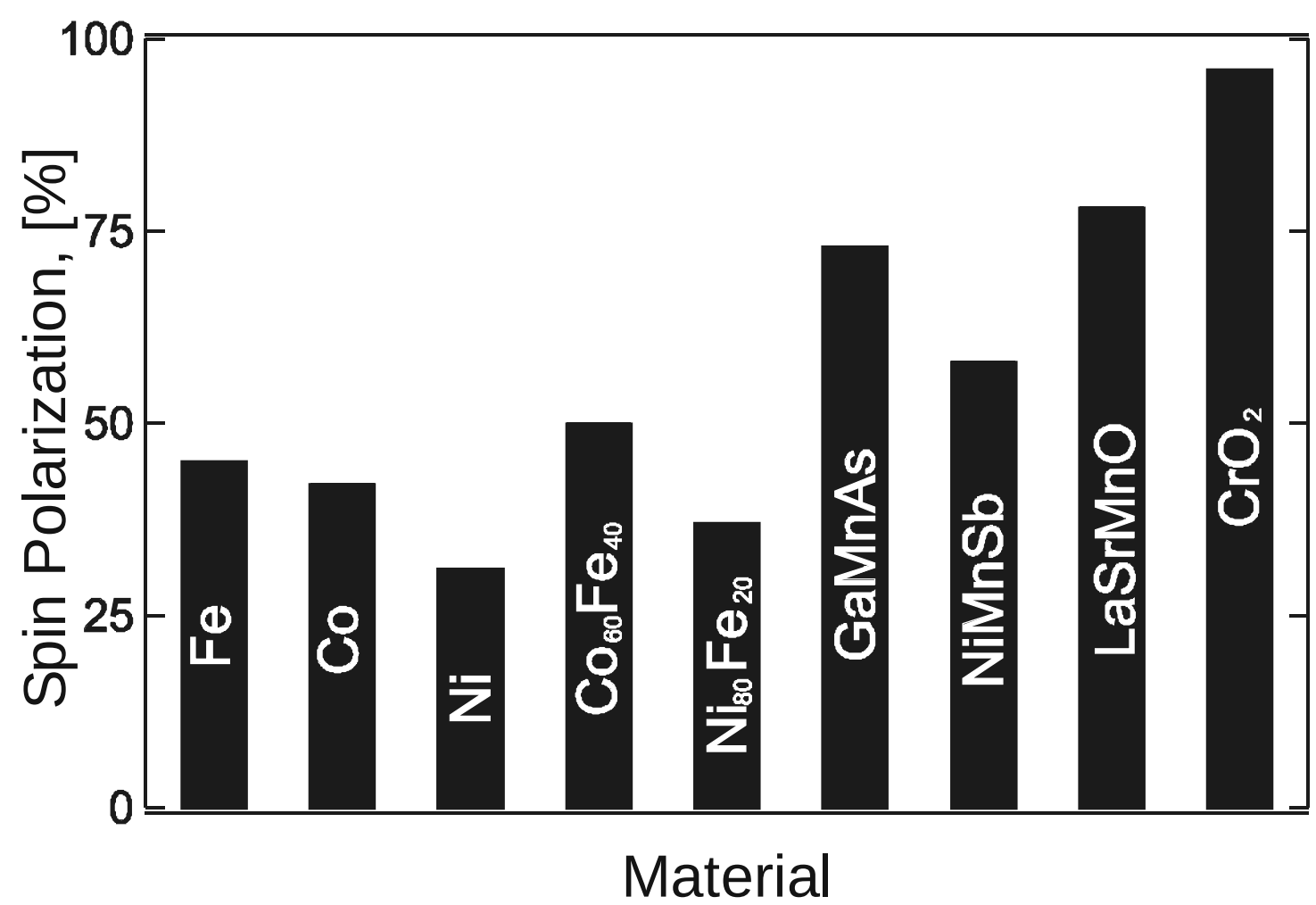

Fig.2.11. The most common ferromagnetic materials vs spin polarization of the charge carriers.

# 3. Spintronics

Spintronics or spin-dependent electronics covers a broad range of devices that utilize electron spin in order to change their electrical properties. In its broad sense it also covers magnetoelectronic passive devices, nevertheless, nowadays there is a strong tendency to detach those two areas. In this thesis the word spintronics is used primary to address semiconductor-based devices whose functionality relies on the electron spin, as well as its charge. Thus, spintronics has a large advantage over magnetoelectronics, as creation of active electrical device should be feasible.

## 3.1. Spintronic Devices and Quantum Computation

The first semiconductor-based device utilizing electron spin in order to change its electrical properties was proposed in 1990 [113] (Fig.3.1).

In its functionality it resembles both, the conventional FET transistor and conventional GMR device. It consists of two ferromagnetic Fe contacts, separated over distance $\lambda < \lambda_{sf}$ , where $\lambda_{sf}$ is spin-flip length in the semiconductor, to the 2DEG formed on the InAlAs/ InGaAs interface. Under application of electrical bias, depending on the relative magnetization and character of the density of states in the left and right ferromagnetic metals, assuming that there is no spin scattering on the metal/semiconductor interface, the spin-polarized electrons originating from the first ferromagnetic metal can be either accepted or be rejected by the second one, resulting in the overall conductivity difference. Moreover, the traditional gating of the channel allows controlling not only the overall carrier concentration, but also the spin precession in the 2DEG [114], resulting in additional degree of freedom controlling the electron spin and, hence, the resistance of the device.

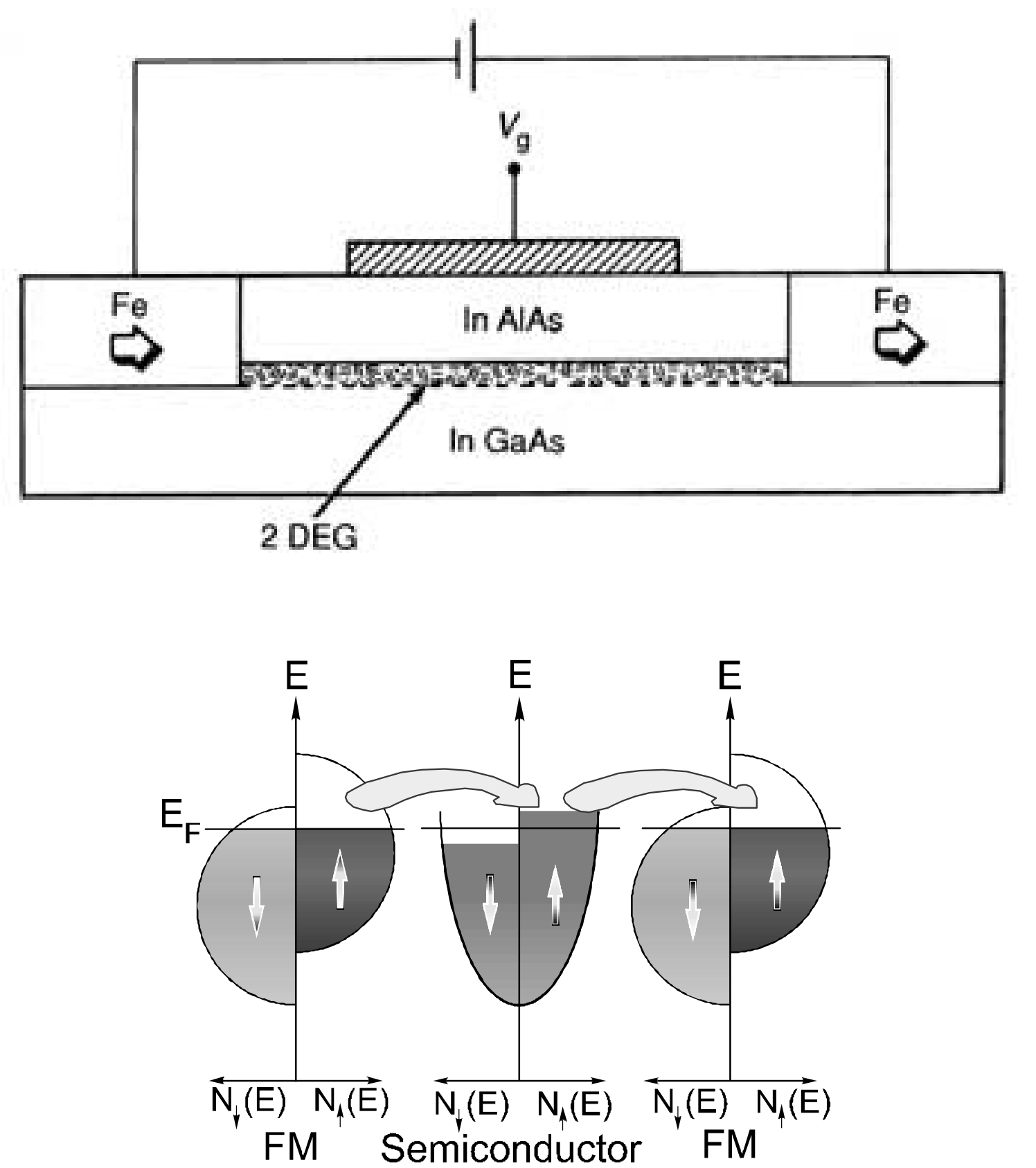

Fig.3.1. Proposed spin-polarized FET, after Datta-Das [113].

Currently the area of spintronics covers other semiconductor based devices, that are the hot electron Spin Valve Transistor (**SVT**) and the spin-polarized Light Emitting Diode (**spin-LED**). Fig.3.2 shows these spintronic devices and their conventional semiconductor counterparts.

The SVT transistor was invented in 1995 [115, 116]. Its operation principle resembles the conventional metal base transistor and, again, the conventional GMR device. The SVT consists of a GMR multilayer base sandwiched between two semiconductors, forming highly resistive emitter and collector Schottky depletion regions in the last ones (Fig.3.2 f). The emitter/base barrier is biased in the forward direction and its barrier height is higher than the reversed biased base/collector barrier. The hot electrons are injected from the emitter over the first barrier into the base, so that some of them can traverse the base ballistically and may overcome the second Schottky barrier, contributing to the collector current. Most of them thermalize, cannot penetrate to the collector and form the base current. The electron thermalization depends on the scattering probability that can be varied by the change of the relative magnetization of the magnetic multilayers forming the GMR-type base. Generally, these devices show very high magnetic field sensitivity, up to two orders of magnitude higher as compared to magnetic tunnel junctions, for example. Today they are the most sensitive electrical devices, the change of the collector current may exceed 1000% [117, 118]. These properties make them almost a perfect magnetic field sensor. However, they suffer from bad scalability, as collector current is low, typically it is in the nano ampere range, and, moreover, it strongly depends over width and cross-section of the metallic base [119, 120],121.

The possibility of a creation of the spin-polarized electrons in a semiconductor by the electrical injection from a ferromagnetic metal using a direct electrical contact was first overseen in 1976 [122]. The spin-LED utilizes the intrinsic properties of III-V semiconductors to efficiently transfer electron angular momentum into angular momentum of the emitted light (see Section 3.3). Thus the change of the magnetization state of the ferromagnetic injector provides control over the polarization state of the injected electrons and, hence, over the polarization of the optical output. Such control increases the complexity of the conventional LED to a level needed for the next generation of the high bandwidth optical communication systems [123], when polarization division multiplexing will be used as wavelength and time division multiplexing are used today.

There is another valuable application, which attracts more and more attention worldwide proportionally or even sub-proportionally to the progressing reduction of the device dimensions. Following Moor’s Law, one should end up with only one electron per device around 2020. The World of Electronics will belong to quantum phenomena, the World of Computation, most probably as well [11-13]. The traditional computer, which is described by the laws of classical physics, operates with bits, namely state $|0\rangle$ and state $|1\rangle$. The quantum computer, which operates accordingly to the laws of quantum physics, operates with quantum bits or qubits, which also have state $|0\rangle$ and state $|1\rangle$. But unlike the conventional bit, the qubit can exist in any superposition of these states. The classical known example of a qubit is a spin of an electron or of a proton. The quantum computer seems already to overperform its classical precursor in certain computational tasks, like the Grover searching algorithm [124], simulating other quantum systems [125] and Shors factoring algorithm [126], which was recently implemented in the enhanced NMR experiment for quantum factoring of number 15 [127].

It is obvious that quantum operations as well as the Datta and Das device rely on certain physical principles. First, the qubits must be initialized into the original state (the spins must be injected into semiconductor). Second, the lifetime of these states must be long enough to perform a quantum operation (to reach the second magnetic contact). And the last one, the final states of the qubits (the spin imbalance or spin polarization) must be easily readable in a straightforward measurement event. The experience shows that circumvention of all these requirements in the case of spin-polarized FET is not a so trivial task to do.

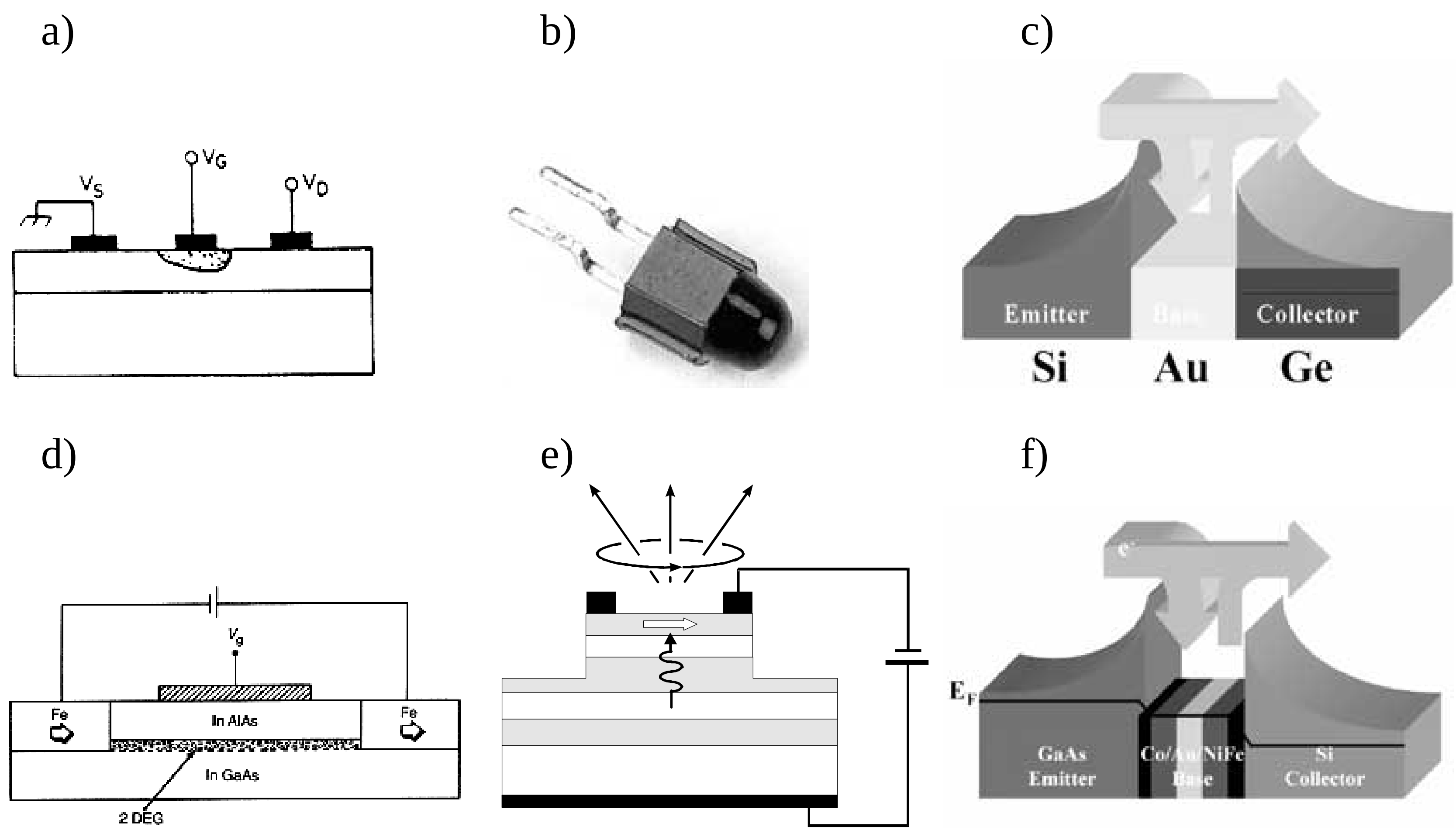

Fig.3.2. Comparison between conventional electronic and spintronic devices. a) conventional FET; b) conventional LED; c) conventional metal base transistor; d) spin-polarized FET; e) spin-LED; f) SVT transistor.

## 3.2. Density of States Matching, Problem of Conductivity Mismatch

The experimental attempts to fabricate and investigate electrical properties of spin-polarized FET have been reported since 1999 [15]. In this experiment a device consisting of NiFe ferromagnetic contacts to the InAs 2DEG was fabricated in order to examine the principles described above. The first results seemed optimistic, the Authors did observe the change of the device resistance of the order $\Delta R/R = 0.8\%$ by changing the relative magnetization of the ferromagnetic contacts. The question is what is causing this change, as there is a number of side effects showing similar behavior in an external magnetic field (like the local Hall effect [128], for example) that mask the real effect and contribute to the experimental change of the resistance.

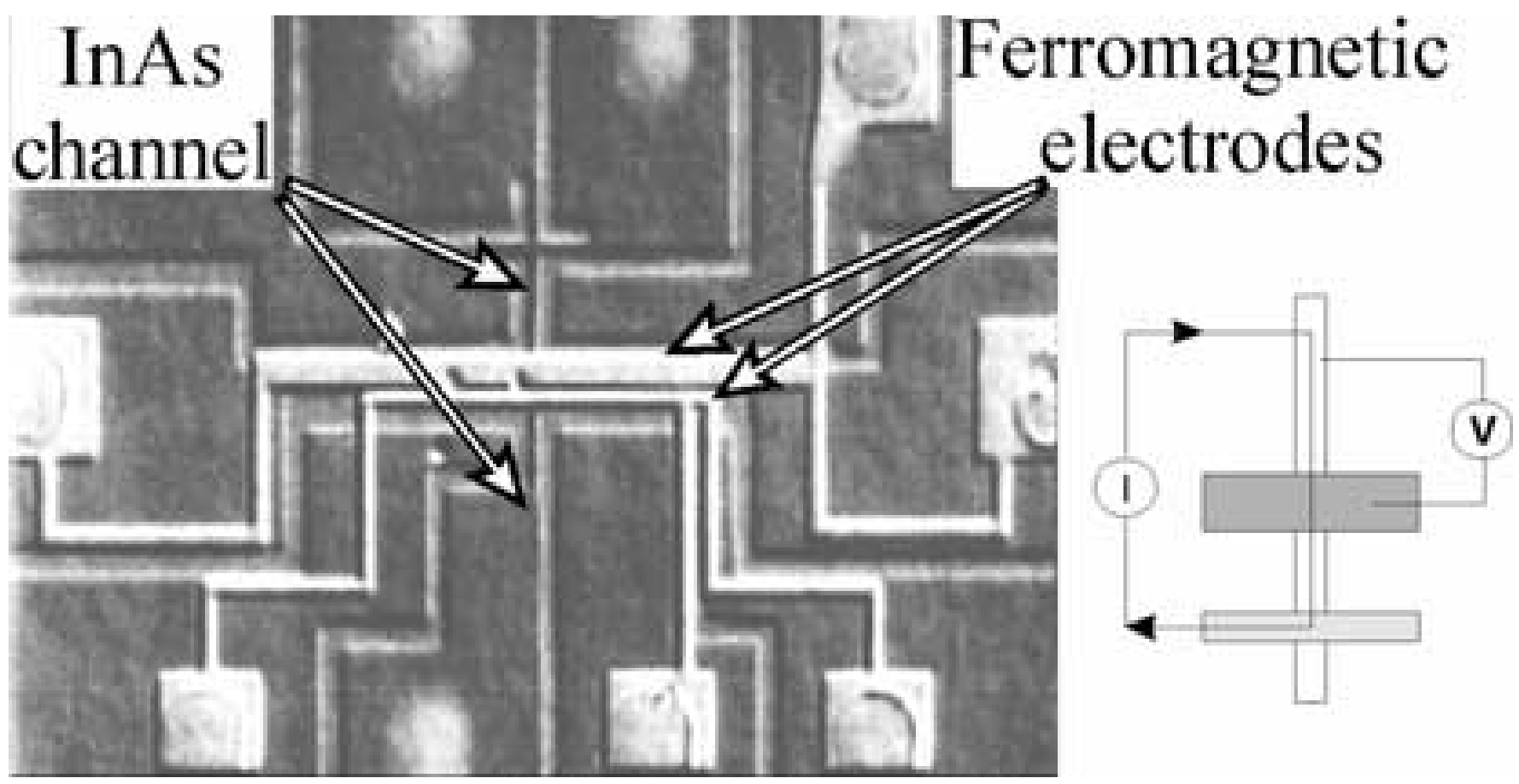

Fig.3.3. SEM image of the fabricated spin-polarized FET and schematic representation of the non-local geometry measurements, after Ref.[16].

The attempt to repeat this experiment, taking special care for the elimination of the different side effects, was performed by different group [16]. In this experiment (Fig.3.3left) devices with different ferromagnetic contacts (Co, Ni, NiFe) to the InAs 2DEG were fabricated. Measurements in the so-called non-local geometry (Fig.3.3 right), pioneered earlier in Ref.[129] for measurements of the spin diffusion length in Au films, when there is no electrical current in the second detecting ferromagnetic metal (and hence no major side effects), has revealed no signal which could be attributed to the spin-polarized transport in any single one of these devices. Moreover, up to now nobody has succeeded in fabrication of such device showing effects caused by spin polarization of the current [130].

So what are the fundamental restrictions that limit the performance of the device ? Is the electron spin lifetime in a semiconductor long enough to ensure the spin polarization of the electrons reaching the second ferromagnetic metal ? Does the high mobility in the 2DEG, as in the original device design, satisfy this requirement ? The answer is Yes. Even traditional semiconductors like Si or GaAs (see next Section 3.3) have electron spin lifetimes long enough to allow fabrication of the device with ferromagnetic contacts separation available to the conventional optical lithography, for example. The most fragile point is the electrical spin injection into a semiconductor in the direct electrical contact. This fact gives rise to all sorts of experiments with spin-LEDs, where electrical spin injection is also a key issue, and, moreover, there is a direct access to the injected spin polarization in the semiconductor by an analyses of the polarization state of the emitted light.

Recently, a number of different theoretical treatments of the problem of electrical spin injection into semiconductors have been reported [17, 131, 132, 133, 134, 135]. All of them came to the same conclusion. It is impossible to inject spin-polarized charges into semiconductor from a ferromagnetic metal in the diffusive ohmic contact. The reason is very different band structure of these materials so that schematic band structure shown in Fig.3.1 is wrong. The density of states available in the semiconductor is couple of orders of magnitude smaller than the one available in the ferromagnetic metal, even for electron spin minority. This leads to complete feeling of both spin channels in the semiconductor and zero spin polarization of the current. The only exceptions are ferromagnetic semiconductors and half-metals (see Sections 2.5.2 and 2.5.1). The first ones have similar to the semiconductors density of states. The half-metals have only one spin channel available on the Fermi level. However, a large half-metallic order is required for the last ones ($\Pi \geq 98\%$) [17, 131] in order to overcome this obstacle.

For the injection from conventional ferromagnetic metals, the only solution proposed, is incorporation of a large interface resistance between metal and semiconductor, like convention tunnel barrier or highly resistive narrow Schottky depletion region. This is not surprising, as electrical spin injection from the ferromagnetic Ni Scanning Tunneling Tip (STM) has been already demonstrated (see Section 3.4.2). Moreover, tunneling effect is known to be proportional to the density of states in both solids, matching the large difference in density of states of a ferromagnetic metal and a semiconductor. Further, in the case of GaAs or Si the highly resistive Schottky barrier is naturally formed on the abrupt interface with a metal. Thus, the only requirement is to ensure high tunnel transmission of such interface by appropriate doping of the semiconductor, for example.

## 3.3. GaAs, Spin Detection and Long Spin Memory

As it was mentioned previously, an electron has an angular momentum, which is known as spin and connected with it magnetic momentum. The quantum of magnetic momentum is the well-known Borr magneton $\mu_B = 9.27 \cdot 10^{-24}$ J/T. An electromagnetic wave also has an angular momentum, being equal to $J = -1,\ 0,\ +1$. The angular momentum $J = \pm 1$ corresponds to the circularly polarized light, while linearly polarized light has the angular momentum $J = 0$. It follows that in solid state an angular momentum of light can be transferred into an electron spin and vice versa. This idea is not new and dates the late 60s [136], when circularly polarized light was used for dynamic polarization of $Si^{29}$ nuclei in the enhanced NMR experiment. Later it was shown that conversion of circularly polarized light into electron spin and, on opposite, the conversion of electron spin into circular polarization of light is very efficient in the III-V semiconductors, like GaAs in particularly [18, 137, 138, 139, 140, 141, 142, 143].

### 3.3.1. Band Structure of GaAs and 'Optical Orientation'

The band structure of GaAs is well known [144]. The bottom of the conduction band and the top of the valence band are located at the center of Brillouin zone (Fig.3.4). The conduction band is formed by s-like atomic states (electron orbital momentum $L = 0$) being twice degenerated in spin. The spin-orbit interaction splits the six fold degenerate valence band formed by p-like atomic states (electron orbital momentum $L = 1$) into fourfold degenerate upper band with total angular momentum J=3/2 and the lower split-off band ( $\Delta_{SO} = 0.34\ eV$ ), being twofold degenerated in spin.

Fig.3.4 (right) shows the degenerate band states, the corresponding magnetic numbers $m_J$, the allowed transitions and their probabilities. Along the quantization axis, only the transitions that follow the rule $\Delta m_J = \pm 1$ are allowed, while transitions with $\Delta m_J = 0$ correspond to the perpendicular direction.

Under optical excitation the quantization axis is defined by the direction of light propagation. Thus, for optical excitation $E_g < h \cdot \upsilon < Eg + \Delta_{SO}$ with circularly polarized light, $\sigma^+$ for example, only transitions with $\Delta m_J = +1$ from upper valance band are allowed. These transitions ( $p_{-3/2} \rightarrow s_{-1/2}$ and $p_{-1/2} \rightarrow s_{1/2}$ ) have different probabilities, hence in the case of equal population of the states in the valence band, for one electron with spin state +1/2, three electrons with spin state –1/2 in the conduction band of the semiconductor are created. As result, the excitation with circularly polarized light $\sigma^+$ creates a spin-polarized electron ensemble in the conduction band of the semiconductor $\Pi_{inj} = \frac{n^\uparrow - n^\downarrow}{n^\uparrow + n^\downarrow} = \frac{1-3}{1+3} = -\frac{1}{2}$ with preferential orientation along the axis of light propagation. Similarly excitation with circularly polarized light $\sigma^-$, accordingly to the same selection rules, creates a spin-polarized electron ensemble in the conduction band of the semiconductor $\Pi_{inj} = \frac{n^\uparrow - n^\downarrow}{n^\uparrow + n^\downarrow} = \frac{3-1}{3+1} = +\frac{1}{2}$.

Further, such excitation creates spin-polarized holes in the valence band as well. However, due to the spin-orbit interaction, the spin lifetime of a hole in the valence band is a few orders of magnitude shorter than the one of an electron in the conduction band. As result, the excitation with circularly polarized light can be thought only as optical orientation of the spin-polarized electron ensemble in the conduction band of the semiconductor.

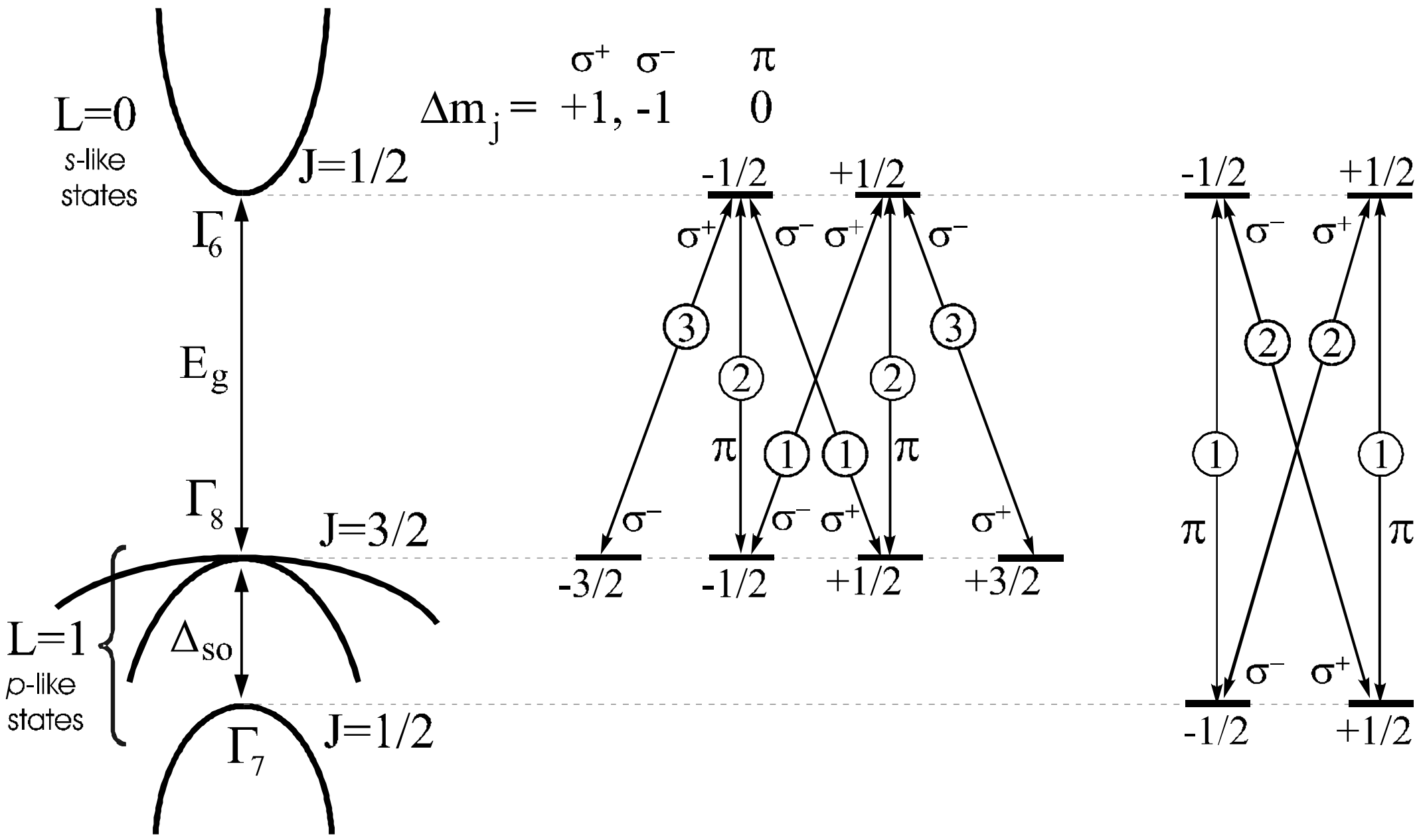

Fig.3.4. Band structure of GaAs in the vicinity of k=0 point, the degenerated band states, the corresponding magnetic numbers $m_J$, the allowed transitions and their probabilities.

It follows that excitation with linearly polarized light does not create any spin polarization of electrons in the conduction band. Linearly polarized electromagnetic wave can be thought consisting of two components having $\sigma^+$ and $\sigma^-$ polarizations, which being equal result in the equal population of the electron spin states and zero overall spin-polarization.

Under radiative recombination of the spin-polarized electron ensemble, in the case of equal population of the states in the valence band, the same transition probabilities are responsible for the emission of circularly polarized light $P = \frac{1}{2} \cdot \Pi$ along preferential direction of the electron spin orientation, being a quantization axis. Here $P = \frac{I^+ - I^-}{I^+ - I^-}$ is the degree of circular polarization, $I^+$ and $I^-$ are the intensities of right ($\sigma^+$) and left ($\sigma^-$) circularly polarized components of light, respectively. In the perpendicular direction such transitions correspond to the linearly polarized light in all sorts of combinations, resulting in overall unpolarized emission.

Following above-mentioned considerations, for optical excitation $E_g < h \cdot \upsilon < Eg+\Delta_{SO}$ with circularly polarized light $P_{excit} = 100\%$, as result of light absorption and radiative recombination, the emission of circularly polarized light $P_{emiss} = \frac{1}{2} \cdot \Pi = \frac{1}{2} \cdot \frac{1}{2} \cdot P_{excit} = 25\%$ is expected. However, during electron lifetime on the bottom of the conduction band, before recombination with holes, some spin scattering may occur. As result, the observed steady state polarization is lower

$$P_{emiss} = \frac{1}{2} \cdot \frac{T_S}{\tau} \cdot \Pi_{inj} = \frac{1}{4} \cdot \frac{T_S}{\tau} \cdot P_{excit} \qquad (3.1)$$

here parameter $\frac{T_S}{\tau}$ describes the spin scattering of electron spins on the bottom of the conduction band, $T_S$ - is spin lifetime ($T_S^{-1} = \tau^{-1} + \tau_S^{-1}$), $\tau$ is electron lifetime and $\tau_S$ is spin relaxation time.

### 3.3.2. Spectral Dependency of 'Optical Orientation'

Fig.3.5 shows the dependency of the steady state polarization of the photoluminescence as function of energy of the exciting light with $P_{excit} = 100\%$ [145, 18]. The circular polarization of the photoluminescence is virtually independent over the excitation energy $E_g < h \cdot \upsilon < Eg+\Delta_{SO}$. For the exciting photon energy exceeding $Eg+\Delta_{SO}$, where $\Delta_{SO}$ is the spin orbit splitting of the valence band, the spin polarization of the photoexcited electrons is strongly reduced, due to the fact that electrons excited from the split-off band have opposite spin orientation as compared to the electrons excited from the upper bands of the light and heavy holes (see Fig.3.4). In the general case, the contribution of the latter ones prevails and the net spin polarization stays positive, but reduced in value. However, the electrons excited from the upper valence bands have much higher kinetic energy and should be thermalized to the bottom of the conduction band before recombination takes place. If these hot electrons loose their spin orientation during thermalization process (DP mechanism see Section 3.3.5), the net spin orientation on the bottom of the conduction band changes sign and becomes negative due to the contribution of "cold" electrons excited from the split-off valence band.

The loss of spin polarization during the thermalization process depends on the speed of electron thermalization, which is different for the samples with different doping levels. In the highly doped samples the thermalization is rapid and polarization losses are small and even insignificant. For the low-doped samples the thermalization process is slow and spin losses during thermalization could be high.

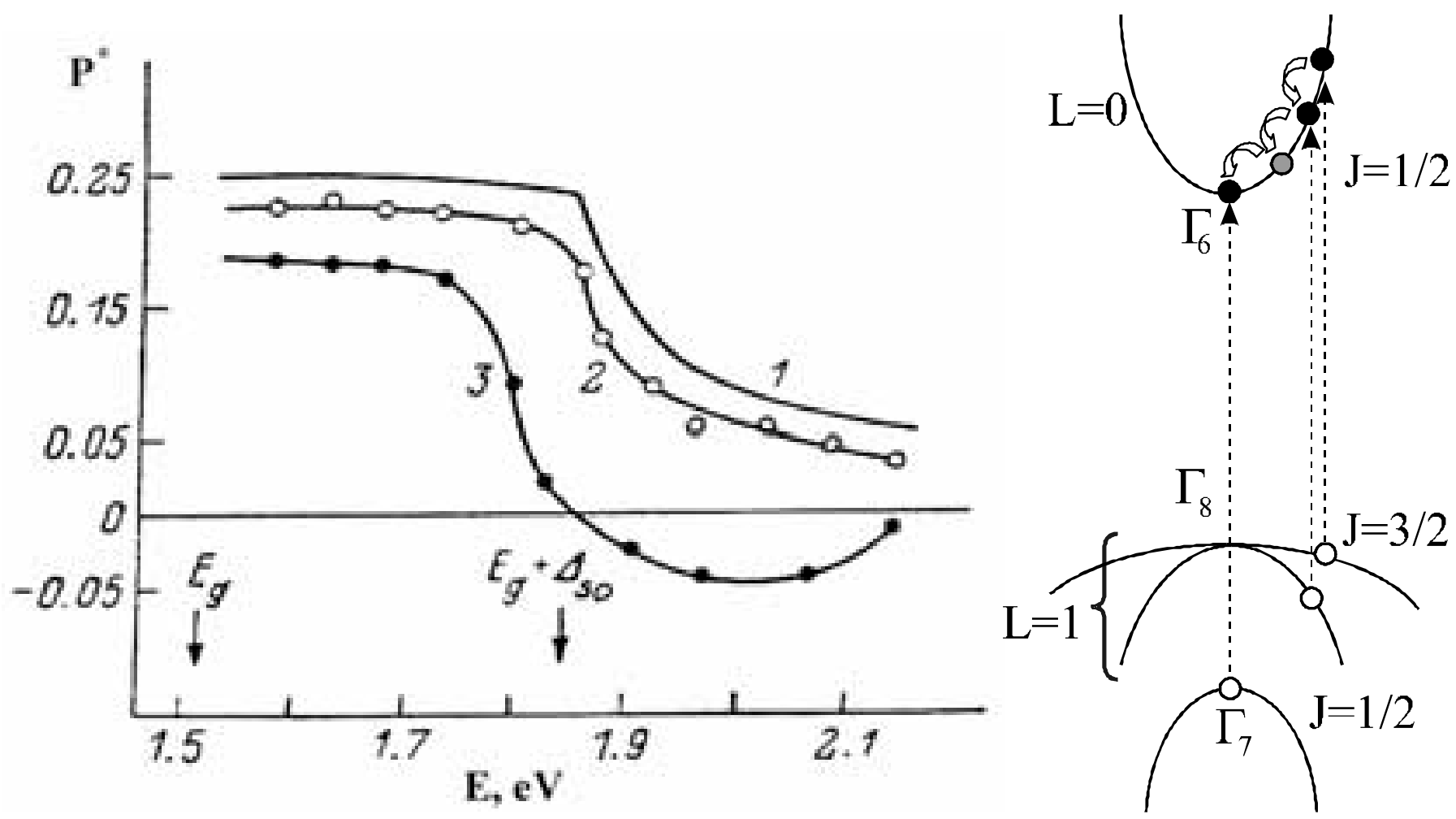

Fig.3.5. The circular polarization of the photoluminescence as function of the excitation energy ( $P_{excit} = 100\%$ ) for GaAs samples with different impurity concentration at 4.2K (left) [145, 18]: 1- theoretical curve; 2- $p=4 \cdot 10^{19} cm^{-3}$; 3- $p=7.8 \cdot 10^{16} cm^{-3}$; and electronic transitions under optical excitation with $h \cdot \upsilon > E_g + \Delta_{SO}$ (right).

### 3.3.3. Influence of Mechanical Stress and Quantum Confinement

The application of uniaxial stress to the GaAs crystal leads to lifting of the fourfold degeneracy of the top of the valence band at k=0. Under compression (stretch) the top of the light-hole (heavy-hole) band happens to be higher than the heavy-hole (light-hole) band. In this case the quantization axis is defined by the axis of applied stress and the emitted light is connected only with the conduction band light-hole (conduction band heavy-hole) transitions. Hence, under optical excitation with circularly polarized light $P_{excit} = 100\%$ along the stress axis, the polarization of the photoluminescence is $P_{emiss} = \Pi = P_{excit} = 100\%$ .

The quantum confinement is known to have a similar effect. The ground state of the energy levels of the charge carriers in the quantum well structure is

$$E_1 \approx \frac{1}{m^* \cdot d^2} \qquad (3.2)$$

where $m^*$ is the effective mass and $d$ is the width of the quantum well. Thus the heavy-hole (**hh**) and light-hole (**lh**) levels in the quantum well have different energy. The splitting of these levels accordingly to Eq.3.2 is proportional to the difference in the effective mass and square of quantum confinement. These splitting causes the change of the transitions probabilities shown in Fig.3.4, as result only conduction band heavy-hole (**c-hh**) transitions contribute to the light emission. Hence, the circular polarization of the emitted light $P_{emiss} = \Pi$. The quantization axis in this case is defined by the quantum confinement and for planar heterostructure it coincides with the growth direction of the semiconductor layers.

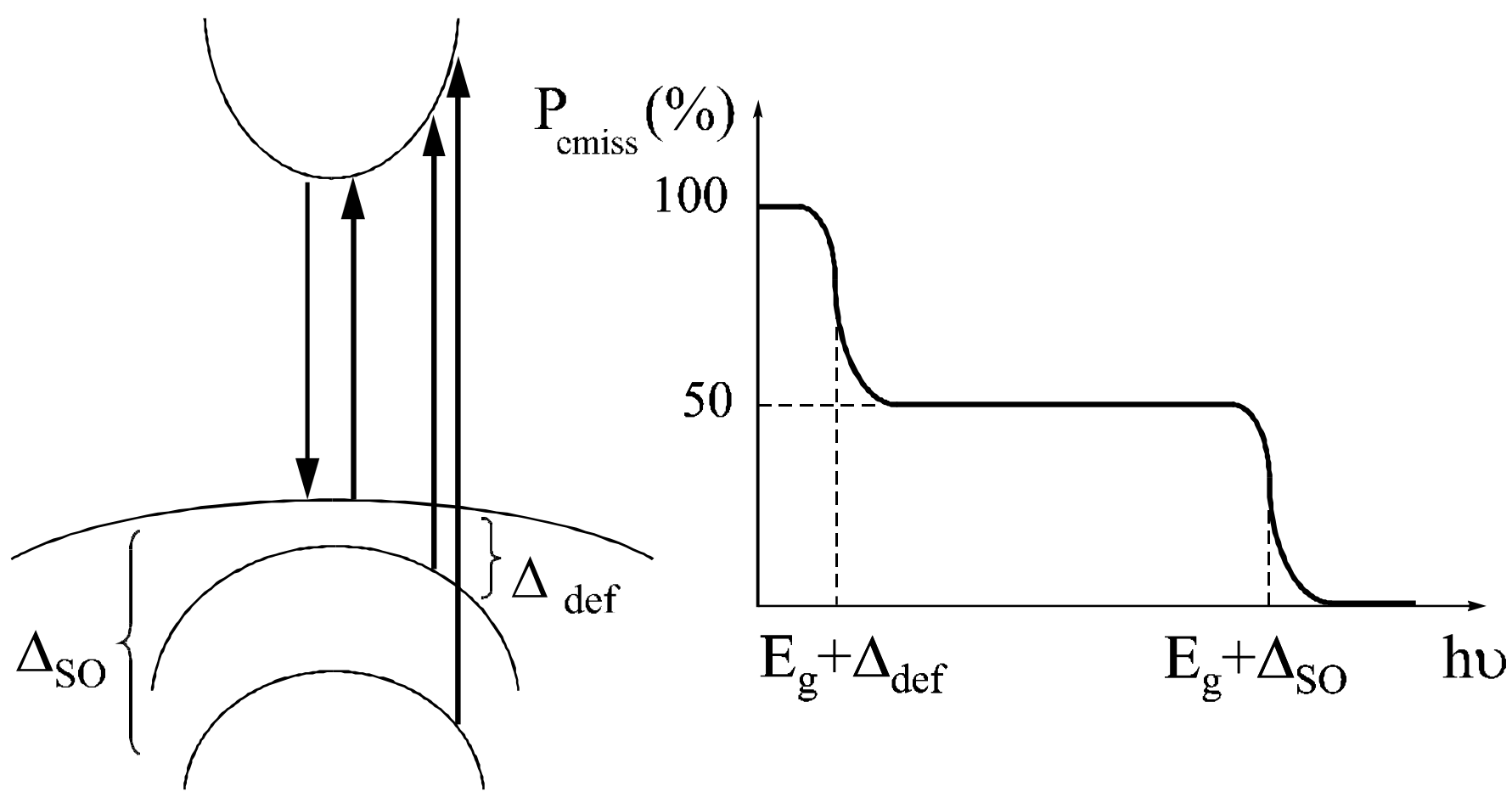

Fig.3.6. Band structure of GaAs under application of uniaxial stress and spectral dependency of optical orientation of electrons. The axis of optical excitation and observation coincides with the axis of stress application.

The more complicated case is when injection and observation axes are perpendicular to the axis of mechanical stress or quantum confinement. In this case the transformation Hamiltonian must be known in order to obtain mathematical expression matching the spin polarization of the electrons and polarization of the emitted light. In the general case unpolarized light is emitted in this direction.

The effect of selection rules damping depends over magnitude of the stress or quantum confinement itself and should be remedied at higher temperatures, when $k \cdot T > \Delta_{lh,hh}$ ($\Delta_{lh,hh}$ is the splitting of the hh-lh levels).

### 3.3.4. Hanle Effect and Optical Investigation of Spin Relaxation

The depolarization of luminescence in a transverse magnetic field is called the Hanle effect. It was discovered in the resonance fluorescence in gases in 1924 [146]. The first observation of Hanle effect in a semiconductor was reported in 1969 [147]. The application of the transverse magnetic field $\vec{B}$ causes the precession of the electron spins with Larmor frequency $\Omega = \mu_B \cdot g^* \cdot B / \hbar$ around $\vec{B}$, were $\mu_B$ - is the Bohr magneton, and $g^*$ is the effective g-factor (Fig.3.7).

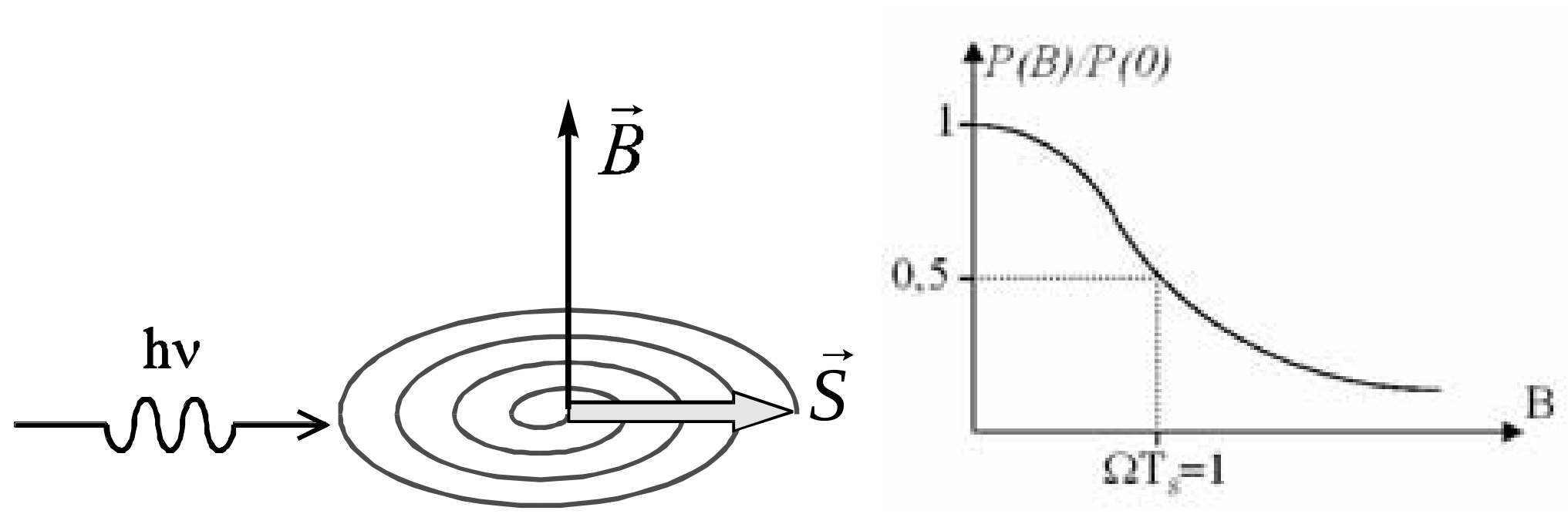

Fig.3.7. Spin injection and depolarization of photoluminescence in the transverse magnetic field.

Under steady state conditions the depolarization of photoluminescence in the transverse magnetic field is described by the well-known expression for Hanle curve

$$P(B) = \frac{P(0)}{1+(\Omega \cdot T_S)^2} = \frac{P(0)}{1+(\frac{\mu_B \cdot g^*}{\hbar} \cdot B \cdot T_S)^2} \tag{3.3}$$

where $P(B)$ is measured circular polarization of the photoluminescence in the external transverse magnetic field, $P(0)$ - is circular polarization of light at $B = 0$ and is given by Eq.3.1. Fig.3.7 shows the calculated Hanle curve after Eq.3.3. The curve has a Lorentzian shape with half-width corresponding to condition $\Omega \cdot T_S = 1$. As one can see from Eq.3.1 and Eq.3.3, the experimental shape of Hanle curve is determined by the two important parameters that are the spin relaxation term $T_S / \tau$ and the spin lifetime $T_S$. Thus experimental investigation of depolarisation of the photoluminescence in the transverse magnetic field allows direct measurements of all characteristic lifetimes of the electrons on the bottom of the semiconductor conduction band (These times are in fact compared to the known period of Larmor precession $\Omega$ ). The $T_S$ is obtained from the fitting of the experimental curves and is giving by the half-width of Hanle curve. The $T_S / \tau$ ratio is given by polarization of the emitted light at $B = 0$, once the circular polarization of the exciting beam is known. The electron lifetime $\tau$ and the spin relaxation time $\tau_S$ could be directly obtained from these parameters. This measurement technique was introduced in 1971 [148, 18] and it has been allowing already for decades an experimental investigation of electron and spin relaxation in III-V compounds.

### 3.3.5. Spin Relaxation in GaAs

In the previous sections it was shown that optical excitation with circularly polarized light creates a spin-polarized ensemble of electrons in the conduction band of the GaAs. During electron lifetime in the conduction band some depolarization of electron spin ensemble may occur due to the following mechanisms of spin scattering.

*D'yakonov-Perel'* (DP) mechanism of electron spin scattering [149, 150, 18] is originating from the lack of inversion symmetry in III-V compounds that leads to spin splitting of the conduction band for k≠0. Electrons with the same wave vector k but with opposite spin orientation have different energies. This splitting is equivalent to the presence in the crystal of an effective magnetic field inducing the precession and mixing of electron spins.

*Elliot-Yafet* (EY) mechanism of electron spin scattering [151, 152, 18] is originating from the mixing of electron wave functions with opposite spin vector orientations due to the spin-orbit interaction. As result, in the process of momentum scattering the disorientation of electron spin becomes possible also. However, the Elliot-Yafet process was shown to play a negligible role for the electron spin relaxation in gallium arsenide [150, 18].

*Bir-Aronov-Pikus* (BAP) mechanism of electron spin scattering [153, 154, 150, 18] is originating from an electron spin scattering on holes. Such spin-flip transitions are caused by electron hole exchange and annihilation interactions. The significant property of this interaction is its dependence on the relative orientation of electron and hole spins.

Fig.3.8 shows the relative efficiencies of DP and BAP mechanisms of spin scattering in p-GaAs as function of temperature and doping concentration [18]. Fig.3.8a shows the variation of spin scattering time $\tau_S$ as function of temperature for GaAs samples with moderate doping level. The dashed and solid lines represent the theoretical calculations for DP and BAP mechanisms, respectively. The DP mechanism gives a good description of experimental data only at high temperatures. At low temperatures BAP mechanism is dominating. Generally the BAP mechanism reveals itself in GaAs at doping concentration $N_A > 10^{17} cm^{-3}$, first in the low-temperature range above the delocalization temperature. As the acceptor concentration increases the BAP starts to dominate at progressively higher temperatures.

Fig.3.8b shows the variation of spin scattering time $\tau_S$ as a function of doping concentration at low and room temperatures. Different behavior of BAP at different doping concentrations is seen at low temperature curve. At acceptor concentrations $N_A < N_0$ ( $N_0$ is critical concentration, which leads to metallization of acceptors)

$\tau_S^{-1} \sim N_A$, while at higher acceptor concentrations $\tau_S^{-1} \sim N_P^{1/3}$ ($N_P$ is free hole concentration).

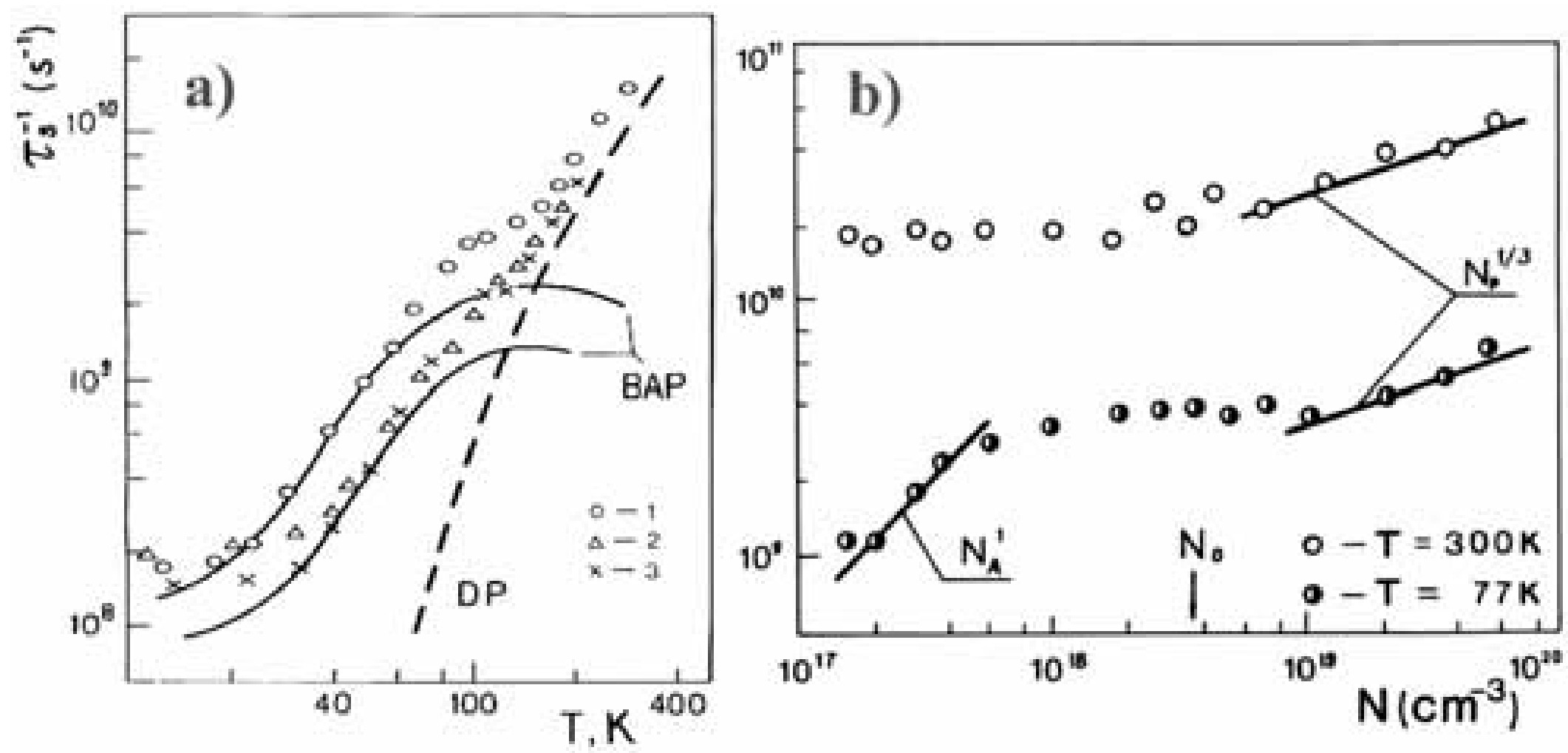

Fig.3.8. Variation of electron spin relaxation time $\tau_S$ in p-type GaAs samples as function of temperature and doping concentration [18]. a) Variation of $\tau_S$ as function of temperature in moderately doped GaAs: 2,3- $N_A$=2.2·10$^{17}$cm$^{-3}$; 1- $N_A$=3.5·10$^{17}$cm$^{-3}$; b) Variation of electron spin relaxation time $\tau_S$ as function of acceptor concentration at 77 and 300K.

### 3.3.6. Pulse-probe Technique and Electron Spin Coherence

Recent developments in the area of ultra short pulse lasers have allowed creation of a new tool for investigation of spin dynamics in semiconductors also. The combination of pulse-probe technique with optical methods for observation of magnetooptical effects allows investigation of all sorts of spin related phenomena in the semiconductors and their interfaces [155, 156, 157, 158, 159, 160]. Moreover, this technique provides new information for the samples where 'optical orientation' methods are not so strong, in particularly, undoped and n-type semiconductors.

Fig.3.9 (left) shows the experimental geometry for the observation of the time resolved Faraday rotation. The ultra short laser pulse having circularly polarization creates the ensemble of spin-polarized electrons in a semiconductor. Such spin-polarized electron ensemble consisting of $n$ electrons creates a local magnetization $M = n \cdot \mu_B$. The application of the transverse magnetic field $B$ leads to spin precession, and, hence, precessin of the connected with it local magnetization $M$ around $B$ with the Larmor frequency. The component of the spin magnetization along the direction of observation creates a transient circular birefringence that can be recorded as the Faraday rotation of a time-delayed, linearly polarized probe-pulse, which passes through the sample. In the Kerr effect geometry the only difference is that probe beam is reflected back from the surface of a semiconductor or its interfaces. So the only spins that are available within the tiny region ~10nm from the refclecting surface can be monitored.

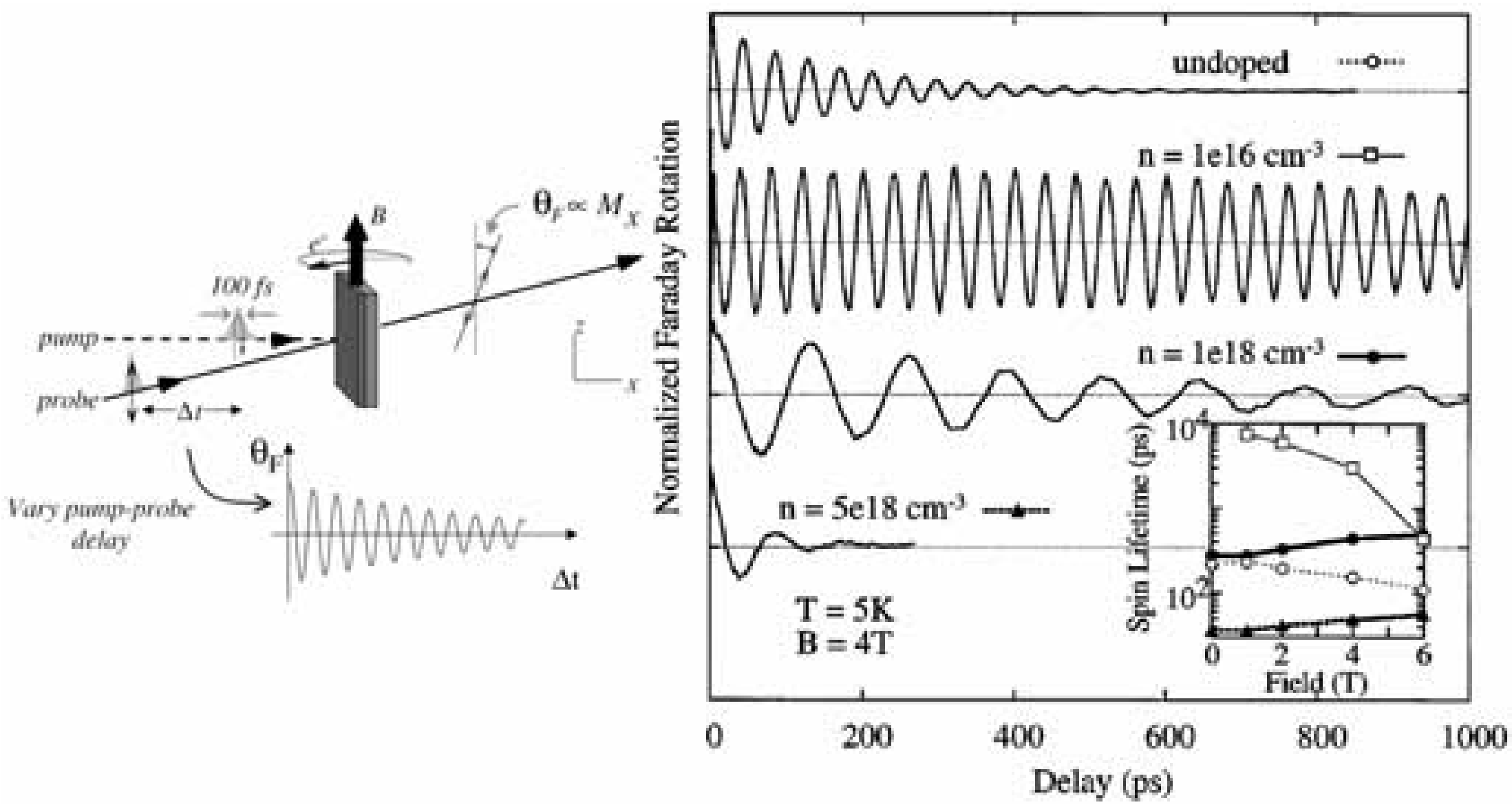

Fig.3.9 Time-Resolved Faraday rotation technique for optical investigation of electron spin dynamics in semiconductors, and experimental results for undoped and n-type GaAs [156].

Fig.3.9 (right) shows the measured spin precession in the time resolved Faraday rotation experiment [156] for GaAs samples with different doping level at low temperature. The exponential decay of the oscillatory behavior directly provides the electron spin lifetime $T_S$. In this experiment an extremely long spin lifetime $T_S = 130ns$ for GaAs sample with donor concentration $n = 1 \cdot 10^{16} cm^{-3}$ was reported.

Further, the spatial separation of pump and probe beams, together with application of the electrical bias has revealed the macroscopic lateral transport of coherently precessing electron spins over distances exceeding 100 μm [157]. Furthermore, electron spins can coherently traverse the interfaces of semiconductors with different g-factors, like GaAs and ZnSe [158]. Finally, very recent experiments involving coherent electron spin manipulation in the parabolic AlGaAs quantum well, where electron g-factor strongly depends over Al concentration, have revealed control of spin coherence via electrical gating, which displaces electron wave function from the center of the quantum well and, hence, changes the effective g-factor of electrons [159, 160].

## 3.4. Electrical Spin Injection into Semiconductors: State-of-the-Art

### 3.4.1. Injection from Magnetic Semiconductors

The concept of using a semi-magnetic diluted semiconductor with small amount of Mn ions as spin aligner was pioneered in 1998 [161]. The idea is to utilize the effect of giant effective Zeeman splitting of the conduction band in the small external magnetic field. This effect is caused by the large effective g-factor (up to 100) appearing as result of interaction of the electron spins with the large spin of localized 3d electrons of Mn ion [162]. The large splitting of the conduction band results in the alignment of the injected electron spins on a picosecond time scale [163].

One of such semi-magnetic semiconductors, BeMnZnSe, was used to demonstrate the feasibility of very efficient spin injection into a semiconductor in the direct electrical contact [164]. In this experiment, initially non-polarized electrons were 'spin aligned' in the semi-magnetic semiconductor and then injected into 15nm GaAs/AlGaAs quantum well, where they radiatively recombined with unpolarized holes supplied by the p-type GaAs substrate (Fig.3.10). The comparison of polarization state of the electroluminescence in the surface emitting spin-LEDs for samples with and without spin aligning layer as function of external magnetic field has revealed the spin polarization of injected electrons in access of 90% at low temperatures (Fig.3.10 right). Although Authors neglected the influence of splitting of heavy- and light-hole levels in the quantum well (see Section 3.3.3) leading to overestimation of the injected spin polarization, this experiment has proven the feasibility of very efficient electrical spin injection into semiconductors in the direct electrical contact.

Another experiment questioning the applicability of magnetic semiconductors for electrical spin injection is presented in the same issue of the same journal [165]. Here, the pioneers of diluted magnetic semiconductors used ferromagnetic GaMnAs to inject spin-polarized holes into non-magnetic active region of the spin-LED consisting of 10nm wide InGaAs/GaAs quantum well (Fig.3.11). Unpolarized electrons were supplied by the n-type GaAs substrate. Analyzes of the edge electroluminescence (the perpendicular to the layers growth direction, see Section 3.3.3) revealed rather small circular polarization, showing clear correlation with magnetization state of the ferromagnetic injector (Fig.3.11 right). Although the observation geometry suggests the damping of selection rules, the supporting test experiments, which include measurements of the polarization of the electroluminescence performed on the identical sample without ferromagnetic layer, and measurements of polarization of the photoluminescence under optical excitation of the active region of the sample with magnetic layer, suggested that such experimental dependencies are connected with spin injection. In this case, the spin injection of spin-polarized holes $\Pi = 1\%$ was reported.

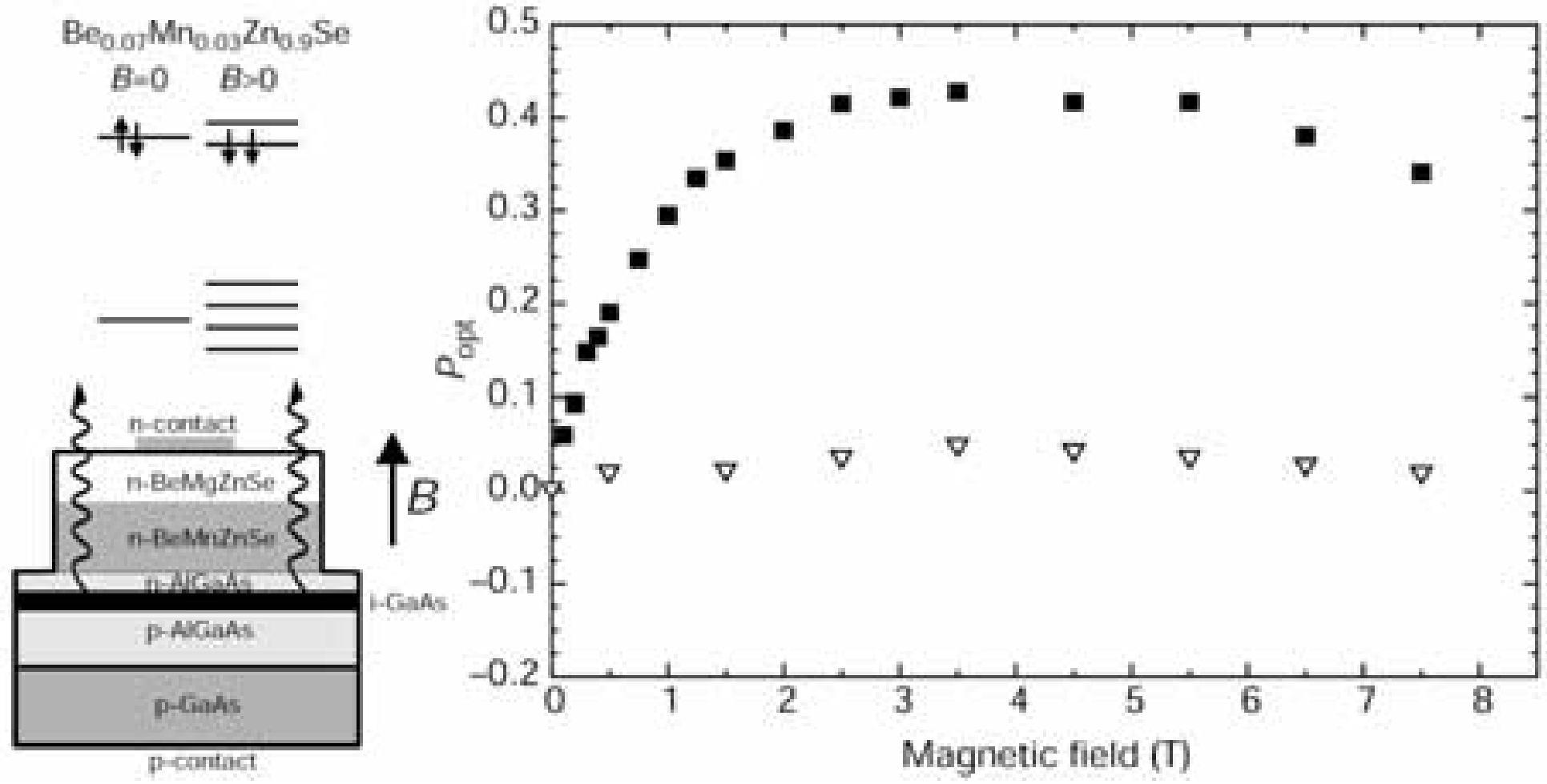

Fig.3.10. Efficient electrical spin injection into a semiconductor from a spin aligning semi-magnetic semiconductor [164]. On the left: conduction and valence band states of BeMnZnSe with and without applied external magnetic field and schematic representation of the spin-LED showing the direction of observation and axis of magnetic field application. On the right: comparison of circular polarization of the electroluminescence for samples with (squares) and without (triangles) spin aligning BeMnZnSe layer.

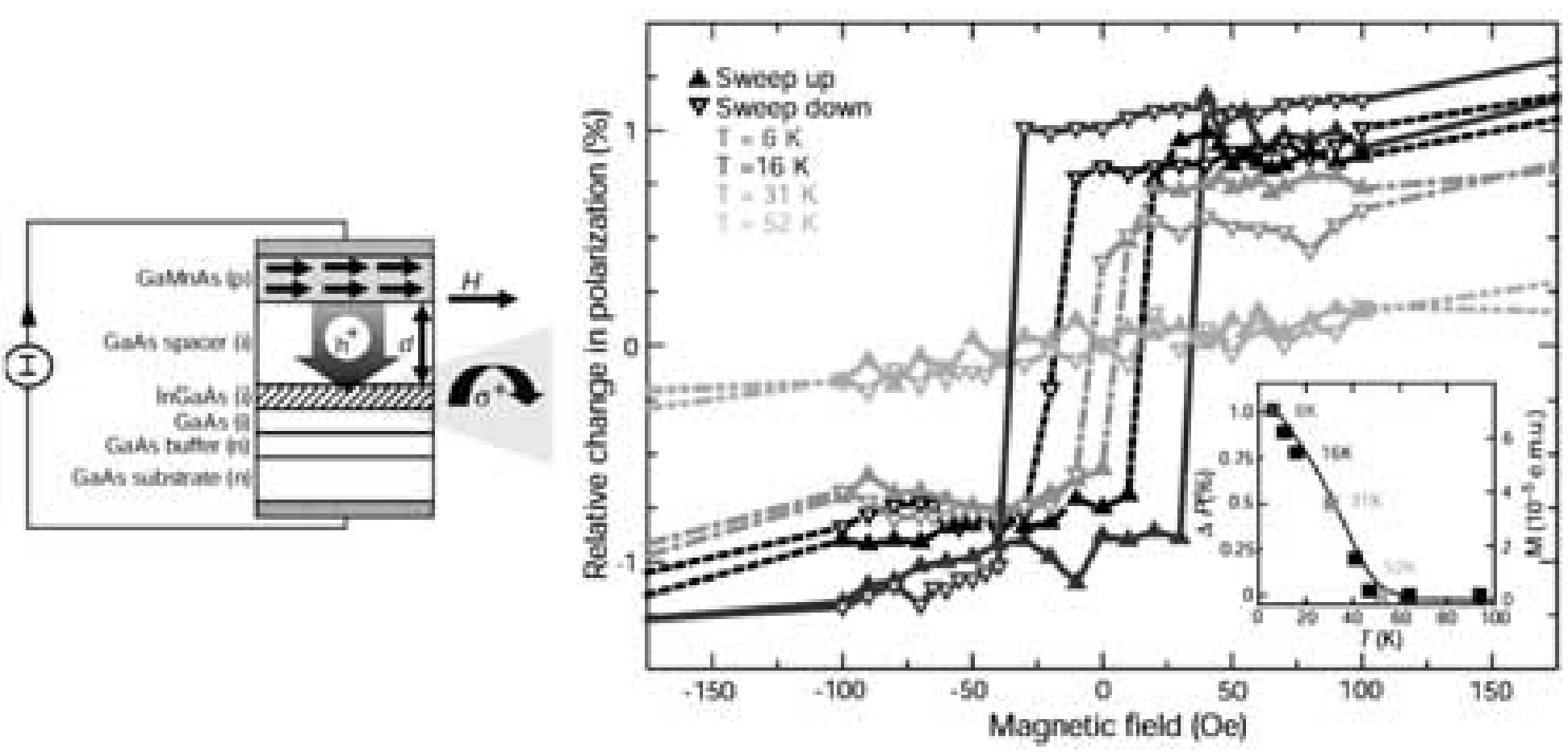

Fig.3.11.Electrical spin injection of spin-polarized holes from ferromagnetic semiconductor GaMnAs into nonmagnetic GaAs [165]. On the left: schematic representation of the device architecture showing the direction of observation and the axis of magnetic field application. On the right: circular polarization of the edge electroluminescence as function of external magnetic field and temperature.

However, the implementation of uniaxial sitem, such as quantum well, for optical detection of electrical spin injection and, moreover, in the complicated experimental configurations (as described in the previous paragraph) may lead to very complicated experimental results as well. Very recently a virtually identical experiment to the one described in the beginning of this section, examining different detection geometries in the case of electrical spin injection from the semi-magnetic semiconductor BeMnZnSe was reported [166]. In this experiment the polarization of the electroluminescence in the surface emission was compared to the one observed in the edge emission configuration at low temperatures. As described above, external magnetic field was used for effective alignment of electron spins. The active region of the device consisted of 15nm AlGaAs/GaAs quantum well. In these devices, the circular polarization of the electroluminescence in the surface emission configuration suggested spin injection in excess of $\Pi = 70\%$, while in side emission configuration, as one can expect (see Section 3.3.3), no signs of spin injection were observed, as circular polarization of electroluminescence remained zero.

Nevertheless, in another experiment, where Authors have compared the polarization of light emitted from 10nm InGaAs/GaAs quantum well active region in the side and surface emission configurations, the measurements have shown almost identical values of circular polarization $P = 6.5\%$ and $P = 8.5\%$, respectively [167]. In this case the electrons, the minority carriers in the valence band of ferromagnetic GaMnAs, in the Zener type diode were injected into the semiconductor.

At present, as it was mentioned in the Section 2.5.2, magnetic semiconductors suffer from low Curie temperature (see also Fig.2.10). Thus low temperatures and, as in the case of semi-magnetic semiconductors, application of a large magnetic field is needed for observation of the spin-dependent effects. Ferromagnetic metals do not have these limitations, offering more freedom for device implementations even above room temperature.

### 3.4.2. Injection from Ferromagnetic Metals

In all senses pioneering experiment questioning electrical spin injection from ferromagnetic metals into semiconductors was reported in 1992 [168]. In this experiment spin-polarized electrons were tunneling in the UHV from ferromagnetic Ni STM tip into cleaved atomically flat surface of GaAs (Fig.3.12). The polarization analyzes of the electroluminescence has revealed emission of circular polarization up to $P = 4.4\%$. While the analyzes of experimental data appears to be wrong, Authors neglected the high refraction index of GaAs leading to overestimation of the injected spin polarization [169] (something like twice in their case) the injection of minority electron spins up to $\Pi = -30\%$ decaying with increase of energy of the injected electrons was reported. Although this approach is not very practical for device implementation, as direct electrical contact is needed, this experiment has proven the feasibility of electrical spin injection into semiconductors from ferromagnetic metals.

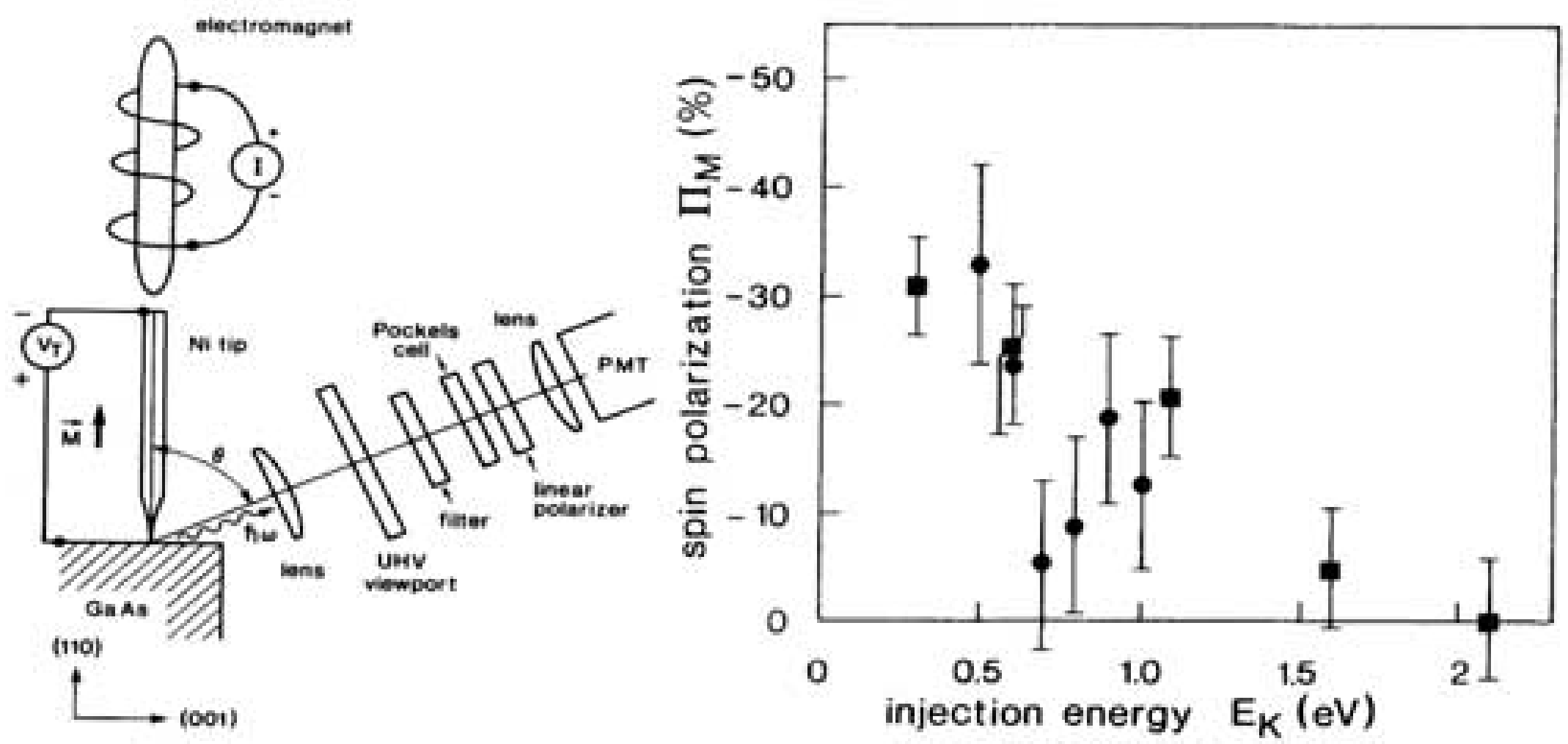

Fig.3.12. Electrical spin injection into GaAs from ferromagnetic Ni STM tip [168].

Very recently a number of experiments questioning electrical spin injection into a semiconductor from a ferromagnetic metal trough highly resistive Schottky [170, 171] and $AlO_X$ tunnel barriers in the direct electrical contact were reported [172, 173]. While the last two are described in this thesis, the first ones represent another interesting approach to the subject.

In the first experiment Ref.[170] the spin-polarized electrons were injected through highly resistive Schottky depletion region formed on Fe/GaAs interface into two 4nm wide InGaAs/GaAs quantum wells (Fig.3.13a), where they radiatively recombined with unpolarized holes supplied by the p-type GaAs substrate. The polarization analysis of light in the backside emission configuration through the transparent on this wavelength GaAs substrate was implemented in order to examine the polarization state of the injected electrons. In such system the polarization of injected electrons is defined by the magnetization axis of the ferromagnetic metal, which is in-plane. So the light emitted in the perpendicular direction is unpolarized. Hence, the magnetization of the ferromagnetic layer had to be forced out-of-the-plane by the application of external magnetic field in order to obtain a non-zero electron spin component along the direction of observation. Fig.3.13b shows the experimental dependency of the circular polarization of the electroluminescence as function of external out-of-plane magnetic field. The solid line represents the out-of-plane component of magnetization of ferromagnetic Fe contact in arbitrary units. These data were compared with the ones obtained on the test sample, where ferromagnetic contact was removed and replaced by the non-magnetic AuGe alloy.

Authors found that circular polarization of electroluminescence follows very closely the magnetization curve of the ferromagnetic metal and is superimposed on the background contribution caused by Zeeman splitting of electron and hole levels in the semiconductor. They concluded that such change of circular polarization is caused by the injection of spin-polarized electrons into semiconductor with $\Pi = 2\%$ . However these experimental data along do not prove the fact of spin injection, as contribution of magnetooptical effects to the polarization of light initially emitted in the opposite to the observation direction and then reflected back from the ferromagnetic film and collected by a photodetector was not evaluated.

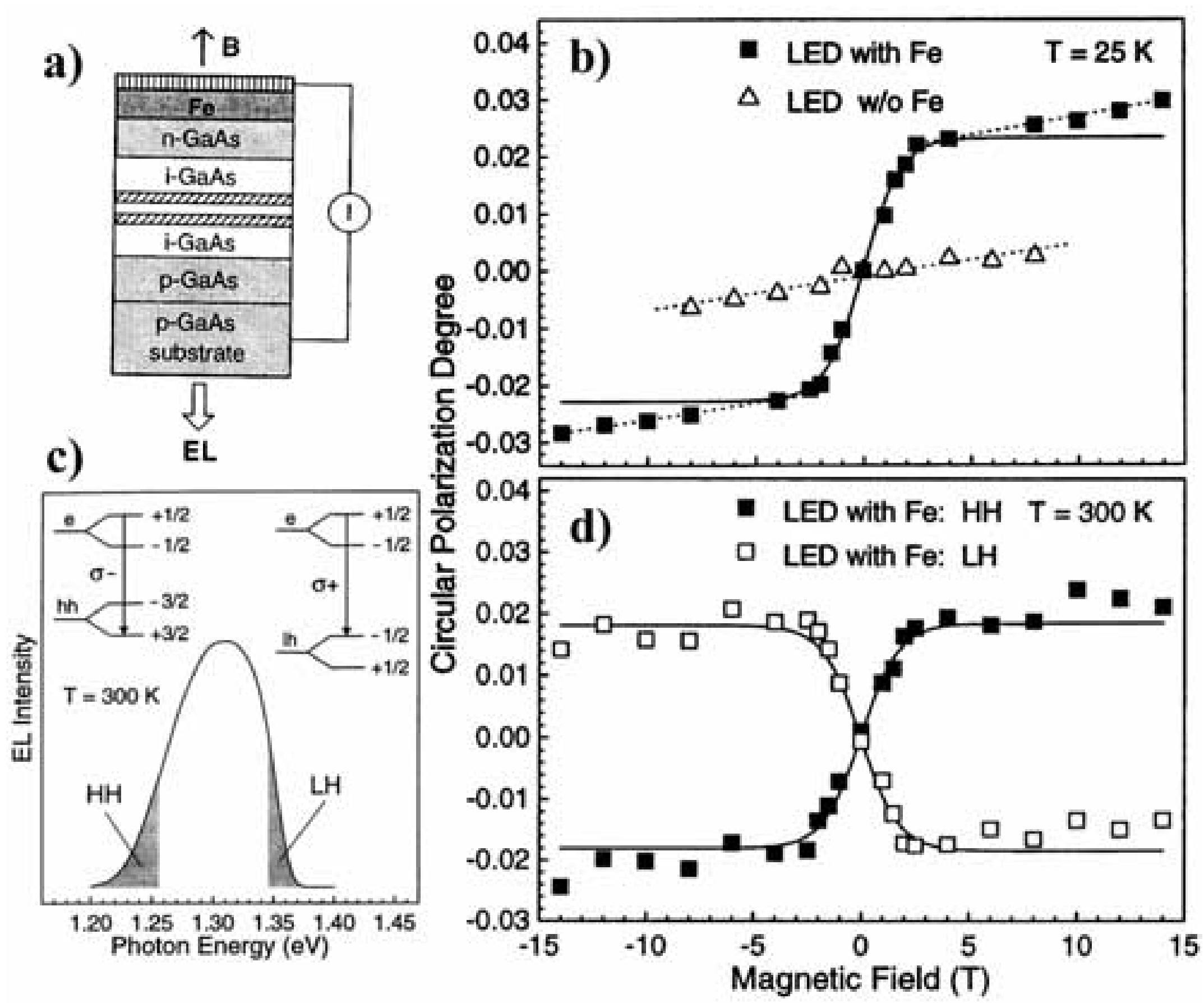

Fig.3.13. Electrical spin injection into semiconductor from ferromagnetic Fe through highly resistive Schottky depletion region formed on the Fe/GaAs interface [170]. a) Device architecture showing the direction of observation and axis of the external magnetic field application; b) Circular polarization of electroluminescence as function of external out-of-plane magnetic field (squares), the out-of-plane component of magnetization of ferromagnetic Fe contact shown in arbitrary units (solid curve) and Zeeman splitting induced spin alignment in the semiconductor as revealed by measurements of the sample without ferromagnetic layer; c) Electroluminescence spectrum of the spin-LED at RT. The shaded areas indicate the integrated intensities used to determine the degree of circular polarization for c-hh and c-lh transitions separately; d) Circular polarization of the c-hh and c-lh transitions as function of external magnetic field at RT.

Further, spectroscopic measurements at room temperature have shown that the width of the electroluminescence peak increases to the value, which is larger than *expected* heavy- and light-hole levels separation (Fig.3.13c). Although different thickness of the quantum wells may contribute to this effect, it inspired Authors to analyze polarization of these transitions separately. The experimental dependencies of circular polarization corresponding to these parts of the electroluminescence spectrum as function of the external out-of-plane magnetic field are shown in Fig.3.13d. The fact that these dependencies show the same absolute value of circular polarization but of opposite sign, while following the magnetization curve of the ferromagnetic layer, very strongly suggests that there is spin imbalance for spin-up and spin-down electrons in the quantum wells influenced by the magnetization state of the ferromagnetic metal. While it is hard to imagine other then electrical spin injection mechanism that would create this imbalance, the fact that the measured circular polarization shows the same absolute value at low and room temperatures is strange. It requires very rapid thermalization of electrons into the quantum wells and as fast recombination with holes. It is very well known that during thermalization, depending on its speed, electrons loose their spin polarization due to spin scattering. Moreover the spin scattering during electron lifetime at the ground level (on the bottom of conduction band), before recombination with holes, is described by the well known parameter $T_S/\tau$, where $T_S^{-1} = \tau^{-1} + \tau_S^{-1}$ is spin lifetime, $\tau$ and $\tau_S$ are electron lifetime and spin scattering time, respectively (see Section 3.3). Although the electron lifetime at the ground level in the quantum well can be very short, first electrons have to be captured and thermalize. Moreover, the efficiency of a LED is known to be lower at room temperature. This requires higher electrical current driven through the LED as compared with low temperatures. As result, electrons injected into semiconductor must have higher energy at room temperature. It is hard to imagine that these factors do not reveal themselves in the measurements at different temperatures.

In the second experiment [171] Authors perform virtually an identical experiment. Spin-polarized electrons are injected through highly resistive Schottky depletion region formed on Fe/GaAs interface, intentionally engineered for high tunnel transparency into 10nm wide AlGaAs/GaAs quantum well. The unpolarized holes are supplied by the p-type GaAs substrate (Fig.3.14a). In the surface emitting spin-LED the polarization of emitted light is examined as a function of an external out-of-plane magnetic field and current driven through the structure (Fig.3.14b). The circular polarization of emitted light follows nicely the out-of-plane component of the magnetization and is substantially lower than the side effects as measured in an identical experiment with photoexcitation of quantum well by linearly polarized light. Moreover, the circular polarization at saturation of the out-of-plane component of ferromagnetic contact magnetization shows clear dependency over electrical current driven trough the device. Authors found injected spin polarization $\Pi = P = 13\%$ at low and $\Pi = 4\%$ at 240K.

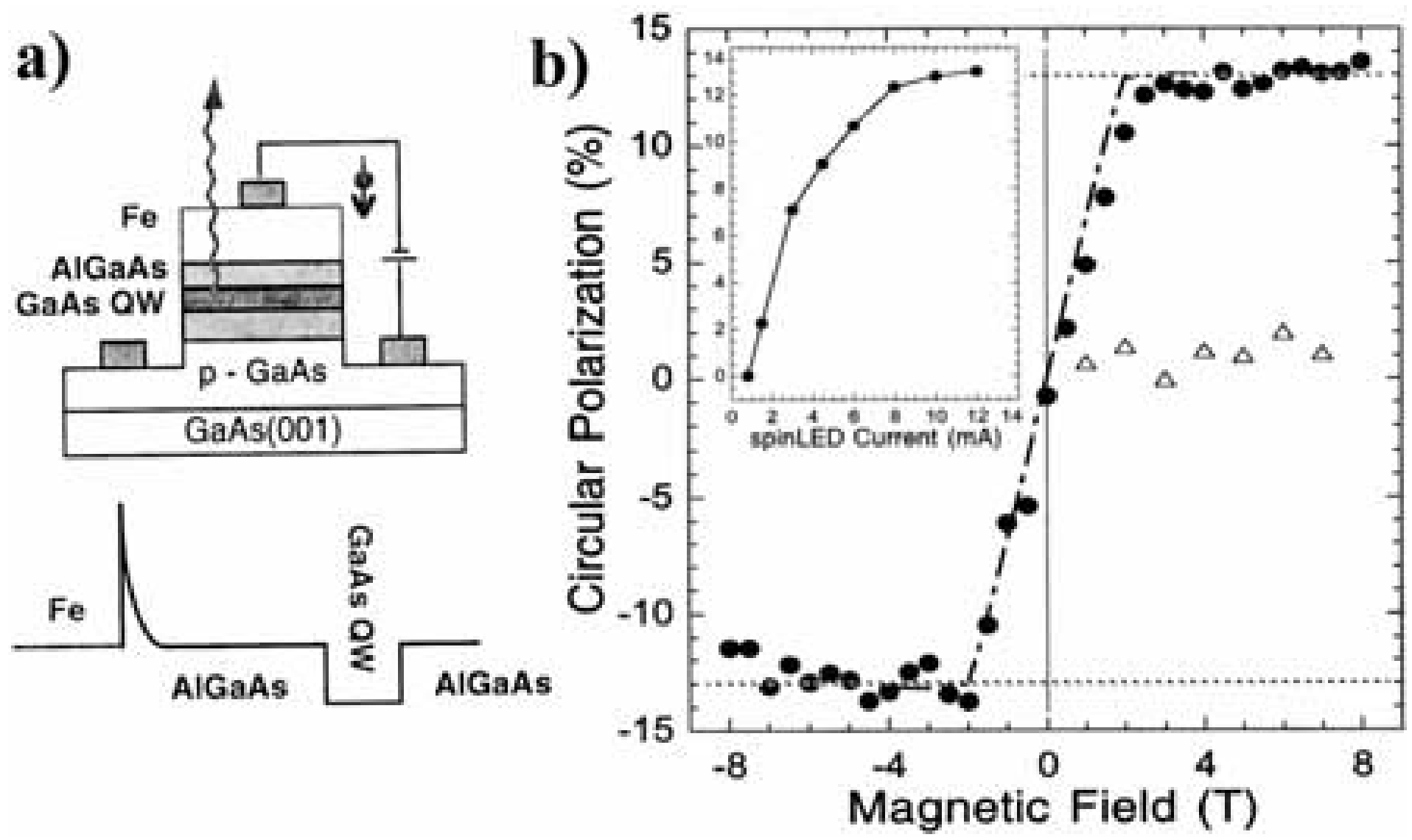

Fig.3.14. Electrical spin injection into semiconductor through highly resistive Schottky depletion region formed on Fe/GaAs interface at 4.5 K [171]. a) The schematic representation of fabricated heterostructure; b) The measured circular polarization of electroluminescence (circles) and photoluminescence (triangles) of a surface emitting spin-LED as function of external out-of-plane magnetic field and driven electrical current (insert). The dashed line represents the out-of-plane component of magnetization of the ferromagnetic contact.

# II. Experimental Investigation of Electrical Spin Injection into Semiconductors

# 4. Experimental Approach

As it was mentioned in Section 3.3.1 the conversion of the angular momentum of light into electron spin and vice versa is very efficient in III-V semiconductors. In these materials and in GaAs in particular, the absorption of 100 % circularly polarized light leads, due to the selection rules for transition probabilities, to the creation of an ensemble of electrons with preferential spin orientation along the axis of the light propagation $\Pi = 0.5$. Where $\Pi$ is the degree of spin polarization of the electron ensemble $\Pi = (n^{\uparrow} - n^{\downarrow}) \big/ (n^{\uparrow} + n^{\downarrow})$, and $n^{\uparrow}$, $n^{\downarrow}$ are numbers of spin-up and spin-down electrons.

In the radiative recombination process, due to the same selection rules of the transition probabilities, light of circular polarization $P = \frac{1}{2} \cdot \Pi$ is emitted along the axis of spin polarization of the electron ensemble. Where $P = (I^{+} - I^{-}) \big/ (I^{+} + I^{-})$ is the degree of circular polarization of light and $I^{+}$, $I^{-}$ are the intensities of right and left circularly polarized components of light.

Under electrical spin injection into a semiconductor the observed emission of circular polarization is in fact the result of a multistep process. Generally at first, the spin-polarized electrons are injected into the conduction band of the semiconductor for the case where the kinetic energy is higher than $k \cdot T$ (hot electrons). Secondly, in the thermalization process and during the electron lifetime at the bottom of the conduction band, before recombination with holes, some loss of spin polarization may occur due to spin scattering. As a result, the measured steady state polarization can be significantly smaller than the originally injected one.

## 4.1. Electron Spin Manipulation in the Semiconductor, Ferromagnetic Film and Related Phenomena

In the optical investigation of spin related phenomena it further appears that, due to the high refractive index of GaAs ($n_{GaAs}$=3.4), only photons emitted within a small angle close to the sample surface normal can escape the solid state. But for most of the thin ferromagnetic films used in spintronic applications (Fe, Co, Ni etc.) the magnetization plane is determined by the shape anisotropy, which is in-plane. Hence, under radiative recombination of spin-polarized electrons injected from the thin ferromagnetic film into the semiconductor, only the light emitted in the direction parallel to the surface of the ferromagnetic film (and hence to the magnetization) carry information about the injected spins. The light emitted in the perpendicular direction is unpolarized.

The edge emitting spin-LED (Fig.4.1 left) suffers from many significant artifacts, like waveguiding and reflections from the surface of ferromagnetic film. Thus emitted light acquires some circular polarization just from the fact of reflection (see Frenel equations) and from Magnetooptical Kerr effect, masking the expected signal and leading to wrong interpretation of the data. The effect of the selection rules damping in the case of quantum confinement further complicates the analyses of experimental data. Moreover, the high absorption rate of the band gap radiation leads to emission only from the surface area of the device (of the order of 1 μm) limiting the total optical output.

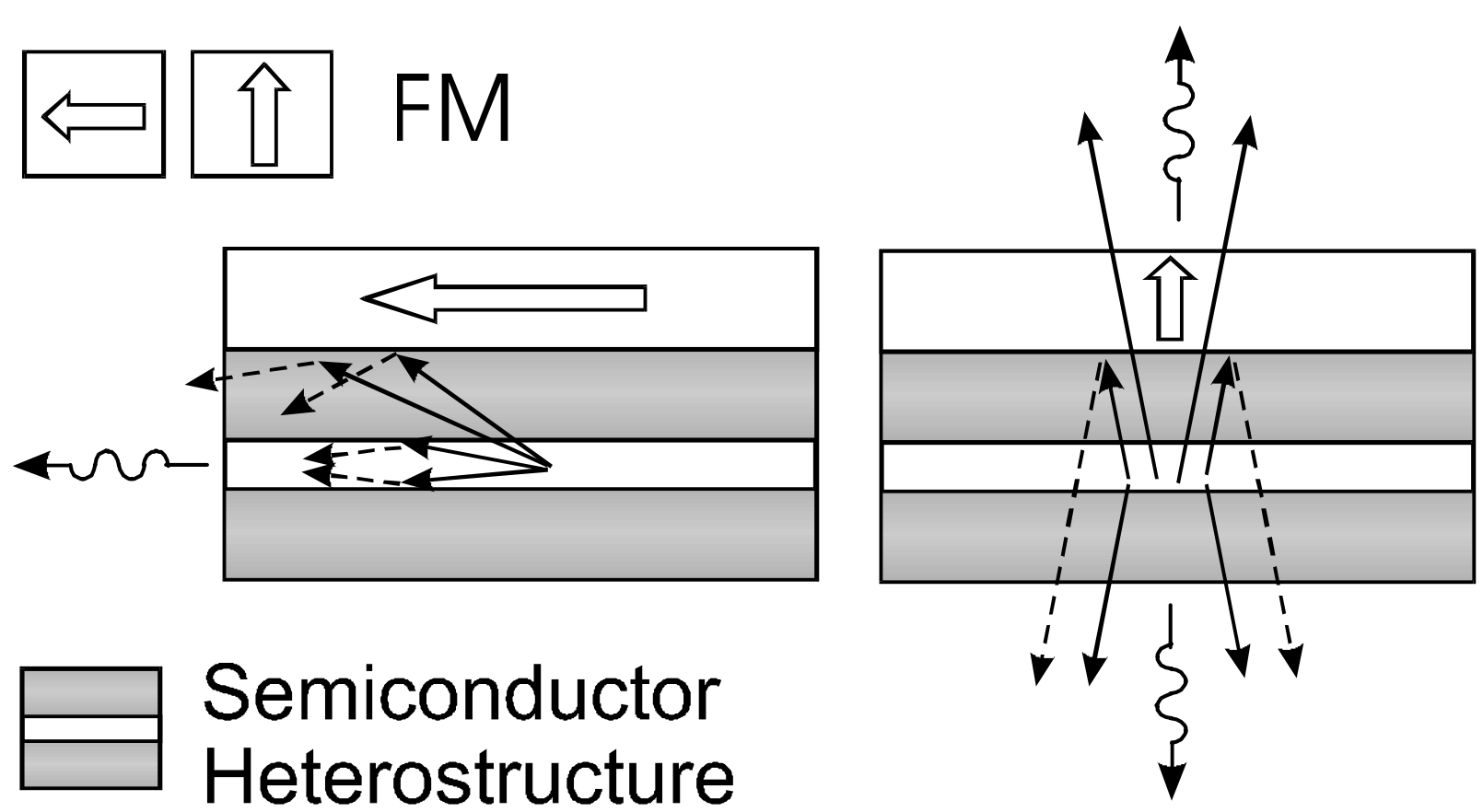

Fig.4.1. Simplified schematic representation of the edge emitting (left), backside and surface emitting (right) spin-LEDs.

The backside and surface emitting spin-LEDs (Fig.4.1 right) may allow higher total optical output, due to the surface multiplication factor, however the absorption in the substrate and in the ferromagnetic thin film must be taken into account. Moreover, as been discussed above, in the normal condition the optical output is unpolarized. In order to optically assess the spin polarization, the spins must be manipulated (in the ferromagnetic film or within the semiconductor) in a way to obtain a non-zero component of the electron spin normal to the surface. One of the common solutions used, consists of applying a strong magnetic field (more than 1T for most common ferromagnetic thin films) perpendicular to the surface, that changes the magnetization of the ferromagnetic film and hence the orientation of the injected spins. This leads to side effects, like Zeeman splitting, the Magnetooptical Kerr Effect (**MOKE**) and Magnetooptical Circular Dichroism (**MCD**). The Zeeman splitting in the semiconductor splits the spin-degenerated electron and hole levels, leading to effective spin polarization of carriers in the conduction and valence bands. Moreover, it further changes the transition probabilities defining the selection rules (see Section 3.3). In the backside emission configuration, the magnetooptical Kerr effect changes the polarization of light initially emitted in the opposite direction, then reflected back from the ferromagnetic mirror and later detected by a photodetector. Further, the influence of different absorption of left and right circularly polarized light that is MCD, in the GaAs substrate itself [174] must be also taken into account. In the surface emitting spin-LEDs, the polarization of emitted light can be affected due to MCD effect in the thin ferromagnetic film. Generally, the MOKE and MCD effects caused by the ferromagnetic film are expected to have similar magnitude due to similar depth of light interaction with the ferromagnetic film (typical thickness of ferromagnetic films in the surface emitting spin-LED is $\sim 10nm$). All these effects scale with the external magnetic field or with the out-of-plane magnetization of the ferromagnetic film, complicating the quantitative assessment of electrical spin injection. Moreover, these side effects can dominate the measured quantities and even be entirely responsible for the observed dependencies.

During work presented in this thesis a new approach based on the oblique Hanle effect (first used for detection of nuclear spin polarization [175]) for optical investigation of electrical spin injection in the surface emitting spin-LED heterostructures was developed (although, the same technique can be implemented for optical investigation of electrical spin injection in any type of spin-LEDs with very minor changes). Here, the application of a small oblique magnetic field enables one to manipulate spins inside of a semiconductor once spin-polarized electrons have been injected, and assess spin injection without considerable change of the magnetization of the ferromagnetic film. In addition, this technique reveals the spin dynamics inside the semiconductor simultaneously, separates the spin injection from the side effects and really proves the nature of the observed effects.

## 4.2. Hanle Effect in an Oblique Magnetic Field

In the solid state, the spin of an electron ensemble is characterized by the average electron spin $\vec{S} = \sum_i^n \vec{s}_i / n$, where $\vec{s}_i$ is the spin of an individual electron and $n$ is the number of electrons. The degree of spin polarization $\Pi$ along $\vec{S}$ is $\Pi = 2 \cdot |\vec{S}|$. Under application of the magnetic field $\vec{B}$ ($\left|\vec{B}\times\vec{S}\right|\neq 0$) the electron spins start to precess around $\vec{B}$ with the Larmor frequency $\vec{\Omega}=\left(g^{*}\cdot\mu_B/\hbar\right)\cdot\vec{B}$, where $g^{*}$ is the effective g-factor and $\mu_B$ is the Bohr magneton.

Within the semiconductor, the evolution of the average electron spin $\vec{S}$, taking into account spin injection, spin scattering and electron recombination processes, is described by the well known Bloch-type equation [18]:

$$\frac{d\vec{S}}{dt}=\frac{\vec{S}_0}{\tau}-\frac{\vec{S}}{T_S}+\left[\vec{\Omega}\times\vec{S}\right] \tag{4.1}$$

where $\vec{S}_0$ is the average injected electron spin, $\tau$ is the lifetime of the electrons (i.e. electron recombination time) and $T_S$ is the spin lifetime ($T_S^{-1}=\tau^{-1}+\tau_S^{-1}$, where $\tau_S$ is the spin scattering time).

In the steady state conditions $d\vec{S}/dt=0$ (the steady state could be thought any conditions with the fastest processes happening on the timescale $t>>\tau,\ T_S,\ \Omega^{-1}\sim 1\ldots 100ns$) the Eq.4.1 – the system containing three linear equations

$$\begin{cases}\dfrac{dS_X}{dt}=\dfrac{S_{0X}}{\tau}-\dfrac{S_X}{T_S}+S_Z\cdot\Omega_Y-S_Y\cdot\Omega_Z=0\\ \dfrac{dS_Y}{dt}=\dfrac{S_{0Y}}{\tau}-\dfrac{S_Y}{T_S}+S_X\cdot\Omega_Z-S_Z\cdot\Omega_X=0\\ \dfrac{dS_Z}{dt}=\dfrac{S_{0Z}}{\tau}-\dfrac{S_Z}{T_S}+S_Y\cdot\Omega_X-S_X\cdot\Omega_Y=0\end{cases} \tag{4.2}$$

has the following solutions

$$S_X=\frac{1}{1+(\Omega_X\cdot T_S)^2+(\Omega_Y\cdot T_S)^2+(\Omega_Z\cdot T_S)^2}\times$$
$$\times\left\{\begin{array}{l}S_{0X}\cdot\dfrac{T_S}{\tau}\cdot\left(1+(\Omega_X\cdot T_S)^2\right)+\\ +S_{0Y}\cdot\dfrac{T_S}{\tau}\cdot\left(\Omega_X\cdot\Omega_Y\cdot T_S^2-\Omega_Z\cdot T_S\right)+\\ +S_{0Z}\cdot\dfrac{T_S}{\tau}\cdot\left(\Omega_X\cdot\Omega_Z\cdot T_S^2+\Omega_Y\cdot T_S\right)\end{array}\right\}$$

$$S_Y = \frac{1}{1+(\Omega_X \cdot T_S)^2 + (\Omega_Y \cdot T_S)^2 + (\Omega_Z \cdot T_S)^2} \times \\ \times \left\{ \begin{array}{l} S_{0X} \cdot \frac{T_S}{\tau} \cdot \left(\Omega_X \cdot \Omega_Y \cdot T_S^2 + \Omega_Z \cdot T_S\right) + \\ + S_{0Y} \cdot \frac{T_S}{\tau} \cdot \left(1 + (\Omega_Y \cdot T_S)^2\right) + \\ + S_{0Z} \cdot \frac{T_S}{\tau} \cdot \left(\Omega_Y \cdot \Omega_Z \cdot T_S^2 - \Omega_X \cdot T_S\right) \end{array} \right\} \tag{4.3}$$

$$S_Z = \frac{1}{1+(\Omega_X \cdot T_S)^2 + (\Omega_Y \cdot T_S)^2 + (\Omega_Z \cdot T_S)^2} \times \\ \times \left\{ \begin{array}{l} S_{0X} \cdot \frac{T_S}{\tau} \cdot \left(\Omega_X \cdot \Omega_Z \cdot T_S^2 - \Omega_Y \cdot T_S\right) + \\ + S_{0Y} \cdot \frac{T_S}{\tau} \cdot \left(\Omega_Y \cdot \Omega_Z \cdot T_S^2 + \Omega_X \cdot T_S\right) + \\ + S_{0Z} \cdot \frac{T_S}{\tau} \cdot \left(1 + (\Omega_Z \cdot T_S)^2\right) \end{array} \right\}$$

In the real experiment one rather deals with the geometries similar to the one shown in Fig.4.2, which allows significant simplification of Eq.4.3. Here the XY-plane is the sample plane, the OZ-axis is pointing along the direction of observation and the OY-axis coincides with the easy axis of ferromagnetic film magnetization (although the case when OX-axis coincides with the easy axis of ferromagnetic film magnetization can be realized in an experiment, the application of external magnetic field, as described below, will result in the presented configuration). The key point is that external oblique magnetic field $\vec{B}(0, B_Y, B_Z)$ is applied under an angle $\varphi$ to the OZ-axis. Below are presented the magnetic field dependencies only for the $S_Z$ component of the average electron spin $\vec{S}$, since only this component can be detected optically.

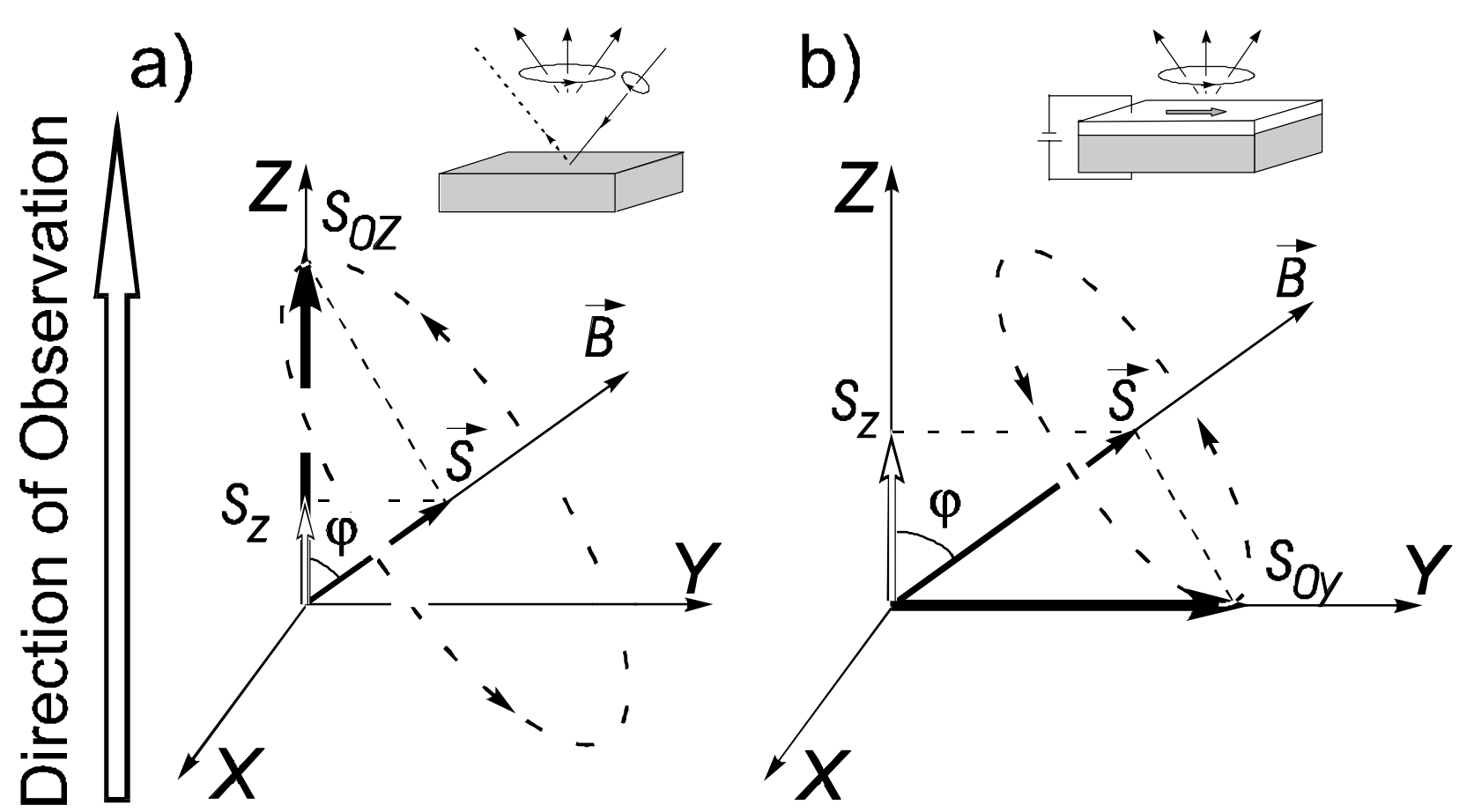

Fig.4.2. Spin precession in the oblique magnetic field in the case of a) optical spin injection and b) electrical spin injection. Under steady state conditions the spin precession leads to averaging and vanishing of the component of $\vec{S}$ perpendicular to $\vec{B}$. The remaining component is parallel to $\vec{B}$, and is accessible in measurements.

In the case of optical spin injection using circularly polarized light $\vec{S}_0(0, 0, S_{0Z})$, the magnetic field dependency of the $S_z$ component is:

$$S_Z = S_{0Z} \cdot \frac{T_S}{\tau} \cdot \frac{1 + (\Omega \cdot T_S)^2 \cdot \cos^2 \varphi}{1 + (\Omega \cdot T_S)^2} = S_{0Z} \cdot \frac{T_S}{\tau} \cdot \frac{1 + (B/\Delta B)^2 \cdot \cos^2 \varphi}{1 + (B/\Delta B)^2} \tag{4.4}$$

where

$$\Delta B = \left( \frac{g^* \cdot \mu_B}{\hbar} \cdot T_S \right)^{-1} \tag{4.5}$$

is the half-width of Hanle curve corresponding to the condition $\Omega \cdot T_S = 1$.

In the case of electrical spin injection $\vec{S}_0(0, S_{0Y}, 0)$ the dependency of the $S_Z$ as the function of the external magnetic field is different:

$$S_Z = S_{0Y} \cdot \frac{T_S}{\tau} \cdot \frac{(\Omega \cdot T_S)^2 \cdot \cos\varphi \cdot \sin\varphi}{1 + (\Omega \cdot T_S)^2} = S_{0Y} \cdot \frac{T_S}{\tau} \cdot \frac{(B/\Delta B)^2 \cdot \cos\varphi \cdot \sin\varphi}{1 + (B/\Delta B)^2} \tag{4.6}$$

The magnetic field dependencies of the $S_Z$ component (normalized to $S_0 \cdot T_S / \tau$) for electrical and optical spin injection, taken after Eq.4.4 and Eq.4.6 for different oblique angles $\varphi$ are presented on Fig.4.3.

In the case of optical spin injection $\vec{S}_0(0, 0, S_{0Z})$, the $S_Z$ component of the average electron spin $\vec{S}$ has a maximum at $B = 0$ and decreases with application of magnetic field. The $S_Z(B)$ dependency has a Lorentzian shape, the value of the asymptotical minimum depends on the angle $\varphi$ and is zero at $\varphi = 90^0$ (ordinary Hanle effect, see Section 3.3.4).

The surprising fact that $S_Z$ is not zero in the case of electrical spin injection from the in-plane magnetized ferromagnetic film $\vec{S}_0(0, S_{0Y}, 0)$ has interesting consequences for optical investigation of electrical spin injection into semiconductors. The $S_Z$ grows from zero at $B=0$ and saturates at higher values of external magnetic field $B \gg \Delta B$. The curve is strongly non-linear. The half-width $\Delta B$ (corresponding again to the condition $\Omega \cdot T_S = 1$) is determined by the effective g-factor and spin lifetime $T_S$ within the semiconductor. This provides a unique signature of spin injection compared to the side effects, which are linear or nearly linear with external magnetic field.

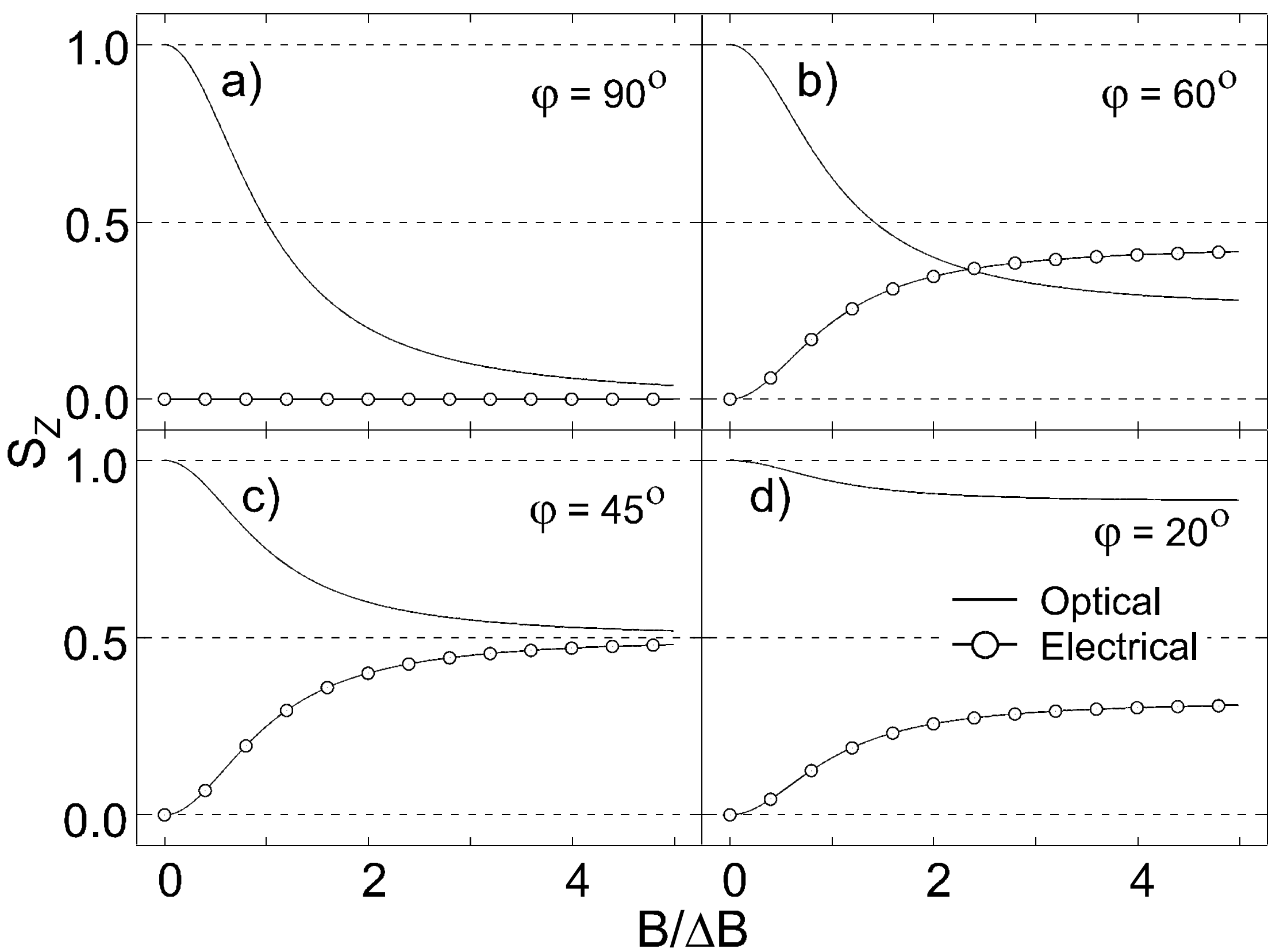

Fig.4.3. The $S_Z$ component of the average electron spin $\vec{S}$ (normalized to $S_0 \cdot T_S/\tau$), in the case of optical $\vec{S}_0(0,0,S_{0Z})$ and electrical $\vec{S}_0(0,S_{0Y},0)$ spin injection into a semiconductor in the oblique Hanle effect geometry (Fig.4.2) for different oblique angles $\varphi$, as revealed by Eq.4.4 and Eq.4.6. The magnetic field is expressed in the units of $\Delta B = \left[\left(g^* \cdot \mu_B/\hbar\right) \cdot T_S\right]^{-1}$, the half-width of the Hanle curve.

The $S_Z$ saturation value is dependent on the angle $\varphi$, having maximum of $S_{Z\,MAX} = \frac{1}{2} \cdot S_{0Y} \cdot T_S/\tau$ for $\varphi = 45^0$. Thus the $S_Z$ value measured from the polarization $P$ of the luminescence ($S_Z = P$) at saturation is lower than the injected one $S_{0Y}$ by the factor $\frac{1}{2} \cdot T_S / \tau$, and the degree of injected spin polarization $\Pi$ is related to the measured degree of circular polarization of light at saturation for $\varphi = 45^0$ as (to be compared with the Eq.2.1)

$$\Pi = 2 \cdot S_{0Y} = 4 \cdot S_{Z\ MAX} \cdot \tau / T_S = 4 \cdot P_{Sat} \cdot \tau / T_S \quad (4.7)$$

The parameter $T_S / \tau$ describes the spin scattering of the electrons during their lifetime on the bottom of the conduction band of the semiconductor and can be measured in the complete optical experiment under excitation with 100% circularly polarized light, when the injected spin polarization is perfectly known $\vec{S}_0(0,0,1/4)$ ([18], See Section 3.3.1).

However, the optical assessment of electrical spin injection in a real device implies consideration of some other effects. Below is shown the influence of magnetization switching in the ferromagnetic film by the in-plane component of the external oblique magnetic field, and how one can take into account the influence of the side effects- the tilting (rotation) of the ferromagnetic film magnetization in the small external oblique magnetic field, the magnetooptical effects caused by the ferromagnetic film and the influence of electron thermalization processes in the semiconductor.

## 4.3. Influence of Magnetization Switching

The switching of the magnetization of the ferromagnetic thin film by the in-plane component of the external oblique magnetic field, as well as coercivity, remanence are natural phenomena characteristic for ferromagnetic order of a solid. In the optical investigation of electrical spin injection in the oblique Hanle effect geometry it reveals itself as change of sign of the emitted circular polarization, once such switching occurred (Fig.4.4).

It is known that change of magnetization in the ferromagnetic film also changes the direction of preferential electron spin orientation and thus of electrically injected electron spins. Accordingly to Eq.4.6 the $S_Z$ component of the average electron spin has linear dependency over injected spin polarization and preferential orientation. This simple phenomenon provides an additional unique signature of electrical spin injection in the oblique Hanle effect experiment.

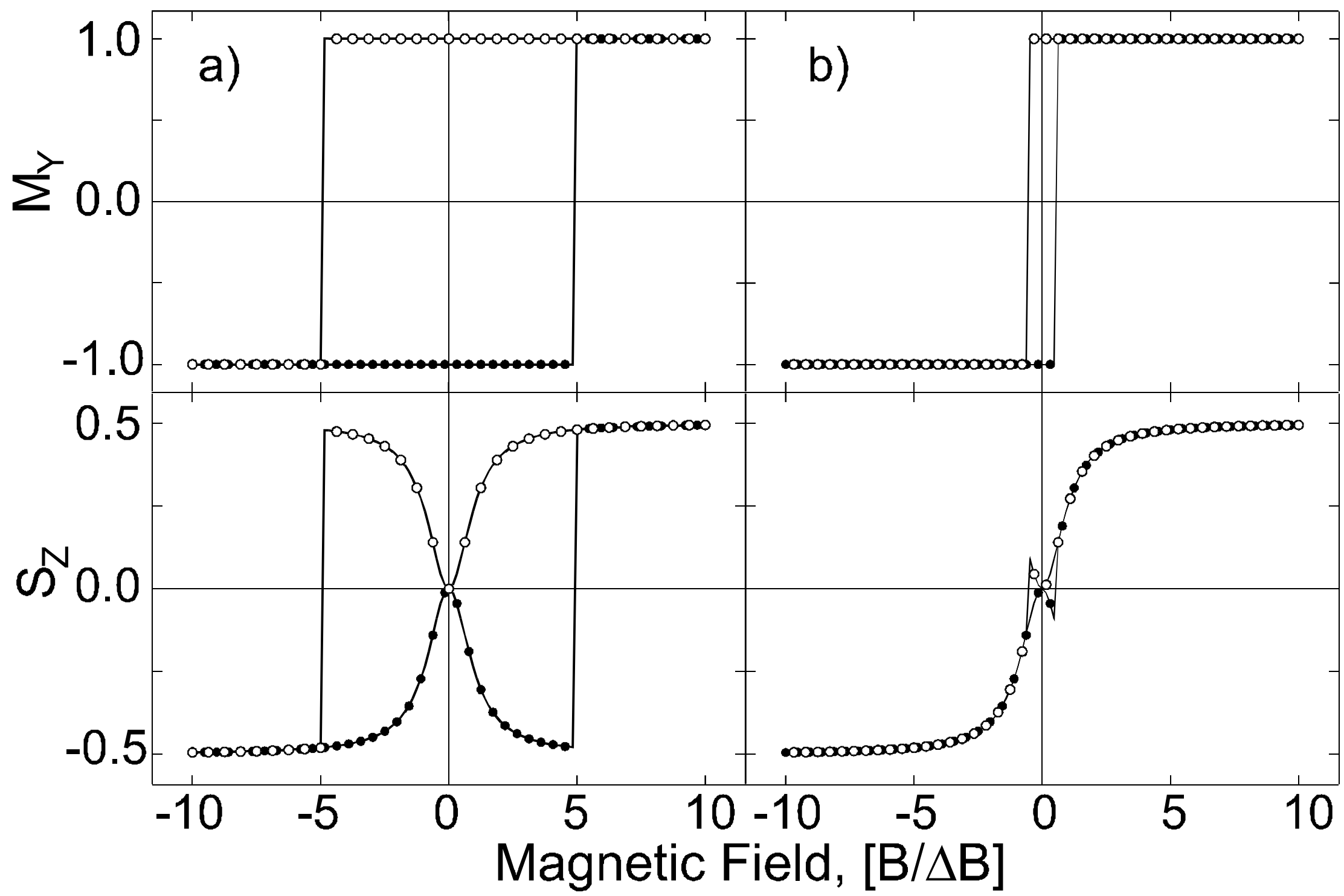

Fig.4.4. The influence of the ferromagnetic film magnetization switching on the $S_Z$ component of the average electron spin $\vec{S}$ in the experimental configuration depicted in Fig.4.2 in the case of: a) the coercive field $M_{Y_C} > \Delta B$ ($M_{Y_C} = 5 \cdot \Delta B$) and b) $M_{Y_C} < \Delta B$ ($M_{Y_C} = 0.6 \cdot \Delta B$). The magnetic field is expressed in the units of $\Delta B = \left[ \left( g^* \cdot \mu_B / \hbar \right) \cdot T_S \right]^{-1}$, the half-width of the Hanle curve.

## 4.4. Tilting of the Ferromagnetic Film Magnetization in an Oblique Magnetic Field

An oblique magnetic field applied under angle $\varphi$ with the magnetization axis of a ferromagnetic film (OZ-axis) will force it to tilt (rotate) out-of-plane with an angle $\psi$ (Fig.4.5). This will lead to the spin injection $\vec{S}_0(0, S^*_{0Y}, S^*_{0Z})$, which is different from the case described in the previous section.

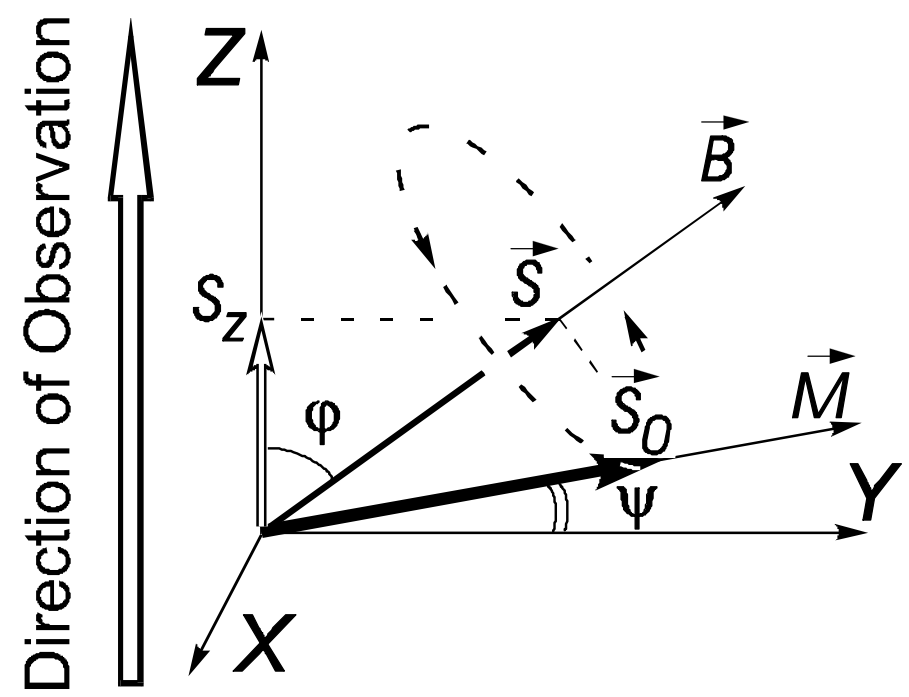

Fig.4.5. Tilting of the magnetization $\vec{M}$ of the thin ferromagnetic film under application of the oblique magnetic field $\vec{B}(0, B_Y, B_Z)$ in the oblique Hanle effect experimental configuration.

Starting from Eq.4.3, for the same applied oblique magnetic field $\vec{B}(0, B_Y, B_Z)$ one can easily obtain an expression for the $S_Z$ component of the average electron spin $\vec{S}$ :

$$S_Z = S_{0Y}^* \cdot \frac{T_S}{\tau} \cdot \frac{(\Omega \cdot T_S)^2 \cdot \cos\varphi \cdot \sin\varphi}{1+(\Omega \cdot T_S)^2} + S_{0Z}^* \cdot \frac{T_S}{\tau} \cdot \frac{1+(\Omega \cdot T_S)^2 \cdot \cos^2\varphi}{1+(\Omega \cdot T_S)^2} =$$

$$= S_{0Y}^* \cdot \frac{T_S}{\tau} \cdot \frac{(B/\Delta B)^2 \cdot \cos\varphi \cdot \sin\varphi}{1+(B/\Delta B)^2} + S_{0Z}^* \cdot \frac{T_S}{\tau} \cdot \frac{1+(B/\Delta B)^2 \cdot \cos^2\varphi}{1+(B/\Delta B)^2} \qquad (4.8)$$

where $S_{0Y}^* = S_0 \cdot \cos\psi$ and $S_{0Z}^* = S_0 \cdot \sin\psi$ .

Now $S_Z$ has two components identical to Eq.4.4 and Eq.4.6 with amplitudes, which are functions of the tilting angle $\psi$ . The magnitude of the angle $\psi$ is dependent on the film saturation magnetization $M$ , angle $\varphi$ and magnitude of the external applied field $H_{ext} = B/\mu_0$ .

For the calculation of the tilting angle $\psi$ one can make the following considerations. The shape anisotropy energy (or demagnetization energy) density is given by $U_d = -\mu_0 \cdot \left(\vec{M} \cdot \vec{H}_d\right)/2$, where $\vec{H}_d = -\vec{M}_Z$ is the demagnetizing field and $M_Z$ is the out-of-plane magnetization of the film. The total energy density of the film in an external field $\vec{H}_{ext}$ is then given by $U = -\mu_0 \cdot \left(\vec{M} \cdot \vec{H}_{ext}\right) - \frac{\mu_0 \cdot \left(\vec{M} \cdot \vec{H}_d\right)}{2} =$

$$= -\mu_0 \cdot M \cdot H_{ext} \cdot \sin(\varphi + \psi) + \frac{\mu_0 \cdot M^2}{2} \cdot \sin^2\psi \qquad (4.9)$$

and the minimum is obtained for the tilt angle ψ given by

$$\frac{B}{\mu_0 \cdot M} = \frac{\sin\psi \cdot \cos\psi}{\cos\varphi \cdot \cos\psi - \sin\varphi \cdot \sin\psi} \tag{4.10}$$

The dependencies of the tilting angle $\psi$ on the external oblique magnetic field $\vec{B}$, taken after Eq.4.10 are shown on Fig.4.6 for two different field angles $\varphi = 45\ ^{o}$ and $\varphi = 60\ ^{o}$. The angle $\psi$ has an almost linear dependency at small values of $B$ with saturation at higher values. The saturation value is $\psi = 90^{o} - \varphi$.

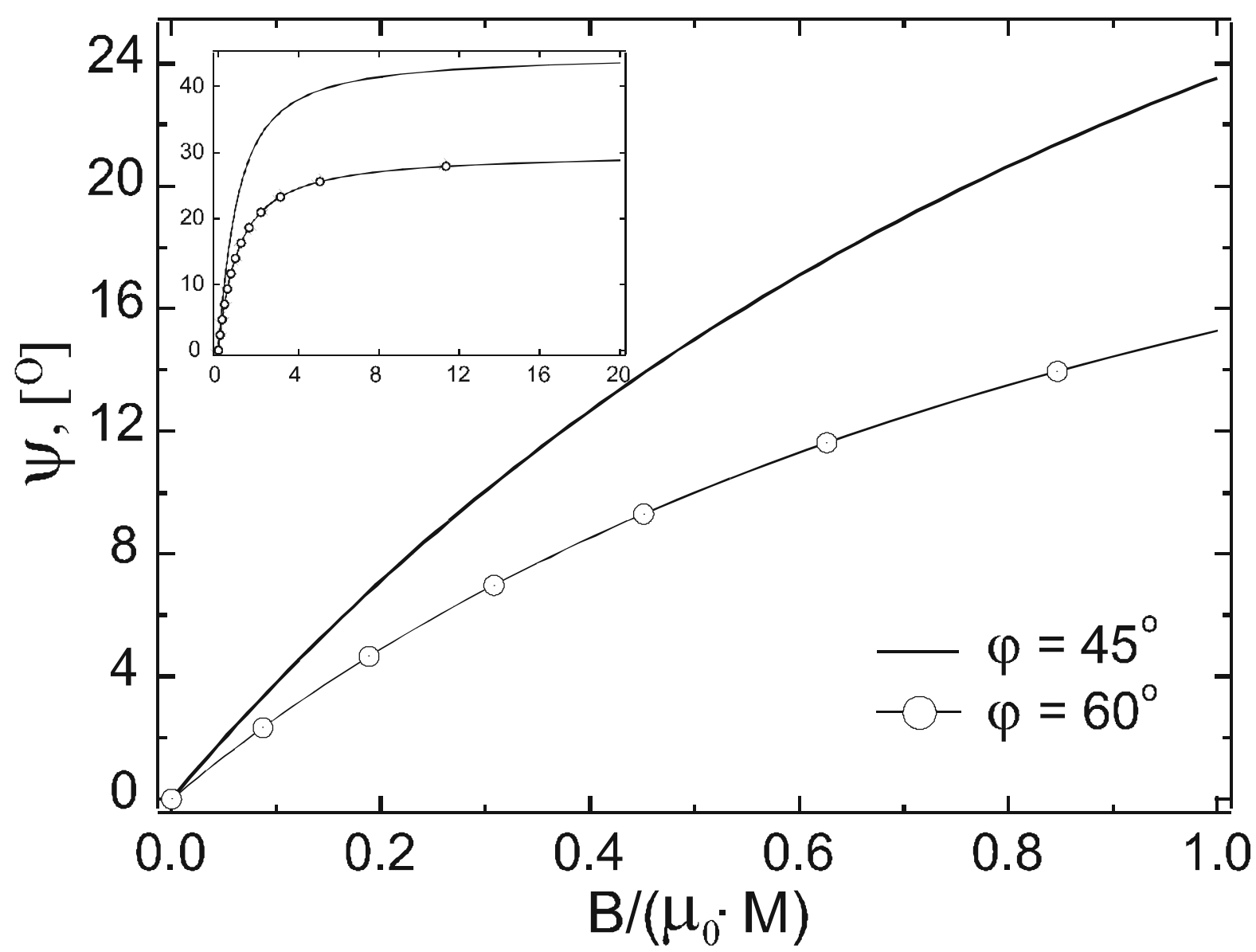

Fig.4.6. The tilting angle $\psi$ as a function of the $B/(\mu_0 \cdot M)$ ratio for different oblique angles $\varphi = 45\ ^{o}$, $\varphi = 60\ ^{o}$, after Eq.4.10. (Inset extended view).

Note that the angle $\psi$ is dependent on the magnetic field $B$ thus the $S^*_{0Y}$ and $S^*_{0Z}$ components of the average electron spin also depend on $B$. This leads to deformation of the Hanle curves described in the Section 4.2. Fig.4.7 shows the calculated dependencies of $S_Z(B)$ resulting from Eq.4.8, Eq.4.10 and Eq.4.6 for oblique angle $\varphi = 45\ ^{0}$. The important parameter is the ratio between the half-width of the Hanle curve $\Delta B$ and the saturation magnetization $\mu_0 \cdot M$ of the film. Fig.4.7a shows the $S_Z(B)$ dependency described by Eq.4.8, Eq.4.10 for various values of the $\mu_0 \cdot M$, while $\Delta B$ is fixed at $\Delta B = 1$.

Fig.4.7b shows the opposite case, when the $\mu_0 \cdot M$ is fixed and the $\Delta B$ takes different values. (A change of $\Delta B$ can be achieved by changing the temperature as well as semiconductor heterostructure, doping level, etc. ([18], see Section 3.3). The bold

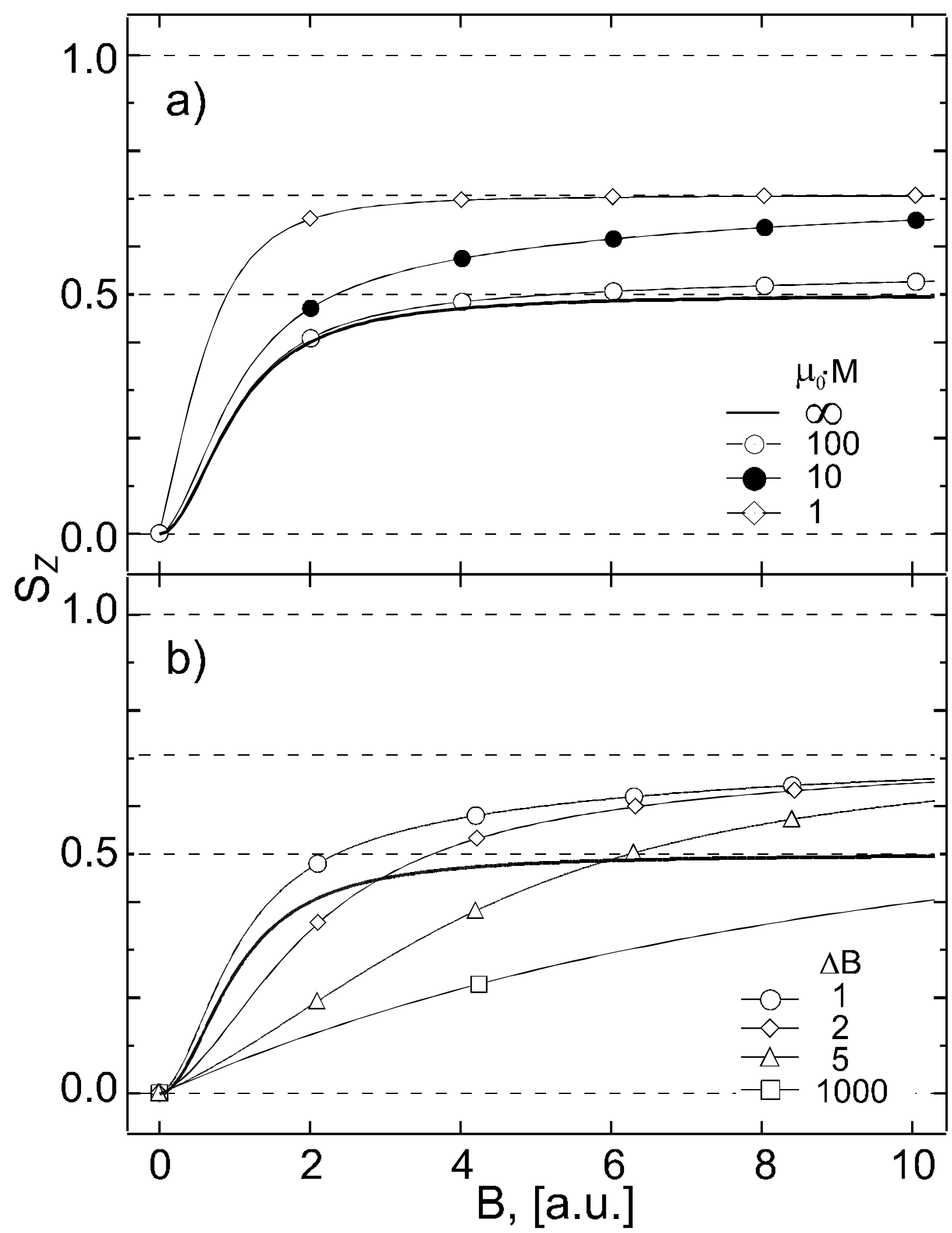

Fig.4.7. The $S_Z$ component of the average electron spin $\vec{S}$ (normalized to $S_0 \cdot T_S/\tau$) as a function of the oblique magnetic field $\vec{B}$ ($\varphi = 45^0$) for different $\Delta B/(\mu_0 \cdot M)$ ratios after Eq.4.8, Eq.4.10 and Eq.4.6. a) The $S_Z(B)$ dependency for different ferromagnetic materials - different values of saturation magnetization $\mu_0 \cdot M$, while $\Delta B = 1$ being fixed. b) The $S_Z(B)$ dependency for different semiconductor spin detectors – different half-width of Hanle curve $\Delta B = \left[\left(g^* \cdot \mu_B/\hbar\right) \cdot T_S\right]^{-1}$, while $\mu_0 \cdot M = 10$ being fixed. The bold line corresponds to the $S_Z(B)$ dependency for in-plane spin injection $\vec{S}_0(0, S_{0Y}, 0)$ (Eq.4.6), when $\Delta B = 1$. The values of $B$, $\Delta B$ and $\mu_0 \cdot M$ are expressed in the same arbitrary units.

line represents $S_Z(B)$ without taking into account the tilting of magnetization in ferromagnetic film (Eq.4.6). One can see that the change of the $S_Z$ caused by the tilting of the ferromagnetic film magnetization in the oblique magnetic field indeed depends on the $\Delta B/(\mu_0 \cdot M)$ ratio and in the general case cannot be neglected. This contribution can be easily taken into account in the fitting of experimental data, when the saturation magnetization of the thin ferromagnetic film is known from independent measurements.

This correction in the fitting procedure decreases the values of $\Delta B$ and $\Pi$. However, the experience shows (see Section 6.3.1) that the fabricated FeCo / $AlO_X$ / Al(GaAs) MIS spin-LEDs this correction does not exceed 10% and 20% of the values of $\Delta B$ and $\Pi$, respectively, obtained from the fitting without taking into account the tilting effect (the measurements have shown the ferromagnetic film saturation magnetization in these devices is $\mu_0 \cdot M = 1.3\ T$ and the external magnetic field in the oblique Hanle effect experiment does not exceed $B \leq 0.6\ T$. See Section 5.4.3).

## 4.5. Influence of Magnetooptical Effects

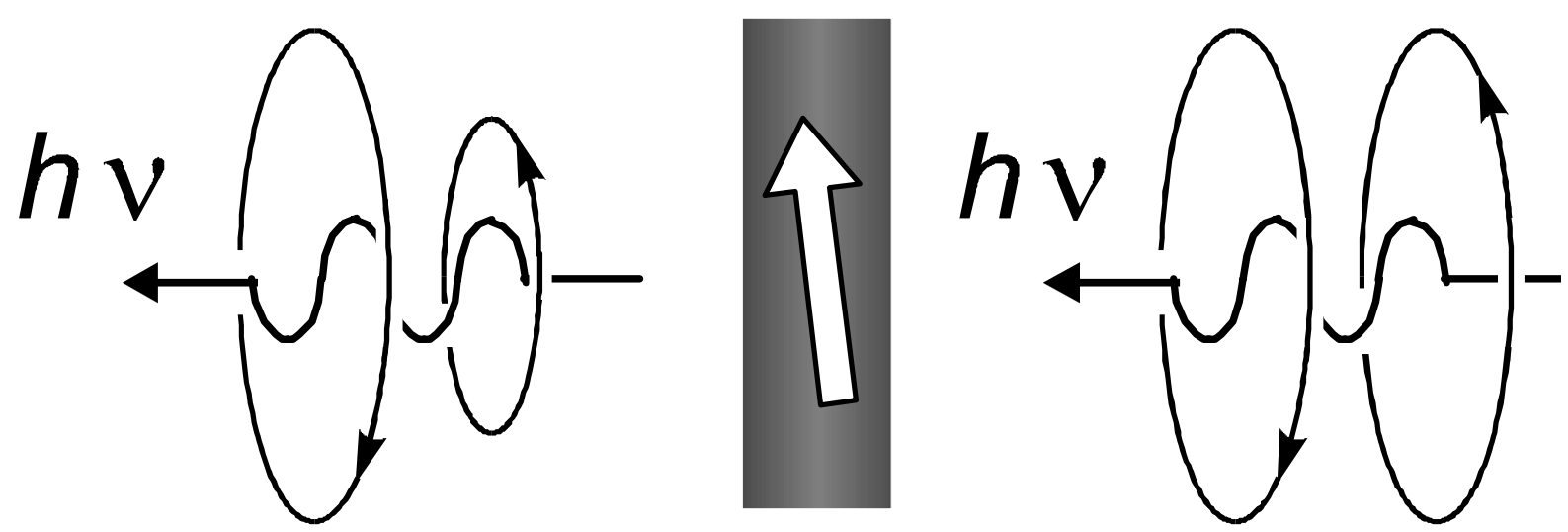

Fig.4.8. Different absorption of left and right circularly polarized light in the thin ferromagnetic film

In the surface emitting spin-LEDs the light emitted in the semiconductor propagates through the thin semi-transparent ferromagnetic layer, which has some out-of-plane magnetization caused by the tilting of the magnetization in the external oblique magnetic field (Fig.4.8). So the Magnetooptical Circular Dichroism (MCD), i.e. the difference in absorption of the light with left and right circular polarization by the ferromagnetic film, can influence the resulting degree of circular polarization. This effect can be easily taken into account.

The MCD contribution to the polarization of the light can be characterized by $D = (T^+ - T^-)/(T^+ + T^-) = \Delta T/T$, where $T^+ = T + \Delta T$ and $T^- = T - \Delta T$ are transmissions of right and left circular polarization components of light in the ferromagnetic film (as result of different absorption). The intensities of right and left circular polarization components of light emitted as a result of the recombination of a spin-polarized electron ensemble inside the semiconductor can be expressed as $I^{+} = I + \Delta I$, $I^{-} = I - \Delta I$. And the degree of circular polarization is then $P_{inj} = (I^{+} - I^{-})/(I^{+} + I^{-}) = \Delta I/I$. The resulting intensities of right and left circular polarization components of light after propagation through the ferromagnetic film are $I^{+}_{meas} = T^{+} \cdot I^{+}$, $I^{-}_{meas} = T^{-} \cdot I^{-}$. So the measured degree of circular polarization of light is

$$P_{meas} = \frac{I^{+}_{meas} - I^{-}_{meas}}{I^{+}_{meas} + I^{-}_{meas}} = \frac{\Delta T}{T} \cdot \left(1 - \frac{\Delta T}{T} \cdot \frac{\Delta I}{I}\right) + \frac{\Delta I}{I} \cdot \left(1 - \frac{\Delta T}{T} \cdot \frac{\Delta I}{I}\right) \qquad (4.11)$$

For typical real structures one generally has $\Delta T/T \ll 1$ and $\Delta I/I \ll 1$ ($\Delta T/T \approx 10^{-3}$, $\Delta I/I \approx 10^{-1} \ldots 10^{-3}$), so the experimentally measured degree of circular polarization can be represented just as

$$P_{meas} \approx P_{inj} + D \qquad (4.12)$$

The contribution of MCD to the resulting circular polarization of the emitted light can be quantitatively characterized in a simple photoluminescence experiment. Excitation with unpolarized or linearly polarized light creates in the semiconductor the population of unpolarized electrons ([18], see Section 3.3.1]. Their radiative recombination with holes results in emission of the unpolarized light with $P_{inj} = 0$. Thus, from Eq.4.12 the measurement of circular polarization of photoluminescence under such excitation provides directly the D value, which characterizes the MCD effect in the ferromagnetic film.

In the same experiment one can also take into account the polarization of luminescence due to Zeeman splitting of electron spin states in the external magnetic field [18]. As it is shown below this effect is very weak, and can be neglected in all measurements presented in this thesis.

In semiconductor under thermal equilibrium the population of spin-up and spin-down states, splitted by external magnetic field, is given by Boltzmann statistics

$$\frac{n^{\uparrow}}{n^{\downarrow}} = \exp\left(-\frac{g^{*} \cdot \mu_B \cdot B}{k \cdot T}\right).$$

The average electron spin $\vec{S}$ corresponding to Zeeman splitting, which is inferior to the thermal energy (it is always the case in the measurements presented in this thesis), is given by the following expression:

$$S_{Zeeman} = \frac{1}{2} \cdot \frac{\mathrm{n}^{\uparrow} - n^{\downarrow}}{\mathrm{n}^{\uparrow} - n^{\downarrow}} \approx \frac{1}{4} \cdot \frac{g^{*} \cdot \mu_B \cdot B}{k \cdot T} \qquad (4.13)$$

Since the lifetime of electrons being optically or electrically injected into semiconductor is comparable to their spin relaxation time, the injected electrons cannot reach the complete thermal equilibrium and Eq.4.13 has to be weighted by the factor $T_S/\tau_S$ ($\frac{T_S}{\tau_S} = \frac{\tau}{\tau + \tau_S} < 1$). In this case the corresponding polarization of luminescence in the direction of observation can be expressed in the following way

$$P_{Zeeman} = S_{Zeeman} \cdot \frac{T_s}{\tau_s} \cdot \cos\varphi \tag{4.14}$$

In the case of GaAs, the estimations for the electron Zeeman splitting contribution to the circular polarization of luminescence at $80\,\mathrm{K}$ gives $P_{Zeeman} \approx 10^{-4}$ in the highest magnetic field used during measurements presented in this thesis ($B_{\max} = 0.6\,\mathrm{T}$, see Section 4.7 and experimental Chapter 6). This contribution is also linear function of $B$, it can be easily measured in the simple experiment under optical excitation with linearly polarized light, exactly the same way as in the case with MCD. Hence, the measurements of $D$ described above include all magneto-optical effects, the MCD in ferromagnetic film and Zeeman splitting induced spin polarization in the semiconductor.

## 4.6. Electron Thermalization in the Semiconductor

In the semiconductor heterostructure under application of sufficiently high bias, the hot electrons are injected into the active region of a spin-LED (Fig.4.9a). It can happen that during their thermalization to the bottom of the conduction band these electrons loose their spin orientation. This effect was studied in detail in all-optical experiments ([18, 145], see Section 3.3) in the 70's. The loss of spin polarization during the thermalization process depends on the speed of electron thermalization, which is different for the samples with different doping levels. In the highly doped samples the thermalization is rapid and polarization losses are small and even insignificant. For the low-doped samples the thermalization process is slow and spin losses during thermalization could be high. Clear indication on the role of thermalization can be obtained from the all-optical experiments with different wavelengths of the exciting light.

For the exciting photon energy (Fig.4.9b) exceeding $E_g + \Delta$, where $\Delta$ is the spin orbit splitting of the valence band, the spin polarization of the photoexcited electrons is strongly reduced, due to the fact that electrons excited from the split-off band have opposite spin orientation as compared to the electrons excited from the upper bands of the light and heavy holes ([18, 145], see Section 3.3). In the general case, the contribution of the latter ones prevails and the net spin polarization stays positive, but reduced in value. However, the electrons excited from the upper valence bands have much higher kinetic energy and should be thermalized to the bottom of the conduction band before recombination takes place. If these hot electrons even partially loose their spin orientation during thermalization process, the net spin orientation on the bottom of the conduction band changes sign and becomes negative due to the contribution of "cold" electrons excited from the split-off valence band ([18, 145], see Section 3.3). Thus, by comparison of circular polarization of photoluminescence under excitation near the band gap $E_g$ and slightly higher than $E_g+\Delta$ the spin loss during electron thermalization can be evaluated.

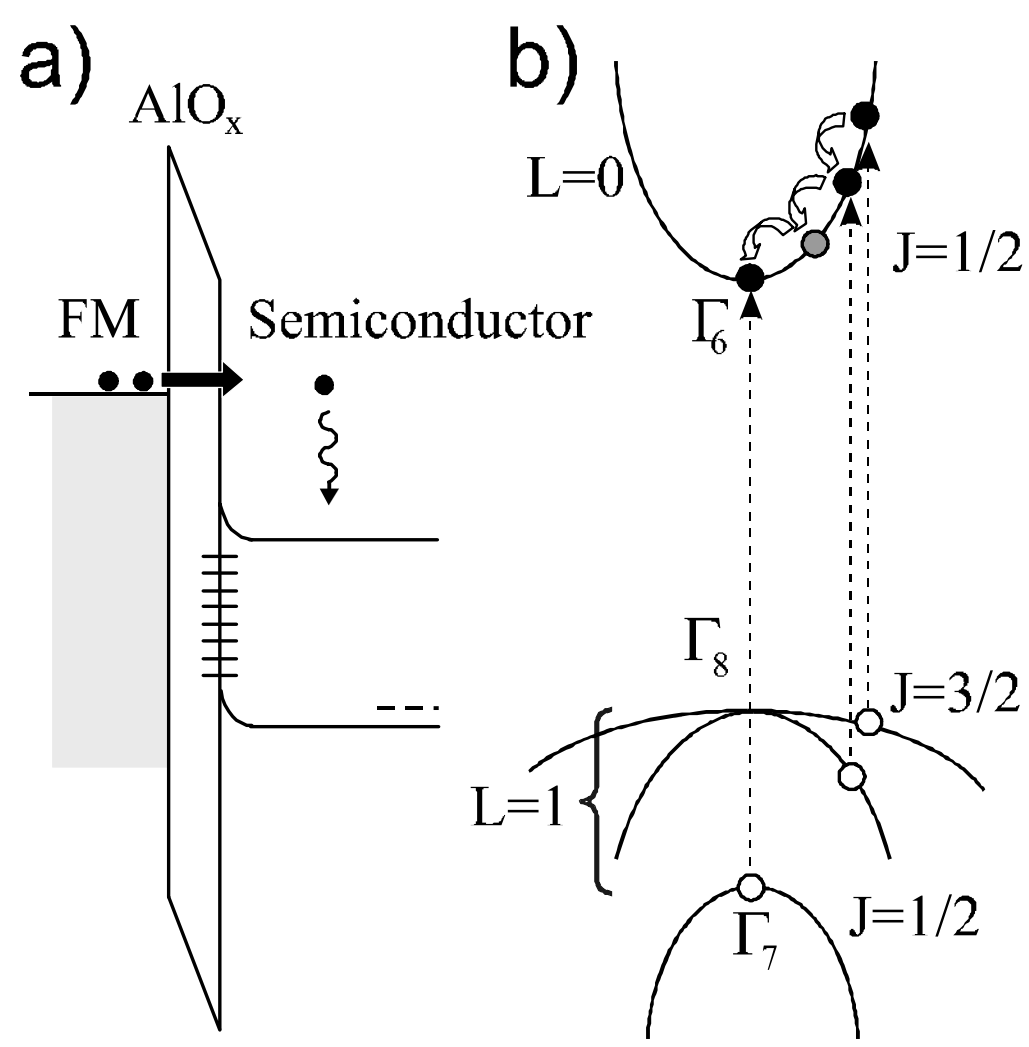

Fig.4.9. Electrical injection of the hot electrons into a semiconductor (a), and electron thermalization under optical excitation with $h\cdot\nu > E_g+\Delta$ (b).

## 4.7. Experimental Setup

During work presented in this thesis the experimental set-up, shown in Fig.4.10, was designed and installed. It allows measurements of the degree of circular polarization of the luminescence under electrical and optical spin injection in an external oblique magnetic field. An amorphous glass liquid nitrogen (home made) and liquid He / N (custom design) optical cryostats allow cooling of the sample to about 80 K and 4.3 K, respectively. The electromagnet (custom design) provides an external oblique magnetic field up to 0.6 T.

The removable mirror (Mr) and video camera (VC) allow precise positioning of the device under investigation into focal plane of the lense $L_1$. The light emitted under electrical spin injection is coupled into an optical fiber by lens $L_2$ and is detected by a photodetector ($PD_1$). In this case no spectral filters, i.e. optical monochromator, were used as EL spectra of fabricated spin-LEDs have shown GaAs interband transitions only. (See Section 5.4.1).

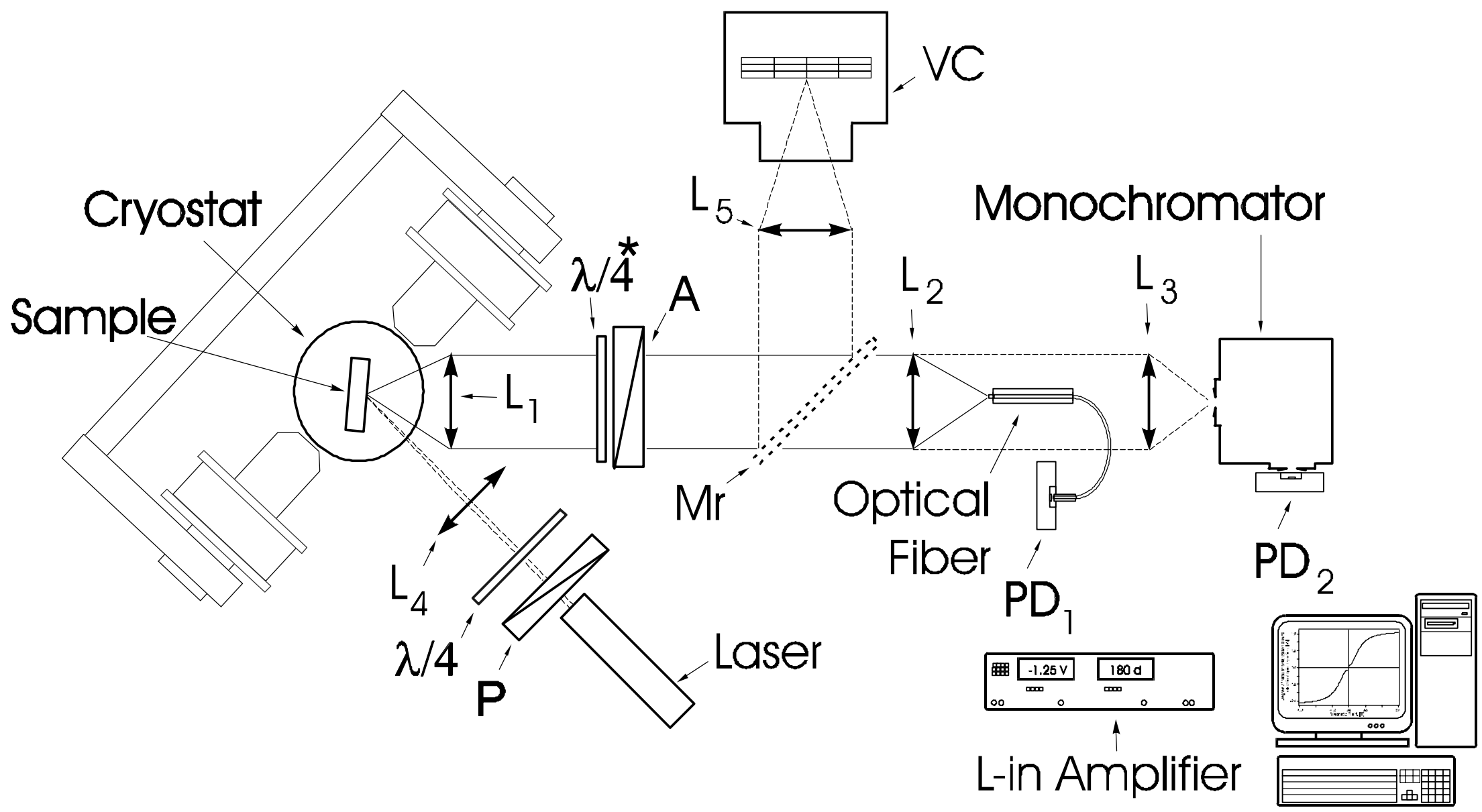

Fig.4.10. Schematic representation of the setup for optical investigation of electrical and optical spin injection into semiconductors under application of external oblique magnetic field.

For the case of optical spin injection and detection (all-optical experiment), semiconductor ($h\nu$=1.58 eV, 50mW) and He-Ne ($h\nu$=1.96 eV, 50mW) lasers (L) together with an optical monochromator and photodetector $PD_2$ are used.

A combination of a rotating quarter waveplate ($\lambda/4^*$) and linear polarizer (A), together with lock-in detection (locked to the double frequency of rotation of the quarter waveplate, as described below) allow precise measurements of the degree of circular polarization of emitted light.

As it follows from any book on optics the combination of quarter waveplate and linear polarizer allows discriminating linear, circular and unpolarized light. While basic working principles of such optical devices could be found there as well, the practical aspects of such analyses remain uncovered. Fig.4.11a shows the conversion of circular polarization into amplitude modulation as result of rotation of the quarter waveplate while higher transmission axis of linear polarizer (analyzer) remains unchanged. The frequency of such oscillations corresponds to double frequency of the quarter waveplate rotation. Fig.4.11b shows the conversion of linear polarization into amplitude modulation as result of rotation of quarter waveplate for two orientations of the linear polarizer. The frequency of such oscillations is four times larger than frequency of the quarter waveplate rotation. In the real electroluminescence or photoluminescence experiment the emission of linearly polarized light is never observed, as electronic transitions have very poor coherency, resulting in emission of unpolarized light which is not effected by quarter waveplate and is added as background DC level into the amplitude modulation observed for the circularly polarized light (Fig.4.11a solid curve).

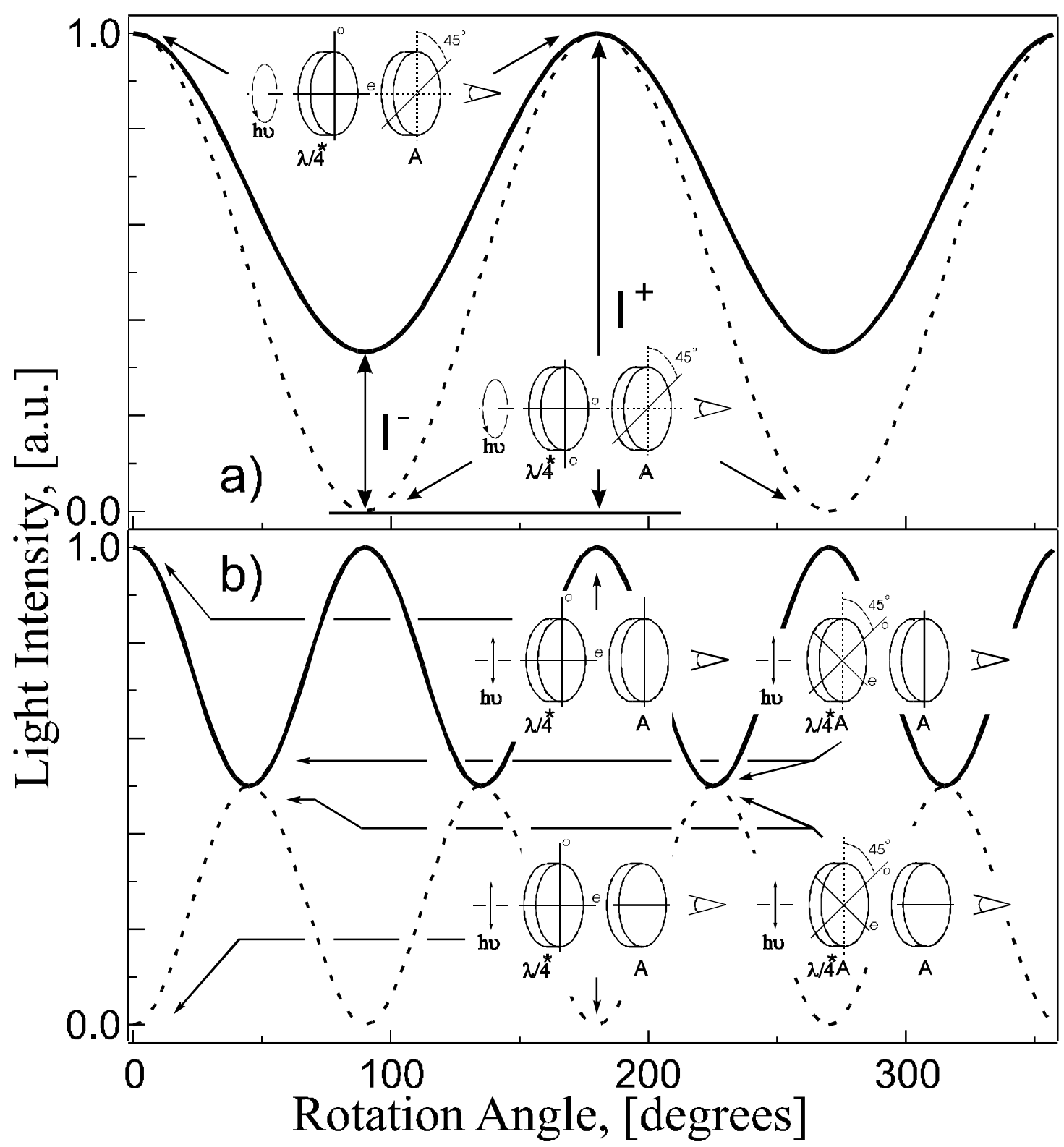

Fig.4.11. The conversion of polarization into amplitude modulation as result of light propagation through rotating quarter waveplate and fixed linear polarizer (analyzer A) as function of rotation angle in the case of a) $P = 100\%$ circularly polarized light (dashed curve) and $P = 50\%$ (solid curve); b) linearly polarized light when axis of light polarization coincides with the higher transmission axis of linear polarizer (solid curve) and in the orthogonal case (dashed curve).

# 5. Design, Fabrication and Characterization of Spin-LEDs

## 5.1. Ferromagnetic Metal/ Insulator/ Semiconductor Spin-LEDs

The studies of Tunnel Magnetoresistance (**TMR**) in metallic structures have shown that high efficiency of spin effects is generally achieved in the systems with very abrupt interfaces, as the interdiffusion on interfaces leads to the "dead layer" formation and a strong reduction of the spin-dependent effects.

In the case of GaAs and other III-V compounds the abrupt metal/semiconductor junction leads to the Schottky barrier formation. The electron injection into the semiconductor conduction band is not so evident for the case of Schottky junctions. Let's consider the example of metal / p-type semiconductor Schottky junction (Fig.5.1a). The use of p-type semiconductor is preferable for optical assessment of electrical spin injection, as in n-type material the presence of strong background of unpolarized majority electrons complicates the quantitative determination of the injected spin polarization. The negative bias applied to the metallic contact of Ferromagnetic Metal (**FM**)/ p-type semiconductor Schottky diode (forward bias) reduces the barrier on the semiconductor side of the junction and induces a strong hole current from the semiconductor into the metal (injection of electrons from the metal into the valence band). No electrons are injected into the semiconductor conduction band from the metal, since the barrier height on the metallic side of the junction is not reduced (Fig.5.1b).

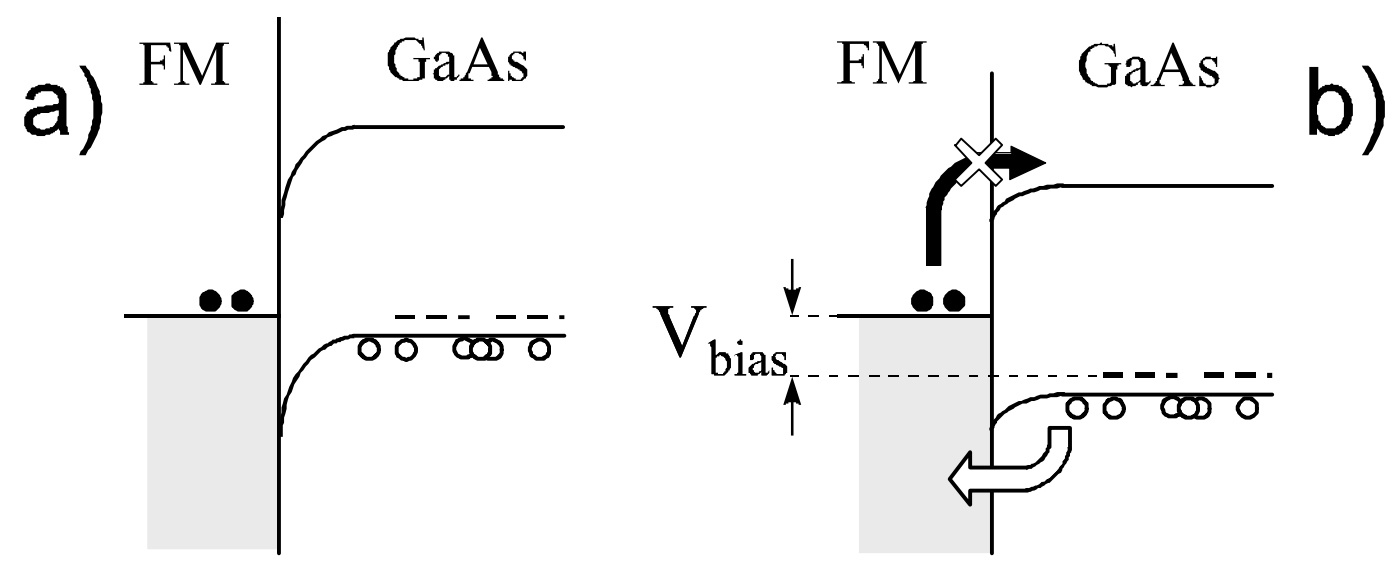

Fig.5.1. Schematic representation of the spin-LED design. a)-b) FM/p-GaAs Schottky diode without and with applied bias: efficient injection of electrons from FM into semiconductor is impossible

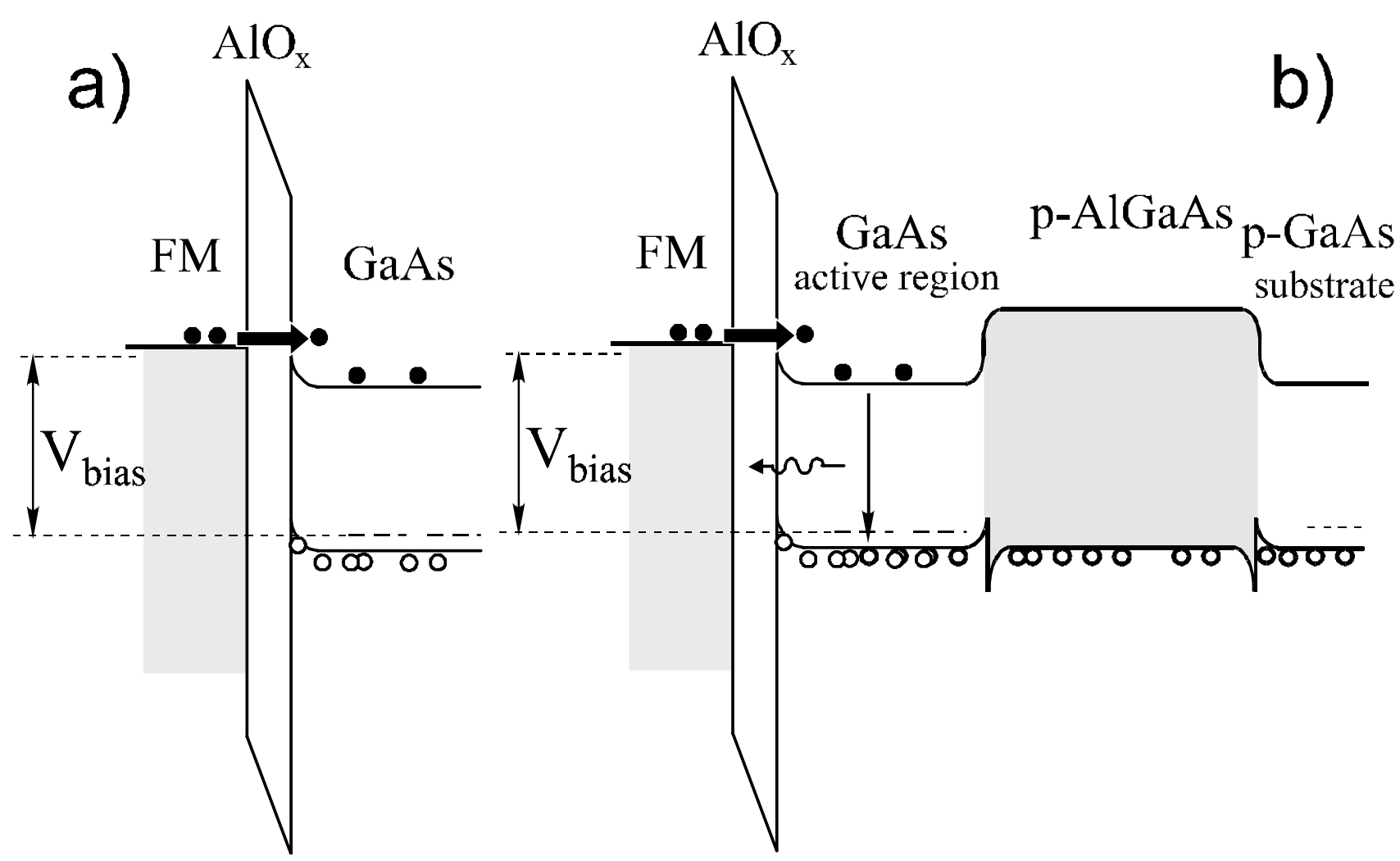

Fig.5.2. Schematic representation of the spin-LED design. a) Introduction of a thin insulator (oxide) layer between the metal and the semiconductor allows alignment of the conduction band edge against Fermi level in the ferromagnetic metal, resulting in electron tunneling directly into the conduction band of the semiconductor; b) Incorporation of the AlGaAs layer does not allow injected electrons to diffuse far from the surface, improving the light extraction efficiency.

This can be remedied by introduction of a thin insulator (oxide) layer between the metal and the semiconductor (Fig.5.2a). The drop of the potential across the tunnel junction reduces the energy separation between the Fermi level of the metal and the conduction band edge. At sufficiently high biases electrons can tunnel from the metal through the oxide layer directly into the conduction band. In addition, the negative bias applied to the metal leads to the formation of the hole accumulation layer at the semiconductor side of the junction, as a result practically all applied bias drops across the tunnel oxide layer. The introduction of the p-AlGaAs layer (Fig.5.2b), like in the conventional LED heterostructures, does not allow injected electrons to diffuse far from the surface, hence enhancing the photon extraction efficiency, as all recombination takes place close to the surface.

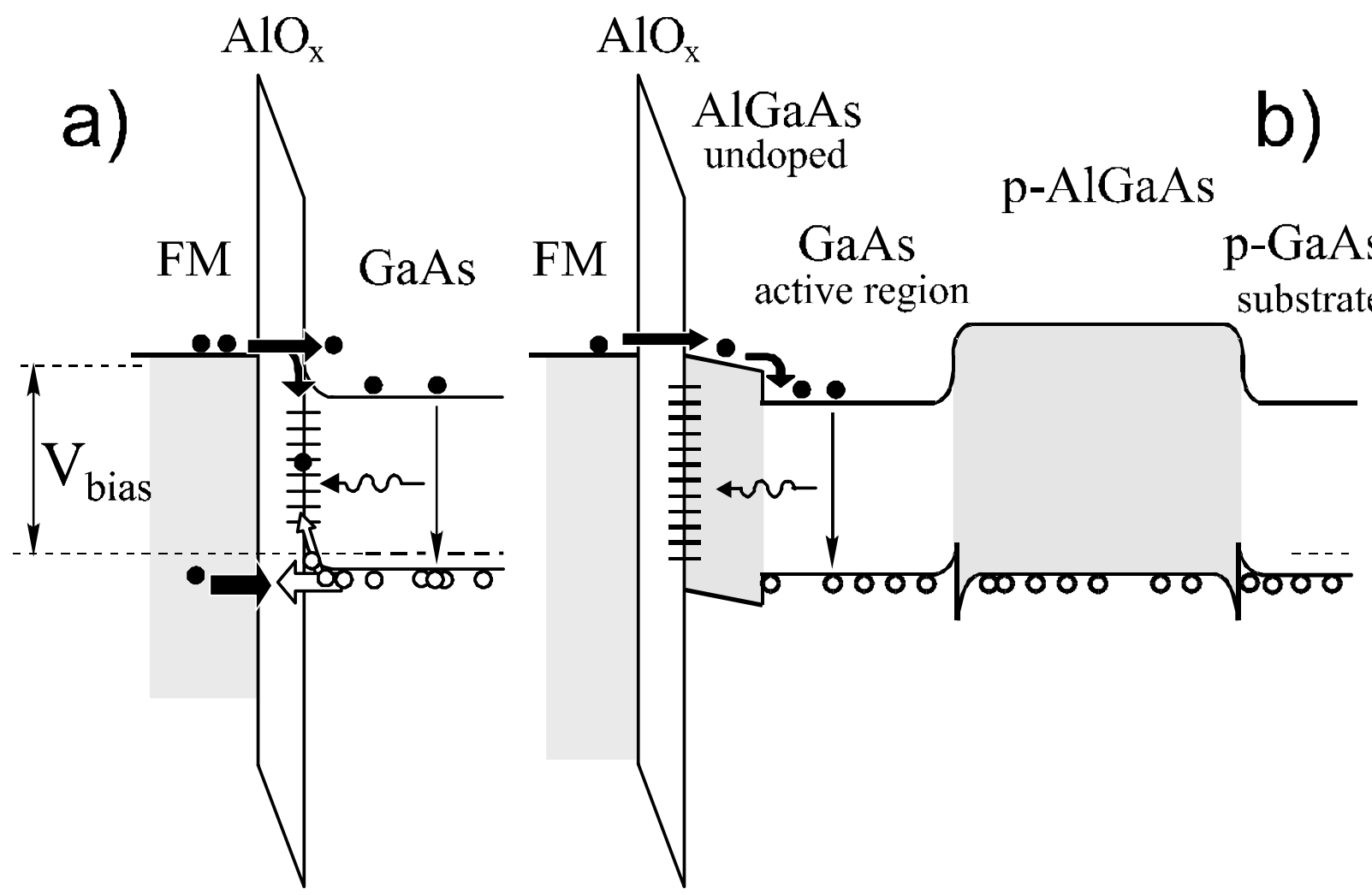

Fig.5.3. Schematic representation of the spin-LED design. a) In a real MIS tunnel junction moderate bias leads to flow of three main currents in the heterostructure: from the FM Fermi level into the conduction band of the semiconductor, from the FM into the valence band of the semiconductor (hole tunneling) and the surface recombination current; b) Incorporation of the two AlGaAs layers leads to formation of the active region preserving GaAs bulk spin detection qualities.

In a real Metal/ Insulator/ Semiconductor (**MIS**) tunnel junction, the application of the forward bias leads to the flow of three main currents (Fig.5.3a). Electrons tunneling from Fermi level of the metal into conduction band of the semiconductor, electrons tunneling from the metal into the valence band of the semiconductor (hole tunneling), and nonradiative recombination of the carriers via interface states. Only the first one results in the spin injection and can be assessed optically. The two others do not reveal themselves in the optical output and result in the local heating of the sample, higher stress (current, bias) applied to the tunnel oxide, device degradation and unreliable operation. The introduction of a thin undoped AlGaAs layer between the tunnel oxide and the GaAs allows substantial reduction of their contributions (Fig.5.3b). It keeps the holes away from the oxide-semiconductor interface and thus reduces the current through the interface states and the hole tunneling current to the metal. Two AlGaAs layers thus create an advanced GaAs active region, where the injected electrons recombine with holes. In the fabricated devices the active region is chosen to be wide enough so that no quantization of electron or hole levels takes place. The quantum well- type active region is completely rejected initially, in order to avoid complications related to the splitting of the valence band and to the partial loss of the spin polarization during the electron trapping into the well. Thus the proposed FM/ Insulator / Semiconductor spin-LEDs consist of two parts: the FM/$AlO_X$ Tunnel Barrier (**TB**) spin injector and III-V heterostructure spin detector.

## 5.2. GaAs and Problem of Conductivity Mismatch

The problem of conductivity mismatch arises in the direct electrical contact of a ferromagnetic metal and a semiconductor having no or almost no interface resistance, diffusive ohmic contact, like ferromagnetic Fe contact to InAs 2DEG for example. In the case of GaAs the abrupt metal/semiconductor junction leads to the Schottky barrier formation, which is very different from the case known as 'basic obstacle for electrical spin injection' or conductivity mismatch (see Section 3.2). Let's briefly reconsider the main phenomena arising on such interface.

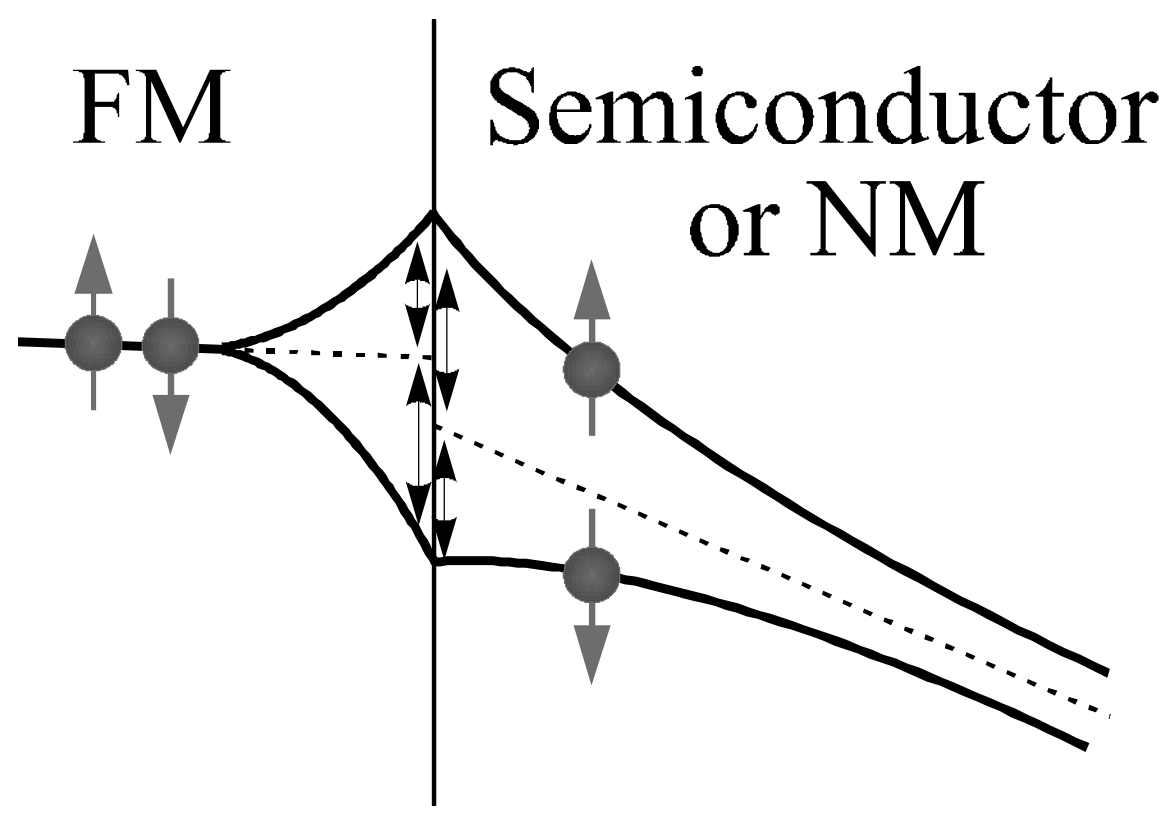

Fig.5.4. Splitting of the electrochemical potentials for spin-up and spin-down electrons under electrical spin injection from ferromagnetic metal into a semiconductor or into a nonmagnetic metal in the diffusive ohmic contact.

In the ferromagnetic metal the current is carried mainly by one of spin channels (spin-up or spin-down, see Sections 2.1, 2.4.2, 2.6 ). On contrary, in the normal metal or semiconductor, the current is carried equally by both spin channels. It is clear that on the interface of ferromagnetic metal and normal metal or semiconductor the conversion of current must occur. It follows that spin polarization of charge carriers in the nonmagnetic part of such junction simply implies the splitting of the quasi Fermi levels for spin-up and spin-down electrons or electrochemical potentials (By definition the electrochemical potential is the energy needed to add a particle to the system $\mu = dE/dn$). Electrons in the different spin subbands in the nonmagnetic part of the junction have different energy (Fig.5.4), starting just from the interface with the ferromagnetic metal ! As it was mentioned above the contact between magnetic/nonmagnetic solids is considered to have no interface resistance, hence spin-up (spin-down) electrons (Fig.5.4) in the ferromagnetic metal must have higher (lower) energy in order to penetrate into nonmagnetic part of the junction. This leads to reduction of spin polarization of electrons injected from the ferromagnetic metal. It follows that magnitude and exponential decay of this splitting is proportional to the electron spin-flip length $\lambda_{sf} = \sqrt{D \cdot \tau_{sf}}$ and conductivities $\sigma^{\uparrow}, \sigma^{\downarrow}$ (hence, density of states) in these materials. Here $D$ is diffusion constant and $\tau_{sf}$ is spin-flip time. In the nonmagnetic solid the conductivity for both spin channels is equal and the exponential decay of the splitting for spin-up and spin-down electrons is also equal. In the ferromagnetic metal the conductivities for spin-up and spin-down channels are different, as result the exponential decay of the splitting is different. This difference gives rise to the interface voltage drop (shown as splitting of total electrochemical potentials on the interface, see Fig.5.4), which is a basis for the non-local geometry measurements mentioned in Section 3.2.

It further follows that in the nonmagnetic metal the spin-flip length and conductivity are comparable to the ones in the ferromagnetic metal. But in a semiconductor the conductivity is much lower. Moreover, the spin-flip length can exceed couple of order of magnitude the one in ferromagnetic metals. As result, the calculations show that spin polarization of electrons injected into a semiconductor rapidly decays and approach zero once spin polarization of electrons in the ferromagnetic metal differ just a little bit from ideal $\Pi = 100\%$ .

However, as was mentioned above, such contact into GaAs samples does not exist, the abrupt metal/semiconductor junction leads to the Schottky barrier formation and surface pinning of the conduction and valance band edges. This fact, as it was shown in the previous Section 5.1, even does not allow electrical injection of electrons in the semiconductor for p-type GaAs samples. The incorporation of a thin tunnel barrier allows bringing the Fermi level in the metal and conduction band edge in the semiconductor into one level, just due to the fact that there is a 'huge' voltage drop across this tiny region. This interface resistance separates the splitting of electrochemical potentials in the semiconductor and ferromagnetic metal and matches the large difference in the density of sates of these materials.

Moreover, the electrical injection of electrons into GaAs from the ferromagnetic metal in the direct electrical contact is possible only in the case of strong n-doping of the semiconductor (Fig. 5.6). Such junction operates in the reverse bias, thus there is a large voltage drop over Schottky depletion region that again matches the splitting of electrochemical potentials in the semiconductor and ferromagnetic metal. The same reality is trough for other semiconductors like Si, for example. Hence, no problem of conductivity mismatch exists for the FM/GaAs, Fe/Si heterojunctions. There is a problem of electrical injection of electrons into conduction band of the semiconductor !

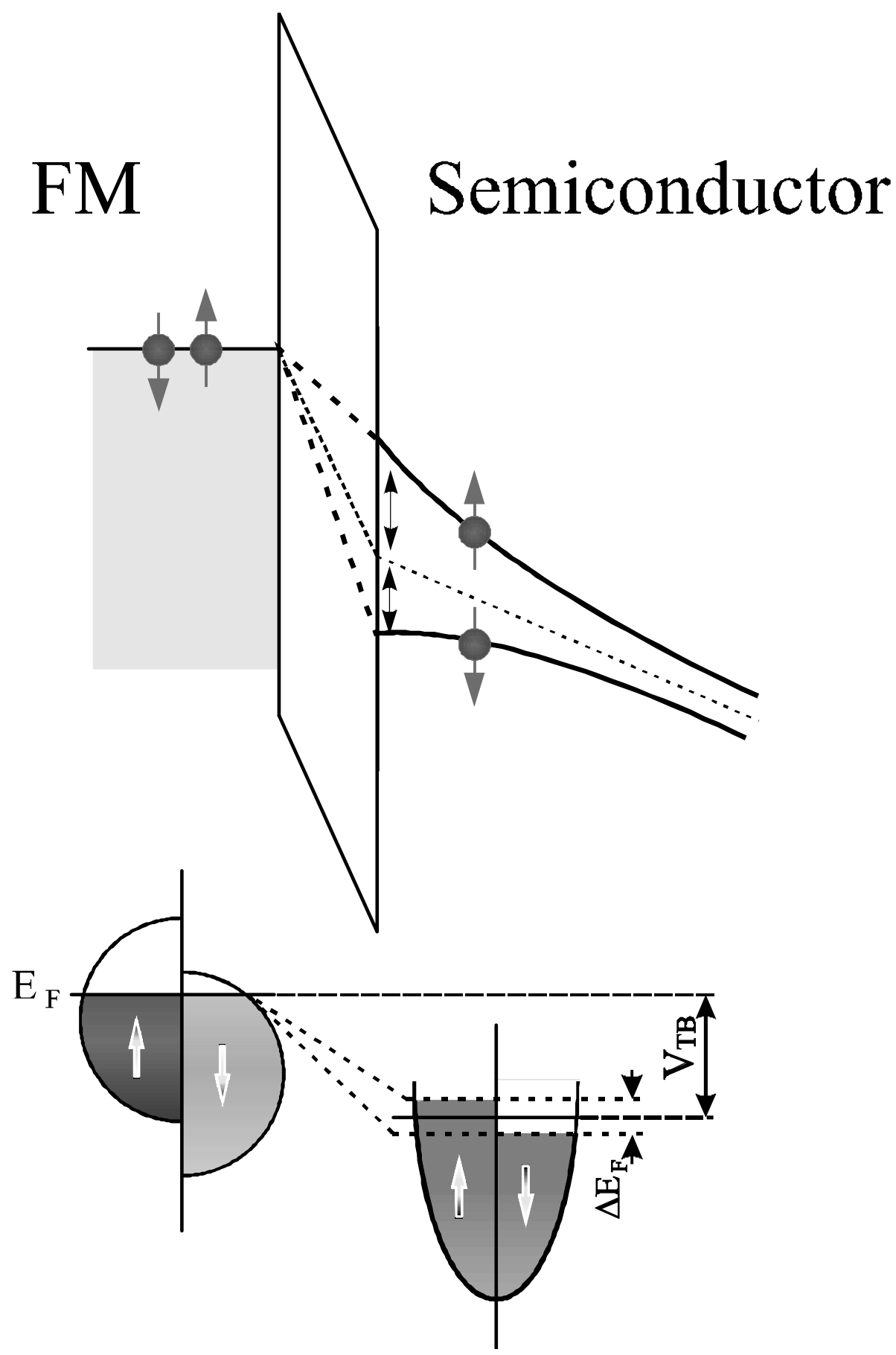

Fig.5.5. The introduction of a large interface resistance allows to overcome the interface splitting of electrochemical potentials.

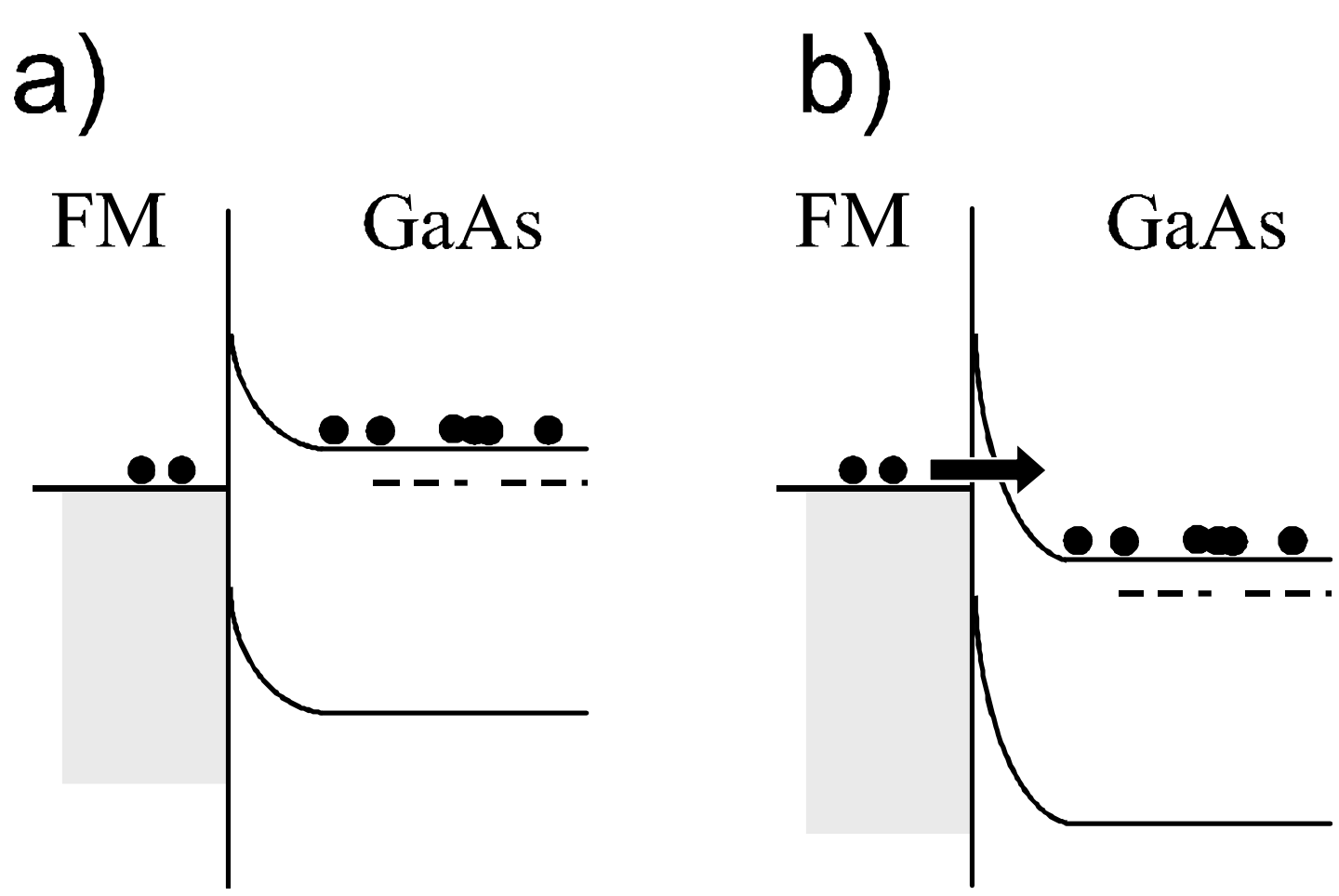

Fig. 5.6. Electrical injection of electrons into GaAs in the ferromagnetic metal/n-GaAs Schottky junction, the only possible case for the direct M/ Semiconductor electrical contact.

The same problem as semiconductor industry challenging during entire era of its existence ! The results presented in this thesis give a strong experimental support to this statement.

Thus, the only question remained untouched, is the question of how large is the splitting of electron levels in the semiconductor for spin-up and spin-down electrons. For this purpose one can open Ref.[177] on the page number 17. In the case of non-degenerate semiconductor $E_F = E_C + k \cdot T \cdot \ln \frac{n}{N_C}$, where $n$ is number of electrons and $N_C$ is effective density of states in the conduction band. It follows that $\Delta E_F = k \cdot T \cdot \ln \frac{n_\uparrow}{n_\downarrow}$. Hence, if the spin polarization of electrons in the conduction band of the semiconductor $\Pi = 0.2$ ($\frac{n_\uparrow}{n_\downarrow} = 1.5$, $\ln \frac{n_\uparrow}{n_\downarrow} = 0.4$), then the splitting of quasi Fermi levels for spin-up and spin-down electrons are $\Delta E_F(80K) = 3{,}2$ meV and $\Delta E_F(300K) = 10$ meV at low and room temperatures, which is negligible comparing with the voltage drop across tunnel barrier or reversed bias Schottky depletion region.

## 5.3. Fabrication of Spin-LEDs

As it was mentioned in Section 5.1 the fabricated MIS spin-LEDs consist of two parts: the FM/$AlO_X$ Tunnel Barrier (**TB**) spin injector and III-V heterostructure spin detector. During work presented in this thesis different III-V semiconductor as well as spin injector heterostructures were fabricated. The overview of fabricated spin-LEDs is given in Section 5.3.3.

All semiconductor heterostructures were grown by Molecular Beam Epitaxy on a (001) $p^+$-GaAs substrate. For fabrication of FM/TB spin injectors, generally immediately after growth the semiconductor heterostructures were transferred in air into sputtering chamber. Where process adopted from TMR junction fabrication technology [176] was implemented.

The thin $AlO_X$ tunnel barrier was fabricated by Al sputtering and subsequent natural oxidation in a controlled oxygen atmosphere of 140 Torr. After fabrication of the $AlO_X$ tunnel barrier, generally the 2 nm $Co_{90}Fe_{10}$ / 8 nm $Ni_{80}Fe_{20}$ / 5 nm Cu ferromagnetic stack was sputtered in the same vacuum chamber. All metals are dc-magnetron sputtered. Magnetic anisotropy was obtained by application of a small in-plane magnetic field of 4 mT. Although the use of such magnetic stack is not critical for electrical spin injection, it allows better control over fabrication process of spin-LEDs, as tunnel magnetic junctions containing exactly the same stack were repeatedly fabricated in the same chamber for other purposes.

The next Section 5.3.1 gives detailed description of the methods used for fabrication and characterization of thin $AlO_x$ tunnel barriers. The processing of the devices is described in the Section 5.3.2.

In general, the process of fabrication of the spin-LEDs is quite reliable and gives very reproducible results. Until now more then ten fabrications starting from 2” GaAs wafer were performed. Each fabrication on the single wafer results in more than 100 spin-LEDs. All of the measured devices on the same wafer have shown similar characteristics.

### 5.3.1. Fabrication of $AlO_X$ Tunnel Oxide

As was mentioned above the spin-LED fabrication process involves transfer of semiconductor heterostructures in air into sputtering chamber for fabrication of spin injectors. Although the ‘time in air’ generally was kept below 10 min, during such transfer some surface contamination by oxygen, carbon and water, for example, may occur. In order to test the influence of this step on the quality of fabricated thin $AlO_X$ TB layers a set of three test samples was fabricated. All samples have identical FM/ TB/ Semiconductor heterostructure. The semiconductor part of the junction contains only GaAs buffer layer grown by MBE. For fabrication of Sample N 1 a $1.4nm$ Al layer was grown further in the same MBE chamber. Immediately after growth, sample was transferred into sputtering chamber for Al layer oxidation. For fabrication of Sample N 2 a $1.4nm$ Al layer also was grown in the same MBE chamber, but before transfer into sputtering chamber for final Al oxidation, it was kept in air for 24 hours. Finally, Sample N 3 contained only GaAs buffer layer grown by MBE. Immediately after growth, sample was transferred into sputtering chamber for fabrication of $AlO_X$ tunnel barrier and ferromagnetic metal deposition. No special cleaning procedures like temperature treatment were used further. For device processing the similar process and mask layout as described in Section 5.3.2 was implemented.

After processing, the fabricated MIS heterojunctions were characterized by electrical means. Fig.5.7 shows the result of I-V measurements only for Sample N 3 (bold curve), where only semiconductor part of the junction was grown by MBE, as only this sample has shown I-V characteristics typical for MIS heterostructures (see inset on the right) [177]. This curve shows strong nonlinearity connected with participation of surface states in the total charge transfer. The saturation part of the curve corresponds to electron tunneling directly into the conduction band of the semiconductor. The dashed curve shows result of measurements of the Schottky diode formed on the GaAs/Au interface on the same wafer, for comparison.

The fact that I-V characteristics for other two samples have shown result similar to conventional Schottky diode may find a simple explanation in terms of chemistry for Al oxidation process. It is known that oxidation of Al layer goes along grain boundaries. In the Al layer grown by MBE the grain size is much larger than in the one in the layers prepared by sputtering. It follows, that complete oxidation of larger Al grains requires more time leading to not complete oxidation and metallic conduction.

As result of this test, the process where Al layer is sputtered and oxidized in the sputtering chamber was chosen for fabrication of all spin-LEDs presented in this thesis, although the samples having more complicated semiconductor heterostructures have never shown the I-Vs characteristics typical for MIS heterojunctions (Fig.5.7, thin solid curve).

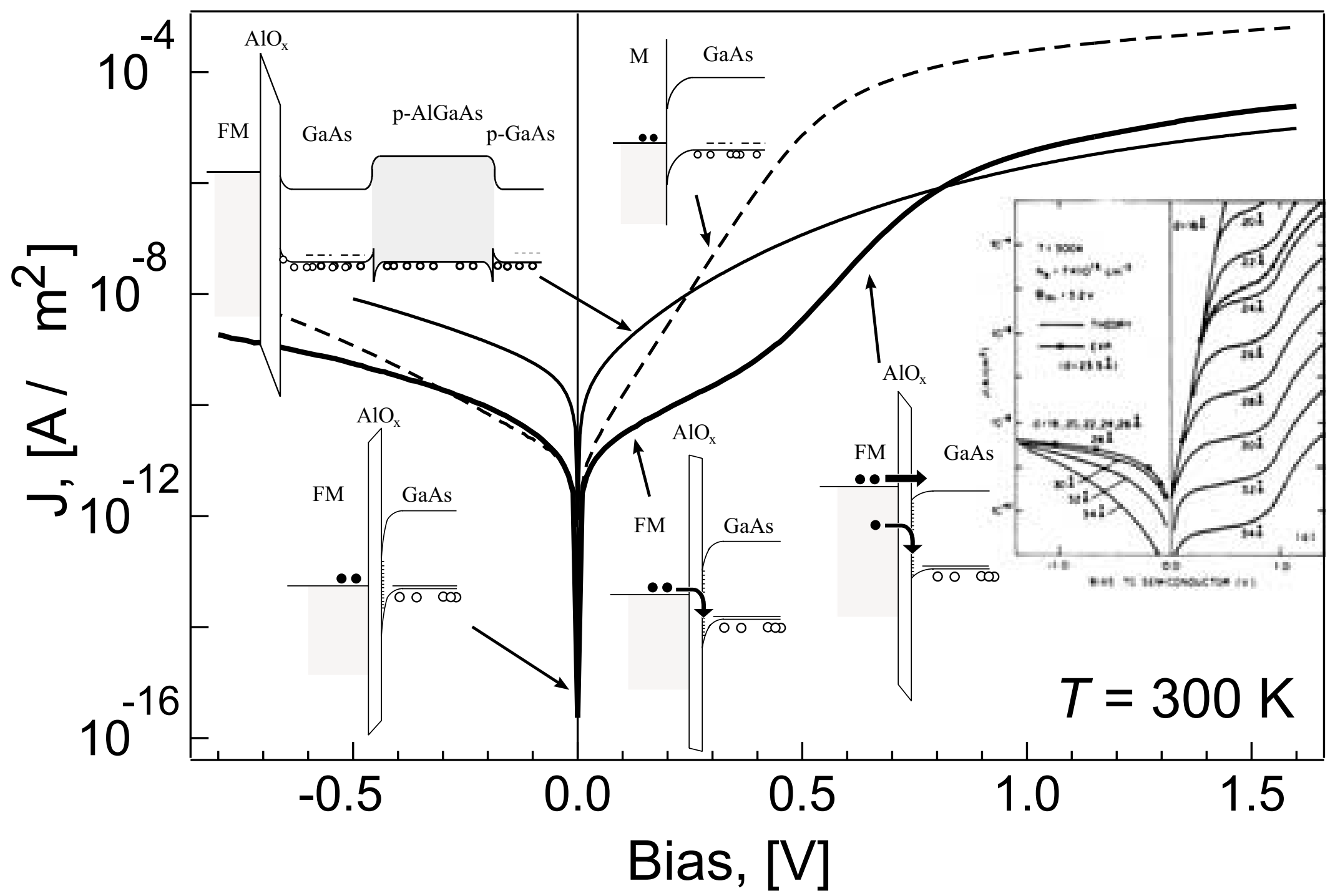

Fig.5.7. Typical result of I-V measurements of: FM/$AlO_X$/GaAs MIS heterostructure (bold), FM/GaAs Schottky diode (dashed) and FM/$AlO_X$/GaAs/AlGaAs MIS heterostructure (thin solid line). The I-V curves for Au/$SiO_2$/Si MIS diodes [177] (Inset on the right).

Recently, thanks to the participation in the research of graduate and postgraduate students (Mayke Nijboer and Pol Van Dorpe) a new process, comprising fabrication of $AlO_X$ tunnel barrier in the double-step process was developed. Here the thinner Al layer sputtered in the first step allows faster oxidation speed. The fabrication of thicker tunnel barriers is achieved by repetition of the first step. Typically one needs at least 7-14 hours for fabrication of $1.4-3nm$ $AlO_X$ tunnel barrier by natural oxidation.

It appears that use of a two-step oxidation process facilitates a full oxidation of the Al, reduces the chance on pinholes [178] and enables the fabrication of thicker barriers. As result, this fabrication process produces an atomically flat, densely packed and pinhole free tunnel barriers (Fig.5.8).

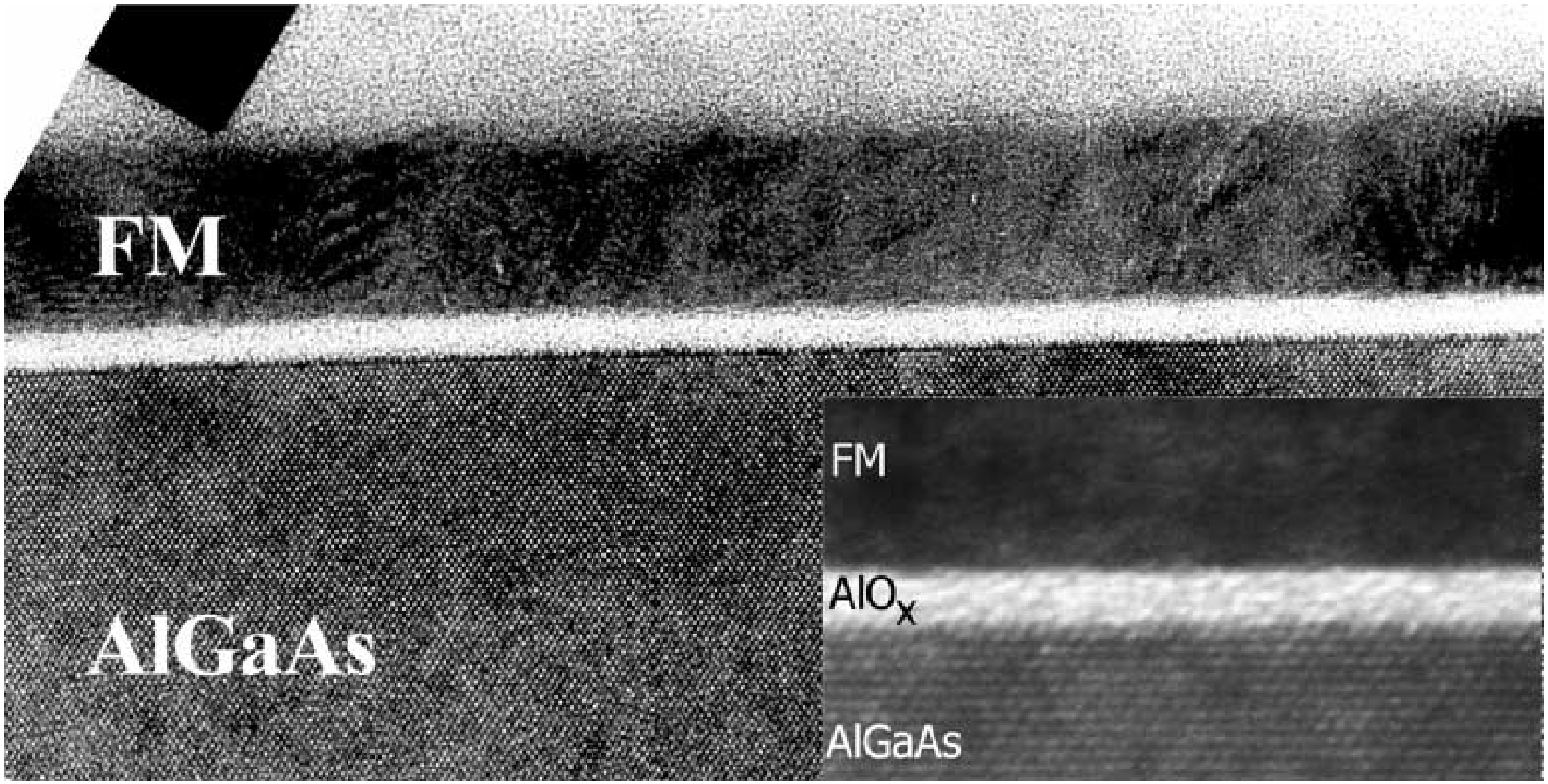

Fig.5.8. TEM characterization of the $AlO_X$ tunnel barrier oxide prepared in the double-step Al natural oxidation process. (to be compared with single step Al oxidation process used for fabrication of magnetic tunnel junctions Fig.5.9). See Section 5.3.1 for details.

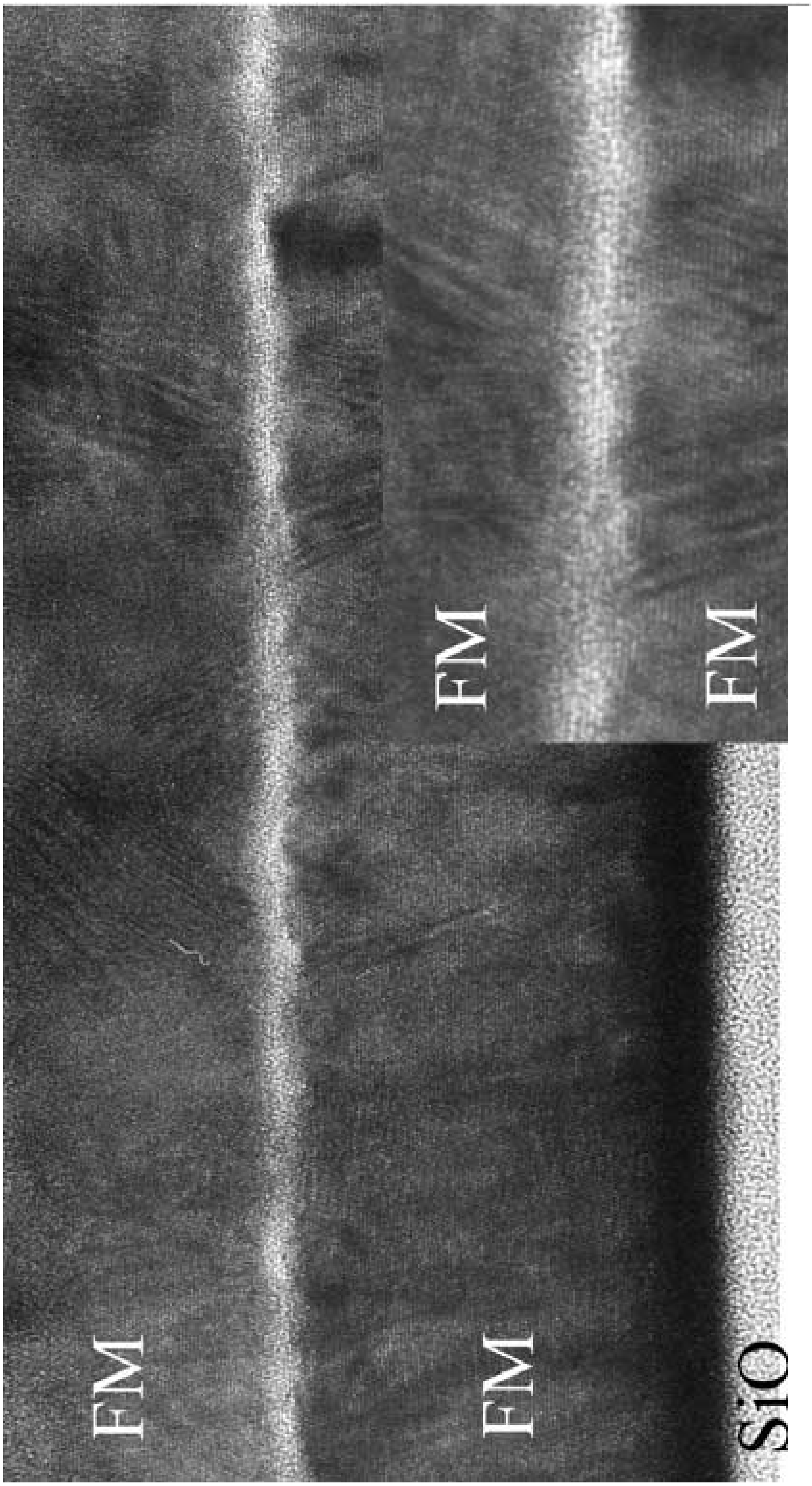

Fig.5.9. TEM characterization of the magnetic tunnel junction with $AlO_X$ tunnel barrier oxide prepared by single step natural oxidation of a thin Al layer. See Section 5.3.1 for details.

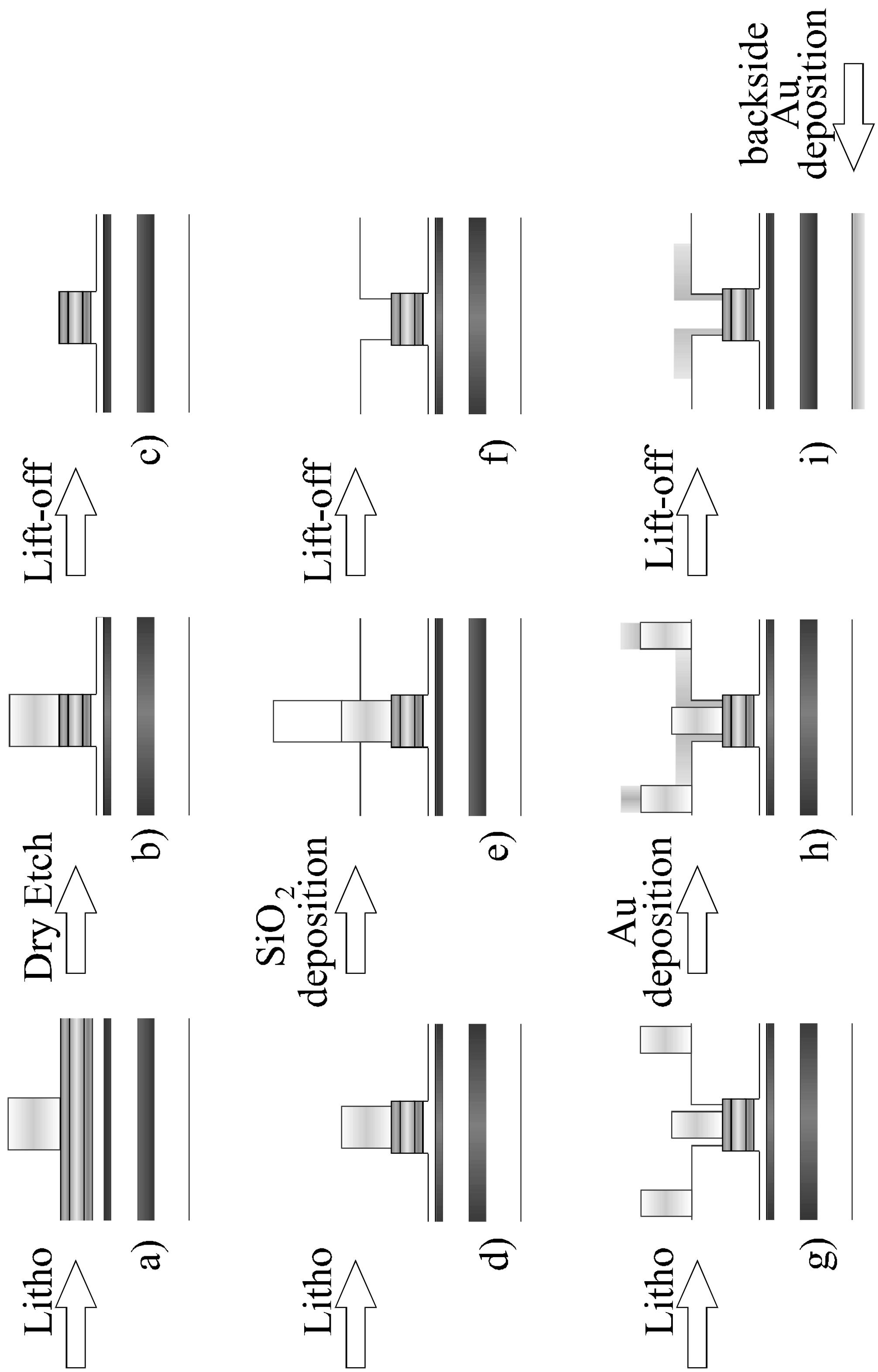

Fig.5.10. Fabrication of the surface emitting spin-LEDs. See Section 5.3.2 for detailes.

### 5.3.2. Device Processing

After fabrication of the detector and injector parts of the devices the surface emitting LEDs were processed using conventional optical lithography, dry and wet processing steps. For this purpose an optical lithography mask and a process sequence itself were developed.

In the first processing step (Fig.5.10a-c) the ferromagnetic metal/tunnel barrier contacts were defined by dry etching of the unprotected by photoresist area of the sample surface, resulting in the magnetic contacts ranging from $20\times20\,\mu m^2$ to $80\times240\,\mu m^2$, with long side of rectangles oriented along the easy axis of the ferromagnetic metal magnetization.

In the second processing step (Fig.5.10d-f) the passivation of the sample surface by sputtering of $SiO_2$ layer and lift-off allows performing the third step (Fig.5.10g-i), where devices were contacted using Au contacts to the backside of the substrate and to the ferromagnetic metal, leaving an optical window (Fig.5.11). For defining the top Au contact and optical window the remaining Au was removed in the lift-off process.

After processing devices were packaged in the 14 pin DIM packages containing no magnetic impurities.

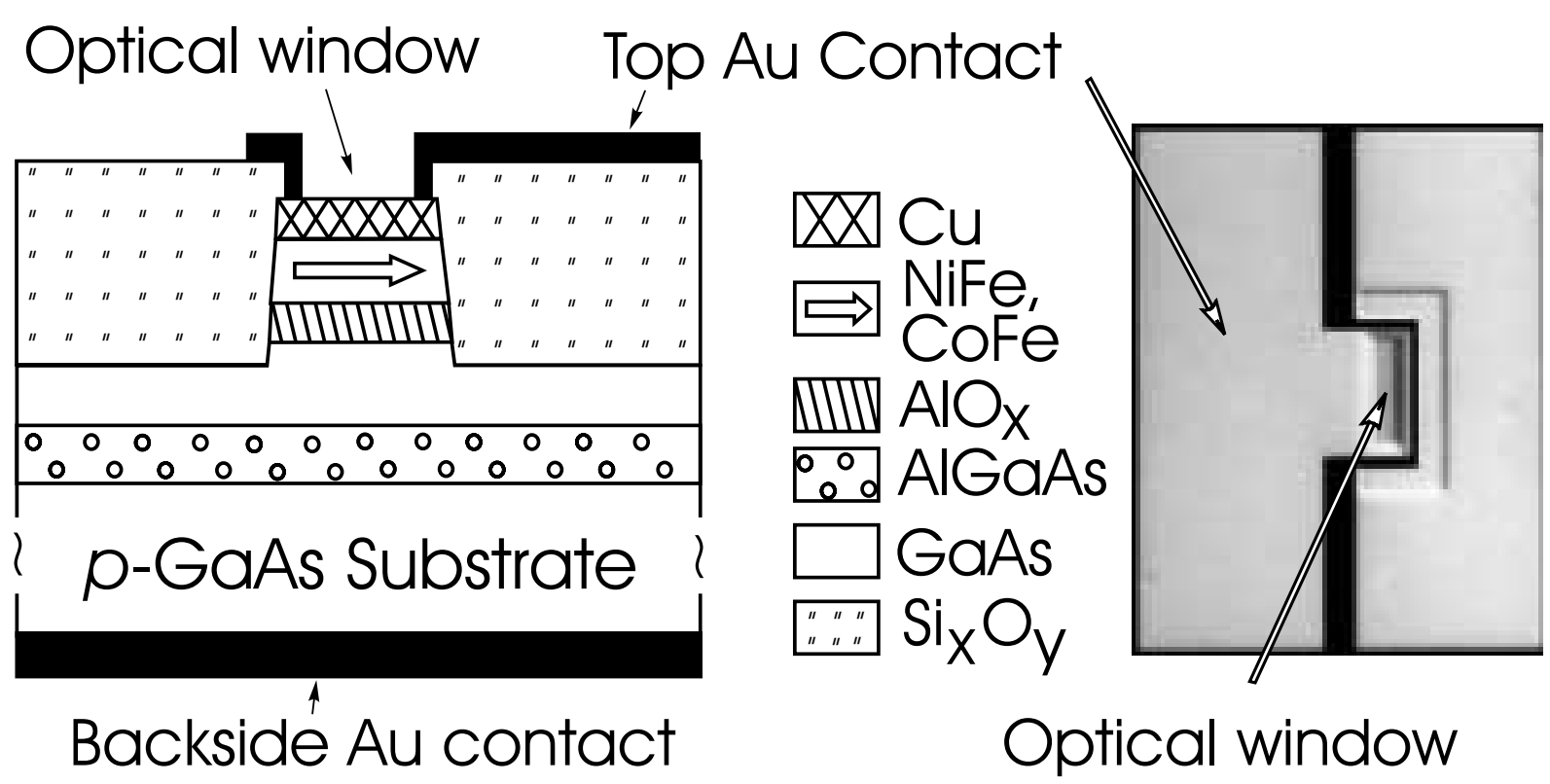

Fig.5.11. The schematic representation of the fabricated MIS spin-LEDs and top view on the processed device, showing top Au contact with optical window.

### 5.3.3. Fabricated spin-LED Heterostructures

Table 5.1 shows an overview of the spin-LED types fabricated during work presented in this thesis. Measurements have shown that the undoped GaAs samples grown by MBE have background doping by carbon $p=10^{14}\ldots10^{15}\ cm^{-3}$. For the different doping the impurities of Si and Be were additionally incorporated into lattice. Optical investigation of electrical spin injection in these samples has led to development of the experimental approach presented in Chapter 4 and Chapter 5.

The samples *Type A* and *Type B* have allowed an experimental confirmation of ideas presented in Chapter 4 and Sections 5.1-5.2. Unfortunately these samples have relatively low optical efficiency due to strong hole current. Even sample *Type B* with surface AlGaAs layer does not allow reliable measurements at room temperature, for example. This is because AlGaAs is not an ideal choice for the 'hole stop' layer, the conduction and valence band offsets with GaAs suppose that it acts better as barrier for electrons [179]. The $Ga_{0.51}In_{0.49}P$ would be a better choice for this purpose, as valence band offset in this compound is much larger than the one for the conduction band.

In addition, the samples with thicker $AlO_X$ layer generally have shown better optical efficiency. The increased tunnel barrier thickness in these samples changes the relative electron and hole tunneling probabilities and, hence, allows correction for the strong hole current.

## Table 5.1 Fabricated spin-LEDS.

| *Sample Type* | *MIS Spin-LED heterostructure* |
|---|---|
| *Type A* | 2 nm $Co_{90}Fe_{10}$ / 8 nm $Ni_{80}Fe_{20}$<br>1.8nm $AlO_X$ (1.4nm Al, single step)<br>100nm GaAs $p=2\cdot10^{18}cm^{-3}$<br>200nm $Al_{0.3}Ga_{0.7}As$ $p=2\cdot10^{18}cm^{-3}$<br>2μm GaAs $p=2\cdot10^{18}cm^{-3}$<br>2μm GaAs buffer $p=2\cdot10^{18}cm^{-3}$<br>GaAs substrate $p=2\cdot10^{18}cm^{-3}$ |
| *Type B* | 2 nm $Co_{90}Fe_{10}$ / 8 nm $Ni_{80}Fe_{20}$<br>1.8nm $AlO_X$ (1.4nm Al, single step)<br>15 nm undoped $Al_{0.2}Ga_{0.8}As$<br>100nm undoped GaAs<br>200nm $Al_{0.3}Ga_{0.7}As$ $p=2\cdot10^{18}cm^{-3}$<br>2μm GaAs buffer $p=2\cdot10^{18}cm^{-3}$<br>GaAs substrate $p=2\cdot10^{18}cm^{-3}$ |
| *Type C* | 2 nm $Co_{90}Fe_{10}$ / 8 nm $Ni_{80}Fe_{20}$<br>2.6nm $AlO_X$ (2x1nm Al, double-step)<br>15 nm undoped $Al_{0.2}Ga_{0.8}As$<br>100nm GaAs $p=2\cdot10^{18}cm^{-3}$<br>200nm $Al_{0.3}Ga_{0.7}As$ $p=2\cdot10^{18}cm^{-3}$<br>2μm GaAs buffer $p=2\cdot10^{18}cm^{-3}$<br>p-GaAs substrate $p=2\cdot10^{18}cm^{-3}$ |

| *Sample Type* | *MIS Spin-LED heterostructure* |
|---|---|
| *Type D* | 2 nm $Co_{90}Fe_{10}$ / 8 nm $Ni_{80}Fe_{20}$<br>2.6nm $AlO_X$ (2x1nm Al, double-step)<br>15 nm undoped $Al_{0.2}Ga_{0.8}As$<br>100nm undoped GaAs<br>200nm $Al_{0.3}Ga_{0.7}As$ $p=2\cdot10^{18}cm^{-3}$<br>2µm GaAs buffer $p=2\cdot10^{18}cm^{-3}$<br>GaAs substrate $p=2\cdot10^{18}cm^{-3}$ |
| *Type E* | 10 nm $Co_{90}Fe_{10}$<br>2.6nm $AlO_X$ (2x1nm Al double-step)<br>15 nm undoped $Al_{0.2}Ga_{0.8}As$<br>100nm GaAs $N_A=4\cdot10^{17}cm^{-3}$, $N_A/N_D=2$<br>200nm $Al_{0.3}Ga_{0.7}As$ $p=2\cdot10^{18}cm^{-3}$<br>2µm GaAs buffer $p=2\cdot10^{18}cm^{-3}$<br>p-GaAs substrate $p=2\cdot10^{18}cm^{-3}$ |
| *Type F* | 7 nm $Co_{90}Fe_{10}$<br>3nm Cu<br>1.8nm $AlO_X$ (1.4nm Al, single step)<br>15 nm undoped $Al_{0.2}Ga_{0.8}As$<br>100nm GaAs undoped<br>200nm $Al_{0.3}Ga_{0.7}As$ $p=2\cdot10^{18}cm^{-3}$<br>2µm GaAs buffer $p=2\cdot10^{18}cm^{-3}$<br>p-GaAs substrate $p=2\cdot10^{18}cm^{-3}$ |

## 5.4. Spin-LED Characterization

### 5.4.1. LED Characterization

Under application of the forward bias, the LEDs emit light corresponding to the GaAs band gap transitions only (Fig.5.12, Fig.5.13). This simple fact has allowed us to simplify the experimental setup to the one presented in Section 4.7.

Generally, as it was mentioned in the previous section the devices on the bottom of the Table 5.1 have higher optical efficiency, nevertheless the typical optical output is in the nanowatt range. The light emission threshold for all fabricated devices is $\sim 1.5\ldots1.7$ V. However due to signal to noise reasoning the measurements were performed at higher biases. The more detailed information concerning applied electrical bias during the measurements can be found in the experimental Chapter 6. From the experince of observation of the spin-dependent effects in magnetic tunnel junctions such electrical biasing is expected to reduce injected electron spin polarization, due to higher stress (current/voltage) applied to the tunnel oxide, for example. Going a little bit further, in the spin-polarized FETs one does not need such high current levels, moreover, the semiconductor can have n-type doping leading to longer spin coherency and better overall performance.

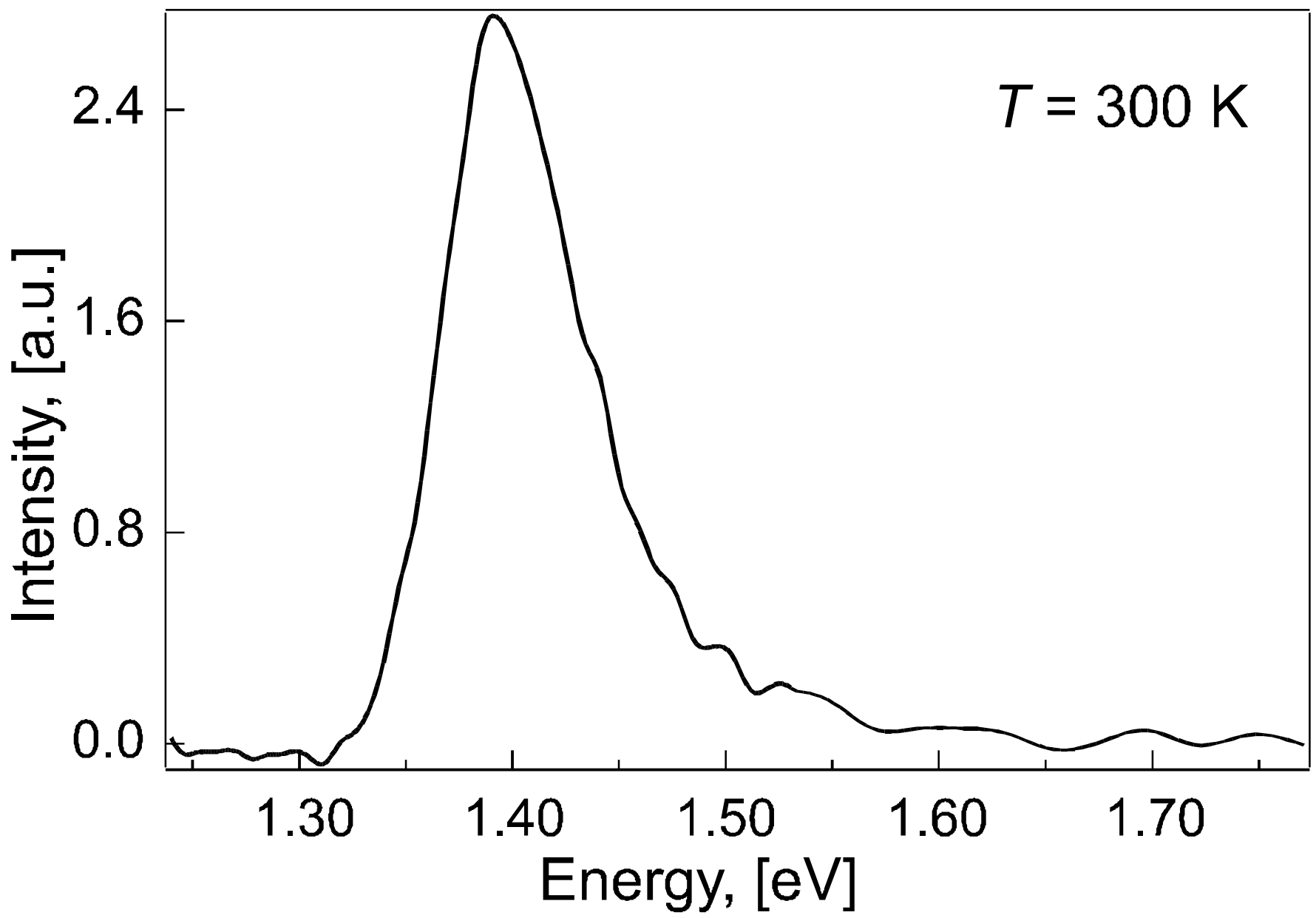

Fig.5.12. Typical spectrum of electroluminescence for sample *Type A* (see Section 5.3.3)

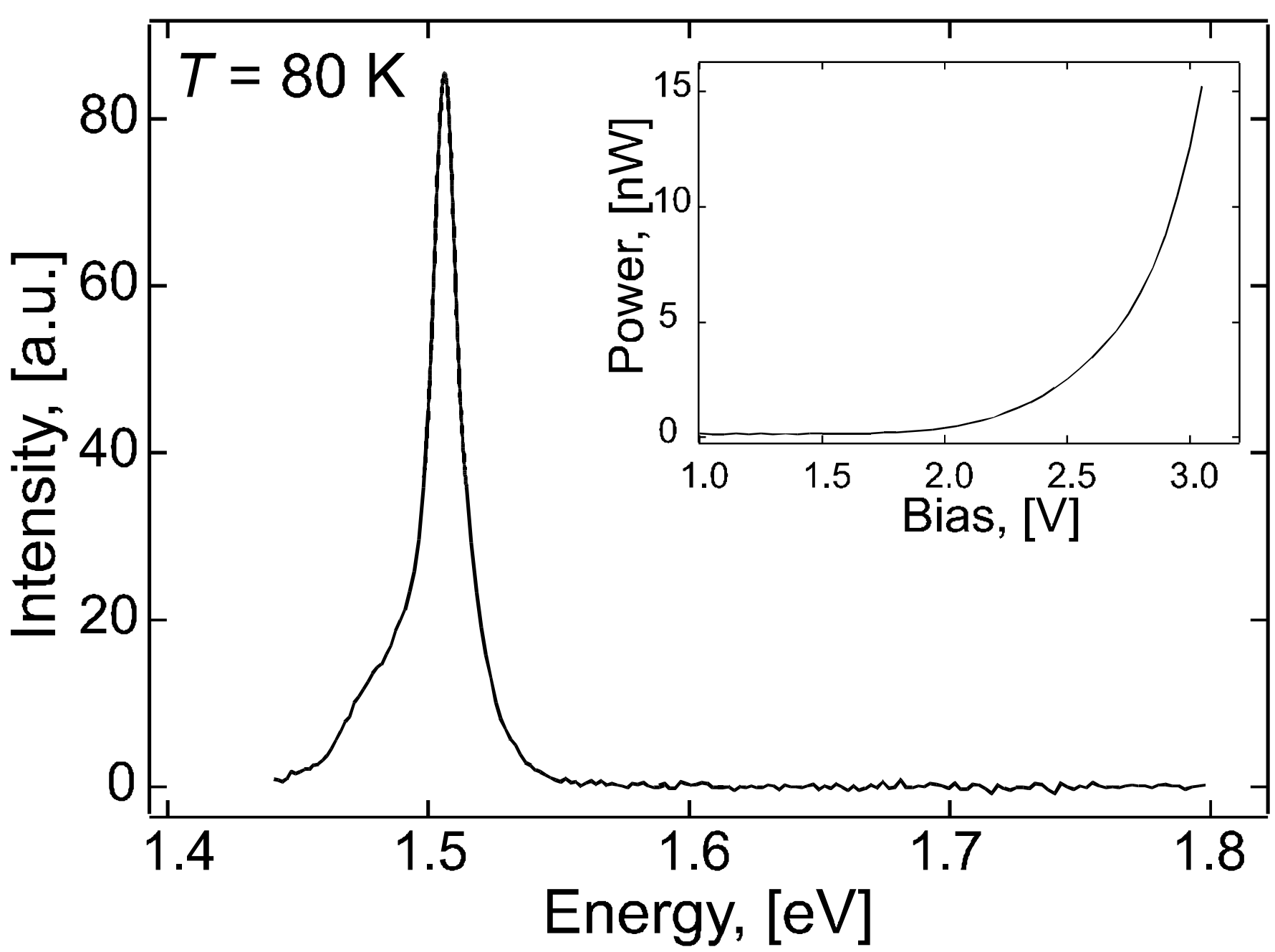

Fig.5.13. Typical EL spectrum and optical output under forward bias (Inset) for the MIS spin-LED *Type B*.

### 5.4.2. Characterization of Al Oxide and Reliability Study

The studies of the Magnetic Tunnel Junctions (MTJ) have shown that fabrication of good structural quality tunnel oxides is needed for observation of large spin-dependent effects. The tunnel barrier imperfections, like pinholes and defect states lead to current channeling and fast device degradation. Moreover, the reliable operation of MIS spin-LEDs requires higher quality of the tunnel oxide, comparing to MTJs. As electrical injection of electrons into the conduction band of the GaAs is achieved only if there is at least 0.7 V (0.8 V for $Al_{0.2}Ga_{0.8}As$) voltage drop across tunnel oxide, while preferential operation of MTJs is in the zero bias regime. Furthermore, the active cross-section of the tunnel barrier (magnetic contact cross-section) in the MTJs generally is much smaller than in the case of spin-LEDs, which allows much higher defect density as compared to the last ones. All these considerations show that extensive study of tunnel barrier reliability issues is needed for reliable fabrication of good quality FM/TB spin injectors. In fact, such study has been carried out in IMEC [180, 181]. Here the spin-LEDs can also serve as experimental tool for visualization of current uniformity across tunnel oxide and hence reveal defect states, like pinholes, etc.

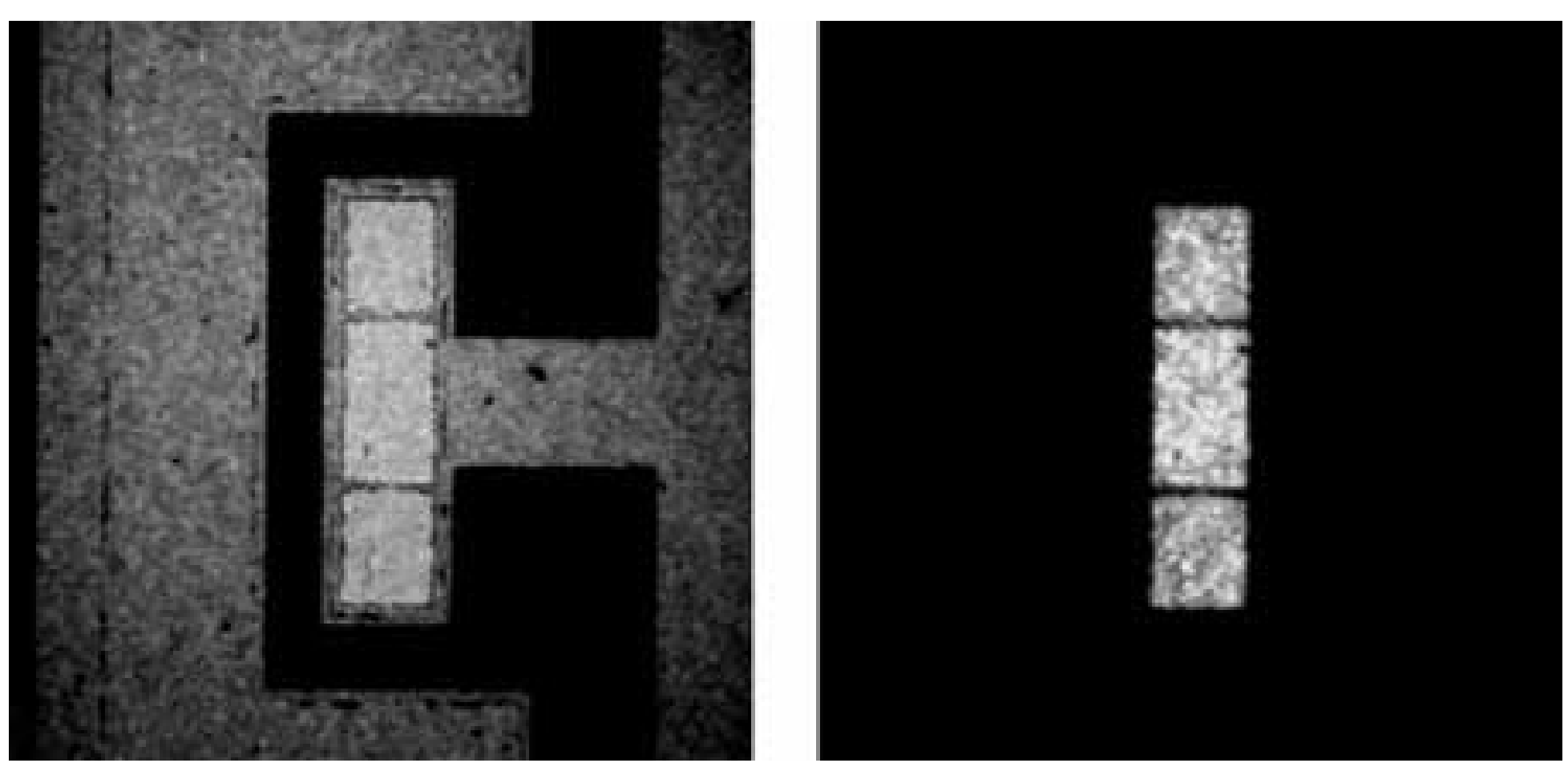

Fig.5.14. Typical emission microscopy measurements of the fabricated MIS spin-LEDs. The lateral dimensions of the ferromagnetic tunnel contact are $80 \times 240\ \mu m^2$.

Fig.5.14 shows a typical result of Emission Microscopy (EMMI) measurement of the fabricated MIS spin-LEDs. Here, both the image from the detector only (right), and a combined image with the detector signal and an optical picture (left) are shown. The EMMI set-up, used at IMEC, consists of an optical microscope and a detector, which is sensitive to a relatively wide spectral range in the visible and near infrared regions, from about 2.5 eV down to 1.2 eV (500 nm to 1100 nm).

These measurements show that the current density through the $AlO_X$ layer is quite uniform. Moreover, there is no enhanced current density along the perimeter of the devices. This means that the dry etch step used for fabrication of spin-LEDs (see Section 5.3.2), doesn't generate defects which may influence the reliability of the devices.

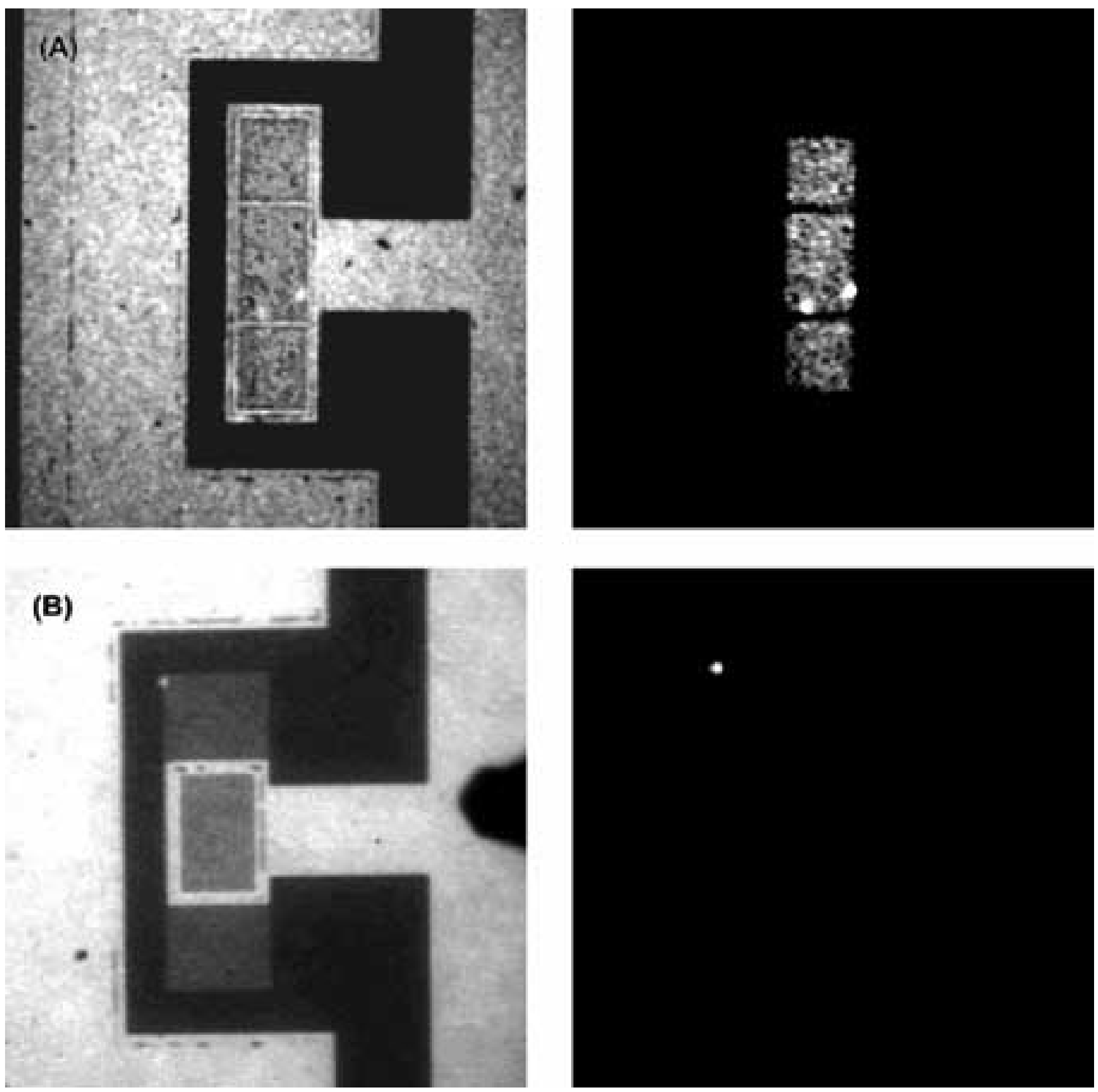

Fig.5.15 Emission microscopy measurements of the fabricated MIS spin-LEDs with bright spots: a) in the middle of the magnetic contact or b) on the perimeter indicating an enhanced current density in these places. The lateral dimensions of the ferromagnetic tunnel contact are $80 \times 240\ \mu m^2$.

Only in very few devices, bright spots were detected in the EMMI measurements. Two examples are shown in Fig.5.15. Here, the bright spots are in the middle of the device (A) or along the perimeter (B). These spots indicate the localized non-uniform current flow, which is most likely to be caused by the damage of the oxide during fabrication of FM/ TB spin injectors or processing. However, since most of the devices show an emission behavior like in Fig.5.14, it indicates that the quality of fabricated $AlO_X$ tunnel barriers is high and device processing most likely does not affect it.

### 5.4.3. Magnetic Characterization

As was mentioned in the Section 5.3, generally the thin magnetic film is composed of two different ferromagnetic materials 2 nm $Co_{90}Fe_{10}$ / 8 nm $Ni_{80}Fe_{20}$. The magnetic in-plane anisotropy was defined by application of small external magnetic field during film deposition. These two layers are expected to show the same magnetic behavior, as domain wall width in such materials is at least of the order of the thickness of entire deposited film [182]. Further, the processing of the spin-LEDs (Section 5.3.2) results in quite large magnetic contacts. Such dimensions can be considered as infinite on the scale of magnetic interactions. Thus no change of magnetic properties of any single contact comparing to the solid ferromagnetic film are expected.

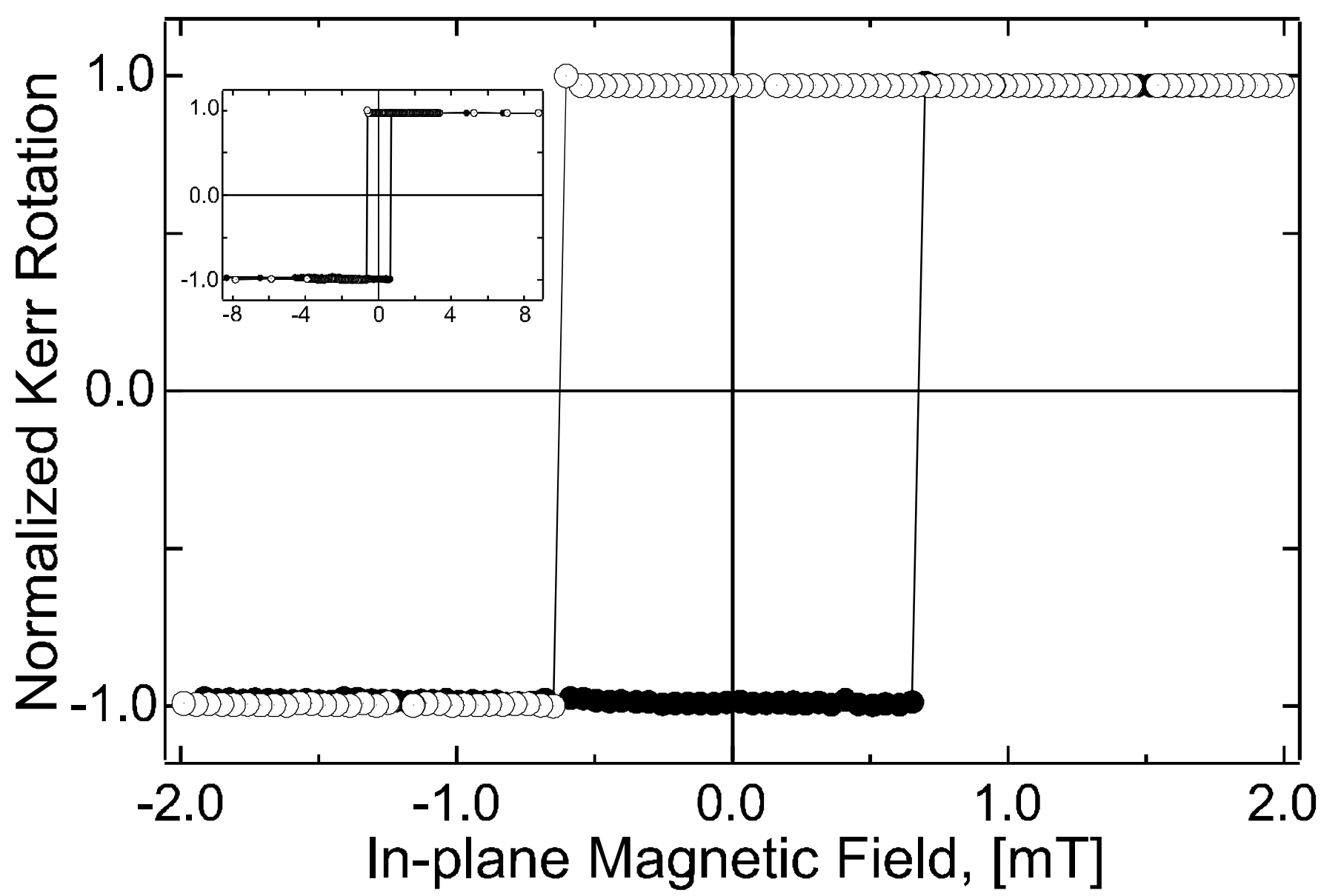

Fig.5.16. Magnetooptical Kerr effect measurements of the easy axis in-plane magnetic reversal. Inset, extended view. The coercive field is 0.65 mT.

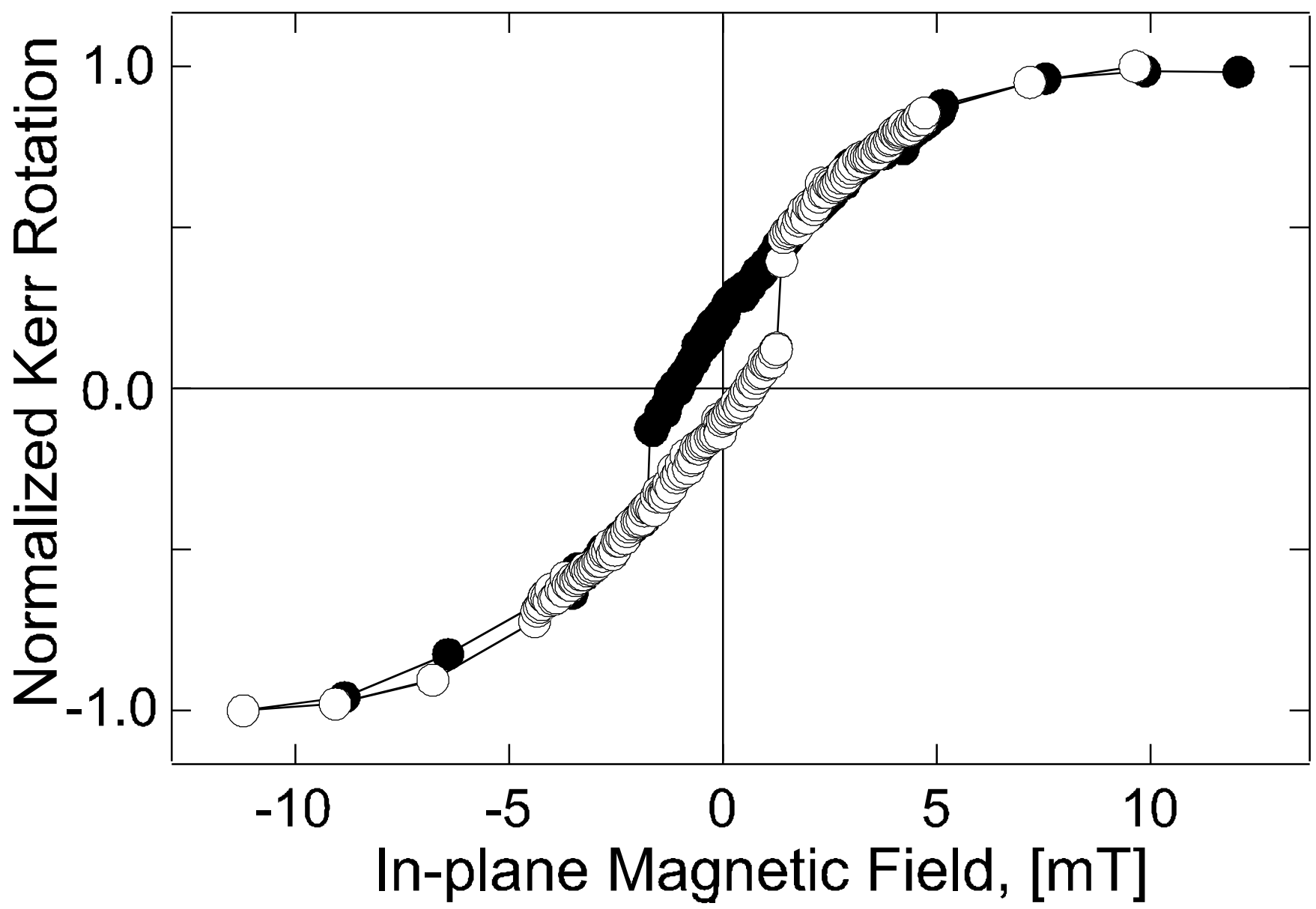

Fig.5.17. Magnetooptical Kerr effect measurements of the hard axis in-plane magnetic reversal.

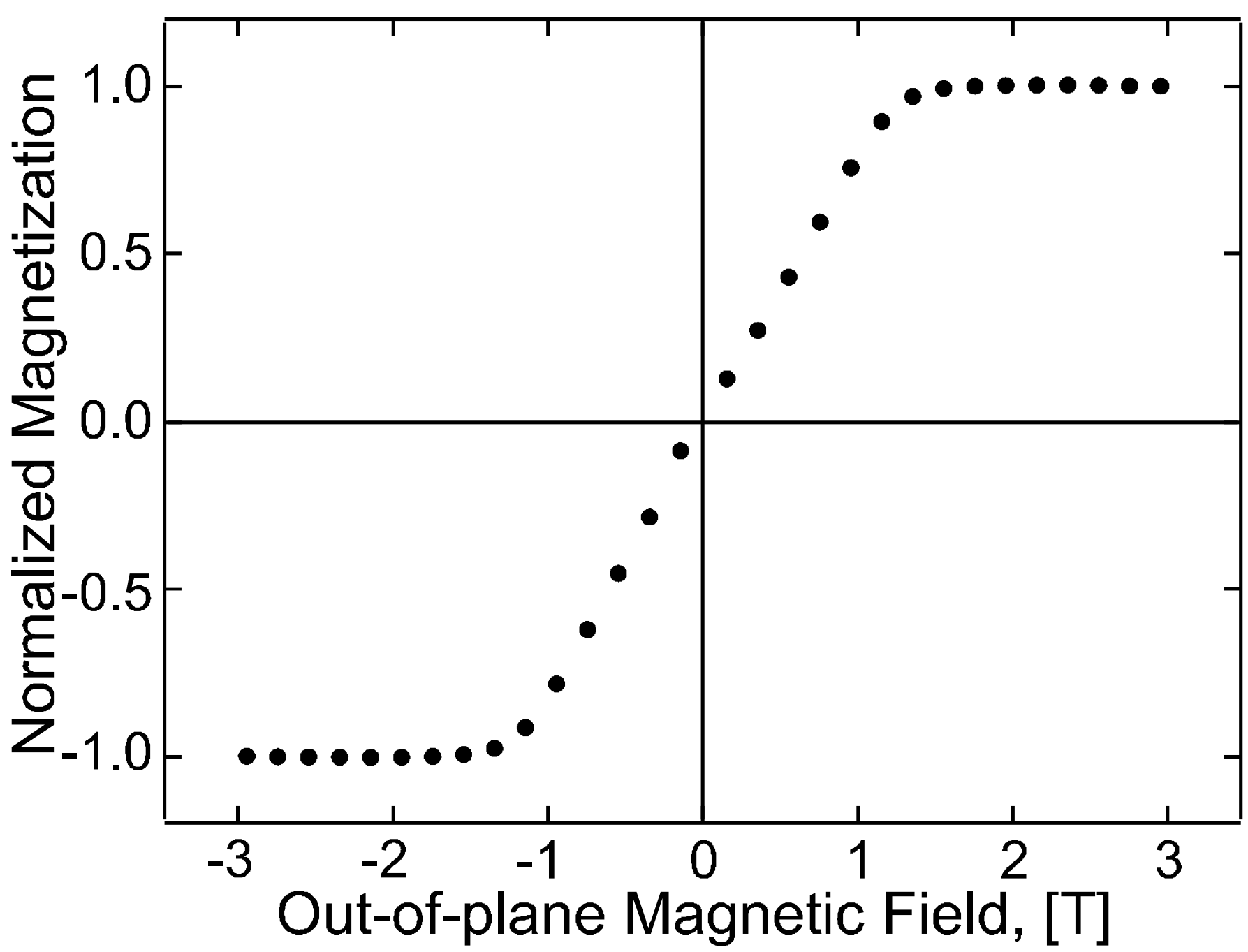

Fig.5.18. Out-of-plane magnetization curve as revealed by extraordinary Hall effect measurements. The saturation field of 1.3 T is a measure for the saturation magnetization $\mu_0 \cdot M$ .

Fig.5.16 and Fig.5.17 show the Magnetooptical Kerr effect measurements of the in-plane magnetization switching in the easy and in the hard magnetic axis, respectively. These measurements show a square hysteresis loop with coercivity of about 0.65 mT in the easy magnetization axis. Some small coercivity in the hard axis measurements corresponds to small sample misalignment in the external magnetic field.

The effective saturation magnetization governing the out-of-plane tilting of the magnetization in the 2 nm $Co_{90}Fe_{10}$ / 8 nm $Ni_{80}Fe_{20}$ ferromagnetic film (see Section 4.4) is $\mu_0 \cdot M = 1.3\,T$. It is determined from the extraordinary Hall effect [183] measurements in the external magnetic field applied in the perpendicular direction to the sample surface (Fig.5.18).

### 5.4.4. Characterization of Spin-injectors

Going a little bit further, it is always interesting to compare the spin polarization of electrons injected into semiconductor with the electron spin polarization in the ferromagnetic metal. In the case of MIS spin-LED heterostructures, due to similar FM/$AlO_X$ interface and the electronic transport involved, the last one can be evaluated from TMR measurements [29]. Moreover it is known that the absolute value of the TMR effect depends not only on the absolute value of the spin polarization within the FM, but also on interface properties and the density of defect states within the tunnel barrier [68, 184]. The quality of the $AlO_X$ interfaces and fabrication of the pinhole free tunnel barriers with good structural and electrical properties is of tremendous importance for both types of the devices. For this purpose, CoFe/$AlO_X$/CoFe TMR junctions were repeatedly fabricated in the same sputtering system. These TMR junctions show 28 % TMR effect at 80 K and 20 % at 300 K. According to the Julliere theory [67] (see Section 2.4.1) these TMR values correspond to the spin polarization in the FM of $\Pi = 40\ \%$ and $30\ \%$ at 80 K and 300 K, respectively.

# 6. Experimental Investigation of Electrical Spin Injection

This chapter gives an overview of the main results obtained during work on the topic of this thesis. The experimental approach is described in the Chapter 4. The MIS spin-LED fabrication and characterization is described in the Chapter 5.

## 6.1. Electrical Spin Injection in the Sample *Type A*

After initial experiments combining different $AlO_X$ tunnel barrier fabrication methods a 8nm NiFe/ 2nm CoFe/ 1.8nm $AlO_X$ (single-step oxidation)/ 100nm GaAs (p=$2\cdot10^{18}$cm$^{-3}$)/ 200nm AlGaAs (p=$2\cdot10^{18}$cm$^{-3}$)/ 2μm GaAs (p=$2\cdot10^{18}$cm$^{-3}$) buffer/ p-GaAs substrate (See Table 5.1) MIS spin-LED heterostructure was fabricated in order to test ideas described in Sections 5.1, 5.2.

Under forward bias conditions spin-LEDs emit light corresponding to GaAs band gap transitions only (see Section 5.4.1). At ~80 K the light emission threshold is about 1.5 V. In order to get sufficient signal to noise ratio the measurements were carried out at about 2 V biasing, with typical current around 90-100 mA.

As it was mentioned before, such MIS spin-LEDs can be considered consisting of two parts: FM/ $AlO_X$ TB - spin injector, and III-V semiconductor heterostructure spin detector. In order to assess the spin injection in such heterostructure, the detector part of the junction has to be calibrated, which can be done in the all-optical experiment with optical spin injection and detection (see Chapter 4). In Fig.6.1 are shown the measurements in the oblique Hanle effect geometry using optical spin injection and detection to determine $T_S$ and $\tau$ for this sample. The closed circles correspond to the measurements of the circular polarization of luminescence under optical excitation with h$\nu$=1.58 eV ($P = 100\%$), i.e., near the GaAs band gap. The solid line represents the fit using Eq.4.4 with the following parameters: $T_S/\tau = 0.48$, $\Delta B = 0.22\ T$ ($T_S = 0.12\ ns$).

An identical experiment with photons of higher energy $h\nu$=1.96 eV (open circles) shows the same sign of circular polarization, indicating ([18, 145], see Section 3.3) that the electrons keep their spin orientation during the thermalization to the bottom of the conduction band. The reduction of the degree of polarization is due to the fact that "cold" electrons excited from the split-off valence band have the opposite spin orientation. The nature of small asymmetricity of these curves is discussed in the Section 6.3.

This result is now used to interpret the measurements of the circular polarization of electroluminescence (Fig.6.2, closed circles) in the oblique Hanle effect geometry. The measured curve does not fit the Eq.4.6 predicted for the oblique Hanle effect, due to the superposition of the spin injection signal and a magnetooptical effect in the ferromagnetic film. In an oblique magnetic field the MCD occurs, which can be measured in a photoluminescence experiment on the same setup with linearly polarized excitation, $\Pi_{inj}=2\cdot S_Z=0$ (see Chapter 4 and Section 3.3). The MCD is proportional to the out-of-plane magnetization component of the ferromagnetic layer induced by the oblique magnetic field. Under photoexcitation with linearly polarized light an unpolarized electron population is created within the semiconductor heterostructure. Radiative recombination with holes produces unpolarized luminescence. Propagation through the ferromagnetic layer induces slight circular polarization caused by MCD. Fig.6.2 (diamonds) shows the resulting circular polarization of photoluminescence with a linear dependence on the oblique magnetic field, as expected for the MCD and the experimental configuration (although the excitation near GaAs band gap is used, see Section 3.3.2, the contribution of MCD on the polarization of exciting light in this experiment can be neglected due to the following reason. If one would imagine that measured polarization of electroluminescence shown in Fig.6.2 is caused by MCD then linearly polarized light will obtain 0.33% circular polarization after propagation through the ferromagnetic film. Selection rules under absorption and emission, as well as measured $T_S/\tau$ ratio, Fig.6.1, will result in the emission of light with only 0.04% circular polarization.). Note that the magnetic field dependency of MCD is quite linear and changes sign when $B$ passes trough zero. It corresponds to the magnetization reversal in the ferromagnetic film. As was mentioned in the Section 5.4.3, the ferromagnetic layer is made from a soft magnetic material, the hysteresis loop is quite narrow and is not seen on the scale of the Fig.6.2.

The difference between the measured degree of circular polarization in the case of electrical spin injection and that, caused by MCD (Fig.6.2, open and closed circles, respectively), gives the net effect of the injected spin polarization (Fig.6.3), which now can be fit (solid line) using the Hanle curve, Eq.4.6, with the following parameters: $T_S/\tau = 0.48$, $\Delta B = 0.22T$ and $S_{0Y} = 0.62\%$. The *only* fit parameter is $S_{0Y}$, the average electron spin of electrically injected electrons. All other parameters were taken from the measurements of oblique Hanle effect with optical spin injection. The change of the sign of the circular polarization is caused by the switching of the magnetic contact at 0.65mT by the in-plane component of the oblique magnetic field (see Section 5.4.3). In this case the degree of spin polarization $\Pi$ of electrically injected electrons (Eq.4.6) was found to be $\Pi_{inj} = (1.2 \pm 0.2)\%$ [172, 185].

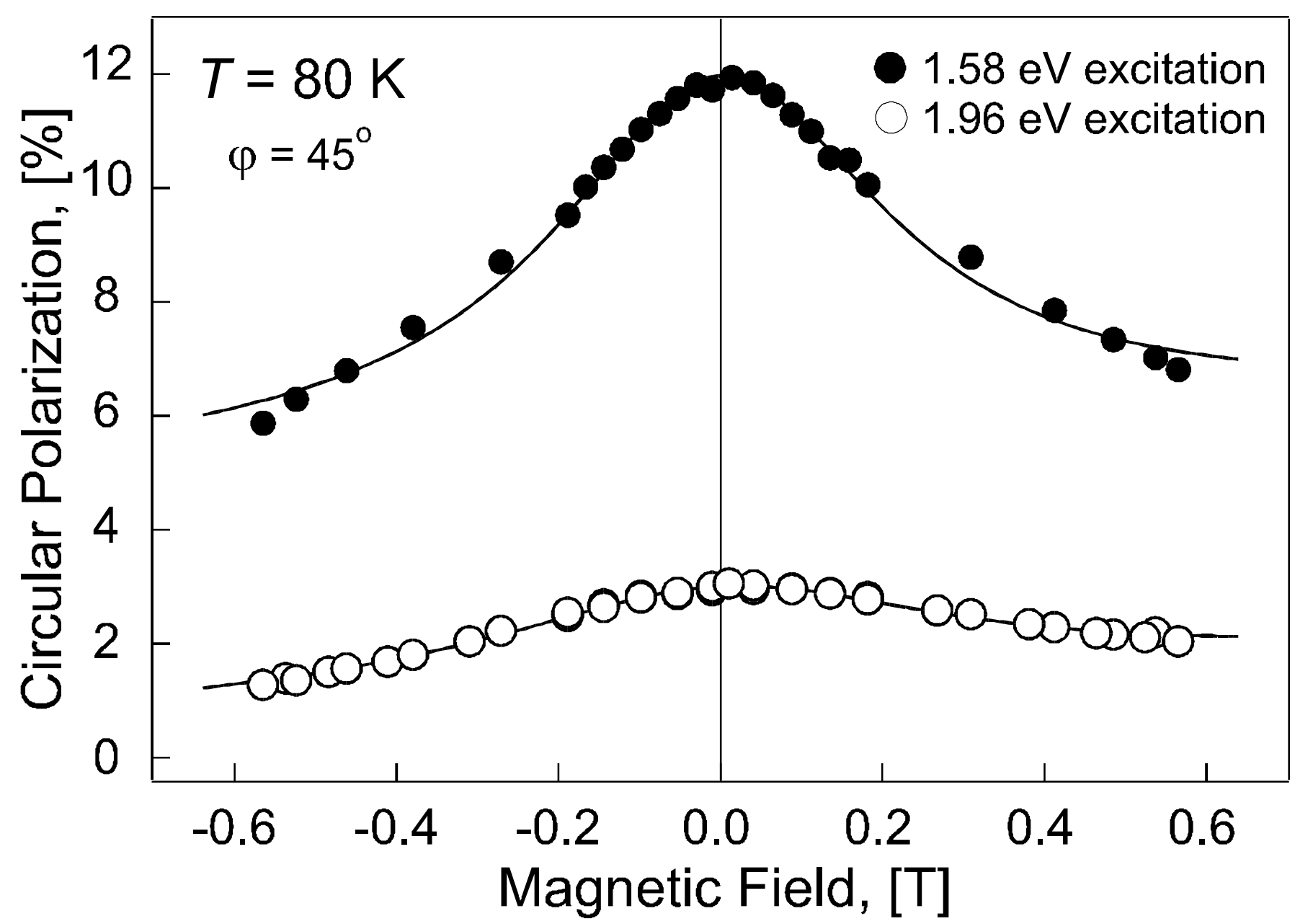

Fig.6.1. Damping of the circular polarization of photoluminescence in the oblique magnetic field under optical spin injection.

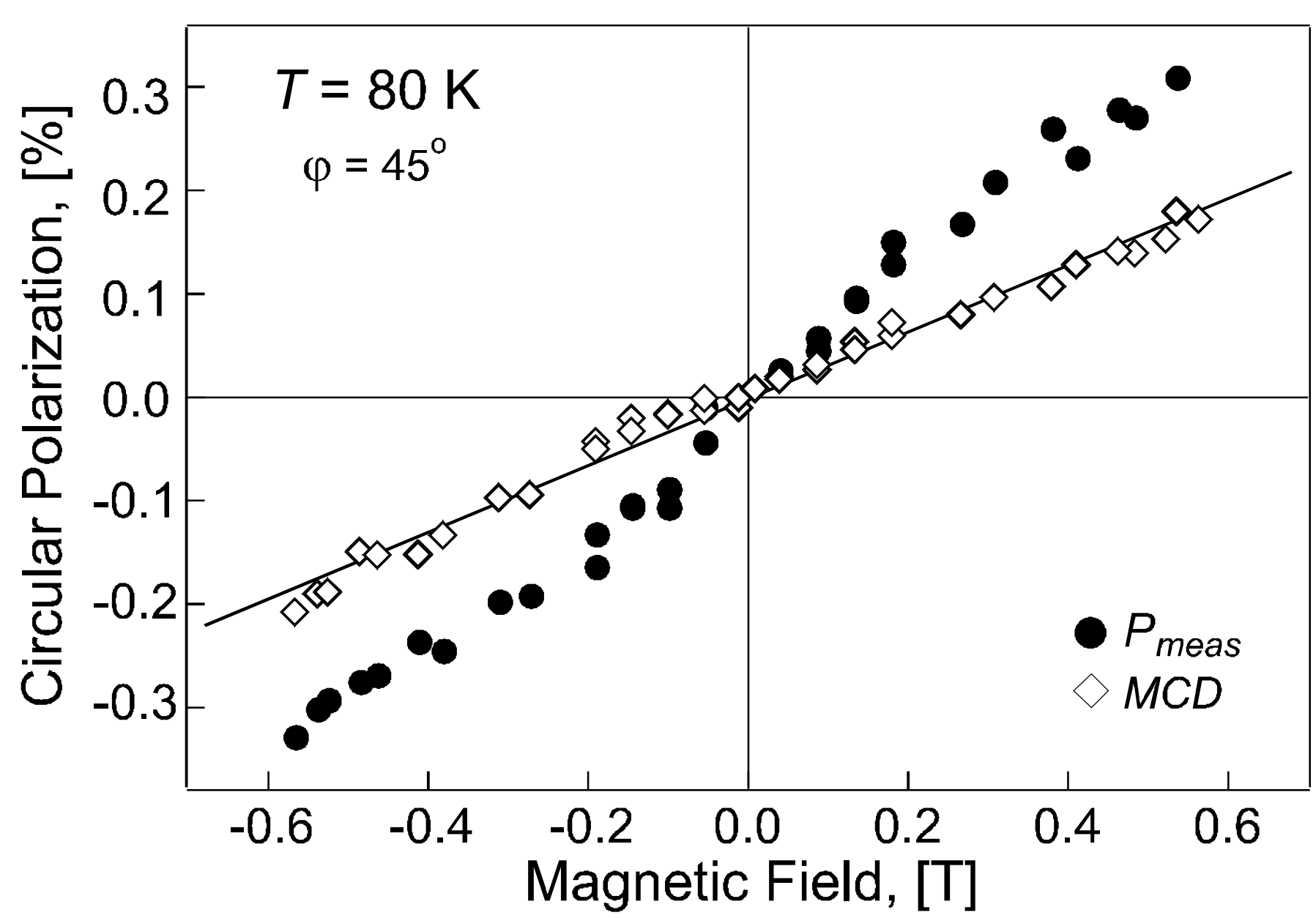

Fig.6.2. Measured degree of circular polarization in the oblique Hanle effect geometry in the case of electrical spin injection (closed circles), and optical linearly polarized laser excitation with hν=1.58 eV (diamonds).

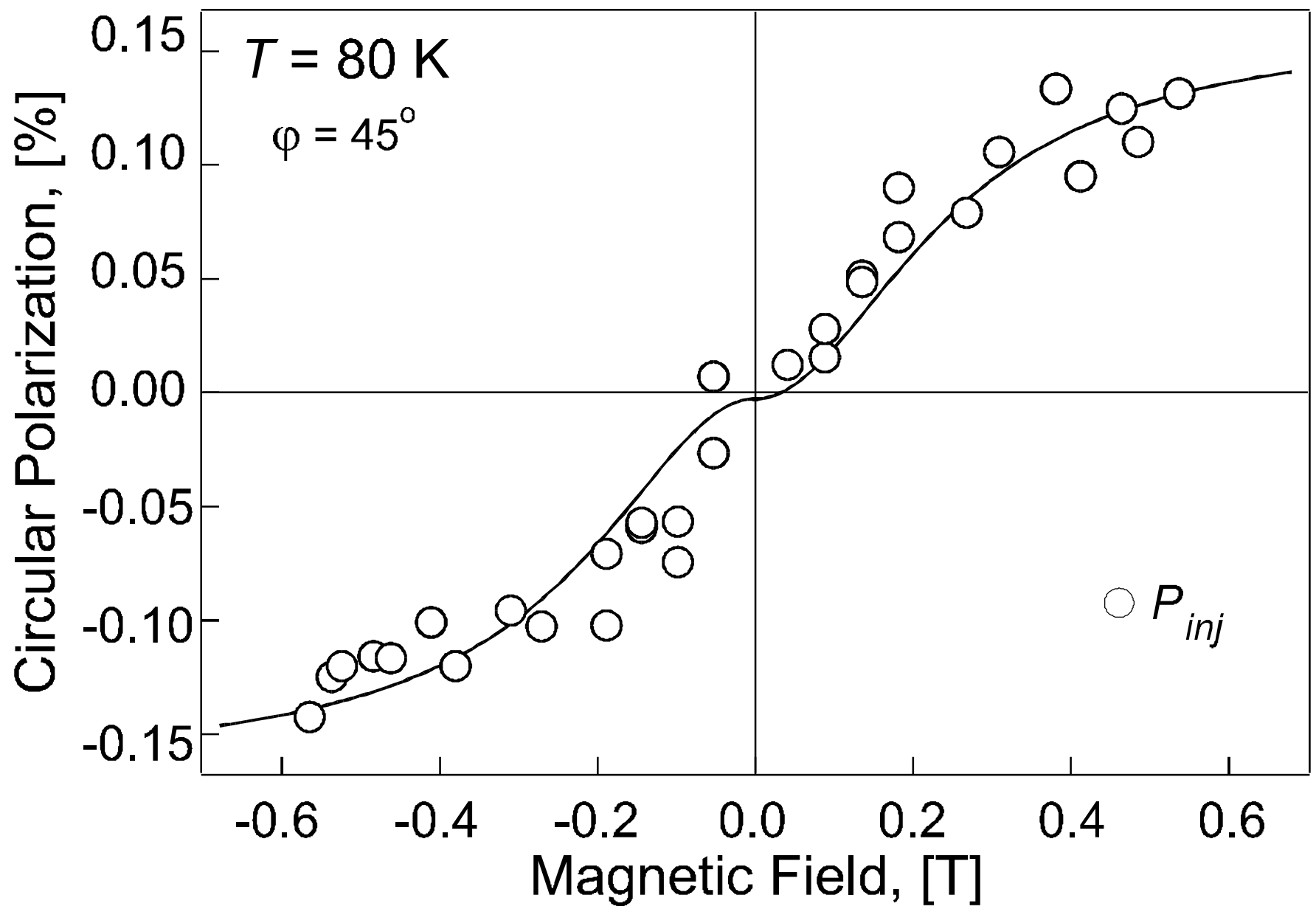

Fig.6.3. Electrical spin injection signal, the difference between circular polarization of electroluminescence and MCD curves on Fig.6.2, and the Hanle curve fit, Eq.4.6.

On behalf of afterword it should be noted that the fitting of the spin injection curves in the case of optical and electrical spin injection presented on Fig.6.1-Fig.6.3 is intentionally made with the smallest half-width of Hanle curve possible. Such fit gives the most conservative number for spin polarization of electrically injected electrons.

## 6.2. Electrical Spin Injection in the Sample *Type B*

In an effort to study the effect of a different $AlO_X$ / semiconductor interface, the sample *Type B* (Table 5.1), consisting of 8nm NiFe/ 2nm CoFe/ 1.8nm $AlO_X$ (single-step oxidation)/ 15nm AlGaAs (undoped)/ 100nm GaAs (undoped)/ 200nm AlGaAs (p=$2{\cdot}10^{18}$cm$^{-3}$)/ 2µm GaAs (p=$2{\cdot}10^{18}$cm$^{-3}$) buffer/ p-GaAs substrate was fabricated. As before, under application of forward bias the fabricated MIS spin-LEDs emit light corresponding to GaAs band gap transitions only (see Section 5.4). At ~80K the light threshold is about 1.6 V. In order to get sufficient signal to noise ratio measurements were carried out at about 3 V biasing and a typical current $70 \ldots 90$ mA .

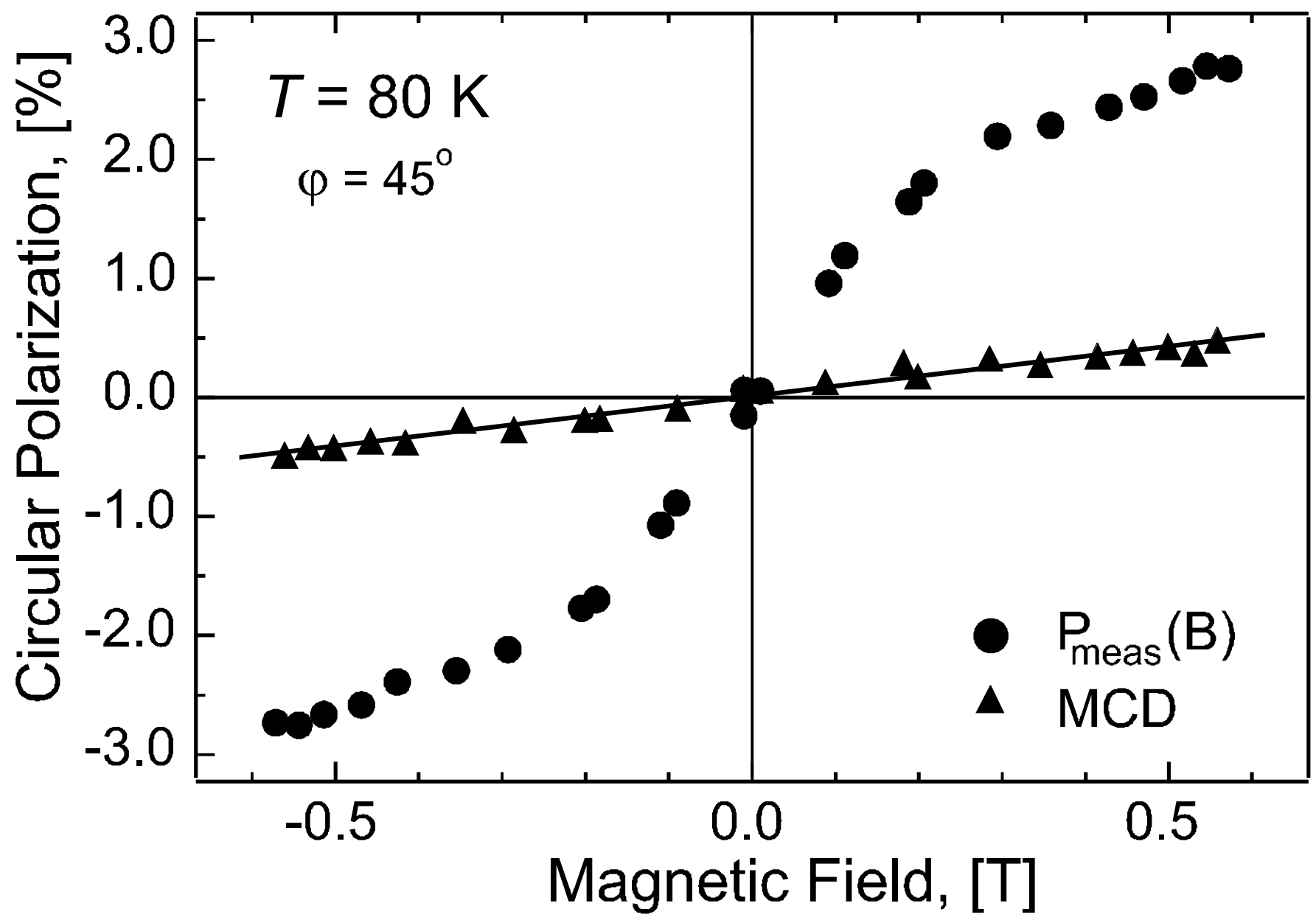

Fig.6.4. Measured circular polarization of the electroluminescence (closed circles) and the GaAs edge photoluminescence (triangles) under optical linearly polarized laser excitation with hν=1.96 eV.

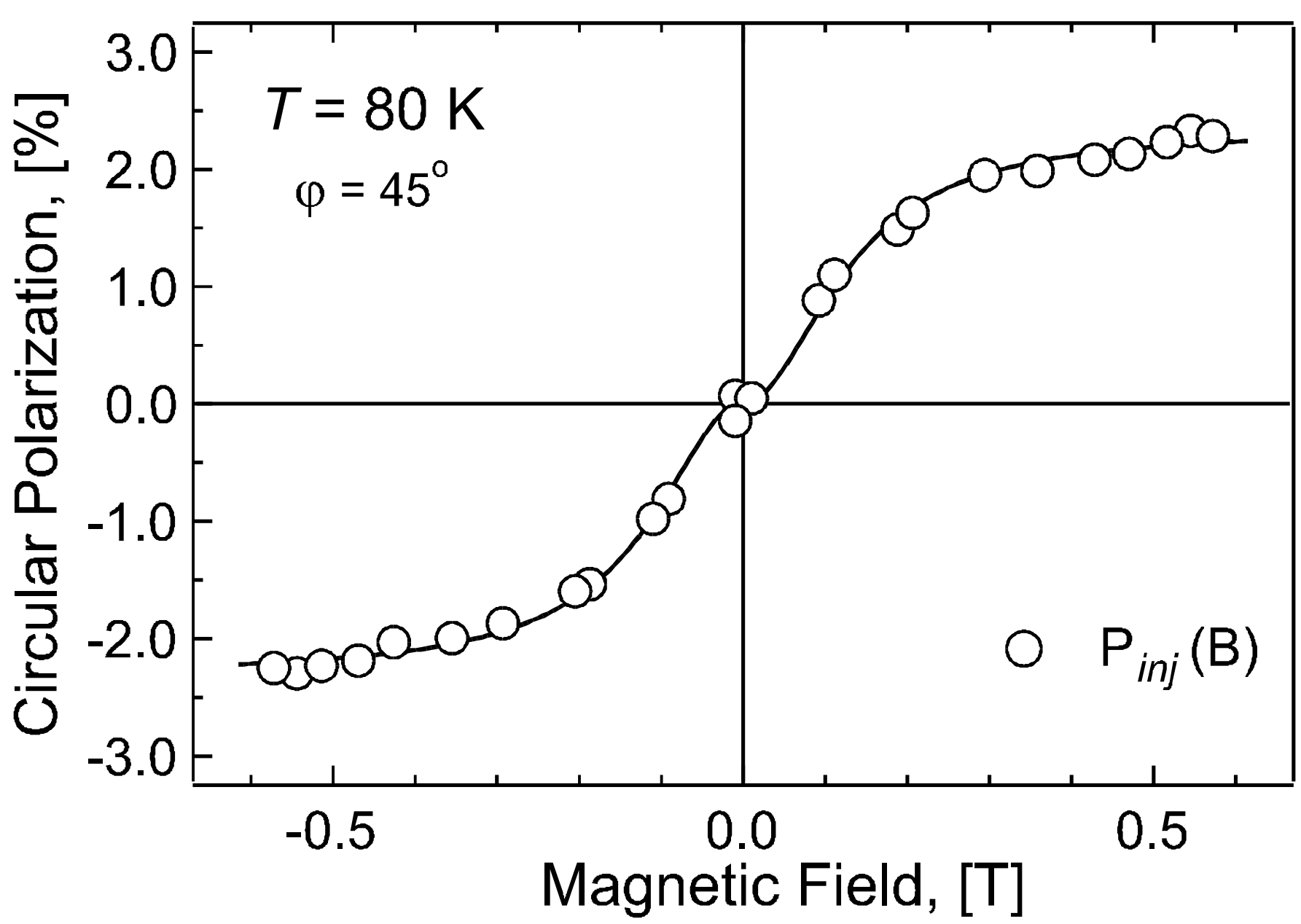

Fig.6.5. Electrical spin injection signal (the difference between $P_{meas}$ and MCD curves on Fig.6.4), and the Hanle curve fit (Eq.4.6). The spin polarization of electrically injected electrons $\Pi_{inj}$ was found to be in access of 9%

The typical evolution of the measured circular polarization of the emitted light as a function of the oblique magnetic field ($\varphi = \pi/4$) is shown in Fig.6.4 (circles). The measured curve is strongly non-linear with tendency to saturation above 0.3 T. This field is too small to induce a saturation of out-of-plane magnetization in the ferromagnetic contact ($M_{Sat} = 1.3T$). On the other hand, the Zeeman splitting induced by this field is very small compared to the thermal energy at 80 K and cannot explain the magnitude of the effect, nor the observed saturation.

Equation 4.6 qualitatively describes the measured data. However, for quantitative determination of the spin injection magnitude, as in the case with the sample *Type A*, the contribution of MCD in the resulting signal must be taken into account. As before this contribution was measured in an all-optical experiment with excitation by linearly polarized light hν=1.96 eV (which is different from the case with the sample *Type A*, for difference see Sections 3.3.2, 6.1).

The measured circular polarization (Fig.6.4, triangles) of the photoluminescence is small and has a linear dependency on the oblique magnetic field, as expected for the MCD and the experimental configuration. The change of the sign of the circular polarization is caused by the switching of the magnetic contact at 0.65mT by the in-plane component of the oblique magnetic field.

Fig.6.5 shows the experimental spin injection signal after subtraction of the MCD contribution, representing the change of circular polarization of the emitted light caused by spin injection only. The Eq.4.6 with parameters $S_{0Y} \cdot \frac{T_S}{\tau} = 2.3\ \%$ and $\Delta B = 0.13\ T$ ($T_s = 0.20\ ns$) is used to fit the data (solid line). In this case, the value $\Pi_{inj} \cdot T_S/\tau$ was found to be (9.2 ± 1.6) % [172, 173, 185]. The value of $T_S/\tau$ ratio, which describes the spin scattering of electrons during their lifetime within the semiconductor, is not known for sample *Type B*. This is because, the undoped GaAs has very poor photoluminescence efficiency and it is difficult to discern this luminescence on the strong background of intense photoluminescence coming from highly p-doped substrate. However the actual spin polarization of electrons that traversed the ferromagnetic metal / semiconductor interface $\Pi_{inj} = \frac{\tau}{T_S} \cdot 9.2\ \%$, is higher than $9.2\ \%$ by the factor $\tau/T_S = (\tau_S + \tau)/\tau_S > 1$.

## 6.3. Electrical Spin Injection in the Samples *Type C* and *Type D*

As was mentioned before the samples *Type A* and *Type B* have limited performance, due comparatively low optical efficiency, which does not allow reliable measurements at room temperature. In an effort to improve the optical efficiency the samples *Type C* and *Type D* were fabricated as described in Chapter 5. The sample *Type C* consists (Table 5.1) of 8nm NiFe/ 2nm CoFe/ 2.6nm $AlO_X$ (double-step oxidation)/ 15nm AlGaAs (undoped)/ 100nm GaAs ($p=2\cdot10^{18}cm^{-3}$) active region/ 200nm AlGaAs ($p=2\cdot10^{18}cm^{-3}$)/ 2μm GaAs ($p=2\cdot10^{18}cm^{-3}$) buffer/ p-GaAs substrate MIS heterostructure. The sample *Type D* has an identical heterostructure, the only difference is that active region is undoped. Under application of forward bias the fabricated MIS spin-LEDs emit light corresponding to GaAs band gap transitions only. At ~80K the light threshold is about 1.7 V for both samples. In order to get sufficient signal to noise ratio, the measurements were carried out at about $1.9\ldots2.5$ V and $2.6\ldots3.6$ V at 80K, $1.6\ldots2.5$ V and $2.0\ldots3$ V at 300K biasing for samples *Type D* and *Type C*, respectively. A typical current driven through the device during measurements was $30\ldots100$ mA. These devices allow better optical output due to increased thickness of $AlO_X$ layer resulting in the change of relative tunnel probabilities for electrons and holes.

### 6.3.1. Low Temperature Investigations

As it was mentioned before, such MIS spin-LEDs consist of two parts: FM/ $AlO_X$ TB-spin injector, and III-V semiconductor heterostructure- spin detector. In order to assess the spin injection in such heterostructure, the detector part of the junction has to be calibrated, which can be done in the all-optical experiment with optical spin injection and detection.

Fig.6.6 shows a typical result of measurements of emitted circular polarization as a function of external oblique magnetic field under 100 % circularly polarized optical excitation with $h\cdot\upsilon=1.58$ eV ($h\cdot\upsilon\geq E_g$) and $h\cdot\upsilon=1.96$ eV ($h\cdot\upsilon\geq E_g+\Delta$) for the sample *Type C* (p-type active region). As it has been discussed in the previous sections the optical measurements under excitation near the band gap of the semiconductor with 100% circularly polarized light allows complete characterization of the semiconductor as spin detector. The fitting of the data ($h\cdot\upsilon=1.58$ eV) with Eq.4.4 ($S_0(0,0,1/4)$) reveals the following parameters of the GaAs active region: spin relaxation term $T_S/\tau=0.67\pm0.08$ and the half-width of the Hanle curve $\Delta B=(0.28\pm0.03)$ T (These parameters correspond to electron lifetime $\tau=0.14$ ns and spin relaxation time $\tau_S^{-1}=3.6\cdot10^9$ $s^{-1}$, see Fig.3.8b, and *Appendix A* for comparison).

For excitation with $h\cdot\upsilon=1.96$ eV, the reduced circular polarization is due to excitation of electrons with opposite spin orientation from the split-off band. The fact that circular polarization does not change sign in this case suggests that electrons do not lose their spin during the thermalization process (see Section 4.6).

On the same figure is presented the typical result of measurements of emitted circular polarization as a function of external oblique magnetic field under excitation with linearly polarized light $h \cdot \upsilon = 1.96$ eV, which does not create any spin polarization in the semiconductor ([18], see Section 3.3.1). The observed polarization of the luminescence is due to the MCD effect in the ferromagnetic layer (see Section 4.5). It varies linearly with the magnetic field and gives the value of $D(B)$ which characterizes the MCD contribution to the observed circular polarization of light emitted by the structure. In order to obtain the real polarization of the emitted light $P_{inj}$ one have to subtract the $D$ contribution from all measured $P_{meas}$ values. Thus the subtraction of $D(B)$ from the curve $P_{meas}(B)$ ($h \cdot \upsilon = 1.96$ eV, Fig.6.6) transforms it into a perfect Lorentzian, typical for the Hanle effect. Note that the $D(B)$ is quite linear and changes sign when $B$ passes trough zero. It corresponds to the magnetization reversal in the ferromagnetic film. As was mentioned in the Section 5.4.3, the ferromagnetic layer is made from a soft magnetic material, the hysteresis loop is quite narrow and is not seen on the scale of the Fig.6.6.

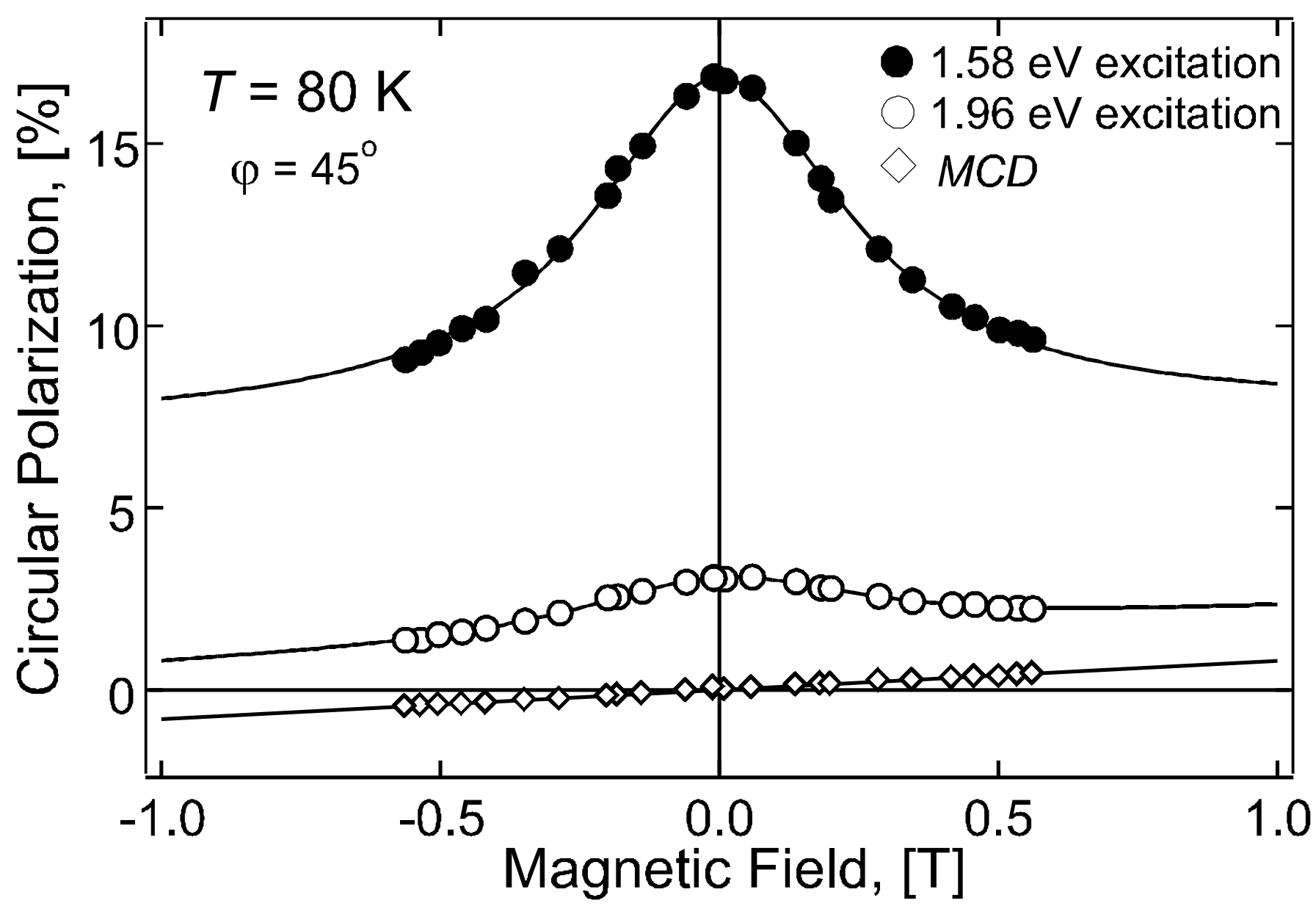

Fig.6.6. Damping of circular polarization of photoluminescence under optical spin injection with $h \cdot \nu = 1.58$ eV and $h \cdot \nu = 1.96$ eV for the sample *Type C* (doped active region), Hanle fits (Eq.4.4), and MCD effect in the ferromagnetic film.

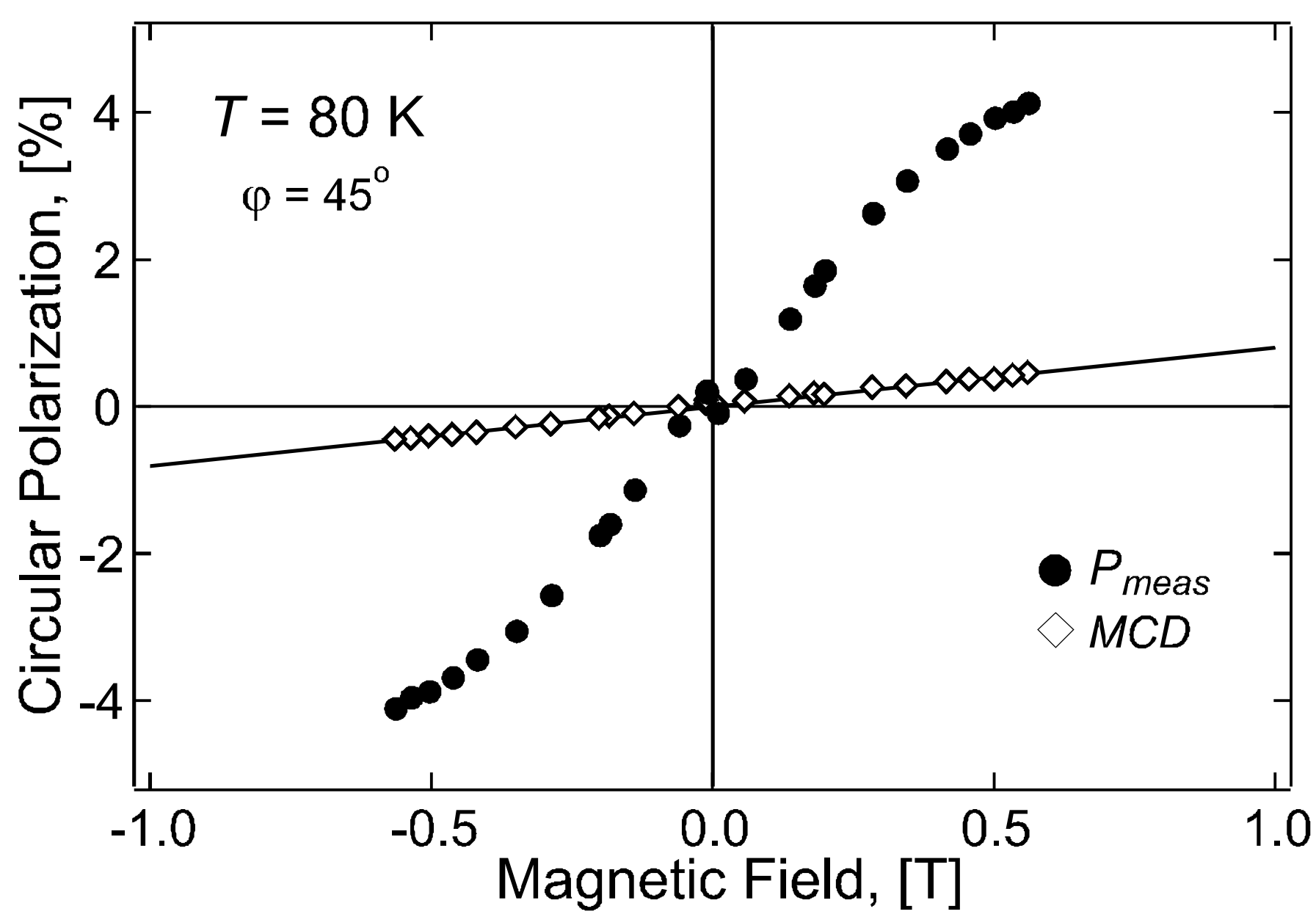

Fig.6.7. Typical result of measurement of the degree of circular polarization of electroluminescence for the sample *Type C* (doped active region) and MCD contribution shown for comparison.

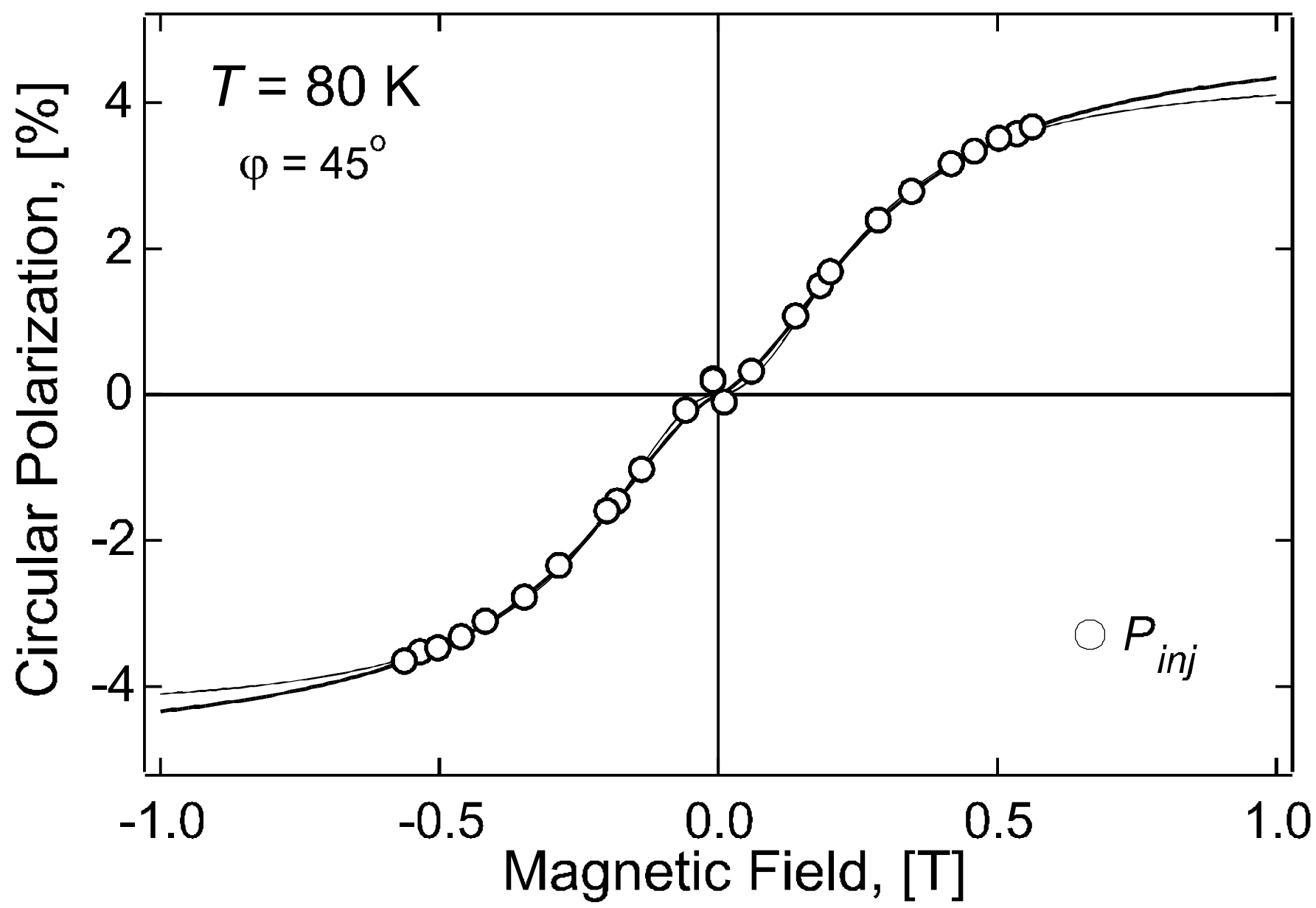

Fig.6.8. The change of circular polarization of the optical output of the device caused by the spin injection and precession only (the difference between spin injection and MCD curves on Fig.6.7) for the sample *Type C* (doped active region), and Hanle fits using Eq.4.6 (thin) and Eq.4.8 (thick). The spin polarization of electrically injected electrons is found to be $\Pi_{inj} = (21 \pm 3)\ \%$.

The typical result of measurements of circular polarization of the emitted light $P_{meas}$ under application of electrical bias for the sample *Type C* is shown in Fig.6.7. The curve is non-linear with tendency to saturation at $B \geq 0.4$ T. The polarization changes sign when $B$ passes trough zero, which again is related to the magnetization reversal in the ferromagnetic film. This clearly indicates that the observed polarization of luminescence is related to the ferromagnetic layer. The $P_{inj}(B)$ variation is obtained from the measured curve by subtraction of $D\,(B)$ measured in the previous experiment.

As it was mentioned in Chapter 4, $P_{inj}(B) = S_Z(B)$ and one can fit the data of Fig.6.7 using Eq.4.6. As before, the situation is quite different from the case of optical excitation near the GaAs band gap (Fig.6.6). In the latter case $S_0$ is known (it is given by selection rules), the $T_S/\tau$ term is directly determined from the $P(B=0)$ value, the only fitting parameter remains the $\Delta B$. In the case of electrical spin injection, all three parameters should be obtained from the $P(B)$ variation: $S_0$, $T_S/\tau$ and $\Delta B$. The first parameter is the most interesting one, since it characterizes the spin injection, i.e. the spin polarization of electrons injected from the ferromagnetic metal into the semiconductor $\Pi = 2 \cdot S_0$. The last two parameters characterize the electron spin evolution in the active region of the MIS spin-LED, i.e. they characterize the spin-detector part of the device. These parameters are known from the all-optical calibration experiment (Fig.6.6), and the experimental data are perfectly fitted with the parameters $T_S/\tau$ and $\Delta B$ derived from these measurements. The thin line on the Fig.6.8 is a fit made using Eq.4.6 and the thick one using Eq.4.8. These two fits give close values of spin injection: $\Pi_{inj} = 26\ \%$ without and $\Pi_{inj} = 21\ \%$ with taking into account the effect of magnetization tilting in the ferromagnetic metal.

Fig.6.9 represents results of measurements of circular polarization of the emitted light $P_{meas}$ under application of electrical bias and MCD contributions for two different orientations of the oblique magnetic field: $\varphi = 45^0$ and $\varphi = 60^0$ for the sample *Type C*. Fig.6.10 shows resulting change of circular polarization of the optical output of the device caused by spin injection and precession only ($P_{inj} = P_{meas} - D$, the difference between measured circular polarization of electroluminescence and MCD curves on Fig.6.9). The solid lines represent the fits obtained after Eq.4.8 with the *same* set of parameters, the only difference is the oblique angle $\varphi$. (The curves were fitted independently, resulting in the same values of spin injection $\Pi$ and half-width $\Delta B$). As one can see, the angular dependence of the effect is perfectly described by the Eq.4.8.

Fig.6.11 shows the typical result of measurements of circular polarization of the emitted light $P_{meas}(B)$ under application of electrical bias for the sample *Type D* (undoped active region) and MCD contribution. The $P_{inj}(B)$ variation is obtained from the measured curve by subtraction of MCD contribution (Fig.6.12). The parameters characterizing spin detecting qualities of the active region of the device are not known for the sample *Type D*. As before this is because, the undoped GaAs has very poor photoluminescence efficiency and it is difficult to discern this luminescence on the strong background of intense photoluminescence coming from highly p-doped substrate. The Hanle curve fit using Eq.4.8 reveals the following parameters, the injected spin polarization normalized to spin scattering parameter $\Pi_{inj} = (21 \pm 3) \cdot \tau / T_S$ % and half-width of the Hanle curve $\Delta B = (0.16 \pm 0.02)$ T . The spin scattering parameter $\tau / T_S$ describes the spin scattering of electrons during their lifetime on the bottom of the conduction band of the semiconductor. Its value is not known for the sample *Type D*, but in any case $\tau / T_S = (\tau_S + \tau) / \tau_S > 1$ and the real value of spin polarization of electrons injected through FM/ semiconductor interface $\Pi_{inj}$ is certainly larger than 21 %.

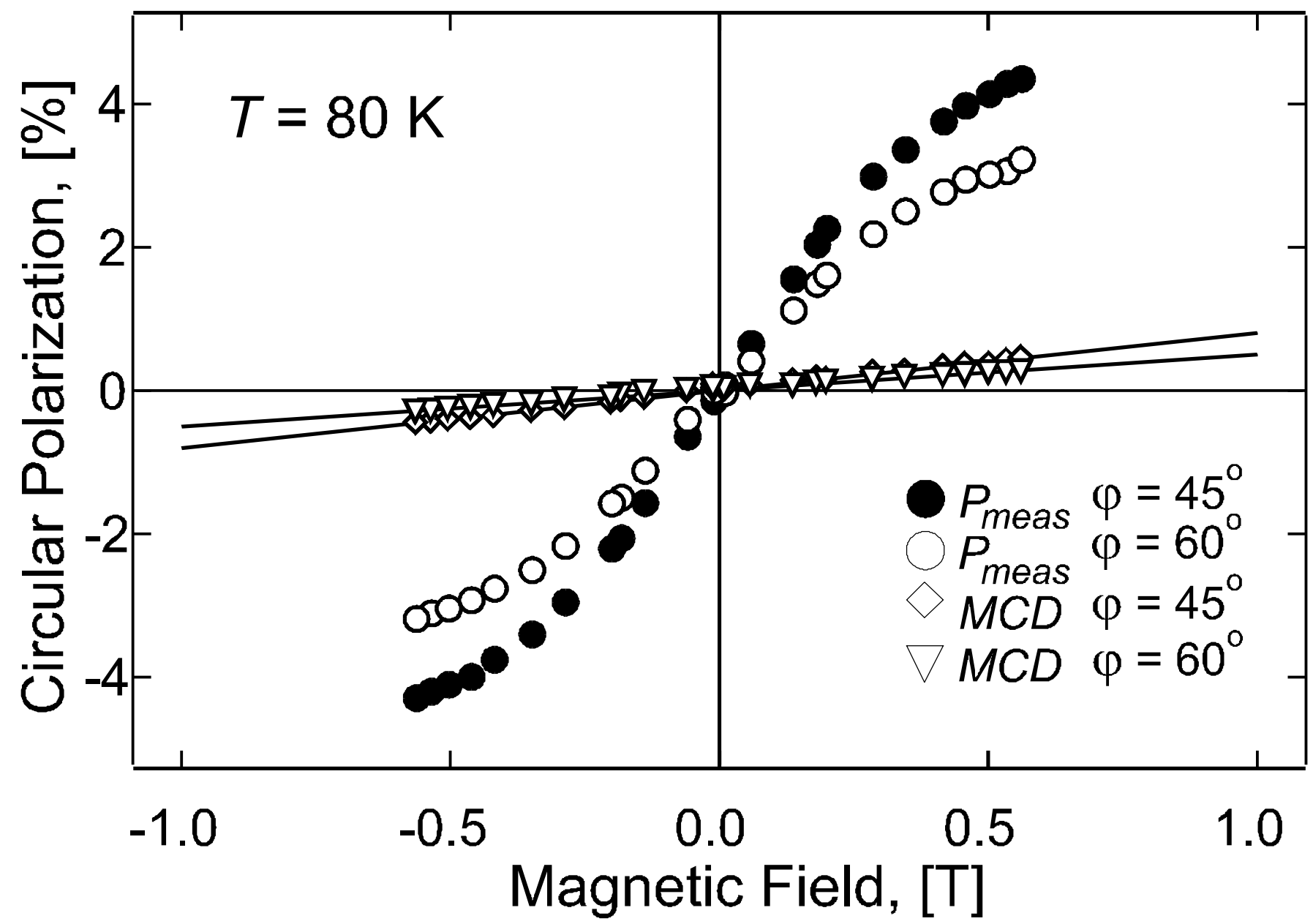

Fig.6.9. Measured degree of circular polarization of the electroluminescence in the external oblique magnetic field and MCD contributions for two different orientations of the oblique magnetic field B for the sample *Type C* (doped active region).

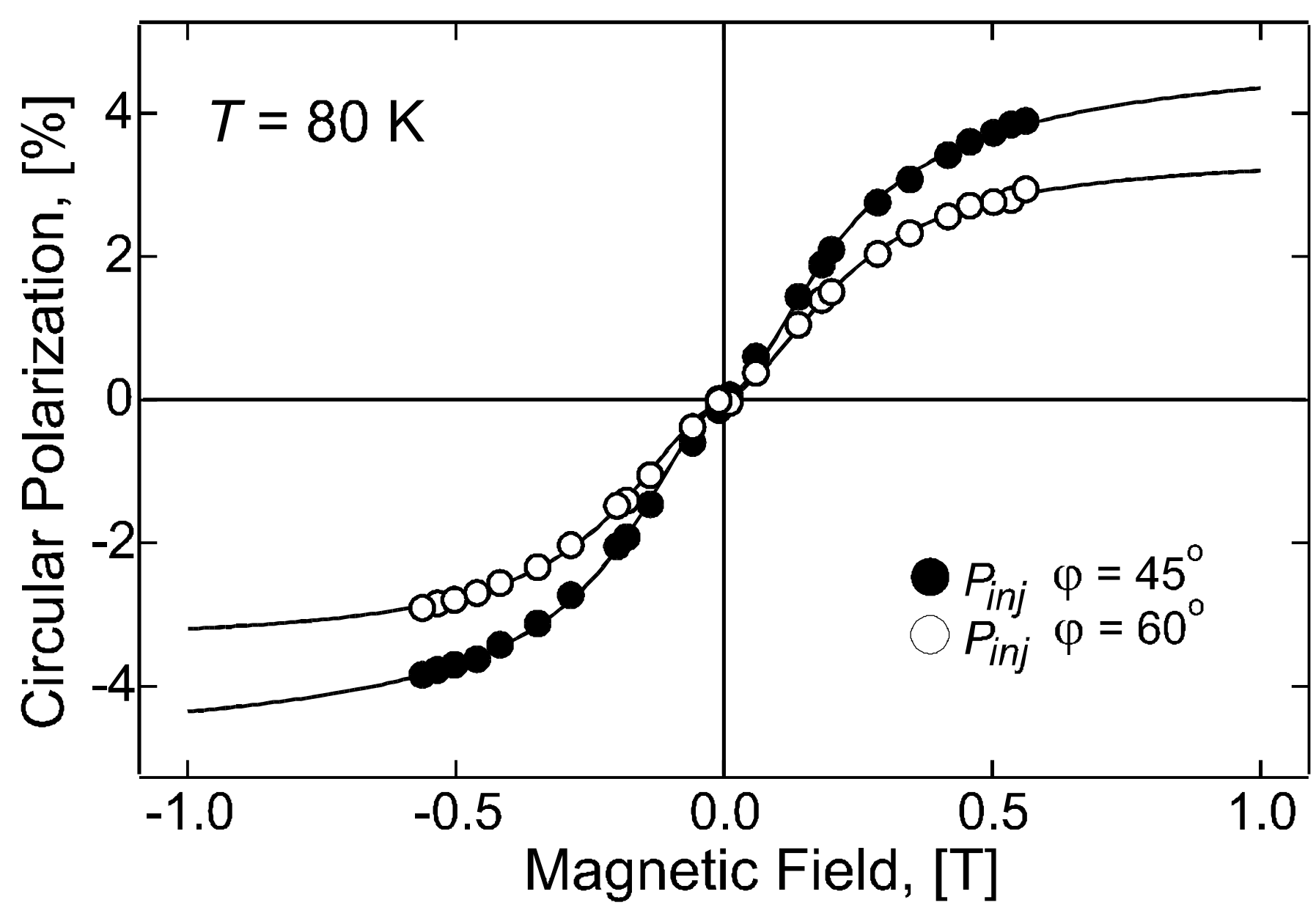

Fig.6.10. The change of circular polarization of the optical output of the device caused by spin injection and precession only (the difference between $P_{meas}$ and MCD curves on Fig.6.9, and Hanle fits (Eq.4.8) with the same sets of parameters: $\Pi_{inj} = (21 \pm 3)\%$ , $\Delta B = (0.23 \pm 0.03)$ T . The only difference is oblique angle $\varphi$

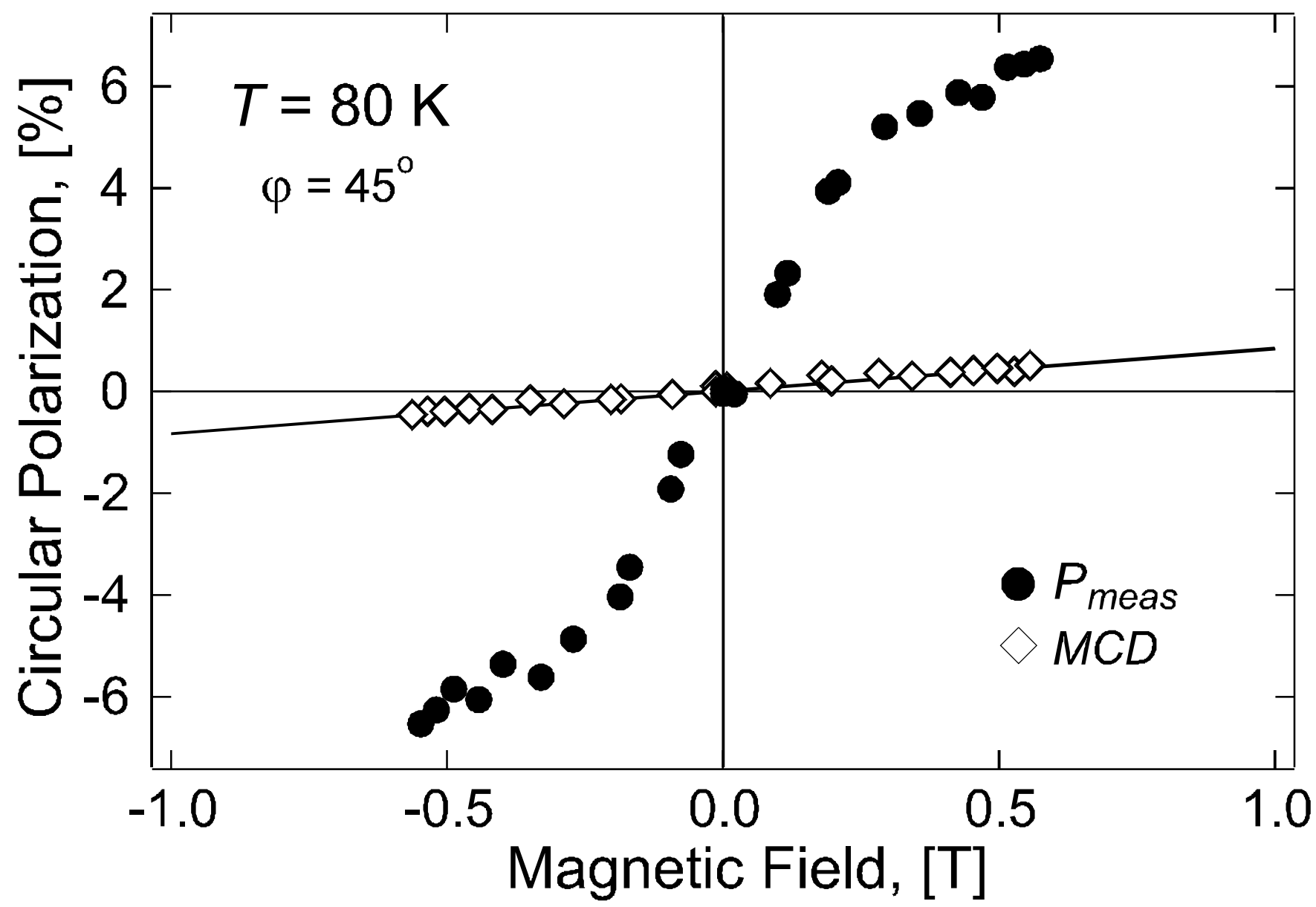

Fig.6.11. Typical result of measurement of the degree of circular polarization of electroluminescence for the sample *Type D* (undoped active region) and MCD contribution.

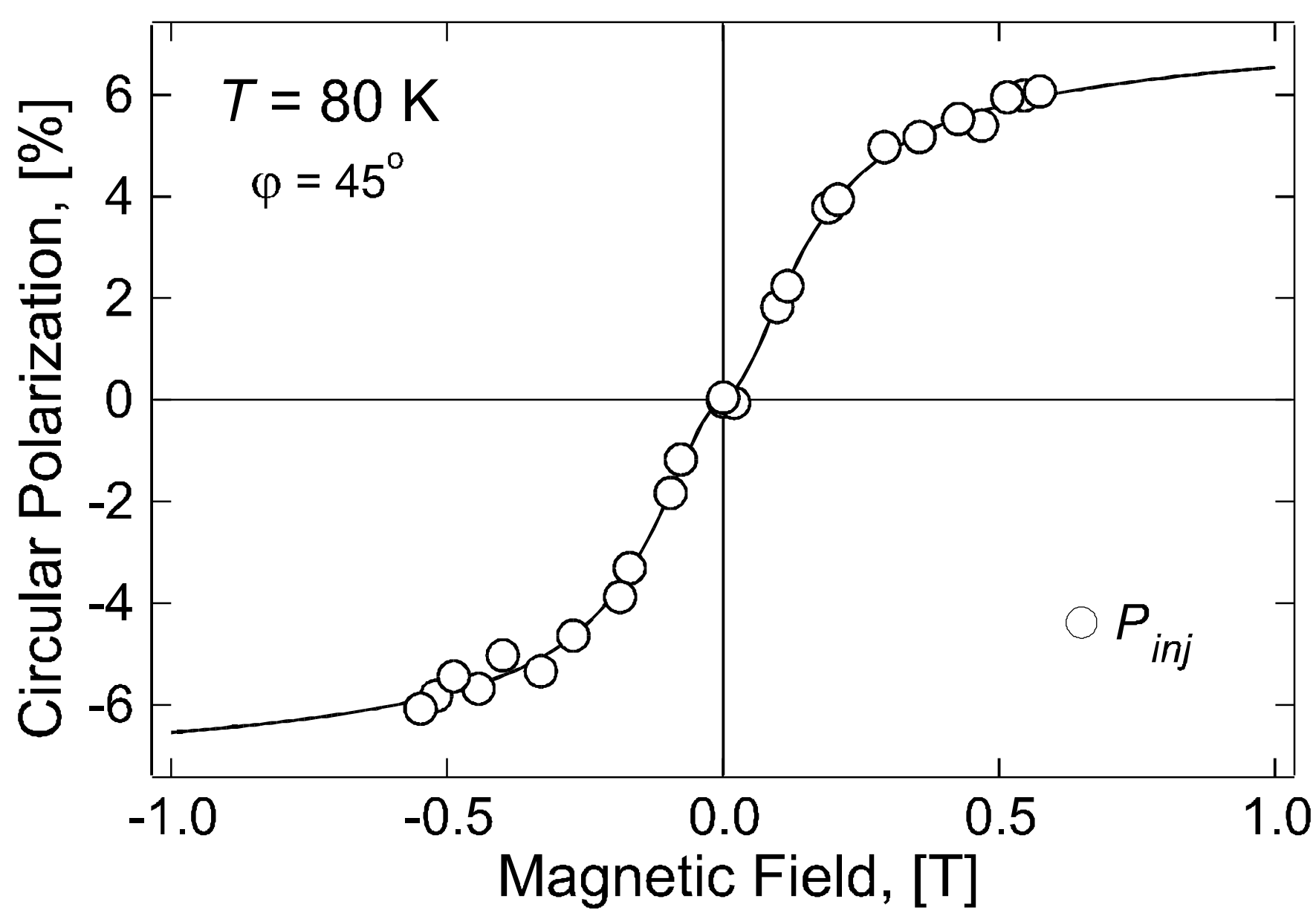

Fig.6.12. The change of circular polarization of the optical output of the device caused by spin injection and precession only (the difference between $P_{meas}$ and MCD curves on Fig.6.11. The Hanle fit (Eq.4.8) reveals the injected spin polarization normalized to the spin scattering parameter $\Pi_{inj} \cdot T_S / \tau = (21 \pm 3)$ % and half-width of the Hanle curve $\Delta B = (0.16 \pm 0.02)\ T$ .

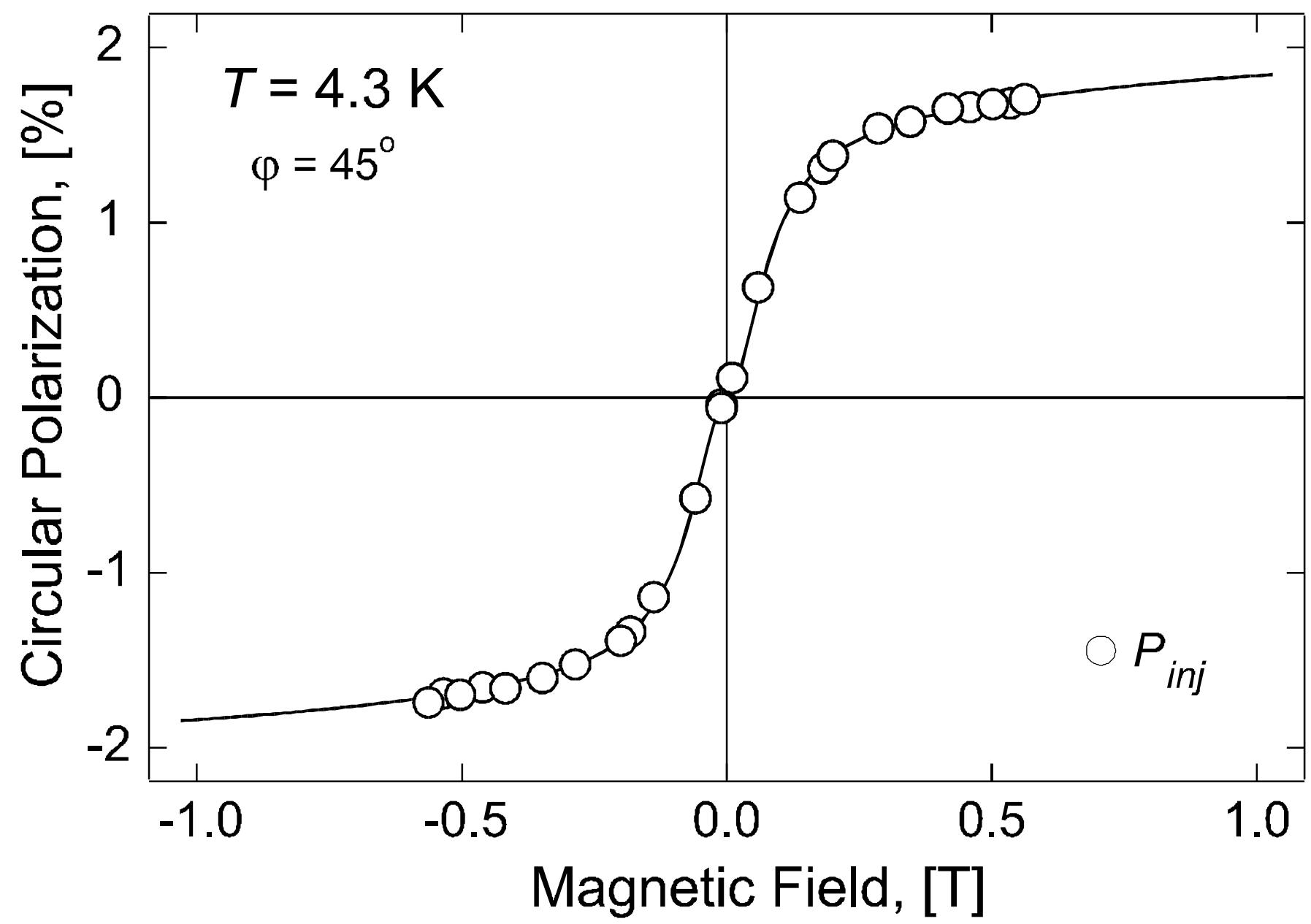

Fig.6.13. Measurements of circular polarization of electroluminescence after subtraction of MCD contribution $P_{inj}$ ( $P_{inj} = P_{meas} - D$ ) for the sample *Type D* (undoped active region) at 4.3 K. The Hanle fit (Eq.4.8) with the following parameters: $\Pi_{inj} \cdot T_S / \tau = (5.8 \pm 0.7)$ % , $\Delta B = (82 \pm 10)$ m$T$ is used to fit the data. The reduced injected spin polarization is caused by Cu contamination of CoFe layer.

In addition, it is interesting to see the influence of temperature on the observed circular polarization of electroluminescence as function of oblique magnetic field. It is known that properties of ferromagnetic film does not change much, due to high Curie temperature $T_C$, while all characteristic electron lifetimes within the semiconductor have strong influence by change of temperature [18]. Such change must reveal itself in the experimental investigation of electrical spin injection in the oblique Hanle effect geometry. While room temperature measurements are presented in the next section, the Fig.6.13 shows the typical result of measurements of circular polarization of the emitted light $P_{inj}$ (after MCD subtraction) under application of electrical bias for the MIS spin-LED heterostructure *Type D*. The narrowing of the experimental Hanle curve is particularly remarkable. Such narrowing is caused by increase of electron lifetime $T_S$ with temperature within semiconductor, which is characteristic for this type of measurements. The reduced circular polarization of emitted light at saturation is caused by the Cu contamination of the CoFe ferromagnetic layer, particularly in this fabrication run. This contamination was found in the supporting measurements of TMR junctions and later confirmed in independent test.

### 6.3.2. Room Temperature Investigations

Fig.6.16 shows the typical results of measurements of the circular polarization of the emitted light under optical excitations with the 100 % circularly polarized light with $h \cdot \upsilon = 1.58$ eV as a function of external oblique magnetic field ($\varphi = 45^{0}$) for the sample *Type C* (doped active region). The fitting of the measured data after Eq.4.4 reveals the following characteristic parameters of GaAs as spin detector: spin relaxation parameter $T_S/\tau = 0.39 \pm 0.05$ and half-width of the Hanle curve $\Delta B = (0.8 \pm 0.1)$ T. These values differ significantly from the ones obtained at 80K. The variation of these parameters is related to the enhancement of spin relaxation with temperature. This effect was studied in details in the 70's [18] and these observations correspond quantitatively to the published data (These parameters correspond to electron lifetime $\tau = 82.8$ ps and spin relaxation time $\tau_S^{-1} = 1.89 \cdot 10^{10}$ s$^{-1}$, for comparison see Fig.3.8b). At room temperature the half-width of the Hanle curve $\Delta B$ becomes very large, it is comparable to the magnetization saturation value of the ferromagnetic film $\mu_0 \cdot M$. As result, the polarization of the electro-luminescence as a function of magnetic field does not show the typical Lorentzian shape (Fig.6.15).

For the sample *Type C*, it is possible to profit from the results of the all-optical experiment on determination of $\Delta B$ and $T_S/\tau$ values. The fitting of the measured data after subtraction of the MCD contribution $P_{inj}(B)$ (Fig.6.16) with these parameters (Eq.4.8) gives an injected spin polarization of $\Pi_{inj} = (16 \pm 2)$ %.

Fig.6.17 shows the typical results of measurements of the circular polarization of the electroluminescence for the sample *Type D* (undoped active region) and MCD contribution. Fig.6.18 shows the $P_{inj}(B)$, the difference between electroluminescence and MCD curves shown on Fig.6.17. As in the previous case, the saturation part of the Hanle curve is not reached in the available magnetic field range. However, the emitted light has substantially higher circular polarization than the MCD contribution. The Hanle fit after Eq.4.8 reveals the following parameters, the minimal injected spin polarization normalized to spin scattering parameter $\Pi \geq (5 \pm 1) \cdot \tau / T_S\ \%$ corresponding to the minimal half-width of the Hanle curve $\Delta B \geq (0.6 \pm 0.2)\ \mathrm{T}$. Again the exact value of the spin scattering parameter $T_S / \tau$ is not known from independent measurements for the sample *Type D*. Comparison of $\Delta B$ measured at low and room temperatures gives the following relative variation of $T_S$ with temperature $\Delta B_{300\,K} / \Delta B_{80\,K} \simeq T_{S\ 80\,K} / T_{S\ 300\,K} \geq 3.8 \pm 1.3$ (see Eq.4.5). Taking into account that the decrease of $T_S$ is entirely due to enhancement of spin relaxation at room temperature (electron lifetime $\tau$ slightly increases while spin scattering time $\tau_S$ decreases drastically [18]):

$$\frac{T_S/\tau\ (300\mathrm{K})}{T_S/\tau\ (80\mathrm{K})} \leq \frac{T_S\ (300\mathrm{K})}{T_S\ (80\mathrm{K})} \Rightarrow\ \tau/T_S\ (300\mathrm{K}) \geq \frac{T_S\ (80\mathrm{K})}{T_S\ (300\mathrm{K})} \cdot \tau/T_S\ (80\mathrm{K})\,.$$

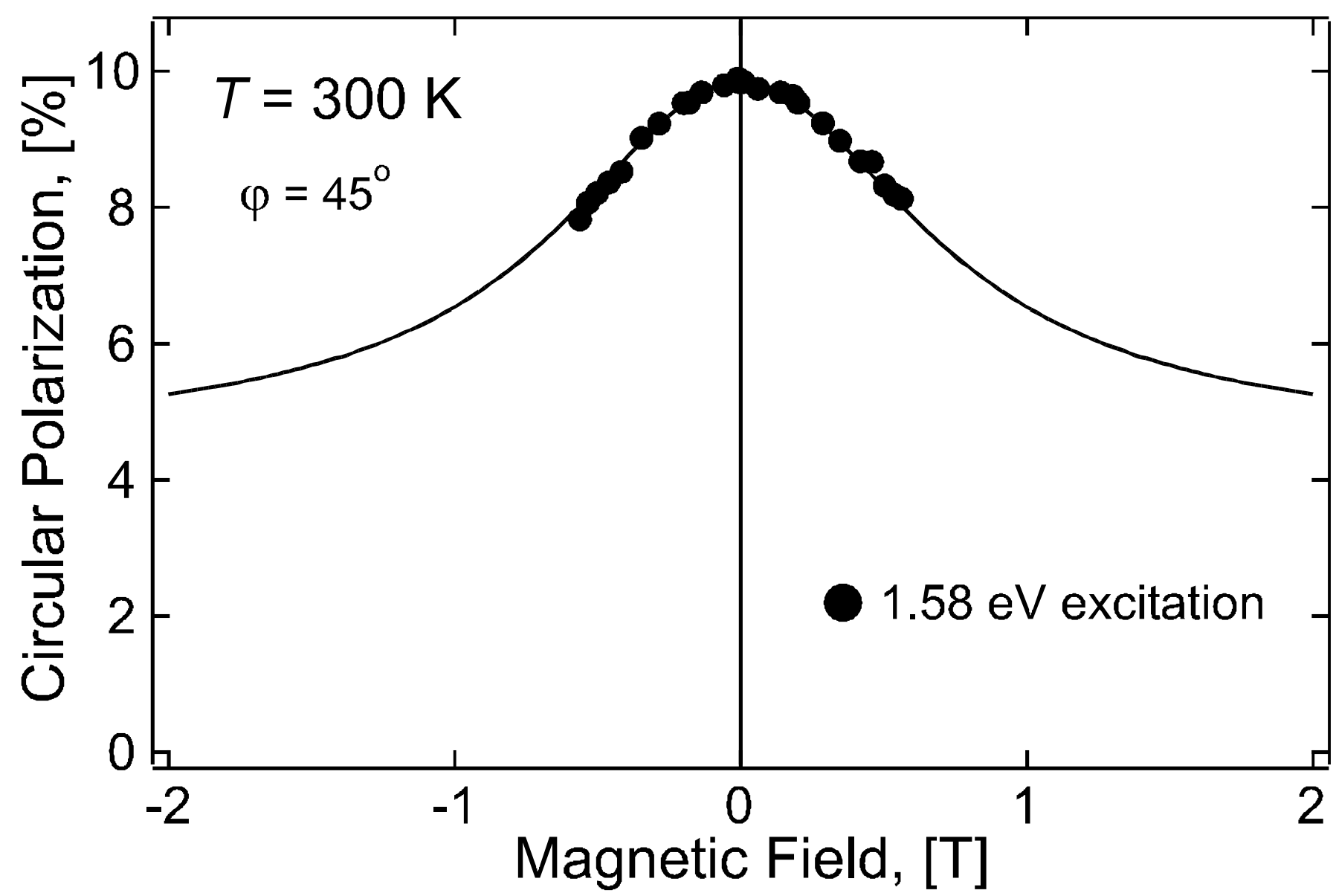

Fig.6.14. Damping of circular polarization of photoluminescence under optical excitation with $h \cdot \nu = 1.58$ eV ( $P = 100\%$ ) for the sample *Type C* (doped active region) and Hanle fit using Eq.4.4.

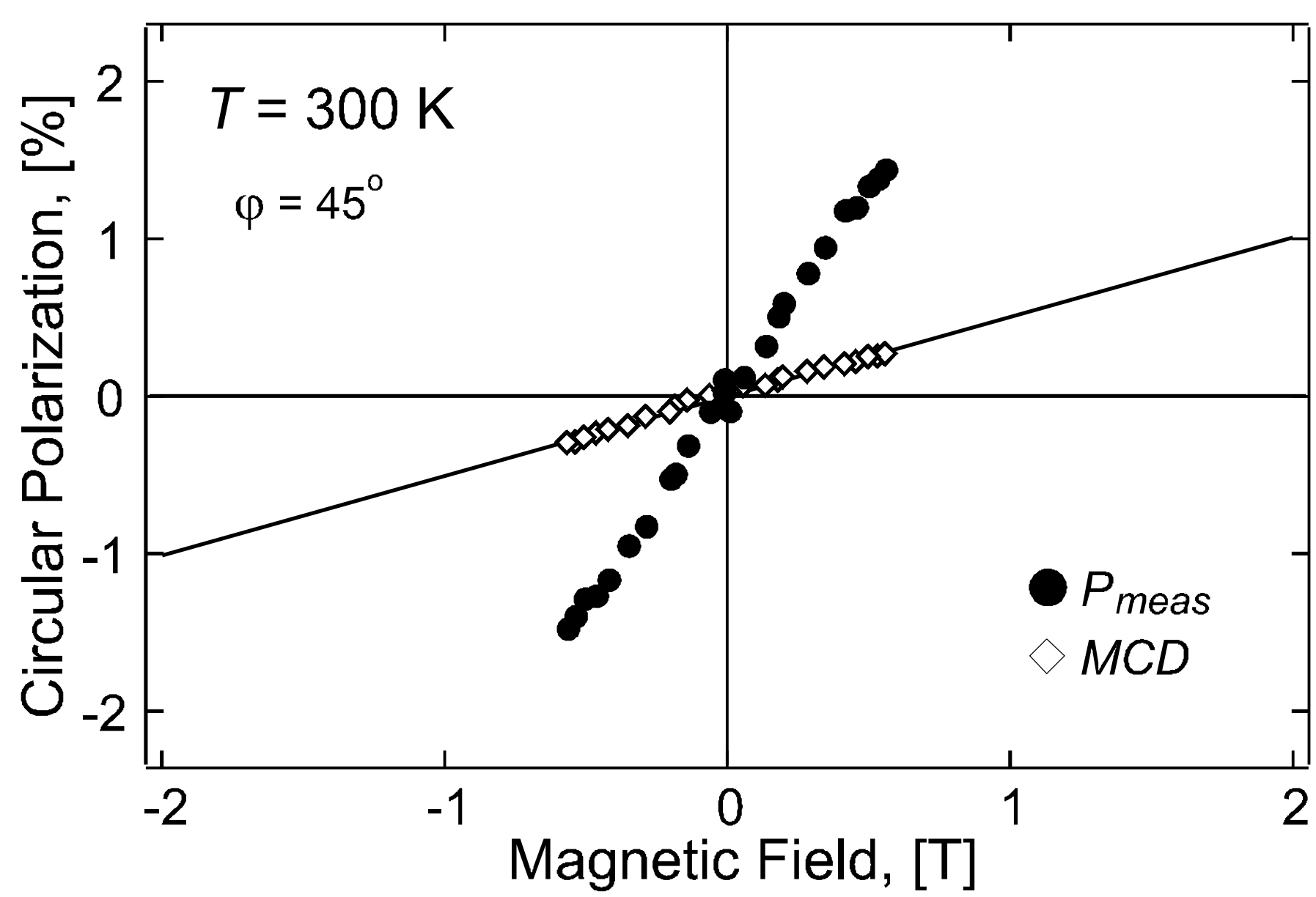

Fig.6.15. Typical result of measurements of the degree of circular polarization of the electroluminescence for the sample *Type C* (doped active region) and MCD contribution.

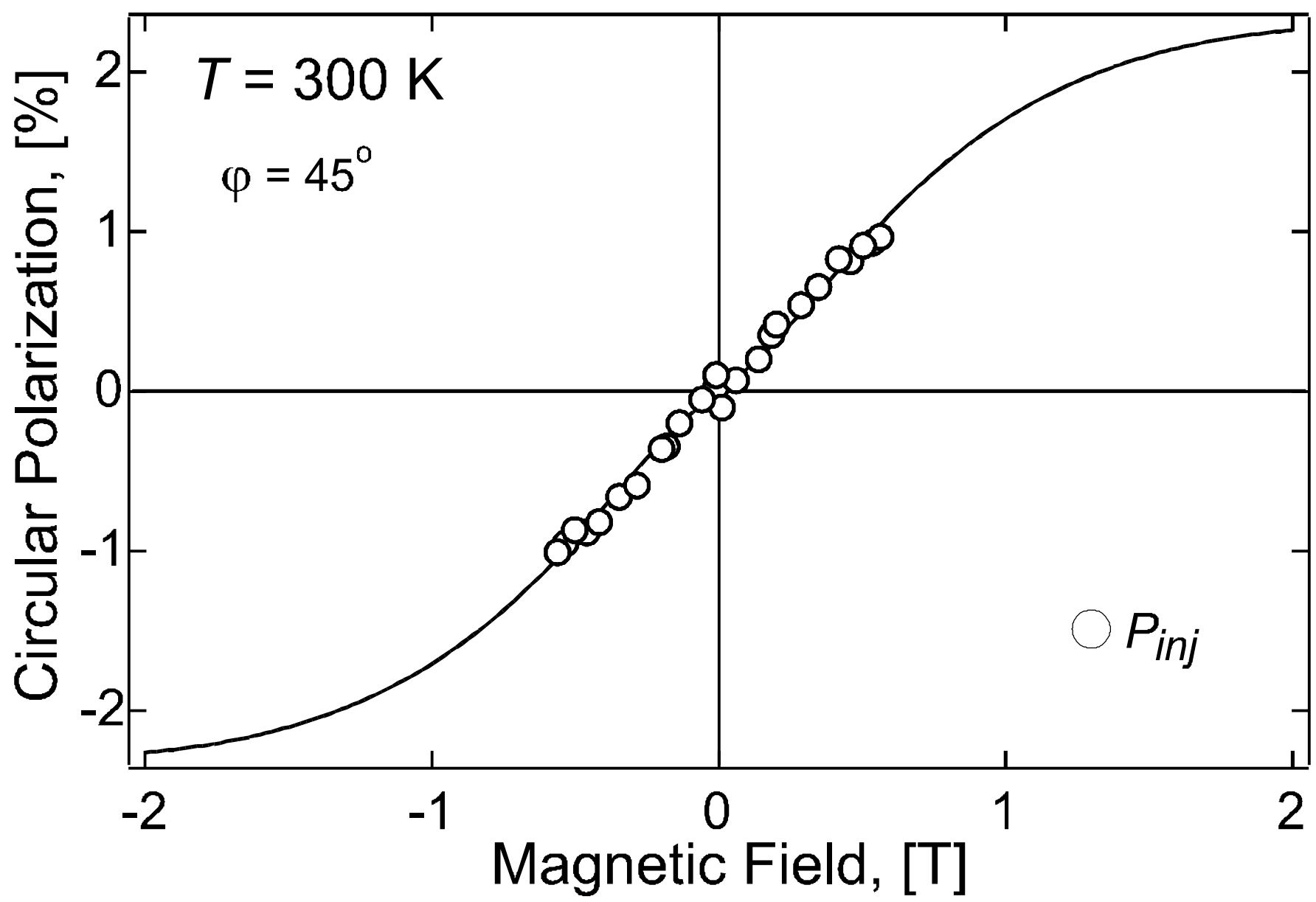

Fig.6.16. The difference between spin injection and MCD curves on Fig.6.15, and Hanle fit using Eq.4.8, from which the degree of injected spin polarization is inferred: $\Pi_{inj} = (16 \pm 2)\ \%$.

Since the spin scattering term $\tau/T_S\,(80\ K) = (\tau_S + \tau)/\tau_S > 1$, one can easily obtain a lowest limit of spin scattering term $\tau/T_S(300K) > 3.8 \pm 1.3$ and injected spin polarization $\Pi_{inj} \geq (19 \pm 7)\ \%$ at room temperature.

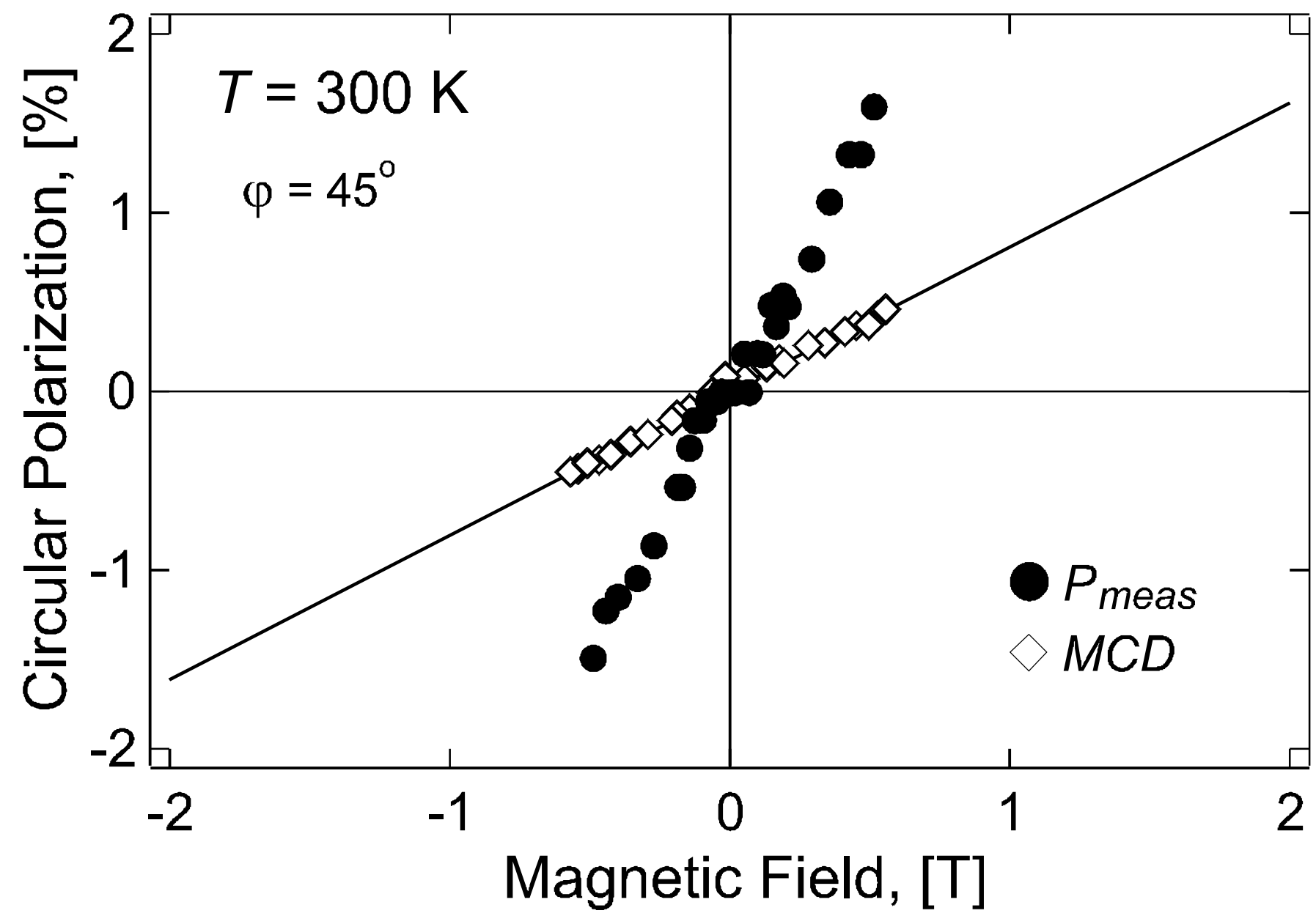

Fig.6.17. Typical result of measurements of the degree of circular polarization of the electroluminescence and MCD contribution in the external oblique magnetic field for the sample *Type D* (undoped active region).

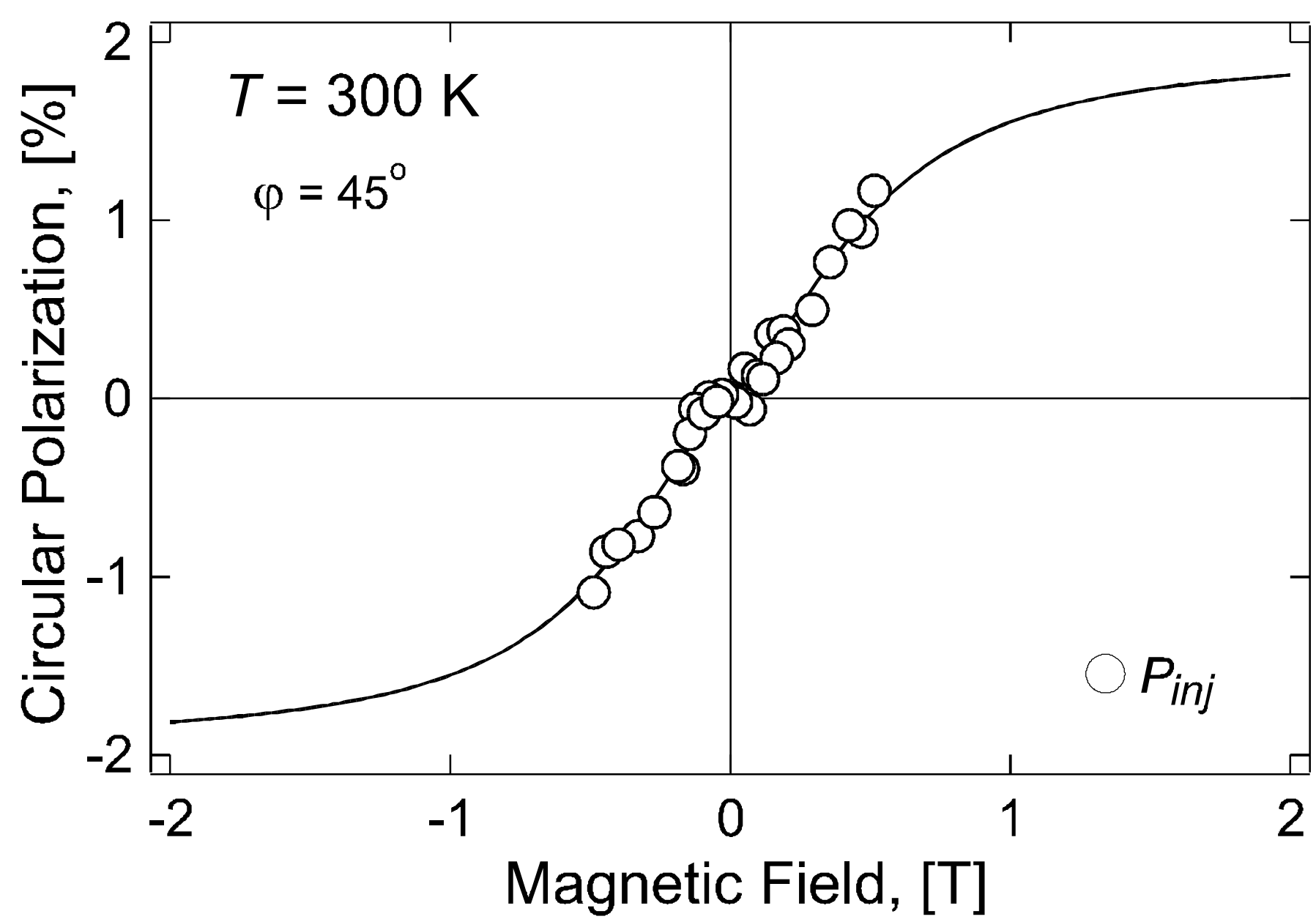

Fig.6.18. The difference between electroluminescence and MCD curves on Fig.6.17. The Hanle fit (Eq.4.8) reveals the degree of injected spin polarization normalized to spin scattering parameter $T_S/\tau$, $\Pi_{inj}\cdot T_S/\tau \geq (5\pm1)$ %.

### 6.3.3. Influence of Electrical Bias

As developed measurement technique allows simultaneous measurements of injected spin polarization and spin dynamics inside of the semiconductor (half-width of the Hanle curve), it is interesting to look how these parameters change as a function of the electrical bias applied to the device (devices are biased using electrical contacts to the ferromagnetic metal and substrate, see Section 5.3.2).

Due to historical reasons the measurements of experimental bias dependencies for the sample *Type D* (undoped active region) are presented first. Fig.6.19 and Fig.6.20 show the typical experimental bias dependencies of the injected spin polarization and half-width of Hanle curve for this sample. Fig.6.21 shows the experimental Hanle curves corresponding to the experimental points marked as triangle and diamond on these figures.

As one can see these measurements show quite strong reduction of the injected spin polarization with bias. Here any possibility of TB degradation or of its interfaces on the result of measurements can be completely excluded, since experimental points were taken with increasing as well as decreasing bias sequence. Moreover, another surprising fact is the narrowing of the Hanle curve with increase of electrical bias. Even though this change is not very important (~1.5 times), going a little bit further, it was observed on all measured devices on all fabricated samples (even for the *Type B* samples). This change cannot be attributed to the heating of the sample at higher biases, when higher Joule energy is dissipated in the device. The heating causes the opposite effect - acceleration of the spin relaxation and increase of the half-width of Hanle curve $\Delta B$.

Such narrowing is very difficult to explain from the fundamental point of view. This effect can be explained in terms of the so-called cascade process during electron thermalization in the semiconductor [18] or taking into account double-step electron tunneling through the intermediate defect states within tunnel barrier (Fig.6.22).

In any case before electron arrives on the bottom of the conduction band (state 2) it can spend some time in the state 1 (defect state within tunnel barrier or multiple change of states during thermalization).

The double-step tunneling was originally presented in Ref [184] in order to explain the bias dependencies of TMR effect in magnetic tunnel junctions. The key point is the uniform distribution of defect states within tunnel barrier. Taking into account the Fermi-Dirac function $f(E)=\left\{1+\exp\left[{(E_C-E)}/{k\cdot T}\right]\right\}^{-1}$ character of the available defect states, their density exponentially increases with the increase of the energy level. Hence, the two-step tunneling increases quickly with increase of electrical bias and is dominant electron transport mechanism at higher biases. The defect states within tunnel barrier are known to be paramagnetic, thus some spin scattering of electrons on this levels occurs. As result the spin polarization of electrons injected into semiconductor is significantly lower. Further, spin precession in the oblique magnetic field during electron lifetime on such level may lead to change of preferential spin orientation. As result, the average electron spin $\vec{S}$ of injected electrons is different from the case described in Chapter 4. Such process may lead to narrowing of the effective halfwidth of the experimental Hanle curves. Numerically, the evolution of average electron spin in this case is described by the two equations similar to Eq.4.1.

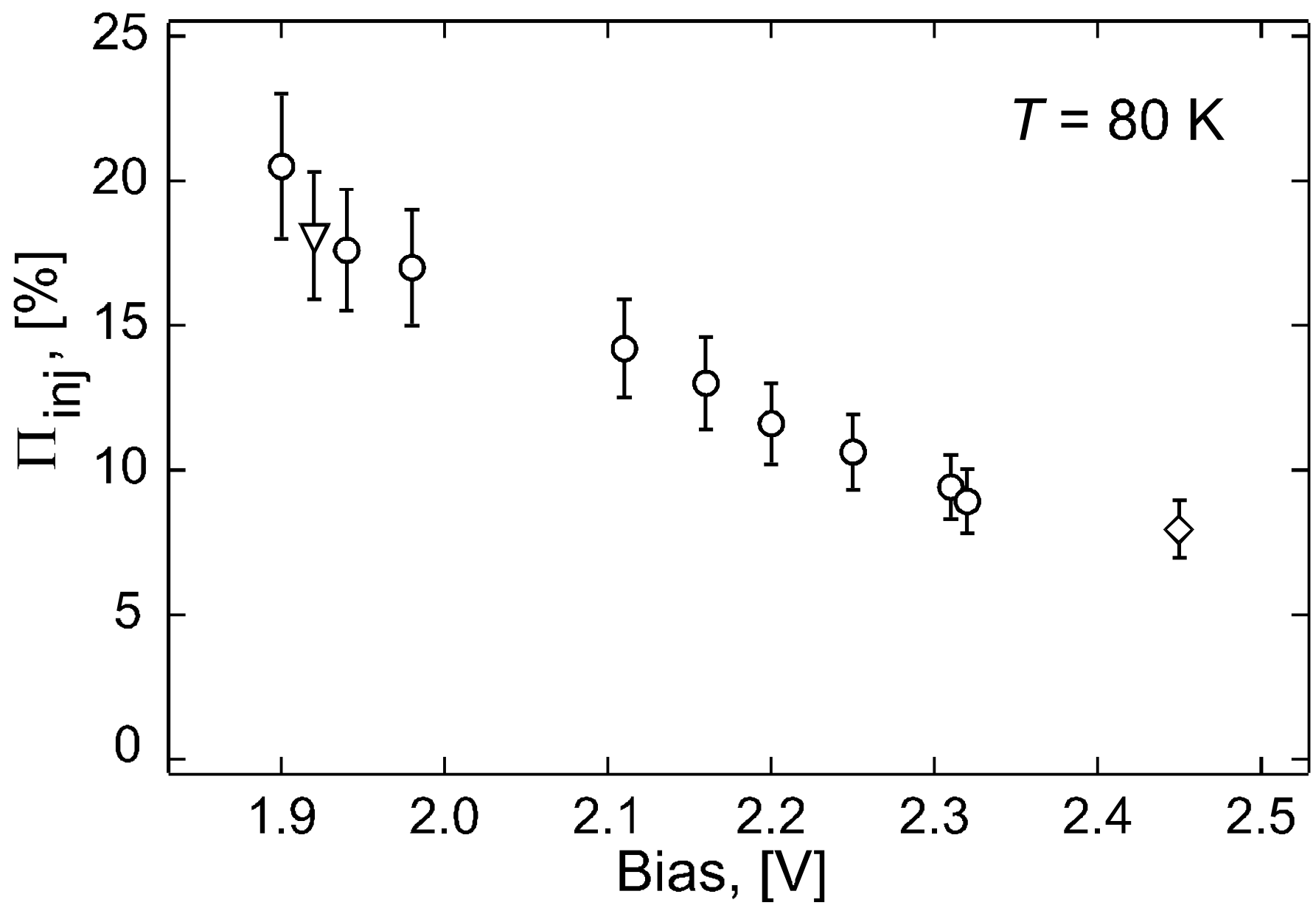

Fig.6.19. The typical experimental bias dependencies of the injected spin polarization for the sample *Type D* (undoped active region). The experimental Hanle curves corresponding to the experimental points marked as triangle and diamond are presented in Fig.6.21.

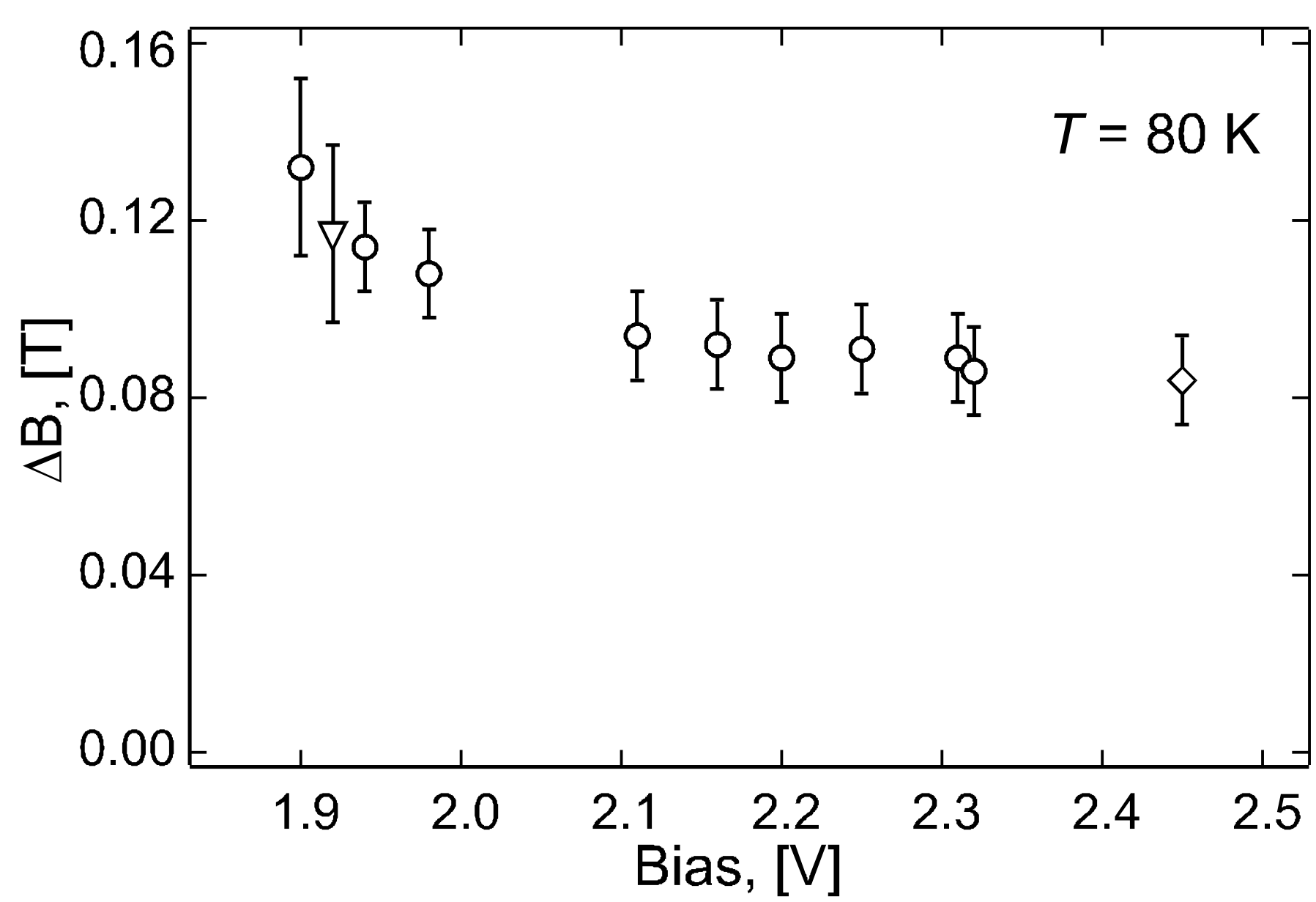

Fig.6.20. The typical experimental bias dependencies of the half-width of Hanle curve for the sample *Type D* (undoped active region). The experimental Hanle curves corresponding to the experimental points marked as triangle and diamond are presented in Fig.6.21.

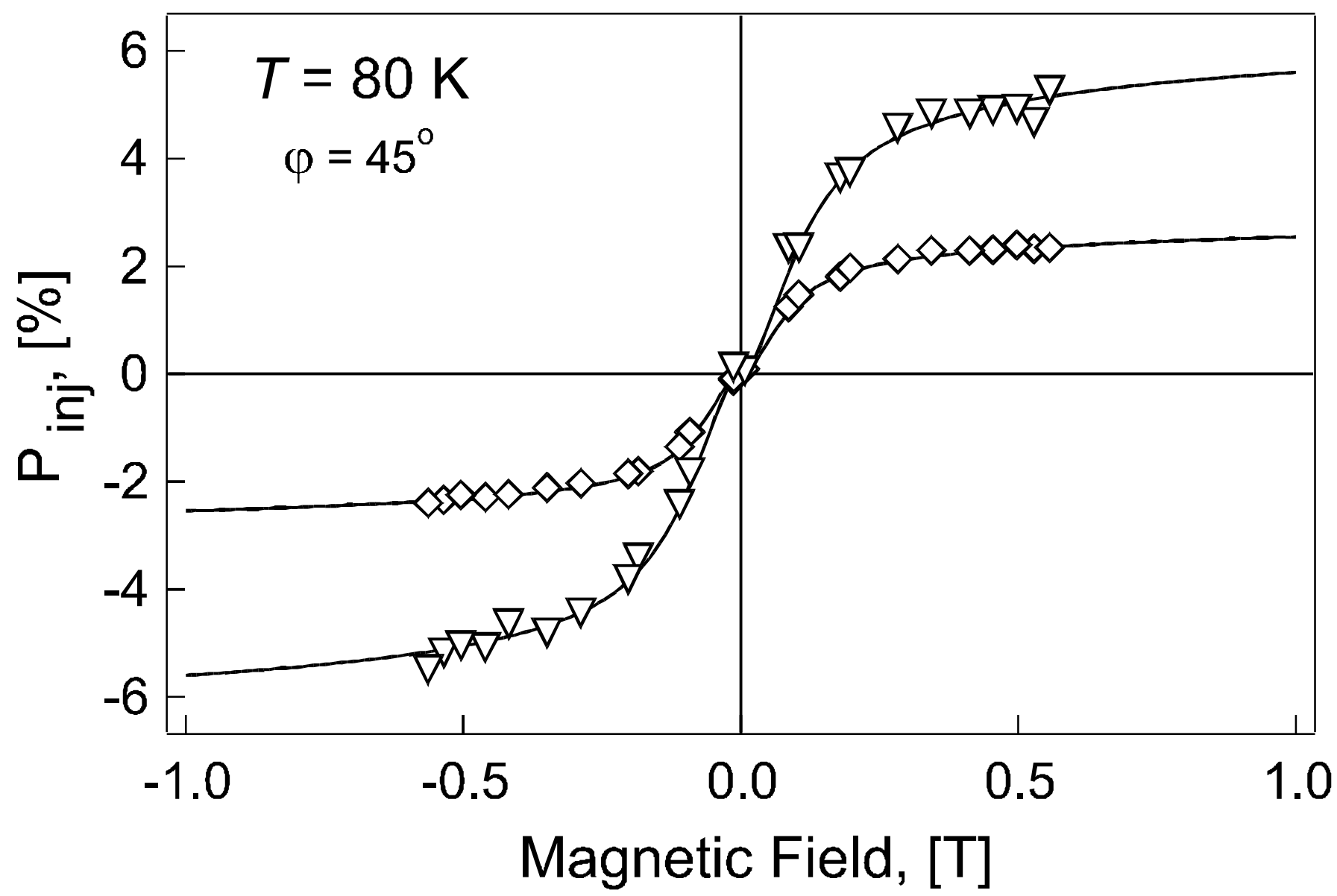

Fig.6.21. The experimental Hanle curves corresponding to the experimental points marked as triangle and diamond on Fig.6.19 and Fig.6.20.

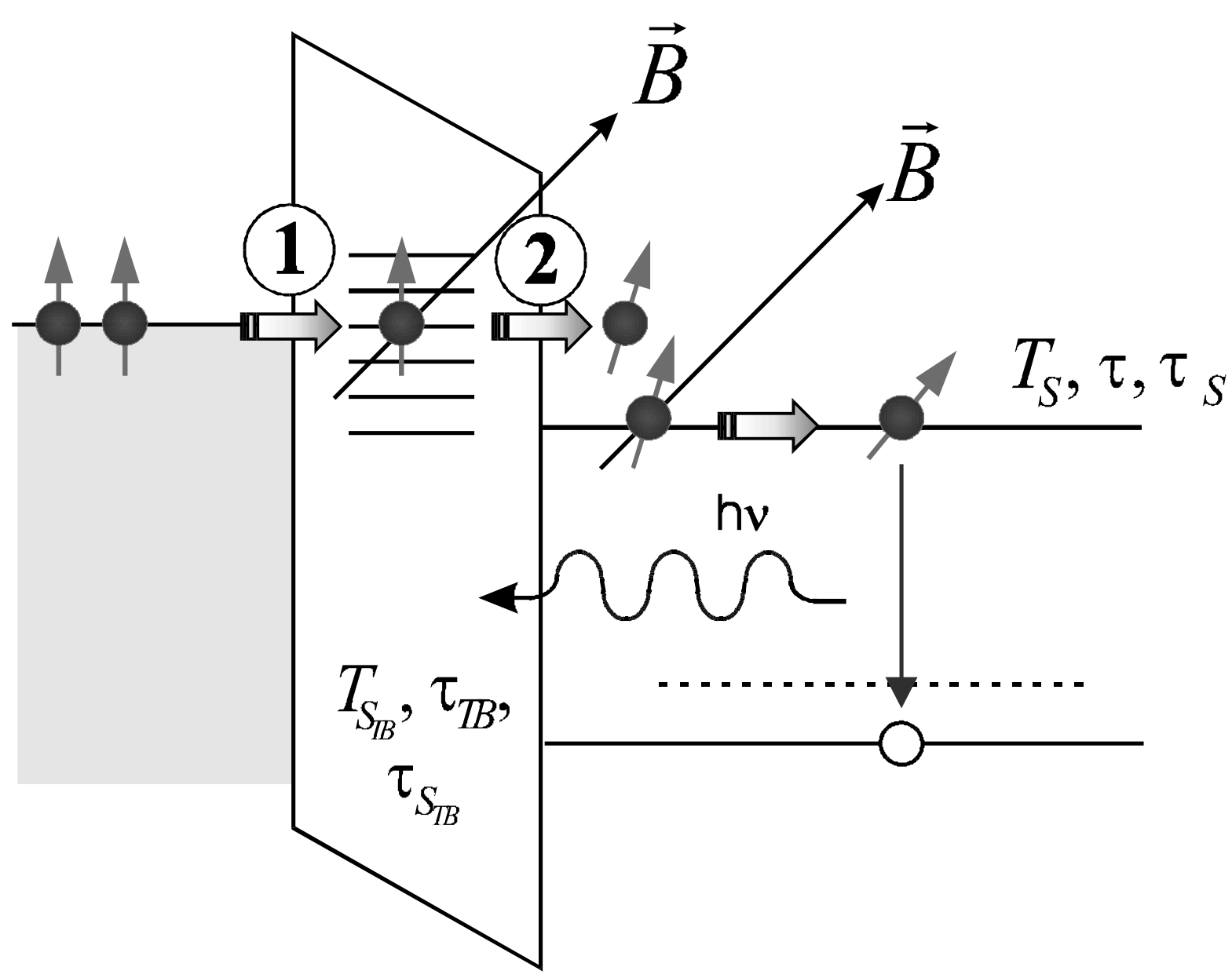

Fig.6.22. Double-step electron tunneling process on the FM/ $AlO_X$/ Al(GaAs) interface. The spin precession on the defect states in the band gap of the insulating layer leads to injection into semiconductor of electrons with the average electron spin $\vec{S}_{0_{TB}}(S_{0X}, S_{0Y}, S_{0Z})$, which is strongly reduced due to the paramagnetic nature of these states.

$$\begin{cases} \dfrac{d\vec{S}_{TB}}{dt} = \dfrac{\vec{S}_0}{\tau_{TB}} - \dfrac{\vec{S}_{TB}}{T_{S_{TB}}} + \left[\vec{\Omega}_{TB} \times \vec{S}_{TB}\right] \\ \dfrac{d\vec{S}}{dt} = \dfrac{\vec{S}_{TB}}{\tau} - \dfrac{\vec{S}}{T_S} + \left[\vec{\Omega} \times \vec{S}\right] \end{cases} \qquad (6.1)$$

Here the first equation describes the evolution of average electron spin on the defect states within tunnel barrier ($T_{S_{TB}}$, $\tau_{TB}$, $g^*_{TB}$) and the second one the conventional evolution of average electron spin within a semiconductor (see Section 4.2). Again under steady state conditions using Eq.4.3 one can easily find the solution for this system. In fact, for experimental verification of this model, sample *Type C* was fabricated.

Fig.6.23 and Fig.6.24 show the typical experimental bias dependencies of the injected spin polarization and half-width of Hanle curve for the sample *Type C*. Fig.6.25 shows the experimental Hanle curves corresponding to the experimental points marked as triangle and diamond on these figures.

As one can see, the spin polarization of electrically injected electrons practically does not change with bias. Such behavior suggests that the effect is associated with properties of the active region of the device and not with the spin injection. The effect can be related to the loss of spin polarization during the thermalization of hot electrons. For higher biases the electrons injected into the active region have higher kinetic energy. From all-optical measurements it is known that the effect of spin scattering and loss of polarization during thermalization is much stronger in samples with lower doping level (DP mechanism, [18], see Sections 3.3.2, 3.3.5, 4.6). This can explain the observed difference between the samples *Type C* and *Type D*.

However, the sample *Type C* shows the same effect of narrowing of the Hanle curve with increase of electrical bias. At low bias the half-width of the Hanle curve corresponds to the one observed in all-optical measurements, and then decreases to lower values at higher biases.

The narrowing of the Hanle curve can also find an explanation in terms of thermalization in cascade process. In the cascade process the evolution of average electron spin can be described by similar to Eq.6.1 system of n equations, where n is number of intermediate states. If the thermalization is slow enough, the spin precession due to external magnetic field during the thermalization cannot be neglected. As was observed in all-optical experiments, this leads to deformation and narrowing of the Hanle curve [186].

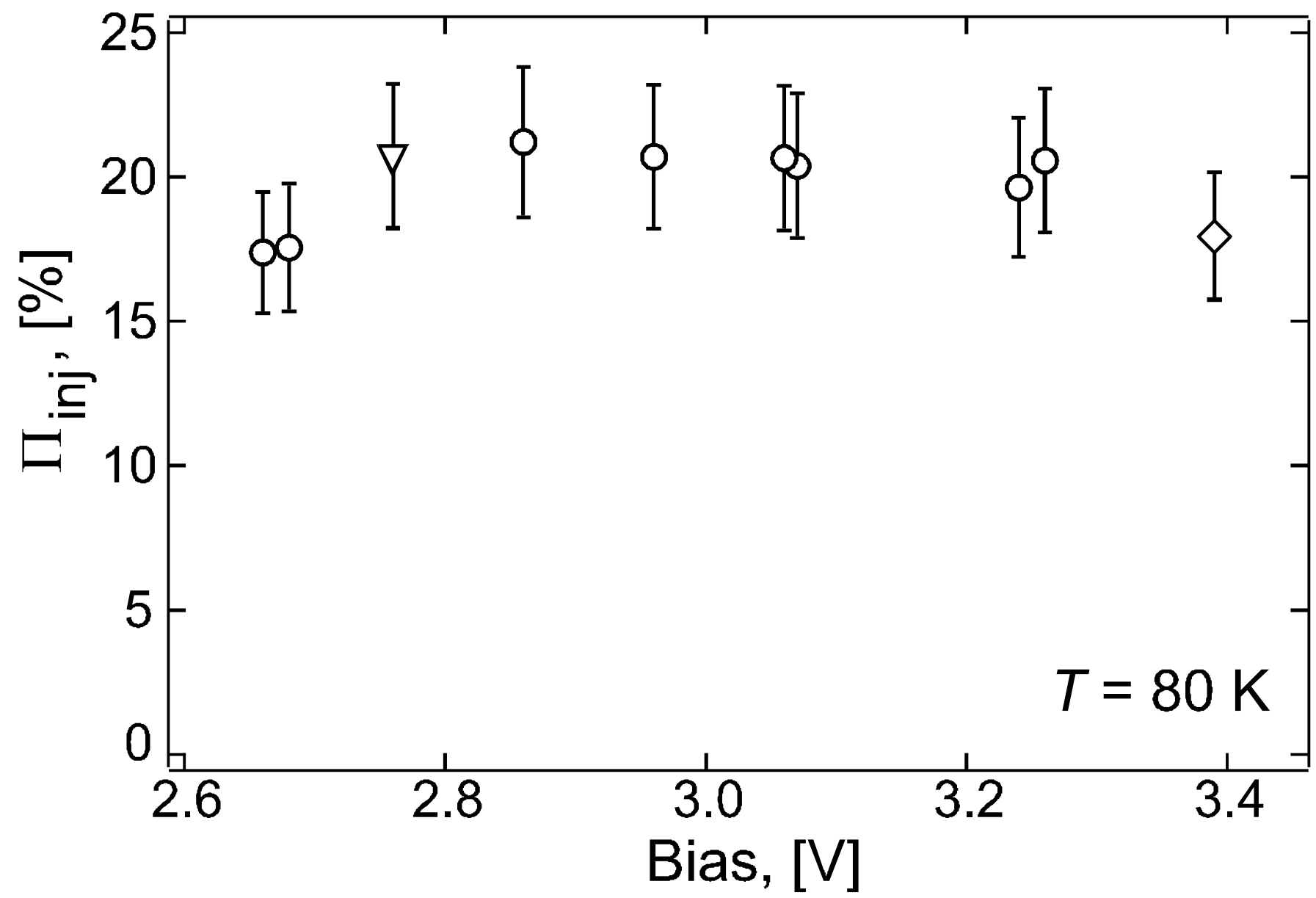

Fig.6.23. The typical experimental bias dependencies of the injected spin polarization for the sample *Type C* (doped active region). The experimental Hanle curves corresponding to the experimental points marked as triangle and diamond are presented in Fig.6.25.

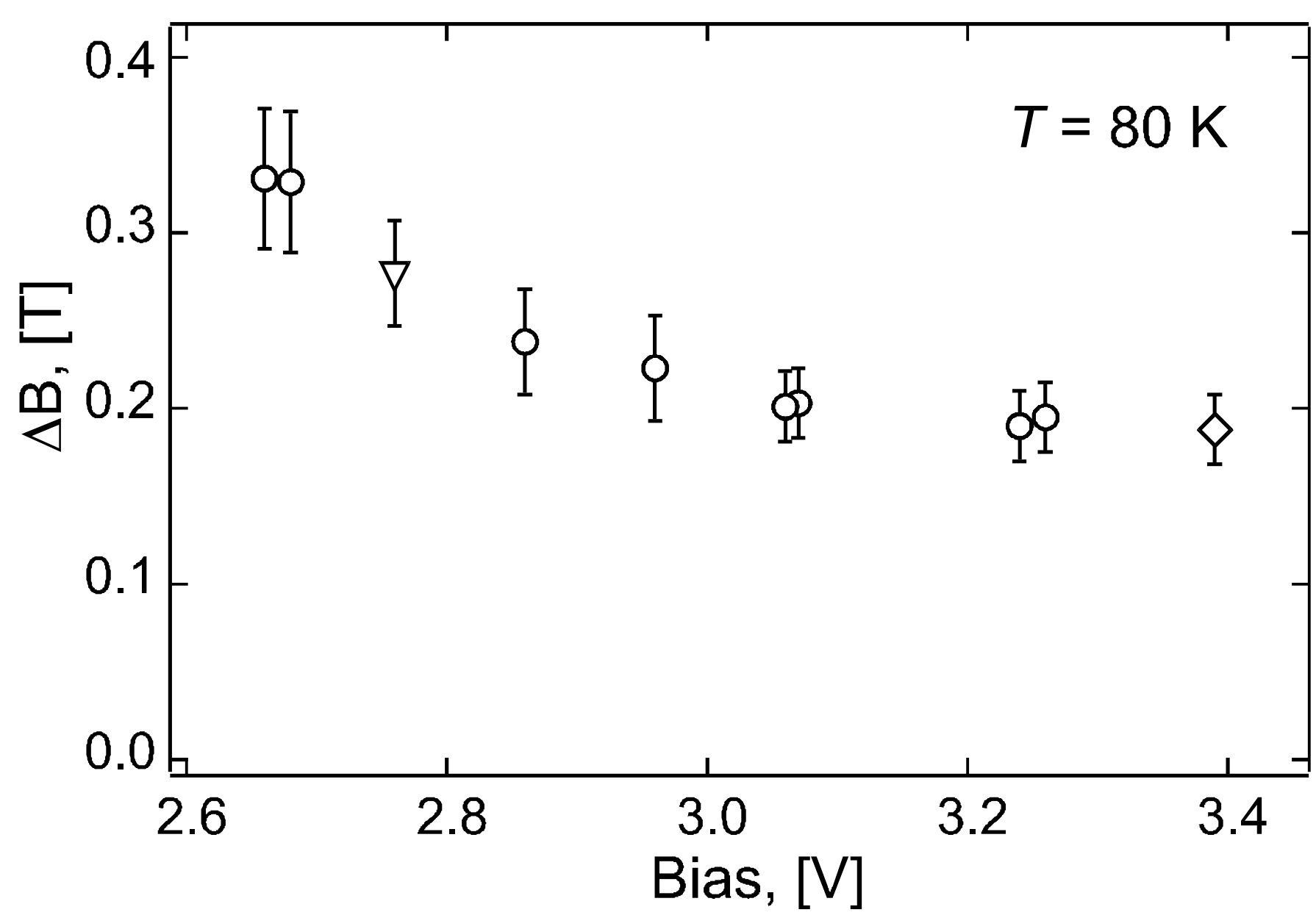

Fig.6.24. The typical experimental bias dependencies of the half-width of Hanle curve for the sample *Type C* (doped active region). The experimental Hanle curves corresponding to the experimental points marked as triangle and diamond are presented in Fig.6.25.

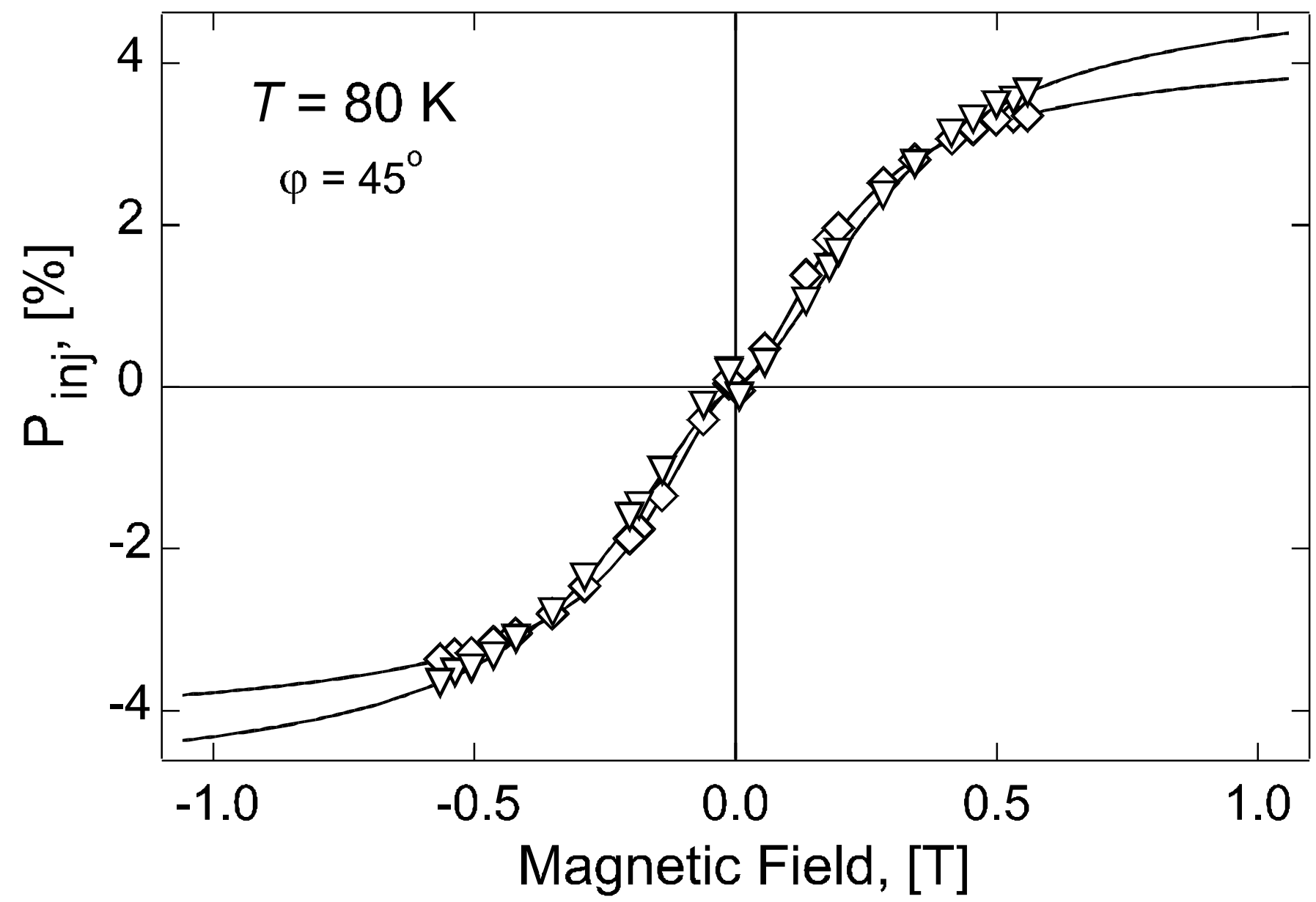

Fig.6.25. The experimental Hanle curves corresponding to the experimental points marked as triangle and diamond on Fig.6.23 and Fig.6.24.

Another factor which may lead to the change of the half-width of Hanle curve is change of effective electron g-factor. The effective g-factor of electrons in the AlGaAs layers, used to form active region of the fabricated spin-LEDs, is different from the one in GaAs- the active region of the device. Moreover, the g-factor in these semiconductors differ even in sign. Hence, displacement of electron wave function in the active region of the device by electrical bias may result in the change of effective g-factor (see references within Section 3.3.6) and thus change of half-width of Hanle curve. However, further investigation is needed for confirmation of any of these models.

Room temperature measurements have shown the similar tendencies of injected electron spin polarization bias dependencies. Fig.6.26 and Fig.6.28 show the typical experimental bias dependencies of the injected spin polarization at room temperature for the samples *Type C* and *Type D*, respectively. Fig.6.27 and Fig.6.29 show the experimental Hanle curves corresponding to the experimental points marked as triangle and diamond on these figures.

Unfortunately, the lack of saturation does not allow reliable experimental investigation of the electrical bias dependencies of the half-width of Hanle curve at 300 K (partly presented in [187,188]).

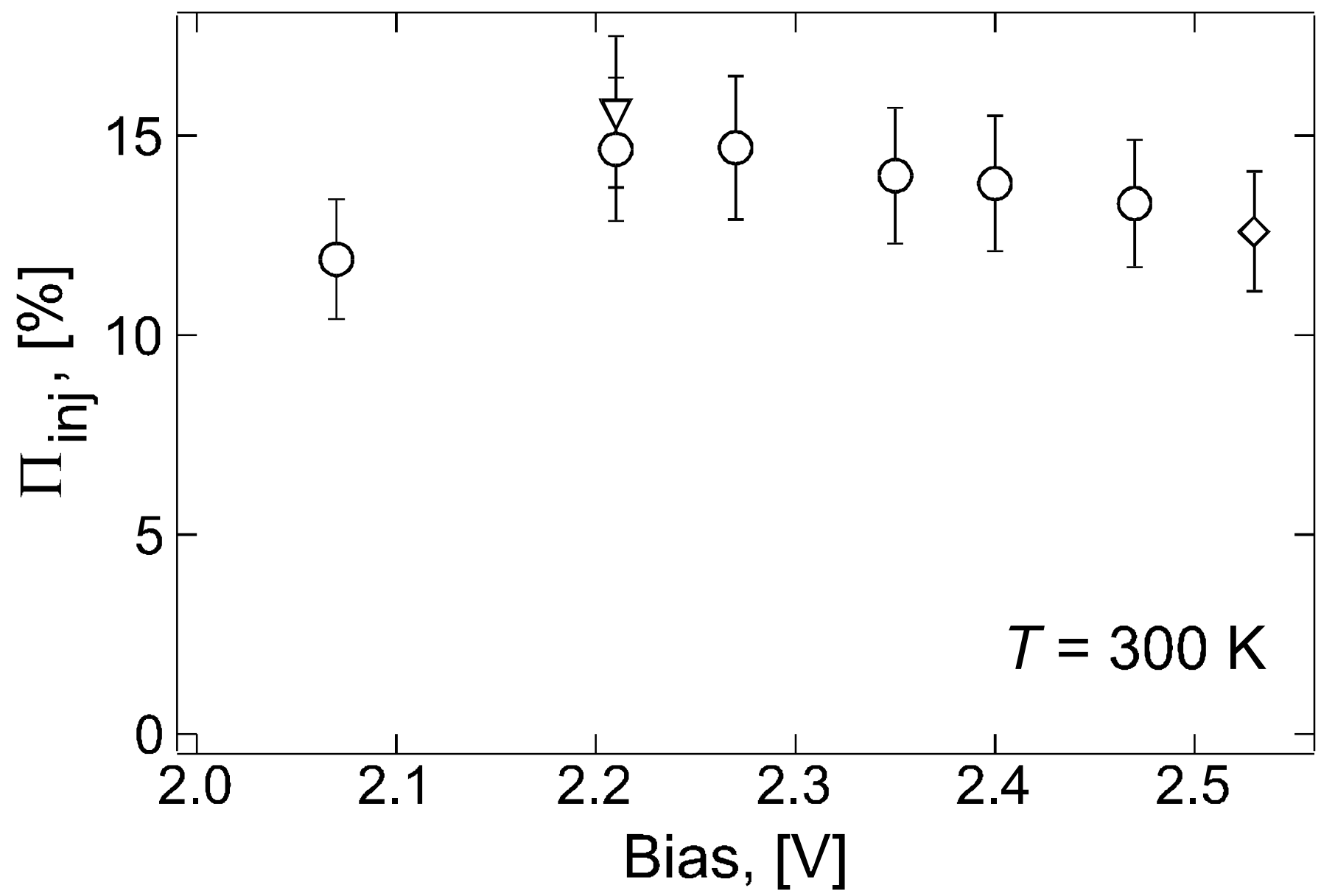

Fig.6.26. The typical experimental bias dependencies of the injected spin polarization for the sample *Type C* (doped active region). The experimental Hanle curves corresponding to the experimental points marked as triangle and diamond are presented in Fig.6.27.

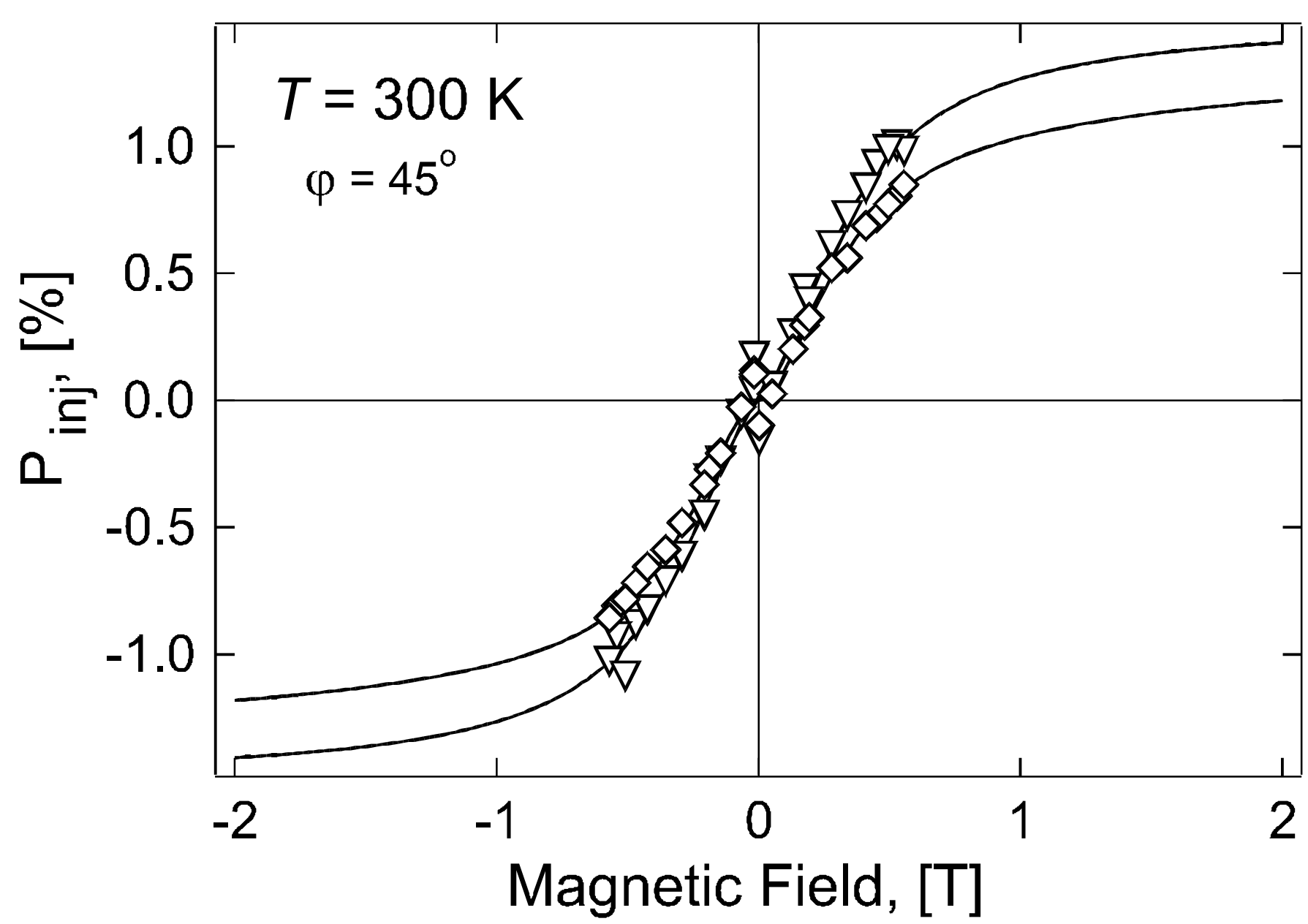

Fig.6.27. The experimental Hanle curves corresponding to the experimental points marked as triangle and diamond on Fig.6.26.

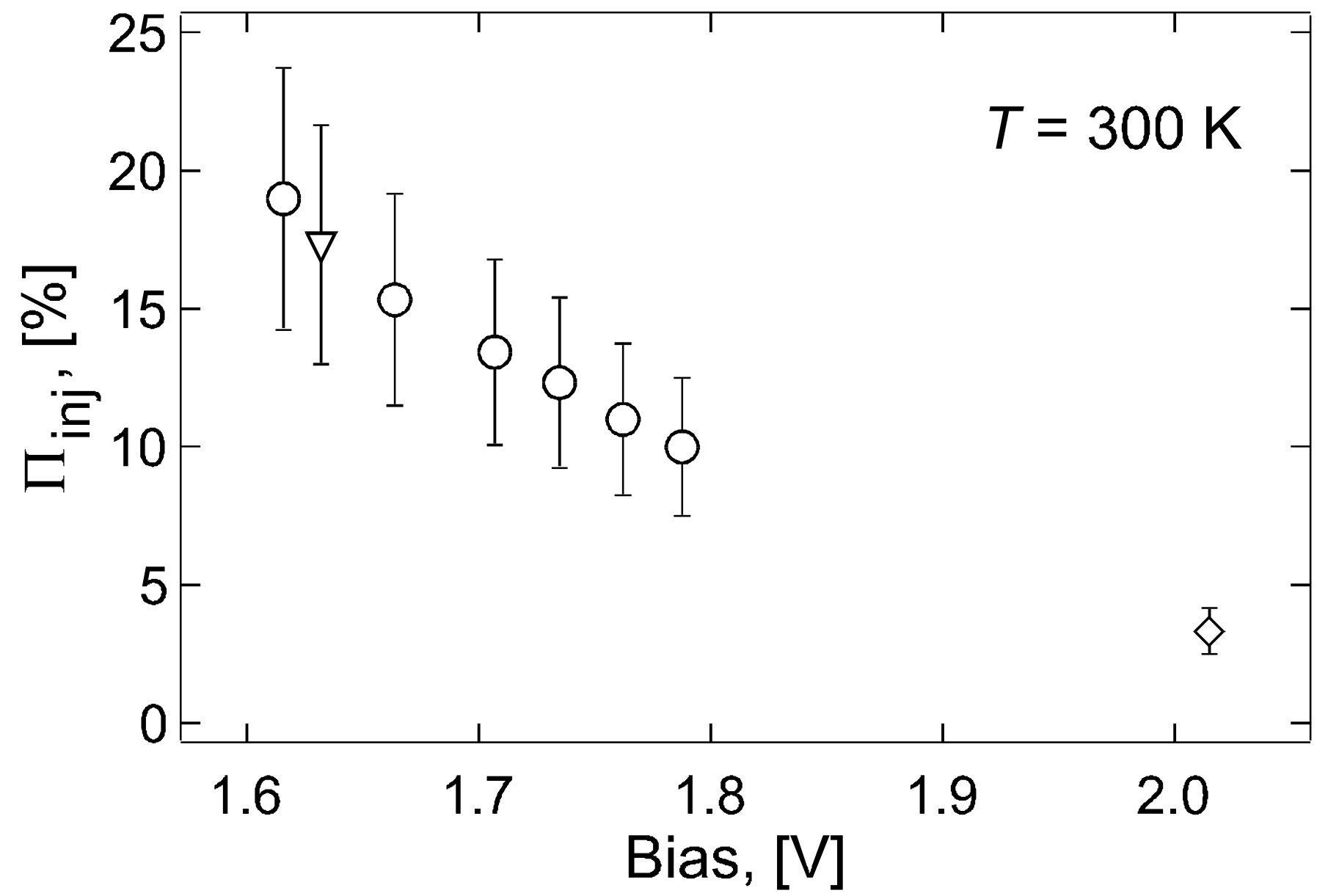

Fig.6.28. The typical experimental bias dependencies of the injected spin polarization for the sample *Type D* (undoped active region). The experimental Hanle curves corresponding to the experimental points marked as triangle and diamond are presented in Fig.6.29.

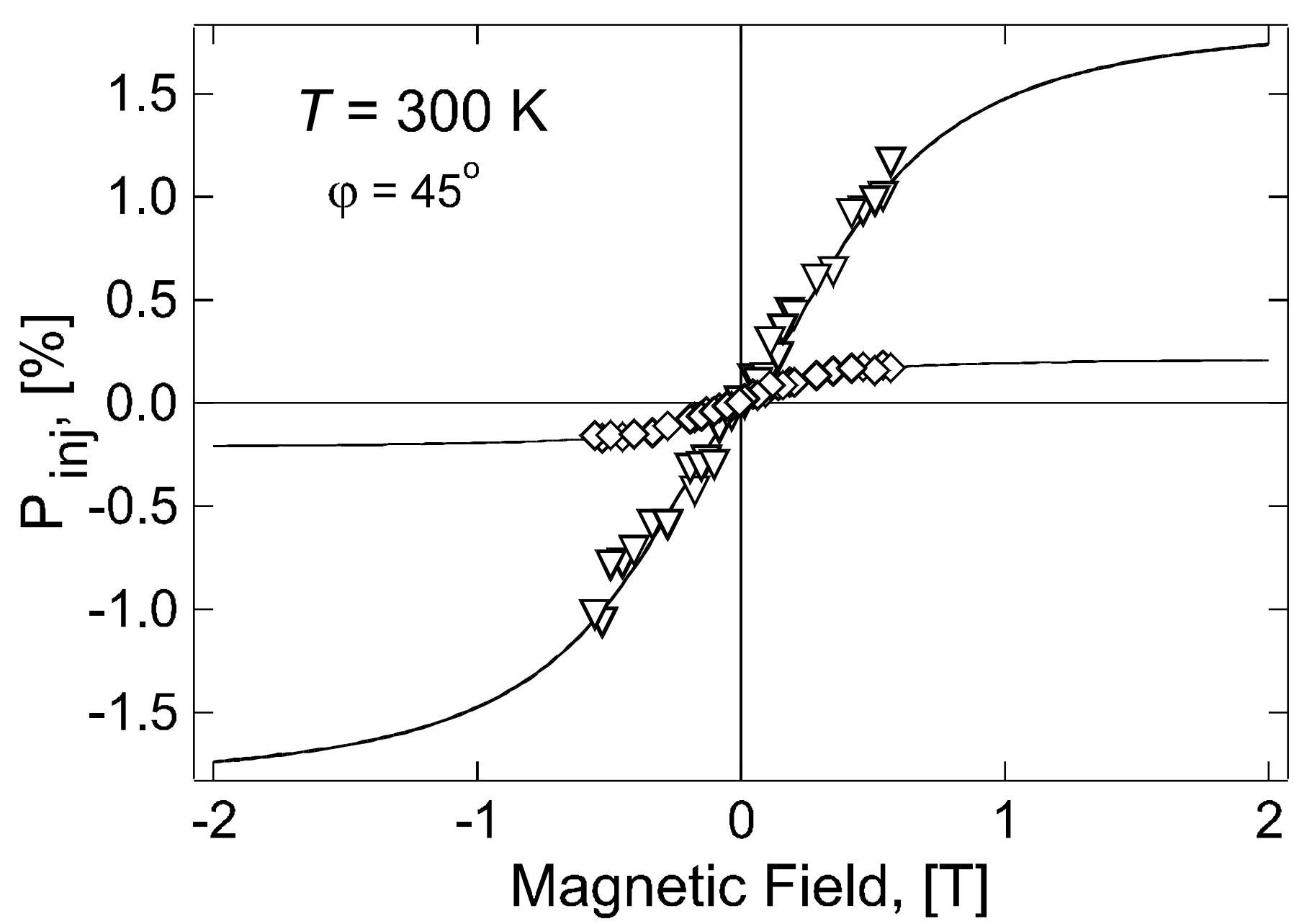

Fig.6.29. The experimental Hanle curves corresponding to the experimental points marked as triangle and diamond on Fig.6.28.

## 6.4. Electrical Spin Injection in the Sample *Type E*: Nuclear Spin Polarization

In the previous sections it was shown that electrical spin injection into semiconductors using ferromagnetic metals indeed looks very promising for utilization in spintronic applications. The high spin polarization of electrically injected electrons has been demonstrated even at room temperature. As it was shown in the all-optical studies of spin-dependent effects in semiconductors, the large electron spin polarization creates a fluctuating local magnetic field, which may lead to dynamic polarization of nuclear spins via hyperfine coupling [18, 136]. This nuclear spin momentum creates a local effective magnetic field acting back on the spins of electrons, which can be detected in the optical measurements [189, 190, 18]. The injection of spin-polarized electrons in an electrical contact opens new possibilities for experimental investigation of effects caused by nuclear and electron spin coupling in the III-V semiconductors. Moreover, the possibility of effective nuclear polarization by electrical current opens a way for practical realization of devices that can use the spin of nuclei in order to process and store information. For the alternative semiconductor- Si, having the most advanced technology available, some of such devices have been proposed already [191, 192]. Here at low temperatures, the combination of fascinatingly long electron spin scattering times (thousands of seconds) combined with as fascinatingly long nuclear spin relaxation times (~$10^{18}$s at millikelvin temperatures, references within Ref.[192]) may lead to creation of prototype of the most advanced logic and memory ever.

However, measurements of the electrical spin injection in the samples described in the previous sections have not revealed any signs of nuclear spin polarization, due to both, high temperature during measurements and relatively short electron spin lifetime $T_S$ ($T_S^{-1} = \tau^{-1} + \tau_S^{-1}$) in the active region of the device. Based upon experience of observation of the effects caused by the spin polarization of nuclei in all-optical experiments [18, 189], the sample *Type E* has been fabricated.

The sample *Type E* (Fig.6.30 right, Table 5.1) consists of 10nm CoFe/ 2.6nm $AlO_X$ (double-step oxidation)/ 15nm AlGaAs (undoped)/ 100nm GaAs ($N_A$=4·$10^{17}$cm$^{-3}$, $N_A/N_D$=2) compensated active region/ 200nm AlGaAs (p=2·$10^{18}$cm$^{-3}$)/ 2μm GaAs (p=2·$10^{18}$cm$^{-3}$) buffer/ p-GaAs substrate MIS spin-LED heterostructure. In this sample the active region was intentionally engineered for high localization of electrons on the local potential fluctuations (band tails, Fig.6.30 left) within the band gap of the GaAs. The localization of electrons on such states leads to very efficient coupling between the spin of nucleus and surrounding electrons, due to localized character of electron wave function, as well as due to longer electron lifetime and spin scattering time [18, 189, 190]. In this section the results obtained for this sample are presented.

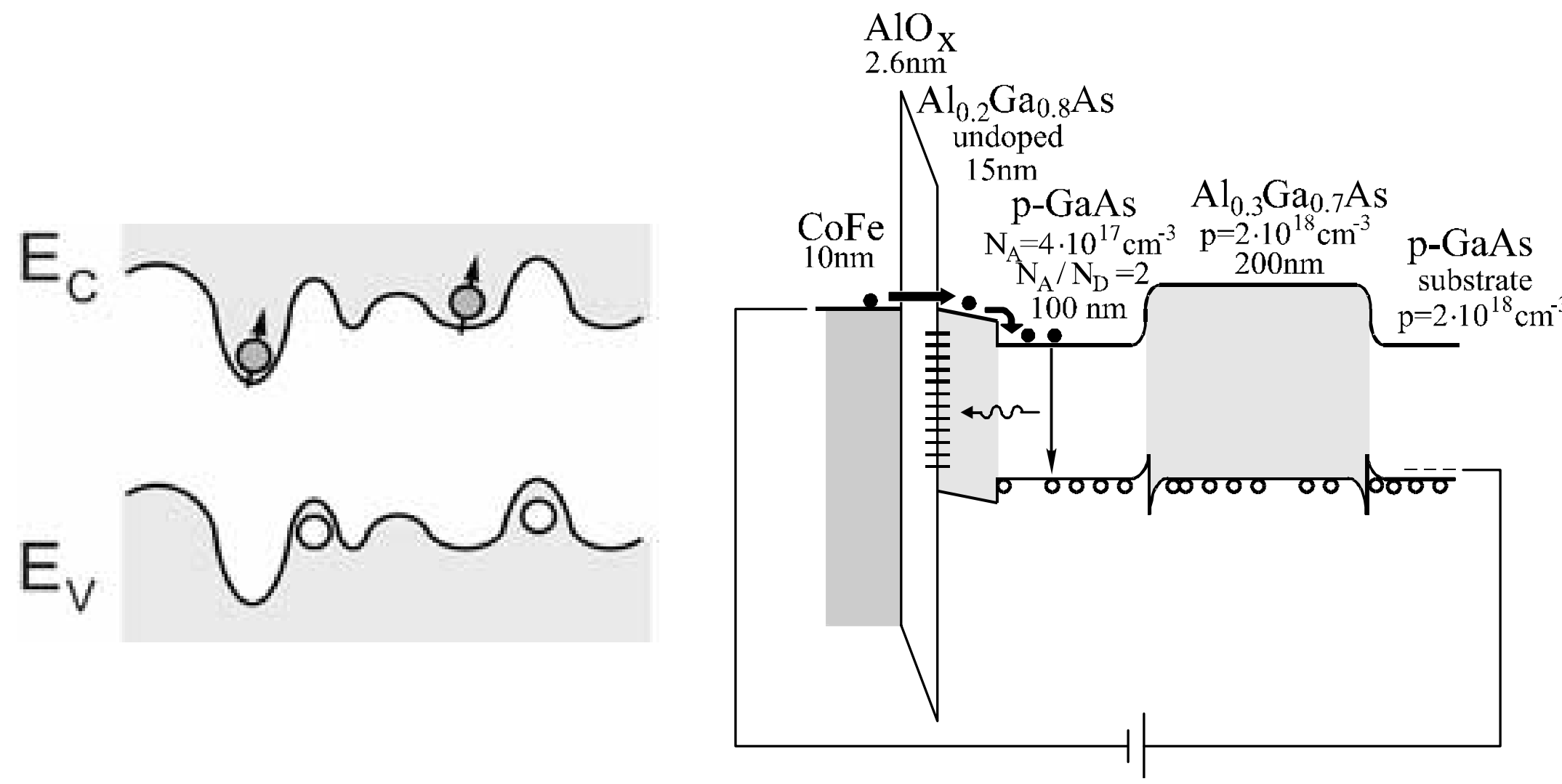

Fig.6.30. Schematic representation of the spin-LED *Type E* (right) and the localized character of the fluctuating potential (band tails) within band gap of the GaAs active region of the device (left).

### 6.4.1. The Manifestation of Nuclear Magnetic Field in the Oblique Hanle Effect Experiment

As described in Ref.[189, 18], the magnetic field $\vec{B}^*$ acting on the electron spin in the presence of a nuclear field consists of an external magnetic field $\vec{B}$ and the effective magnetic field of the nuclei $\vec{B}_N$. The hyperfine interaction tends to polarize the nuclei along the direction of the average electron spin $\vec{S}$. However, the nuclear spin component that is transverse with respect to the external field $\vec{B}$ is destroyed by this field. Hence, with the exception (as explained below) of the weak external magnetic field region

$$B^* = B + B_N \tag{6.1}$$

Fig.6.31 shows the calculated (Eq.4.4 and Eq.6.1) magnetic field dependency of the $S_Z$ component of the average electron spin $\vec{S}$ in the case of optical spin injection (see Fig.4.2, Section 4.2) taking into account the nuclear spin polarization [189, 18]. The maximum of the curve is simply shifted, so that it corresponds to the magnitude of external magnetic field $B = -B_N$.

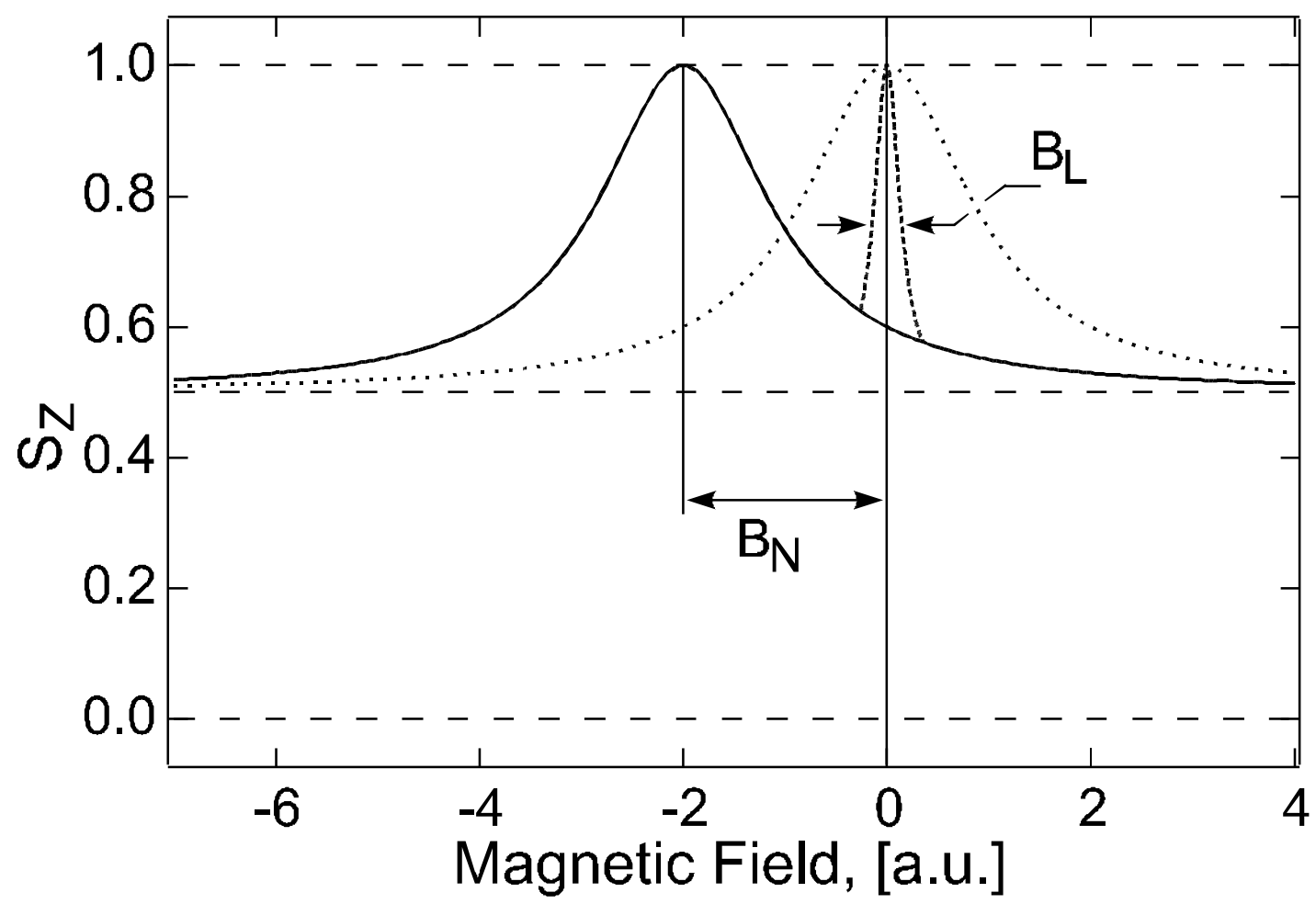

Fig.6.31. The $S_Z$ component of the average electron spin $\vec{S}$ (normalized to $S_0 \cdot T_S/\tau$) under optical spin injection in the oblique Hanle effect geometry ($\varphi = 45^o$, Fig.4.2). The dotted curve is calculated accordingly to Eq.4.4 ($B_N = 0$). The solid curve takes into account the effective nuclear magnetic field $B_N = const$ acting on electrons, Eq.4.4 and Eq.6.1. The dashed curve show the peak in a weak external magnetic field ($B \sim B_L$), where $B_N$ vanishes abruptly.

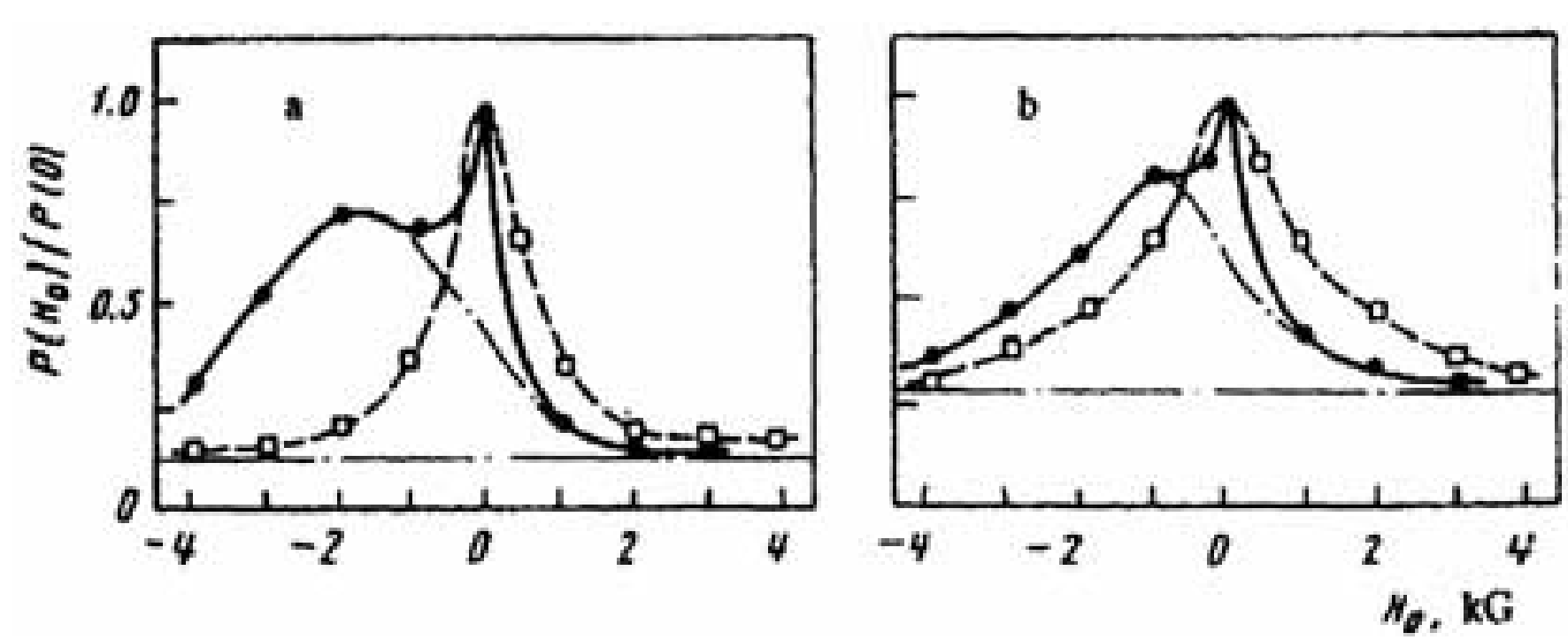

Fig.6.32. Measurements of the circular polarization of photoluminescence as function of external oblique magnetic field in GaAs crystals at: a) 4.2K and b) 77K [189]. The squares correspond to measurements in the absence of nuclear spin polarization (under optical excitation with alternating $\sigma^+$, $\sigma^-$ polarization). The circles correspond to measurements under presence of effective nuclear magnetic field (under optical excitation with $\sigma^+$). The Hanle fit Eq.4.4 (dashed curve) and Eq.4.4, Eq.6.1 (dash-doted curve) are used to fit the data.

In the region of weak external magnetic field $B \sim B_L$, the $B_N$ is not constant, it is small (at $B = 0$ the $B_N = 0$) and is not collinear with $B$. The $B_L$ is characteristic local fluctuating magnetic field due to the dipole-dipole interactions of the nuclei. This field acts on the individual nucleus and is caused by the nuclear spin moment of its neighbors. The nuclear spin precession in this local field leads to overall zero nuclear spin polarization. As result, the $S_Z$ has a sharp maximum (dashed curve, Fig.6.31) in this region. The experimentally observed manifestation of the nuclear field in the all-optical experiment in the oblique Hanle effect geometry is shown in (Fig.6.32).

Following the considerations mentioned above and Eq.4.6, the magnetic field dependency of the $S_Z$ component of the average electron spin $\vec{S}$ in the case of electrical spin injection from the in-plane magnetized ferromagnetic metal (see Fig.4.2), is shown in Fig.6.33.

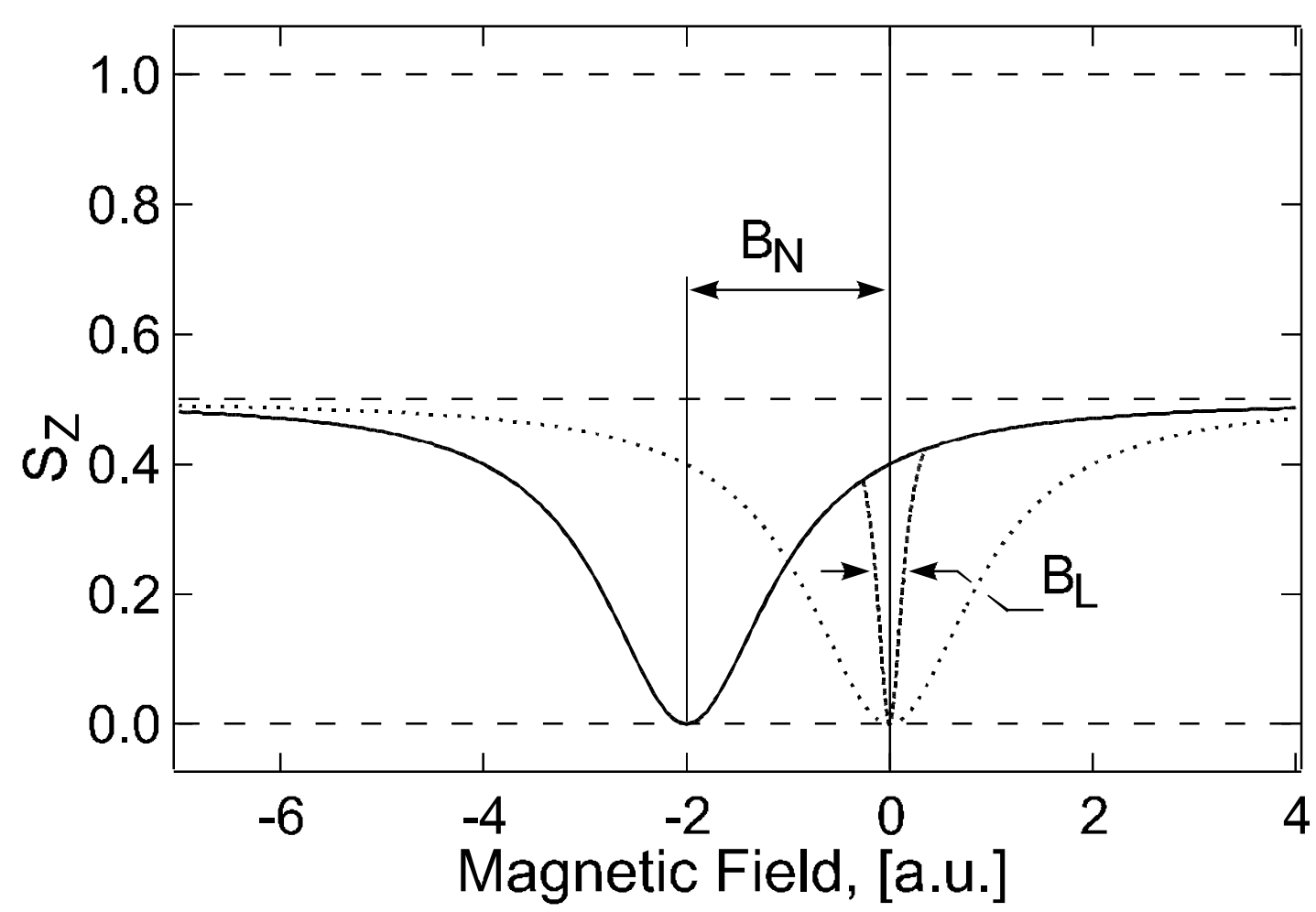

Fig.6.33. The $S_Z$ component of the average electron spin $\vec{S}$ in the case of electrical spin injection in the oblique Hanle effect geometry ($\varphi = 45^{\circ}$, Fig.4.2). The dotted curve is calculated accordingly to Eq.4.6 ($B_N = 0$). The solid curve takes into account the influence of the nuclear magnetic field $B_N = const$, Eq.4.6 and Eq.6.1. The dashed curve shows the dip at weak external magnetic fields ($B \sim B_L$), where $B_N$ vanishes abruptly.

In this case the minimum of the Hanle curve is shifted, so that it corresponds to the $B = -B_N$. Moreover, there is a dip at $B \sim 0$ corresponding to vanishing of the nuclear magnetic field in this region. Hence, the experimental Hanle curve has two minima. The 'broad' one, shifted on $B = -B_N$ with half-width $\Delta B = \left( \frac{g^* \cdot \mu_b}{\hbar} \cdot T_S \right)^{-1}$, corresponding to the 'true' Hanle curve. And the 'sharp' one around $B = 0$ with half-width $\Delta B_L \approx B_L$, corresponding to vanishing of nuclear spin alignment in this region. It should be pointed out that in p-GaAs samples $\Delta B \gg \Delta B_L$ [18].

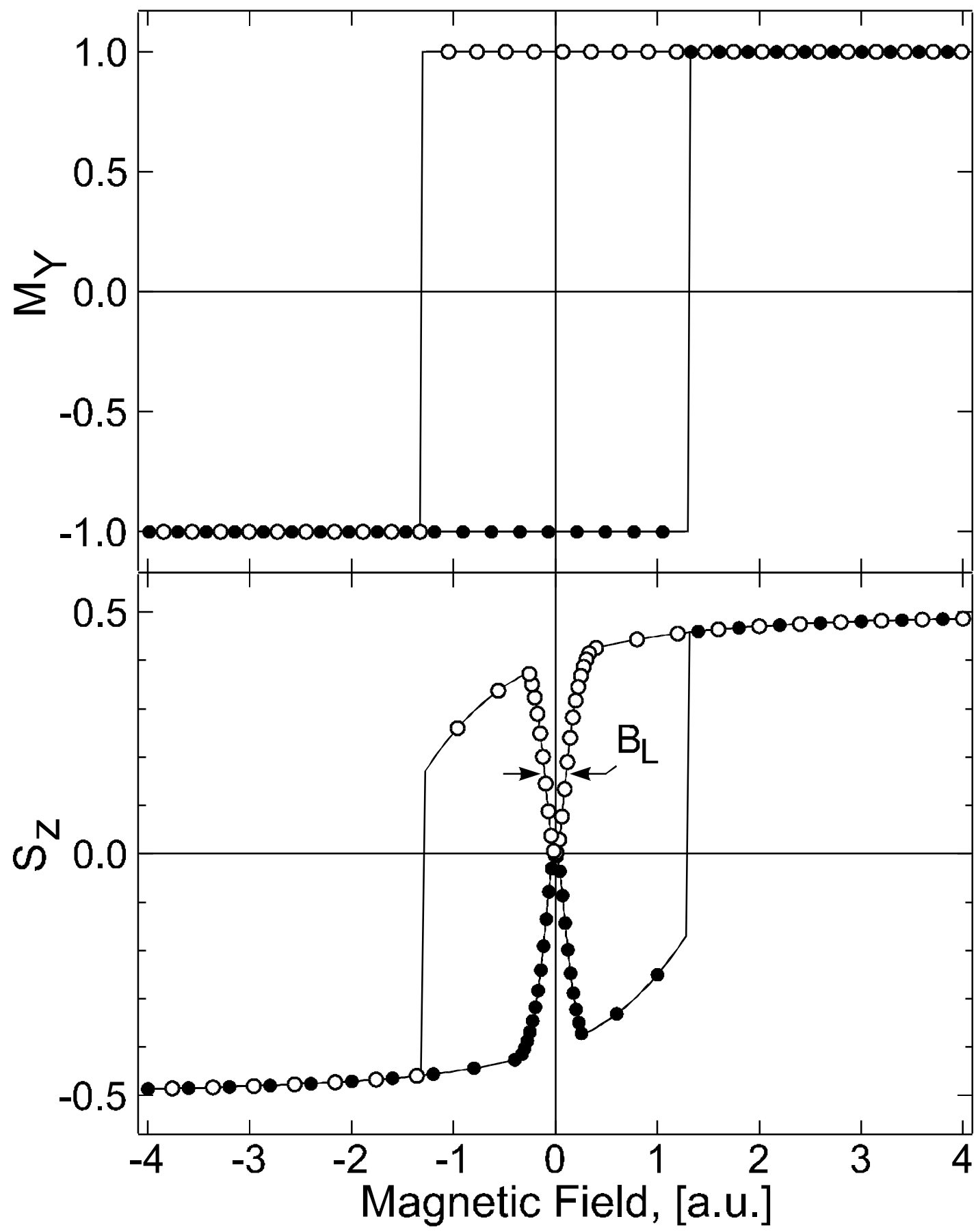

Fig.6.34. The influence of the ferromagnetic film magnetization switching on the $S_Z$ component of the average electron spin $\vec{S}$ in the experimental configuration depicted in Fig.4.2 ($\varphi = 45^o$) under dynamic nuclear spin polarization due to hyperfine interaction with spins of electrically injected electrons (to be compared with Fig.4.4).

In addition, it is clear that in the case of electrical spin injection from a ferromagnetic solid, the switching of the magnetization orientation by the in-plane component of the external oblique magnetic field will result in the change of the orientation of the injected spins. As it was discussed in Section 4.3, such change leads to sign reversal of the average electron spin $\vec{S}$ and its $S_Z$ component. Further, the hyperfine interaction tends to dynamically align the spins of nuclei along the new orientation of $\vec{S}$. Under steady state conditions, almost all nuclear spins are aligned along the new direction of the parallel to $\vec{B}$ component of $\vec{S}$. Fig.6.34 shows schematically the Hanle curve for the relative values of $\Delta B$, $B_L$ and $M_C$ expected for the CoFe/Al(GaAs) heterostructure. In this case only a very narrow dip is expected around zero external magnetic field with highly asymmetrical 'shoulders' of the Hanle curve. Again, this 'sharp' minimum corresponds to the vanishing of nuclear field in this region. The magnetization switching of the ferromagnetic spin source would not allow observation of the 'true' minimum of the Hanle curve and the experimental determination of the nuclear magnetic field magnitude $B_N$.

### 6.4.2. Experimental Results

Fig.6.35 shows the photoluminescence spectra for the sample *Type E* measured under optical excitation (through the ferromagnetic film) with $h \cdot \upsilon = 1.58$ eV at low temperatures. The spectral analyses allow to conclude that this PL is coming mainly from the active region of the device, since high doping level of the substrate supposes the emission of the optical radiation with $h \cdot \upsilon \sim 1.481$ eV [193] (see also Fig.A.1). This simple observation allows performing the all-optical characterization of the active region of the device as spin detector. As before, the measurements were performed under weak optical excitation with 100 % circularly polarized light with $h \cdot \upsilon = 1.58$ eV, e.g., just above the GaAs band gap.

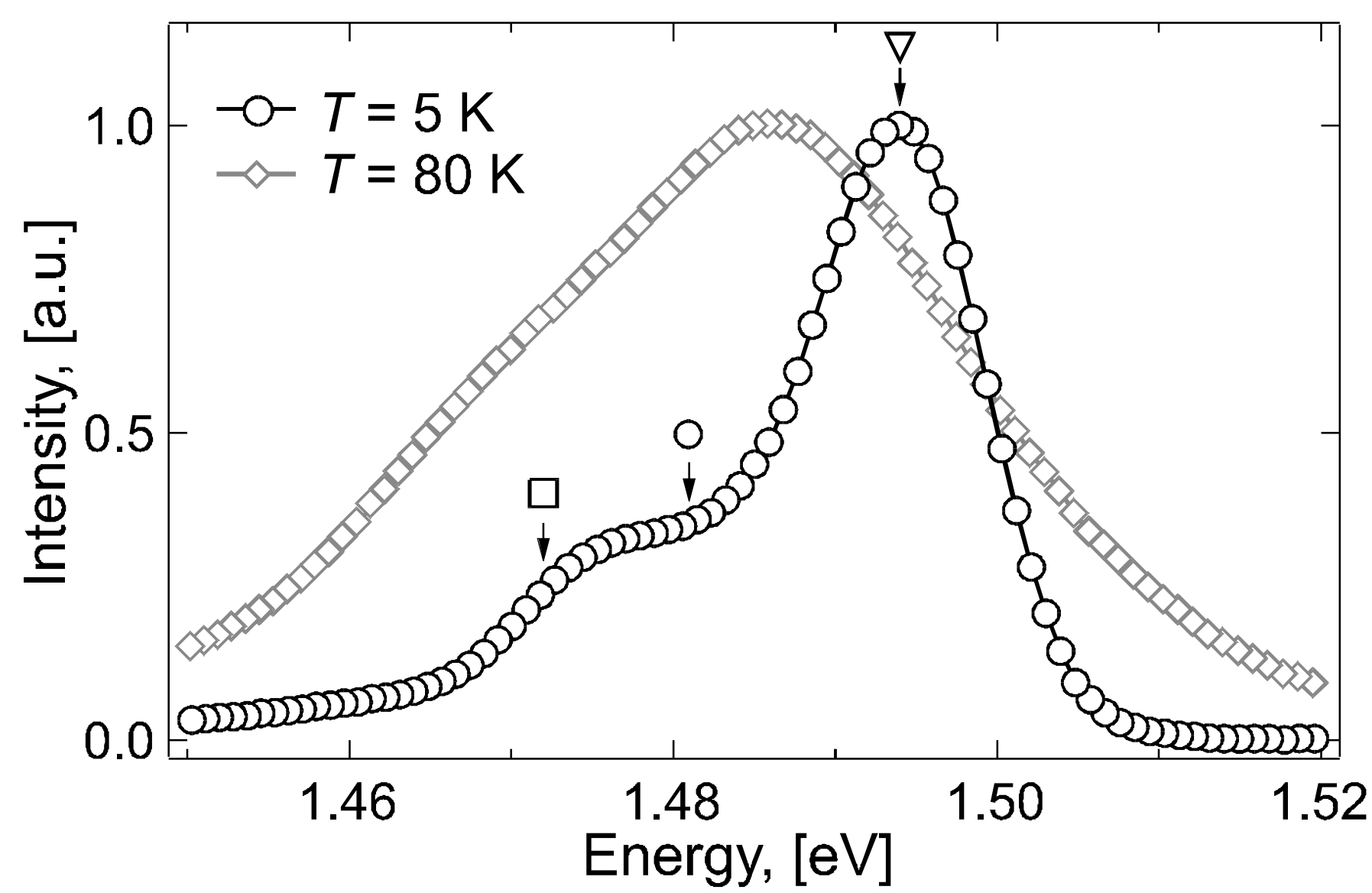

Fig.6.35. Normalized photoluminescence spectra for the sample *Type E* under optical excitation with $h \cdot \upsilon = 1.58$ eV at low temperatures. The square, circle and triangle indicate spectral regions, where the all-optical spin injection and detection Hanle effect measurements were performed (see Fig.6.36).

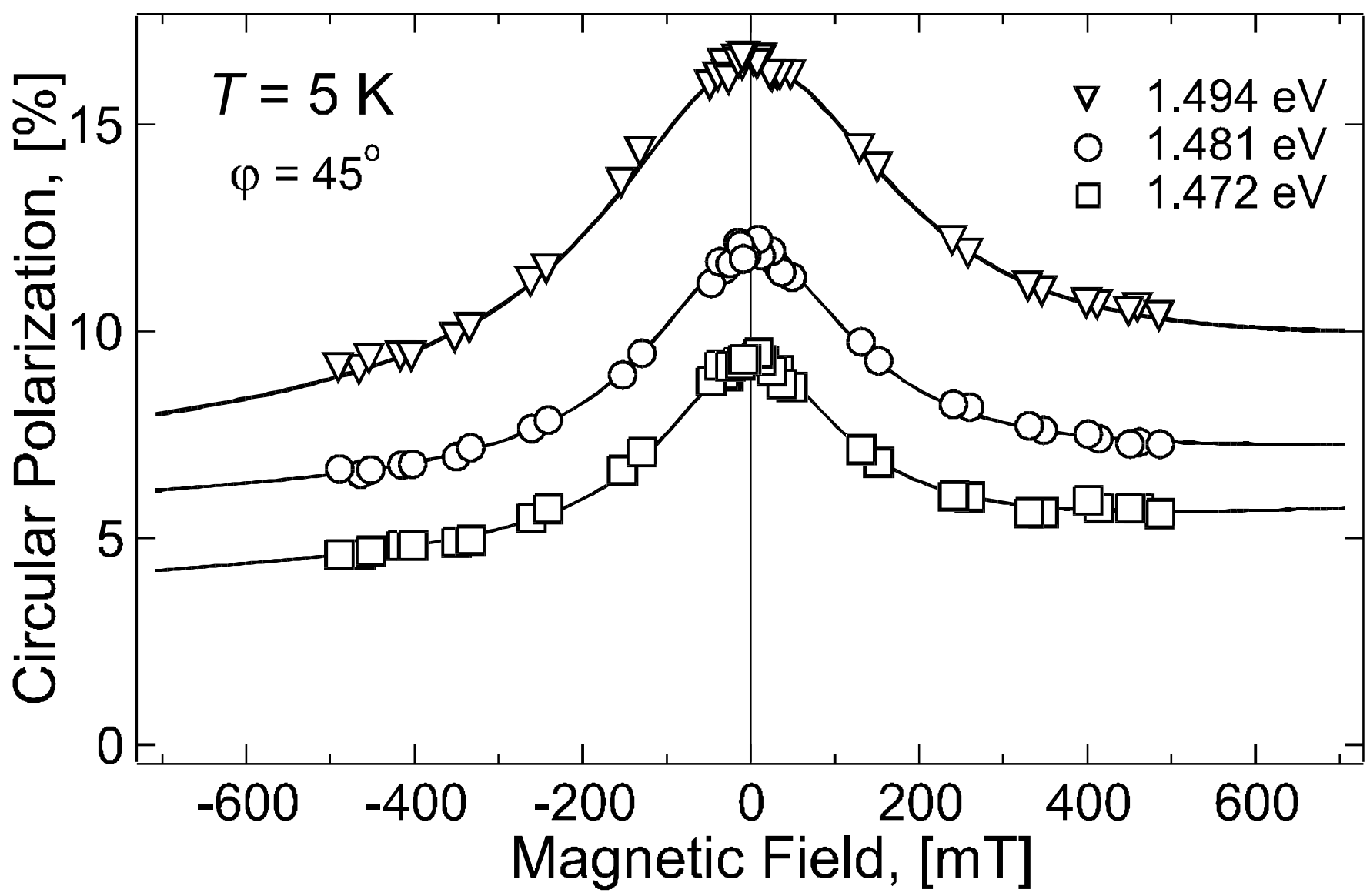

Fig.6.36. Damping of the circular polarization of photoluminescence in the oblique magnetic field ($\varphi = 45^{o}$, Fig.4.2) under optical spin injection in the sample *Type E*. The Hanle curves were measured at the spectral regions marked as square, circle and triangle on Fig.6.35.

Fig.6.36 (triangles) shows the damping of the circular polarization of the photoluminescence as function of external oblique magnetic field measured at the spectral region of 1.494 eV. This energy corresponds to the fundamental band gap transitions in the active region of the sample *Type E*. The Hanle curve (Eq.4.4) with the following parameters: $T_S/\tau = 0.7 \pm 0.1$, $\Delta B = (220 \pm 30)$ mT ($T_S = (0.12 \pm 0.02)$ ns) is used to fit the data (which are reasonable parameters, as the BAP mechanism is dominating mechanism of the spin scattering at this doping concentration and temperature [18]. See Section 3.3.5 and Fig.A.3). The Hanle measurements at other spectral regions of the photoluminescence curve have shown similar experimental half-width of the Hanle curve (see Fig.6.36 circles and squares). As one can see, no nuclear spin alignment was observed in the all-optical spin injection and detection measurements. As it was mentioned before the available optical excitation is weak, which does not allow creation of the reasonable electron spin concentration in the active region of the device. However, it appears that this is not the case in the electrical spin injection experiment.

The spectra of electroluminescence measured at different electrical bias for the sample *Type E* are shown on Fig.6.37. These spectral dependencies show pronounced blue shift at higher electrical biases, characteristic for highly compensated samples. Such a shift is caused by change of the carrier concentration in the active region of the device, which leads to filling of the electron and hole localized density of states (see Fig.6.30 left) with increase of electrical bias. This is the typical behavior of the semiconductors with this kind of doping.

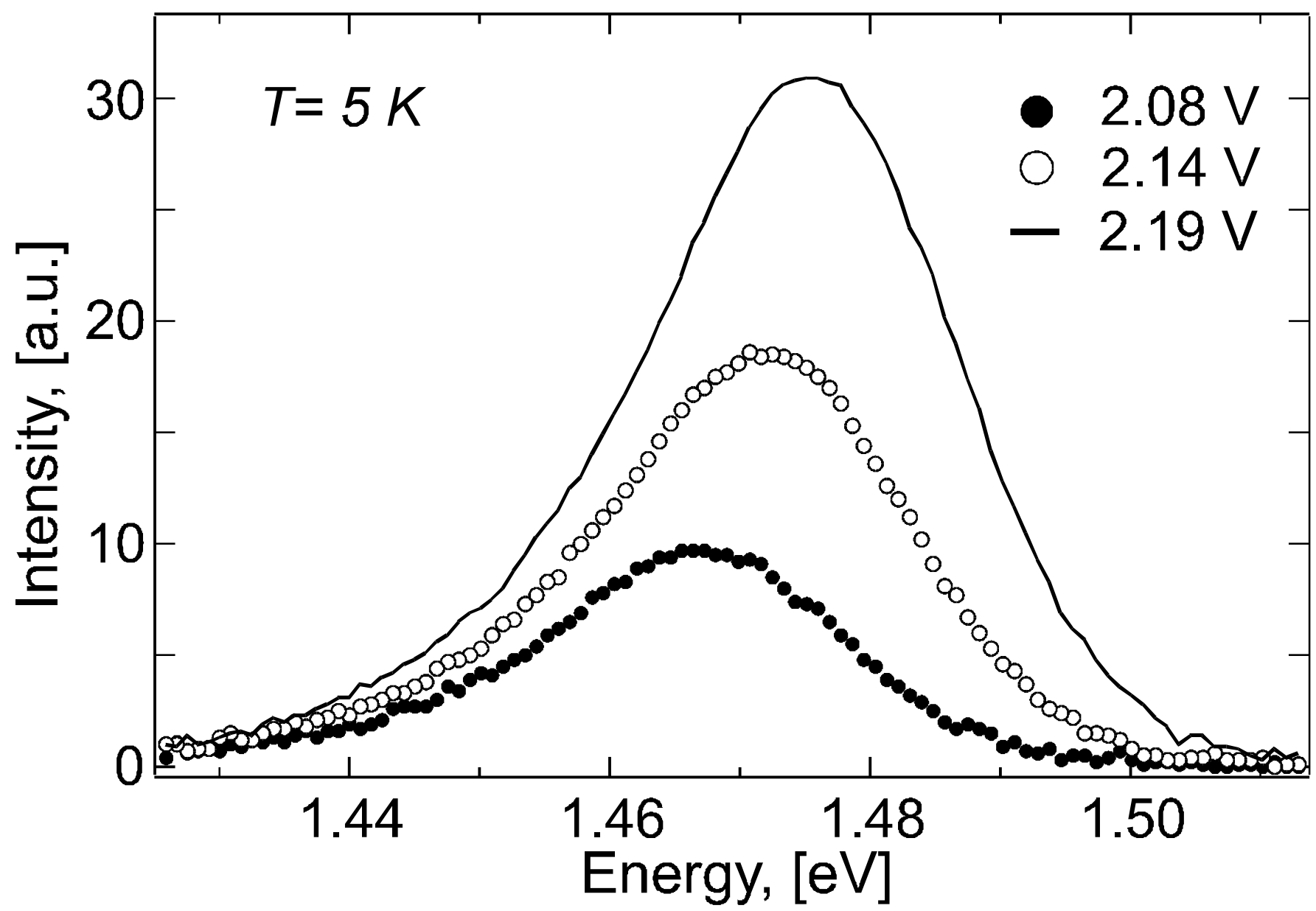

Fig.6.37. Electroluminescence spectra for the sample *Type E* at different electrical biases.

The typical evolution of the circular polarization of the electroluminescence measured in the oblique Hanle effect geometry ($\varphi = 45^{o}$, Fig.4.2) for the sample *Type E* is shown on Fig.6.38 (circles). The electrical bias during measurements was in the 2.02-2.26 V range with typical current 25-45 mA. The time delay between measurements of the different experimental points was ~15 s. All measured curves show quite large values of circular polarization, while the high out-of plane magnetization saturation value for CoFe ($\mu_0 \cdot M \sim 1.8\ \mathrm{T}$) supposes vanishing influence of the effects caused by the ferromagnetic film (MCD and tilting of the magnetization, see Section 4.4 and Section 4.5) in the magnetic field range needed for the saturation of $S_Z$, the out-of-plane component of the average electron spin $\vec{S}$. Taking into account the spin relaxation term obtained in the all-optical measurements, the Hanle fit (Eq.4.8 and Eq.4.10) gives a very high value of injected spin polarization $\Pi_{inj} = (31 \pm 4)\%$. The obvious striking observation is the very narrow Hanle curve seen in the electrical measurements, as compared to the all-optical experiment. The dashed curve on Fig.6.38 shows the calculated Hanle curve with $T_S/\tau$, $\Delta B$ parameters obtained from the all-optical measurements and $\Pi_{inj}$ parameter obtained from the electrical spin injection experiment. Indeed, the difference is striking ! As it is shown below, this difference is due to the large effective magnetic field of spin-polarized nuclei acting on the electrically injected electrons.

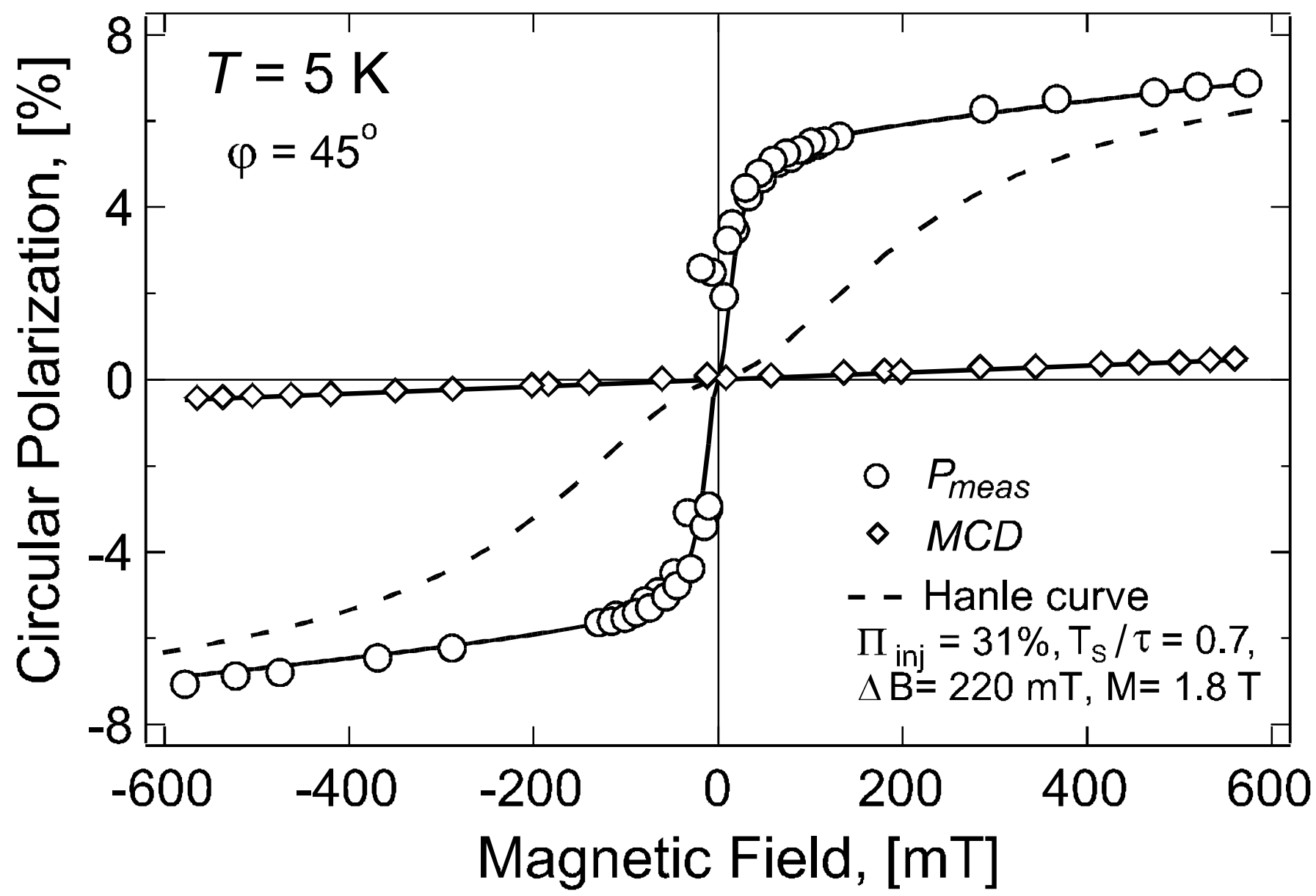

Fig.6.38. Typical measurements of the circular polarization of electroluminescence for the sample *Type E* (circles), MCD contribution (diamonds), and Hanle curve (dashed curve) calculated using Eq.4.8 and Eq.4.10 with the following parameters: $\Pi_{inj} = 31\%$, $T_S/\tau = 0.7$, $\Delta B = 220$ mT, $M = 1.8$ T.

The measurements of the circular polarization of the electroluminescence as function of external oblique magnetic field with increased magnetic field sensitivity for the sample *Type E* at different temperatures are shown in Fig.6.39-Fig.6.43. The insets show extended magnetic field range. The filled and open circles correspond to measurements with increasing and decreasing magnetic field sequence, respectively.

As one can see, the increase of temperature leads to increase of the half-width of the experimental Hanle curves (to be compared with Fig.4.7b) and decrease of the absolute value of the circular polarization at saturation. This change is caused by the enhancement of the spin relaxation, leading to decrease of the electron spin lifetime $T_S$ ($T_S^{-1} = \tau^{-1} + \tau_S^{-1}$) and spin relaxation term $T_S/\tau$ with increase of temperature. Again, this effect was studied in details in the 70's and observed dependencies correspond qualitatively to the published data [18].

In addition, the measurements at low temperatures show the hysteresis-like behavior with shift to the negative values of the external magnetic field. It appears that this effect has a magnetic nature (see Section 4.3). It is caused by the shift of the in-plane magnetization curve of the ferromagnetic spin injector, like in the exchange biased MTJs [194]. Fig.6.44 shows two sets (open and filled circles) of measurements of the same device at 5 K. These measurements were performed on different days, so that during cooling the sample was unintentionally exposed to the nonzero ($\sim 5$ mT) external magnetic field of different sign, caused by coercivity in the ferromagnetic frame of the electromagnet. The oxides of Fe or Co are known to be antiferromagnetic (generally speaking, with their own Curie temperature of transition in the antiferromagnetic state $T_C < RT$ ), so that coupling between ferromagnetic CoFe and antiferromagnetic oxides of Fe or Co may result in the shift of the in-plane magnetization curve as schematically shown in Fig.6.44 (this actually make sense, as Magnetooptical Kerr Effect (MOKE) measurements of the in-plane easy axis magnetic reversal (Fig.6.44 inset) have shown the coercivity ( $2 \cdot M_C$ ) similar to that observed in the spin injection experiment, Fig.6.39-Fig.6.41).

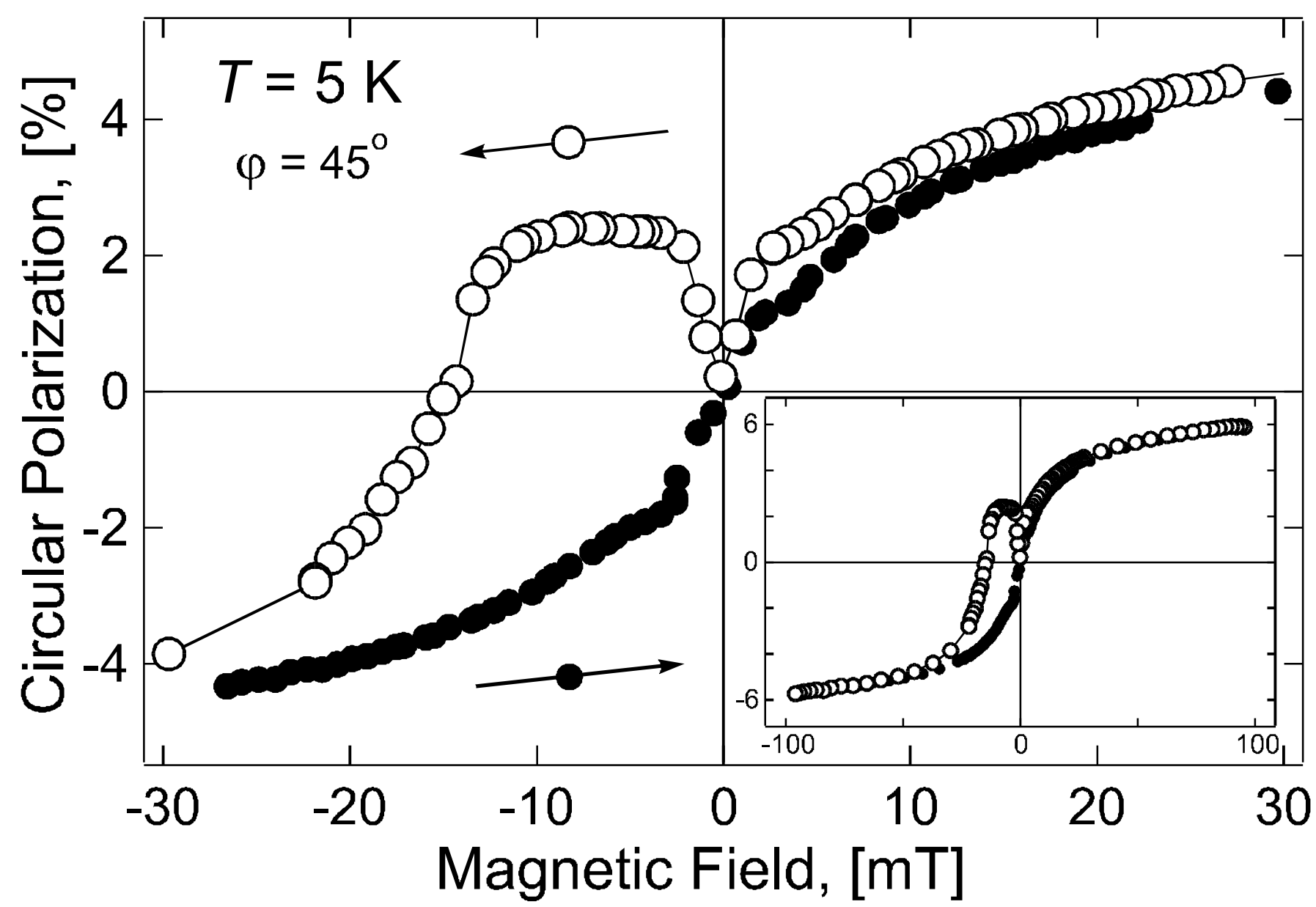

Fig.6.39. Measurements of the degree of circular polarization of electroluminescence for the sample *Type E* (compensated active region). Inset: extended field range.

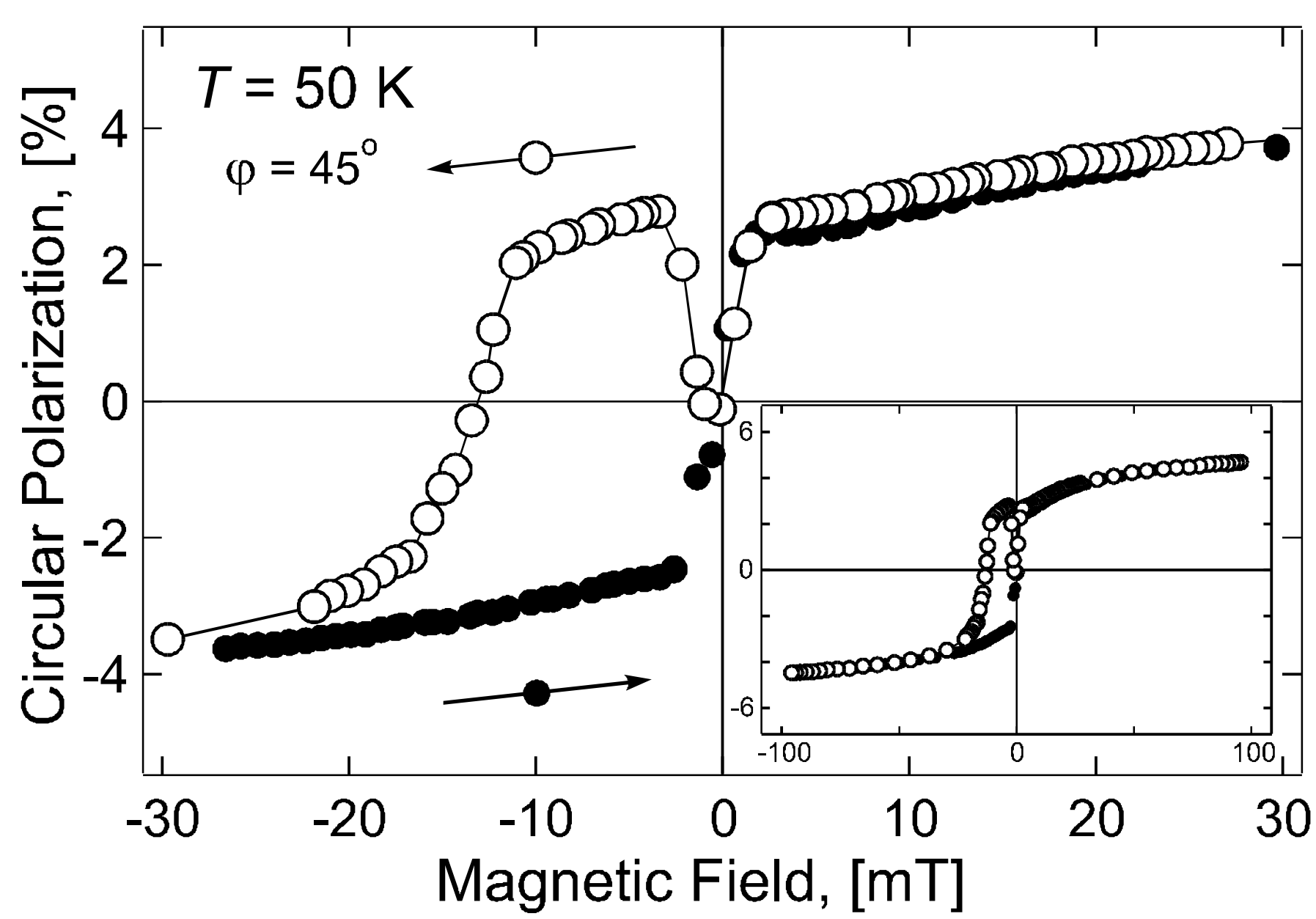

Fig.6.40. Measurements of the degree of circular polarization of electroluminescence for the sample *Type E* (compensated active region). Inset: extended field range.

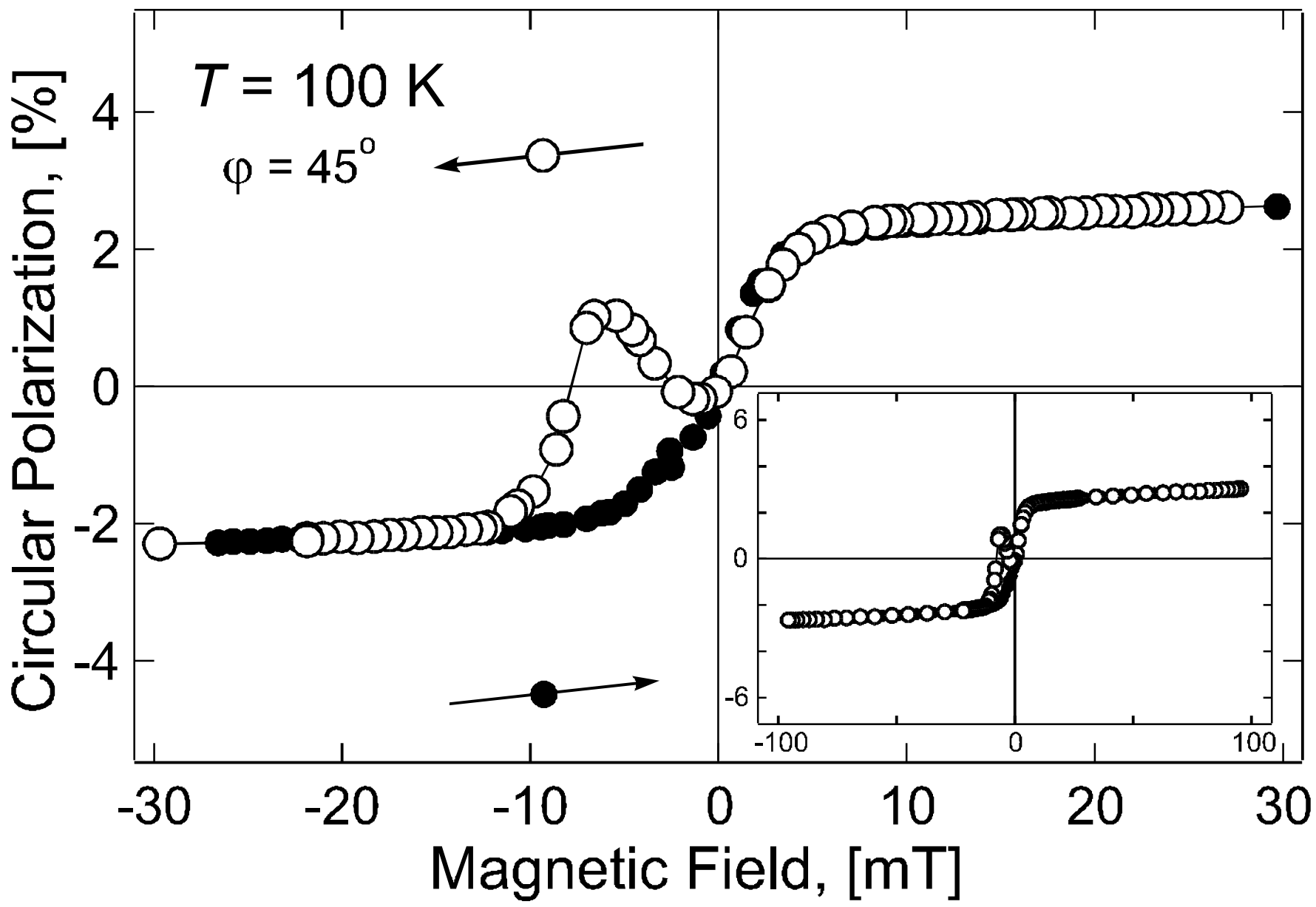

Fig.6.41. Measurements of the degree of circular polarization of electroluminescence for the sample *Type E* (compensated active region). Inset: extended field range.

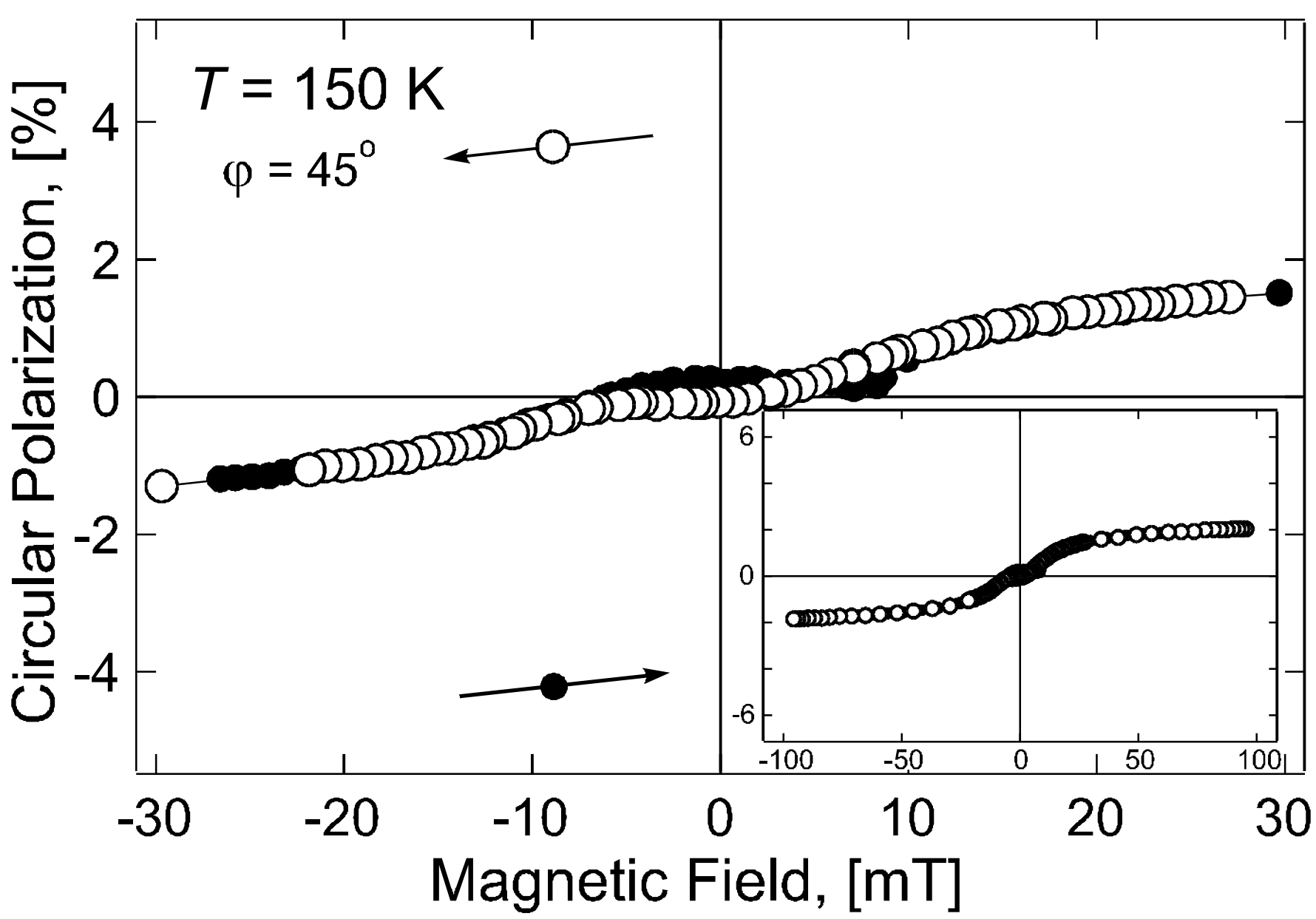

Fig.6.42. Measurements of the degree of circular polarization of electroluminescence for the sample *Type E* (compensated active region). Inset: extended field range.

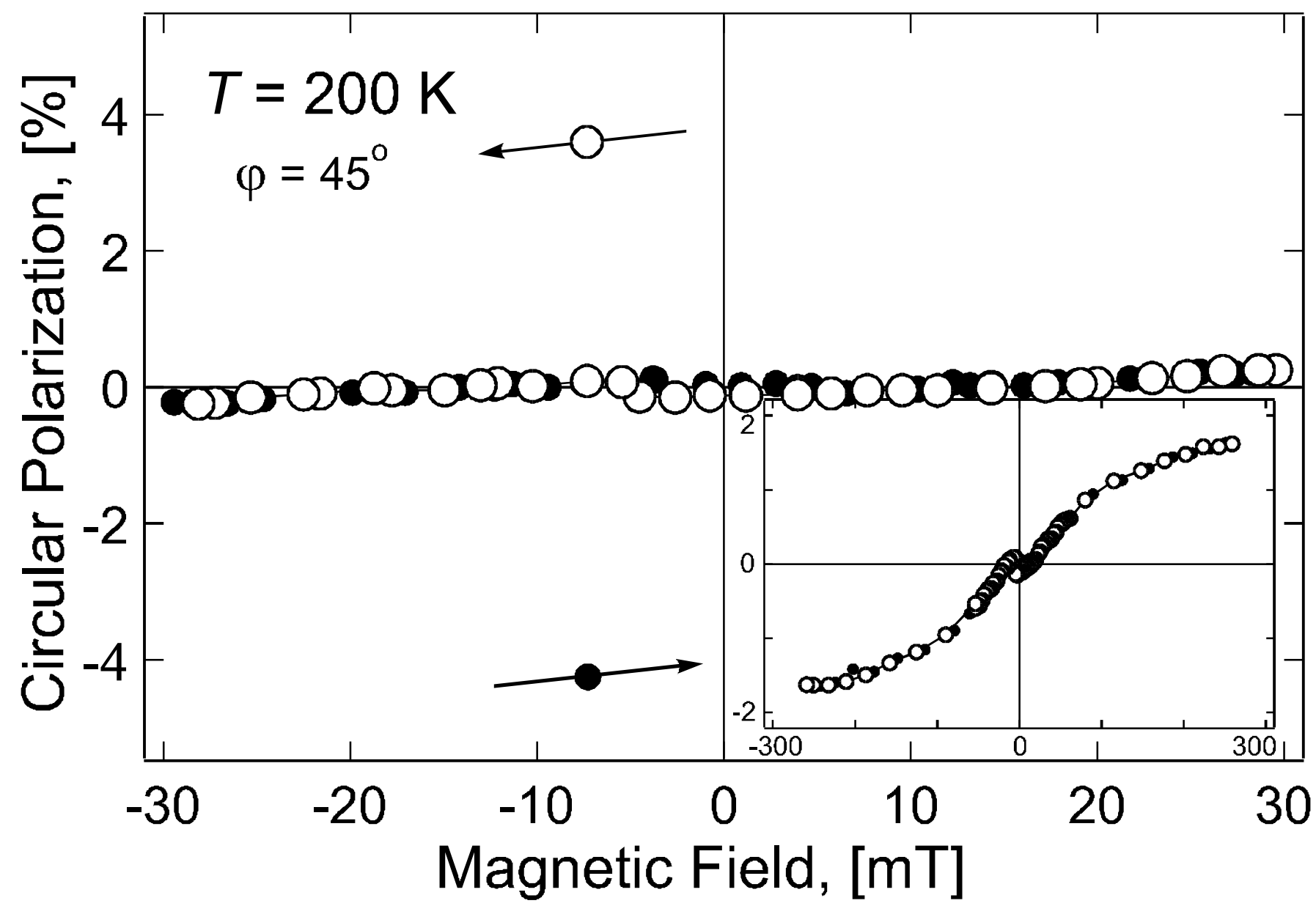

Fig.6.43. Measurements of the degree of circular polarization of electroluminescence for the sample *Type E* (compensated active region). Insets: extended field range.

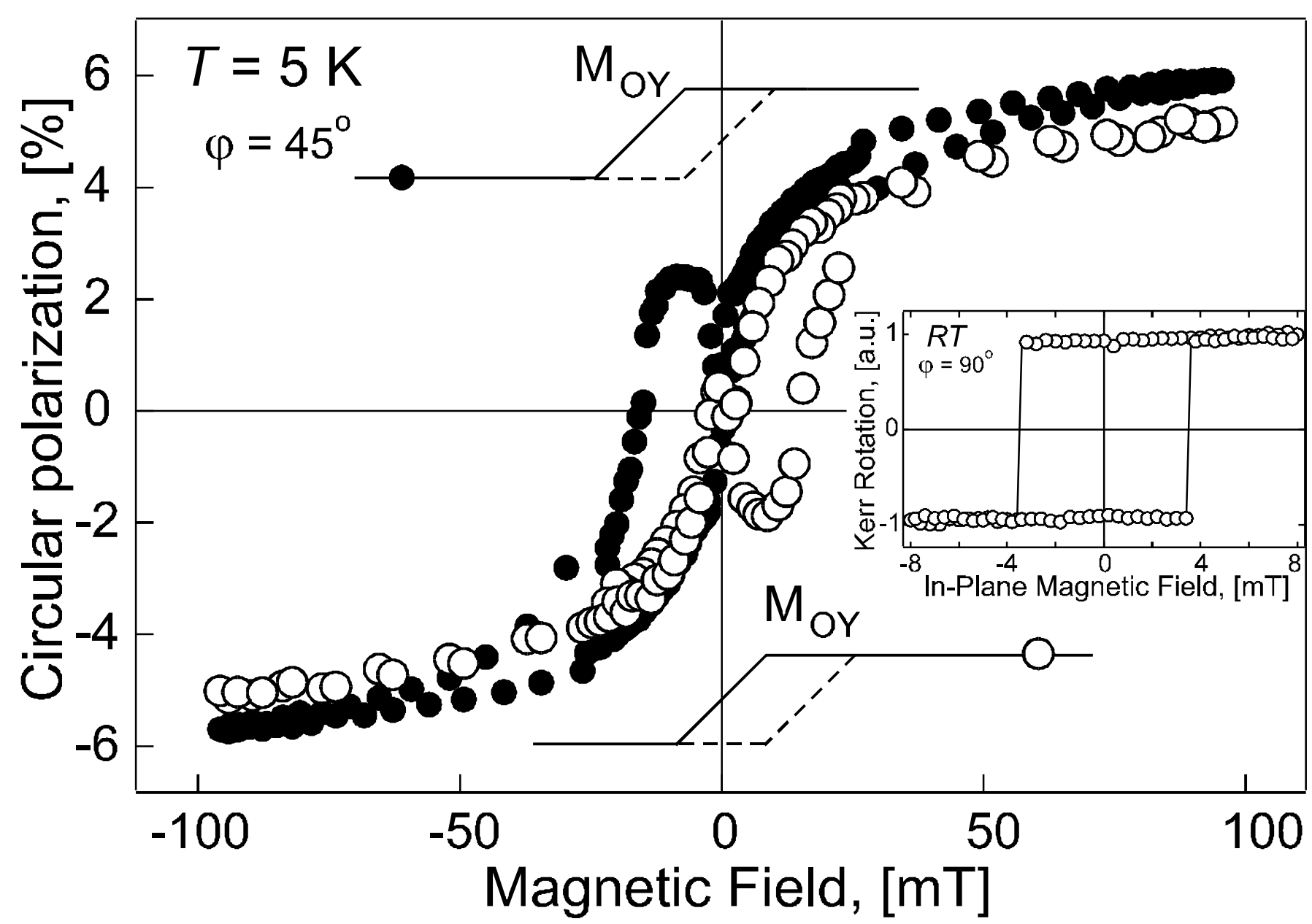

Fig.6.44. Measurements of the degree of circular polarization of electroluminescence for the sample *Type E* (compensated active region). The curves were measured on the same device, the only difference is the sign of the external small magnetic field in which the sample was exposed during cooling. The schematic in-plane magnetization curves ( $M_{OY}$ ) would result in the observed dependencies. Inset: the easy axis in-plane magnetic reversal, as revealed by MOKE measurements.

Moreover, the Hanle curves at 5 K and 50 K show a very abrupt dip around zero external magnetic field with the half-width ~1.2 mT ( $(1.3 \pm 0.1)$ mT at 5 K and $(1.1 \pm 0.1)$ mT at 50 K). The detailed view on this feature is shown on Fig.6.45. Here the word dip is used, as such small half-width of Hanle curve is never observed in the p-type GaAs samples. It exactly corresponds to the value expected for the local fluctuating magnetic field $B_L$ [189, 18]. Hence, the 'strange' narrowing of the Hanle curve in the electrical measurements, as compared to all-optical ones (see Fig.6.36 and Fig.6.38 for example), is in fact caused by the dynamic polarization of nuclei due to hyperfine interaction with electrically injected electrons. Indeed, the experimental shape of the Hanle curves measured at 5 K and 50 K resembles very closely the theoretical curve presented on Fig.6.34.

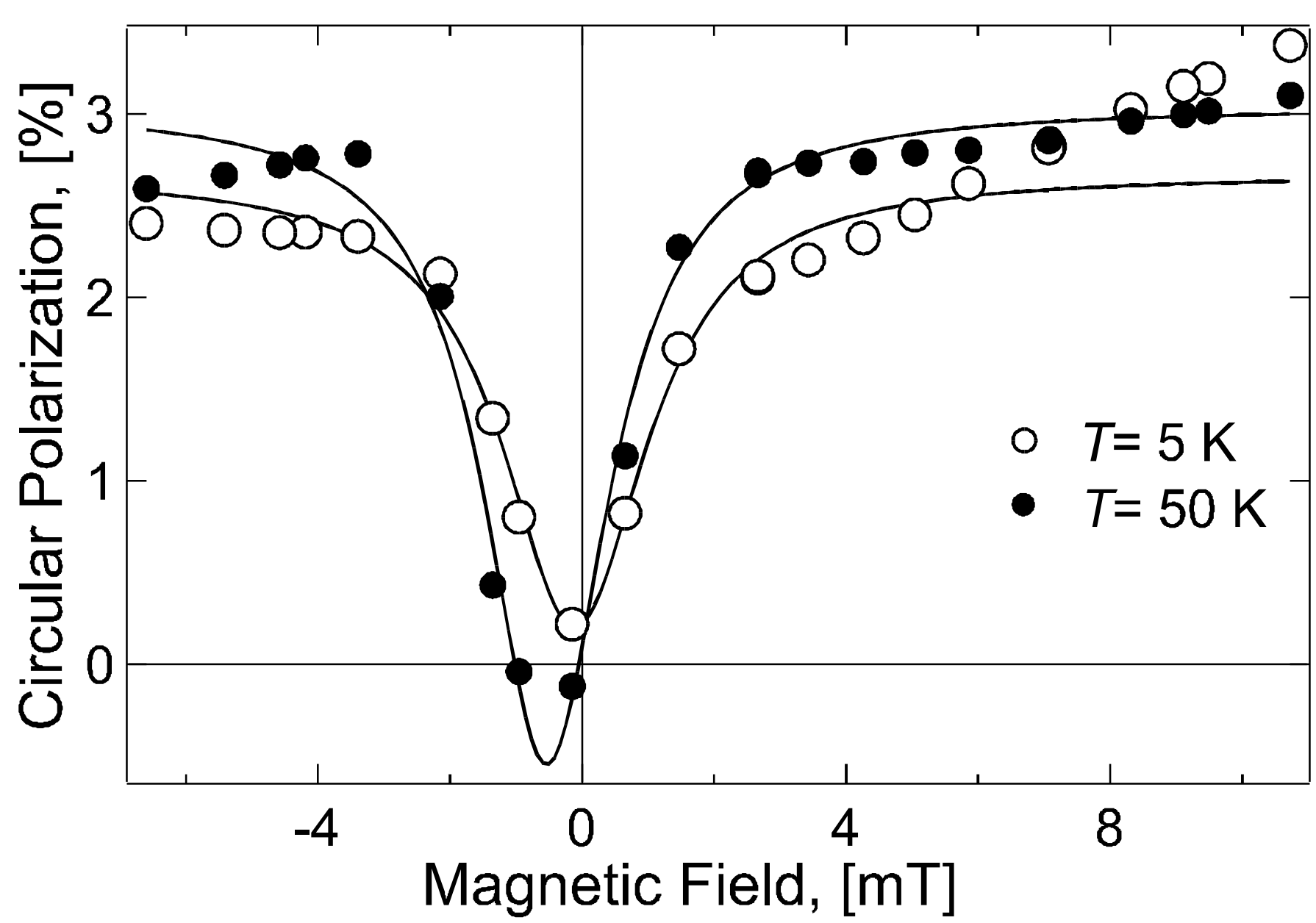

Fig.6.45. The detailed view on the dip observed around zero external magnetic field for the sample *Type E* (compensated active region) at different temperatures. The corresponding experimental Hanle curves are presented in Fig.6.39 and Fig.6.40.

Unfortunately, as it was argued before (see previous Section 6.4.1) the switching of the magnetization of the ferromagnetic spin-injector does not allow independent measurements of the magnitude of the effective nuclear magnetic field as it can be done in all-optical experiment. However, the comparison of the data obtained in the electrical and all-optical experiments suggests that the effective nuclear magnetic field is exceeding 500mT (5kG). At least, this field is needed for the observation of the saturation of the out-of-plane component $S_Z$ of the average electron spin $\vec{S}$ in the all-optical measurements. This nuclear field appears as 'missing' excess field in the measurements presented on Fig.6.38.

## 6.5. Electrical Spin Injection in the Sample *Type F*

The sample *Type F* was fabricated in the fabrication procedure similar to the sample *Type B*. It consists of the 7nm CoFe/ **3nm Cu** / 1.8nm $AlO_X$ (single-step oxidation)/ 15 nm AlGaAs (undoped)/ 100nm GaAs (undoped)/ 200nm AlGaAs ($p=2\cdot10^{18}cm^{-3}$)/ 2µm GaAs ($p=2\cdot10^{18}cm^{-3}$) buffer/ p-GaAs substrate (See Table 5.1). Here the incorporation of 3 nm thin Cu layer between the ferromagnetic metal and the semiconductor allows to reduce the spin polarization of electrically injected electrons to undetectable values [195, 196, 197]. Hence, the circular polarization of electroluminescence should show only the MCD effect in the ferromagnetic film (see Eq.4.12).

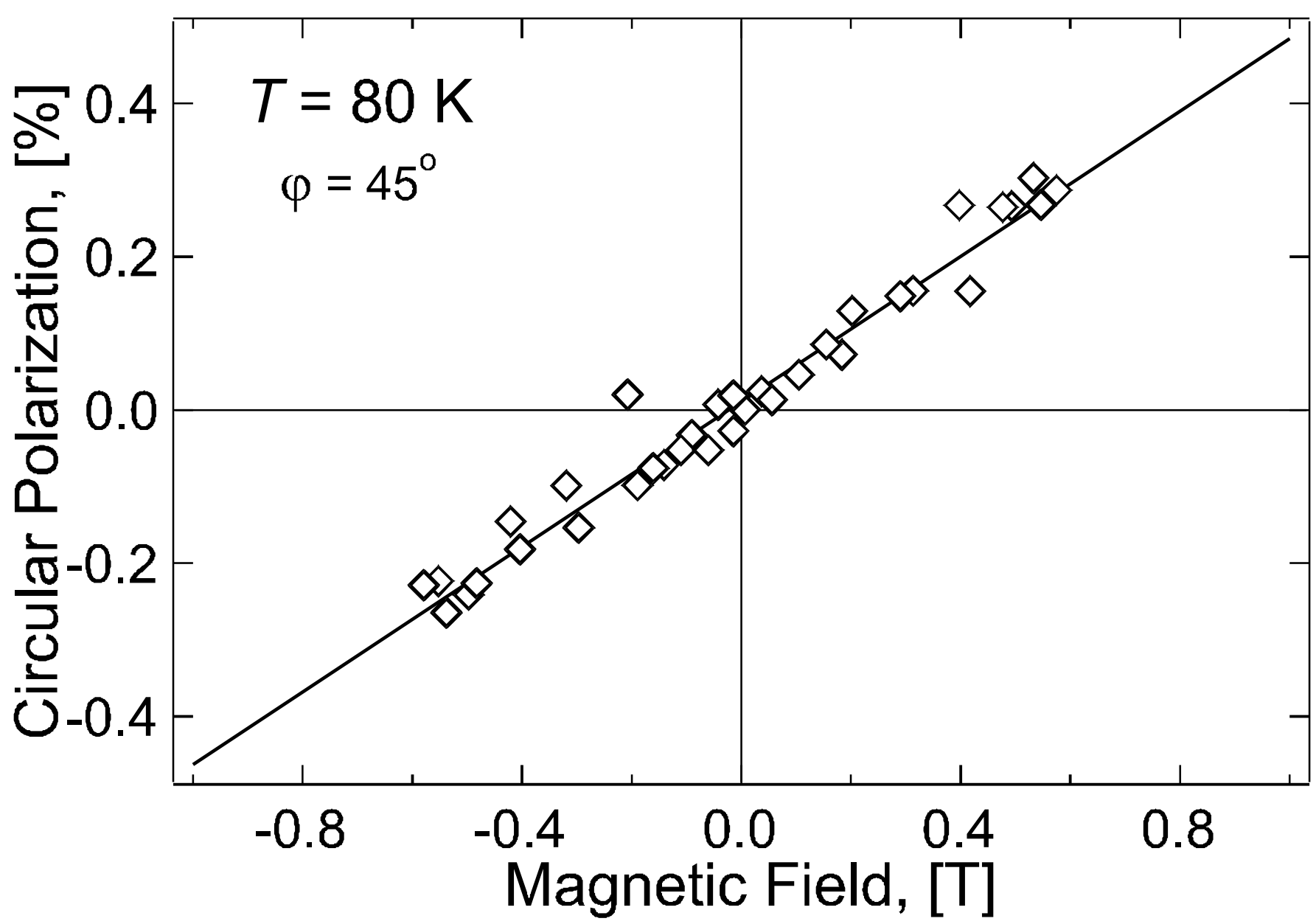

Fig.6.46. Typical result of measurements of the circular polarization of electroluminescence for the sample *Type F* (sample with intentional Cu 'dusting' of the FM/semiconductor interface, resulting in the zero spin polarization of injected electrons, $P_{inj} = 0$ ).

Fig.6.46 shows the typical experimental oblique magnetic field dependency ($\varphi = \pi/4$) of the circular polarization of the electroluminescence. As expected, the curve shows only the linear dependency caused by MCD effect in the ferromagnetic film.

# Summary

As it was pointed out in the introduction chapter of this thesis, the traditional materials and device architectures have serious limitations on the way of further increase of device integration and chip functionality. Moreover, these fundamental physical limits are going to be approached in the couple upcoming years already. At the same time, the recent developments in the areas of mobile communications and multimedia applications, computing and networking, etc., create new demands for further increase of data access and storage, computational power and multifunctionality. It follows that total digitalization does not require analogous signal amplification anymore. What is needed is reliable definition of states $|0\rangle$ and $|1\rangle$, which could be accessed and processed on extremely short timescales. All these circumstances create perfect starting conditions for emerging of new technologies and device architectures.

One of newly emerged technologies relies upon intrinsic property of electron – a spin (see Chapter 2). A classical example of such device, the Giant Magnetoresistance (GMR) junction is currently revolutionizing the world of magnetic recording. Another technology with very promising future is Magnetic Random Access Memories, which is based on other device utilizing electron spin the Magnetic Tunnel Junction (MTJ). All these devices are passive as they contain only metallic multilayers. From other side, the use of electron spin in the semiconductor-based device with relatively simple architecture may bring considerable advantages (see Section 3.1). The advantages of such architecture, the architecture that relies upon quantum mechanical phenomena, become more and more clear as traditional downscaling of device dimensions start to reveal further the quantum character of the nature.

It appears that for successful operation of such device an efficient way of creation of spin-polarized charge ensemble within a semiconductor is needed. Optical methods have proven to satisfy these needs (see Section 3.3), but they do not look very promising, once large integration scales are targeted. It further appears that traditional ferromagnetic metals are currently the most favorable candidate for such mission, as ferromagnetic order makes them almost never lasting source of spin-polarized electrons even at room temperature (see Chapter 2). Moreover, their fabrication and physical properties are well-known.

Unfortunely, the preliminary experiments combining three-terminal geometry for electrical spin injection and detection of spin-polarized electrons into semiconductor (InAs) from a ferromagnetic metal have shown no effects that could be attributed to the presence of electron spin imbalance in the semiconductor (see Section 3.2). The followed theoretical investigation of this problem has shown that electrical spin injection from a ferromagnetic metal into a semiconductor in the diffusive ohmic contact is practically impossible. This is due to the large difference in the density of states in these materials, so that both spin channels (spin-up and spin-down) are completely filled in the semiconductor, resulting in the zero overall spin polarization.

Fortunately, GaAs as other III-V semiconductors provide unique possibilities not only for optical injection of spin-polarized charges, but also for optical detection of their spin-polarization through the polarization state of the light emitted as result of electron and hole recombination (see Section 3.3). This allows direct optical investigation of electrical spin injection into a semiconductor from a ferromagnetic metal in a so-called spin-LED, across a single ferromagnetic metal/ semiconductor interface only.

It follows that traditional problem of ohmic contacts in the case of GaAs, as well as Si, significantly differentiate the problem of electrical spin injection into these semiconductors comparing to InAs, studied in the preliminary experiments (see Sections 3.2, 5.2). In these semiconductors the diffusive ohmic contact simply does not exists, since the abrupt ferromagnetic metal / semiconductor interface leads to Schottky barrier formation. In this case the electrical injection of electrons into conduction band of the semiconductor is possible only in the case of strong n-type doping of the semiconductor, for reverse-biased Schottky contact. In the case of p-type doping, electrical injection of electrons into conduction band of the semiconductor implies incorporation of a thin tunnel barrier on the ferromagnetic metal / semiconductor interface. In both cases the tunneling mechanism is involved in the electronic transport, which is known to be dependent over the density of states in both solids, hence matching the large difference in the density of states for these solids. Moreover, there is a large drop of potential across such interface, which is not the case for the FM/InAs contact.

It further appears that under electrical spin injection into a semiconductor the observed emission of circular polarization is in fact a multistep process. Generally at first, the spin-polarized electrons are injected into the conduction band of the semiconductor for the case where the kinetic energy is higher than $k \cdot T$ (hot electrons). Secondly in the thermalization process and during the electron lifetime at the bottom of the conduction band, before recombination with holes, some loss of spin polarization may occur due to spin scattering. As result, the measured steady state spin polarization of injected electrons determined from the polarization state of the emitted light can be significantly smaller than the injected one (see Section 3.3 and Chapter 4).

The electrical spin injection in the spin-LED type of heterostructures significantly differs from the case of all-electrical devices (like spin-polarized field effect transistor, for example. See Sections 3.1). In the last ones, carriers of only one type, namely electrons, are participating in the electrical current. In the case of a LED, the efficient supply of holes is also needed. Ideally, the electron and hole currents in such heterostructures should be equalized.

As it follows from the mentioned above considerations the design of the tunnel barrier and / or semiconductor heterostructure must be optimized not only for enhanced injection of electrons from a ferromagnetic metal into the conduction band of a semiconductor and sufficient supply of holes for recombination, but also for advanced spin conservation within semiconductor, in order to minimize spin scattering during electron thermalization and lifetime on the bottom of the conduction band. Moreover, the electrical bias across TB in the case of MIS type heterostructure must be minimized, as studies of TMR effect in the magnetic tunnel junctions (MTJ) have shown that high bias applied to the tunnel oxide leads to drastic decreases of spin-dependent effects. This decrease is attributed to electron tunneling via intermediate defect states within tunnel barrier, which act as additional channel for spin scattering ([184], see Section 6.3.3).

The experience of design and fabrication of semiconductor LEDs shows that high optical efficiency of electroluminescence is achieved in the devices having the active region (region where recombination takes place) close to the surface. The typical thickness of the active region in the LEDs is $\sim 4 \ldots 200\ \text{nm}$. In the devices fabricated during work presented in this thesis, the active region was chosen to be wide enough ($100\ \text{nm}$), so that no quantization of electron and hole levels takes place. This allows to avoid partial loss of spin polarization during electron trapping into the well (see Chapter 4). Moreover, it facilitates the quantitative analyzes of measured data, at different temperatures in particular. This is because conduction band to valence band (heavy and light hole) transitions, or by other words selection rules (the rule defining correlation between polarization of emitted light and spin polarization of charge carriers, see Section 3.3.1) strongly depend on energetic splitting of these levels and their population. In the quantum well-type heterostructure, for example, the spin polarization of electrons at low temperatures is equal to the degree of circular polarization of electroluminescence. But at room temperature, the splitting of electron and hole levels is generally smaller than the thermal energy $k \cdot T$ and the spin polarization of electrons is two times larger than the circular polarization of electroluminescence.

However, the optical detection methods in the case of electrical spin injection into semiconductor from ferromagnetic metal cannot be applied as straightforward, as in the case of optical spin injection. In particularly, thin ferromagnetic films typically have in-plane magnetic anisotropy, while high refractive index of the semiconductor allows probing only the component of electron spin normal to the surface (see Section 3.3). It follows that side-emitting geometry (light being detected propagates along the FM/semiconductor interface) is least suitable for this type of experiments, as waveguiding effects in the semiconductor heterostructure and reflections from the FM/semiconductor interface significantly complicates the analyses of experimental data. Moreover, the selection rules damping in the case of quantum confinement (see Sections 3.3.3, 3.4.1) raise serious doubts concerning validity of the measured data.

Further, the light emitted in the surface (light being detected propagates across the FM/ Semiconductor interface through the semitransparent ferromagnetic film, see Chapter 4) and backside (light being detected propagates in opposite direction comparing to surface emitting geometry) emitting configuration is unpolarized, since the preferential spin orientation of electrons injected into semiconductor is orthogal to the direction of observation. The traditional approach of applying strong out-of-plane magnetic field (more than 1 T for most common ferromagnetic metals), which pulls out the magnetization of ferromagnetic metal and hence, changes the orientation of electrically injected spins, leads to significant side effects (Magnetooptical Kerr and Circular Dichroism, Zeeman splitting of electron and hole levels, etc.). These side effects mask the expected spin injection signal and could be entirely responsible for measured quantities.

During work presented in this thesis a new approach for optical assessment of electrical spin injection into a semiconductor from a ferromagnetic metal having magnetic anisotropy orthogonal to the direction of observation was developed. It is based on spin manipulation within a semiconductor (oblique Hanle effect), once spin-polarized charges have been injected. Generally, such measurements can be performed in the relatively week external oblique magnetic field, which does not affect the magnetization orientation of the ferromagnetic film significantly. Moreover, the spin manipulation within semiconductor caused by such magnetic field differs significantly from the case of spin manipulation in the ferromagnetic film (change of the magnetization direction). It provides a unique signature of spin injection, as the experimental magnetic field dependency has the Lorentzian shape, while side effects are linear or nearly linear with the external magnetic field. In addition, the oblique Hanle effect approach reveals the important information on spin kinetics within a semiconductor simultaneously. This is the case, since the change of the out-of-plane average electron spin component due to spin precession (the experimental shape of the Hanle curve) is determined by the longest process, i.e. mainly by the spin precession during the electron lifetime on the bottom of the conduction band (after thermalization) of the semiconductor. The timescale of electron injection or thermalization is a couple of orders of magnitude shorter. Moreover, the oblique Hanle effect approach combined with all-optical characterization of the spin detector part of the device (the semiconductor heterostructure) represents a powerful tool for quantitative evaluation of the spin injection.

Finally, following consideration mentioned above a set of different MIS spin-LEDs was fabricated at IMEC (see Chapter 6). The optical investigation of electrical spin injection in these devices has allowed the experimental demonstration of very efficient electrical spin injection into semiconductor from ferromagnetic metal in the direct electrical contact, even at room temperature. Something that had been thought to be completely impossible before. Moreover, the presented results show the way to increase the efficiency of electrical spin injection in such devices from 2% at 80 K (typically observed by most of the groups around the World) up to more than 60% at low and room temperatures (achieved in this thesis). In addition, the importance of electron thermalization effects and the impact of the doping level of the semiconductor are demonstrated for practical investigation of electrical spin injection by optical means. Further, in the spin-LEDs specifically engineered for high electron localization in the bulk-type active region of the device, measurements at low temperatures have revealed an existence of nuclear magnetic field. This nuclear spin polarization appears due to hyperfine interaction of nuclear spins with the spins of electrically injected electrons. It reveals itself in the experiment as additional magnetic field added to the external magnetic field acting on the spins of electrically injected electrons.

# Conclusions and Outlook

The presented results show that the oblique Hanle effect approach represents a useful tool for optical assessment of electrical spin injection into semiconductors. It discriminates spin injection from side effects, magnetooptical and Zeeman splitting induced spin polarization, for example. In addition, it provides very valuable information about spin kinetics within semiconductor. Combined with all-optical characterization of the spin detector part of the device it represents a powerful tool for quantitative evaluation of the spin injection.

To get a fast feedback on the quality of spin injectors used for spintronics applications it is very important to have an independent characterization tool. In the case of MIS heterostructures the TMR junctions fabricated in the same sputtering system provide essential feedback on the quality of the ferromagnetic metal / tunnel barrier spin injectors. Moreover, such TMR data can be used to estimate the spin injection efficiency in the MIS-type heterostructures. The polarization of injected electrons measured in the fabricated spin-LEDs is already quite close to these values.

It can appear that GaAs is not even the best material for the optical investigation of electrical spin injection in the eoblique Hanle effect geometry, since the electron g-factor is very low ($g^*$=-0,44 [198]). For other materials with higher g-factors [199], for example GaSb ($g^*$=-9,3 [200]), the same value of spin scattering time $T_S$ will give a much narrower (in the case of GaSb – ~20 times narrower) Hanle curve. In this case the Hanle measurements (even at room temperature) can be performed in very low external magnetic field, where the influence of external magnetic field on the magnetization of ferromagnetic film can be totally neglected.

At present, it seems that the spin injection in the hybrid ferromagnet / oxide/ semiconductor devices can be increased the same way as in the TMR junctions, by improving the quality of the oxide barrier and its interfaces, and by using ferromagnetic materials with higher spin polarization.

The presented results indicate that the use of a tunnel barrier injector is indeed an interesting route to inject spins into a semiconductor. The introduction of oxide layer allows obtaining more stable and robust spin injectors. A large variety of ferromagnetic materials can be deposited on top of the oxide layer, forming a universal spin source.

Although, only electrical spin injection into semiconductor from a ferromagnetic metal in the direct electrical contact have been experimentally investigated in the presented research, the electrical spin detection using second ferromagnetic metal in such system does not seem to be a problem. Since the same considerations are valid for the electrical spin detection on such interface, moreover, it seems to be demonstrated in the similar systems already [201, 202, 203, 204].

These results look very promising for future room temperature spintronic devices using stable tunnel barrier injectors, such as $Al_2O_3$ or AlN on III-V (e.g. GaAs, GaN) or state-of-the-art $SiO_2$ for Si/SiGe devices.

Moreover, the possibility of dynamic nuclear spin polarization by electrically injected spin-polarized electrons opens a new way for practical realization of a large scale integration solid state quantum computation, using principles proposed already [192, 205] or entirely new 'the one's we're not thinking about'[10].

# Samenvatting

Zoals wordt aangeduid in het inleidende hoofdstuk van deze thesis, hebben de traditionele materialen en component-concepten aanzienlijke problemen bij verdere toename van component-integratie en chipfunctionaliteit. Meer nog, de fundamentele fysische limieten gaan reeds in de komende jaren benaderd worden. Tegelijkertijd creëren de recente ontwikkelingen in de gebieden van mobiele communicatie, multimedia toepassingen, netwerken, etc. steeds nieuwe eisen wat de verdere toename van data toegang en opslag, rekenkracht en multifunctionaliteit betreft. Hieruit volgt dat een totale digitalisatie geen analoge signaal versterking meer vereist. Een betrouwbare toestandsdefinitie $|0\rangle$ en $|1\rangle$, dewelke op extreem korte tijdsschaal kan uitgelezen en verwerkt worden, is een noodzaak. Deze omstandigheden creëren perfecte begincondities opdat nieuwe technologieën en component architecturen zouden ontstaan.

Eén van de nieuw ontstane technologieën is gebaseerd op een intrinsieke eigenschap van een elektron – de spin (zie Hoofdstuk 2). Een klassiek voorbeeld van zo'n component, de Grote Magnetoweerstand (GMR) junctie, heeft een revolutie veroorzaakt in de wereld van magnetische data-opslag. Een andere technologie met een veelbelovende toekomst is het Magnetisch Random Acces Geheugen (MRAM), dat gebaseerd is op een andere component die de elektron spin benut, namelijk de Magnetische Tunnel Junctie (MTJ). Deze componenten zijn passief, daar ze enkel uit metallische multilagen bestaan. Anderzijds kan het gebruik van de spin in een halfgeleider component met relatief eenvoudige architectuur aanzienlijke voordelen met zich meebrengen (zie Sectie 3.1). De voordelen van een dergelijke architectuur, die berust op quantummechanische fenomenen, worden duidelijker nu de traditionele verkleining van de component-dimensies meer en meer het quantumkarakter van de natuur blootleggen.

Voor een succesvolle werking van een dergelijke component is er een efficiënte manier nodig om spingepolariseerde ladingsdragers in een halfgeleider te creëren. Optische methodes hebben bewezen aan deze vereiste te voldoen (zie Sectie 3.3), maar ze lijken weinig veelbelovend wanneer men naar integratie op grote schaal evolueert. Het blijkt verder dat traditionele ferromagnetische materialen momenteel de beste kandidaat zijn voor deze missie, daar hun ferromagnetische orde hen tot een bijna onuitputtelijke bron van spin gepolariseerde elektronen maakt, zelfs op kamertemperatuur (zie Hoofdstuk 2). Des te meer daar hun fabricatie en fysische eigenschappen goed gekend zijn.

Jammer genoeg hebben voorafgaande experimenten, die een drie-poort geometrie voor elektrische spin injectie en detectie van spin gepolariseerde elektronen vanuit een ferromagnetisch metaal in een halfgeleider (InAs) combineren, geen effect getoond dat kan toegeschreven worden aan een elektronspin-onevenwicht in de halfgeleider (zie Sectie 3.2). Het gevolgde theoretische onderzoek van dit probleem heeft aangetoond dat elektrische spin-injectie van een ferromagnetisch metaal in een halfgeleider in het diffuse ohmse contact praktisch onmogelijk is. Dit is te wijten aan het grote verschil in toestandsdichtheid in deze materialen, zodat beide spinkanalen (spin-op en spin-neer) gelijk gevuld zijn in de halfgeleider, wat resulteert in geen algemene spinpolarisatie.

Gelukkig bieden GaAs en andere III-V halfgeleiders unieke mogelijkheden, dit niet enkel voor optische injectie van spin gepolariseerde ladingen, maar eveneens voor optische detectie van hun spinpolarisatie door de polarisatietoestand van het uitgezonden licht, resulterend uit elektron - gat recombinatie (zie Sectie 3.3). Dit laat een directe optische studie van elektrische spininjectie in een halfgeleider vanuit een ferromagnetisch metaal in een zogenaamde spin-LED toe, met slechts één enkel ferromagnetisch metaal/halfgeleider contact.

Hieruit volgt dat het traditionele probleem van ohmse contacten in het geval van GaAs, zowel als van Si, het probleem van elektrische spin injectie in deze materialen aanzienlijk verschillend maakt in vergelijking met InAs, zoals bestudeerd in voorafgaande experimenten (zie Secties 3.2, 5.2). In deze halfgeleiders bestaat het ohmse contact gewoonweg niet, daar het abrupte ferromagnetische metaal/ halfgeleider grensvlak leidt tot de vorming van een Schottkybarrière. In dit geval is elektrische injectie van elektronen in de conductieband van de halfgeleider enkel mogelijk wanneer er een sterke n-type dopering in de halfgeleider is, en voor een omgekeerd ingestelde Schottky barrière. Bij een p-type dopering is voor elektrische injectie van elektronen in de conductieband van de halfgeleider een dunne tunnelbarrière nodig aan het ferromagnetisch metaal/ halfgeleider grensvlak. In beide gevallen is het tunnelmechanisme betrokken in het elektronisch transport, wat gekend is afhankelijk te zijn van de toestandsdichtheid in beide vaste stoffen. Bijgevolg wordt het grote verschil in toestandsdichtheid voor deze vaste stoffen op elkaar afgestemd. Verder is er een grote potentiaal verval over een dergelijk grensvlak, wat niet het geval is voor het FM/InAs contact.

Het blijkt verder dat bij elektrische spin injectie in een halfgeleider, de geobserveerde emissie van circulaire polarisatie in feite een meerstaps proces is. Algemeen worden er eerst spingepolariseerde elektronen geïnjecteerd in de conductieband van de halfgeleider, hete electronen in het geval dat de kinetische energie groter is dan $k \cdot T$. Vervolgens kan er in het thermalisatieproces en gedurende de spinlevensduur beneden in de conductieband, vóór de recombinatie met gaten, wat verlies van spinpolarisatie optreden omwille van spinverstrooiing. Dit betekent dat de gemeten stationaire-toestand-spinpolarisatie van geïnjecteerde elektronen die bepaald wordt uit de polarisatietoestand van het uitgezonden licht beduidend kleiner kan zijn dan de werkelijk geïnjecteerde (zie Sectie 3.3 en Hoofdstuk 4).

De elektrische spin injectie in het spin-LED type van heterostructuur verschilt beduidend van het geval van volledig elektrische componenten (zoals bijvoorbeeld de spin-gepolariseerde veld-effect-transistor. Zie Sectie 3.1). In deze laatstgenoemden dragen slechts ladingsdragers van één enkel type, namelijk elektronen, bij aan de elektrische stroom. In het geval van de LED is er eveneens een efficiënte toevoer van gaten vereist. In het ideale geval moeten de elektron en gat stromen in dergelijke heterostructuren gelijk zijn.

Zoals volgt uit de hierboven vermelde overwegingen, moet het ontwerp van de tunnel barrière en / of halfgeleider heterostructuur niet enkel voor verhoogde injectie van elektronen van het ferromagnetisch metaal in de conductieband van de halfgeleider en voldoende toevoer van gaten voor recombinatie geoptimaliseerd worden, maar eveneens voor een verbeterd spinbehoud in de halfgeleider , om zo spinverstrooiing gedurende elektronthermalisatie en -levensduur beneden in de conductieband te minimaliseren. Verder moet de elektrische spanning over de TB in het geval van een MIS type heterostructuur geminimaliseerd worden, aangezien studies van het Tunnel-Magneto-Resistief (TMR)-effect in magnetische tunneljuncties (MTJ) aangetoond hebben dat een hoge elektrische spanning over het tunnel oxide, leidt tot een drastische verlaging van spinafhankelijke effecten. Deze verlaging wordt toegeschreven aan elektron-tunneling door via intermediaire defecttoestanden in de tunnel barrière, dewelke zich als een extra kanaal voor spinverstrooiing gedragen ( [184], zie Sectie 6.3.3).

De kennis van het ontwerp en de fabricatie van halfgeleider LED's toont aan dat hoge optische efficiëntie van elektroluminescentie bereikt kan worden in componenten die een actief gebied (gebied waar de recombinatie plaatsvindt) dicht bij het oppervlak hebben. De typische dikte van het actief gebied in LED's is $\sim 4\ldots200$ nm. In de componenten gefabriceerd in het kader van deze thesis, is het actief gebied breed genoeg gekozen, zodat er geen kwantisatie van elektron en gat niveaus plaatsvindt. Dit laat toe gedeeltelijk verlies van spin polarisatie, doordat elektronen gevangen worden in de kwantumput, te vermijden (zie Hoofdstuk 4). Verder vergemakkelijkt het de kwantitatieve analyses van gemeten data, voornamelijk op verschillende temperaturen. Dit komt omdat overgangen van de conductieband naar de valentieband (zware of lichte gaten), of met andere woorden selectieregels (de regel die de correlatie tussen de polarisatie van het uitgezonden licht en de spin polarisatie van de ladingsdragers definieert, zie Sectie 3.3.1), sterk afhangen van de energetische scheiding van deze niveaus en hun bezetting. In de kwantumput-type heterostructuur, bijvoorbeeld, is de spinpolarisatie van elektronen op lage temperatuur gelijk aan de graad van circulaire polarisatie van de elektroluminescentie. Maar op kamertemperatuur is de scheiding van de elektron en gat niveaus kleiner dan de thermische energie $k \cdot T$ en is de spin polarisatie van de elektronen twee keer groter dan de circulaire polarisatie van elektroluminescentie.

Het blijkt verder dat optische detectiemethoden, in het geval van elektrische spininjectie in een halfgeleider vanuit een ferromagnetisch metaal, niet zo vanzelfsprekend kunnen aangewend worden als in het geval van optische spin injectie. Dunne ferromagnetische films hebben namelijk typisch een magnetische anisotropie in het vlak, terwijl de hoge brekingsindex van de halfgeleider enkel toelaat de component van de elektron spin loodrecht op het oppervlak te bekijken (zie Sectie 3.3). Hieruit volgt dat een zijdelings uitzendende geometrie (het gedetecteerde licht propageert langs het FM/ halfgeleider grensvlak) het minst geschikt is voor dit type experimenten, aangezien golfgeleidende effecten in de halfgeleider heterostructuur en reflecties van het FM/ halfgeleider grensvlak de analyses van de experimentele data beduidend compliceren. Verder veroorzaken de selectie regels die dempen in het geval van kwantum begrenzing (zie Secties 3.3.3, 3.4.1), serieuze twijfels omtrent de geldigheid van de gemeten data.

Tevens is het licht dat door het oppervlak (het gedetecteerde licht propageert langs het FM/ halfgeleider grensvlak door de semi-transparante ferromagnetische film, zie Hoofdstuk 6) en de achterkant wordt uitgezonden (het gedetecteerd licht propageert in de tegengestelde richting in vergelijking met de oppervlak uitzendende geometrie) niet gepolariseerd, aangezien de geprefereerde spin oriëntatie van de elektronen die geïnjecteerd zijn in de halfgeleider, loodrecht staat op de observatierichting. De traditionele aanpak, die een sterke magnetische veld uit het vlak aanlegt (meer dan 1T voor de meest gangbare ferromagnetische metalen), welke de magnetisatie van het ferromagnetisch metaal uit het vlak trekt, en dus de oriëntatie van de elektrisch geïnjecteerde spins verandert, leidt tot significante neveneffecten (Magneto-optisch Kerr en Circulair Dichroïsme, Zeeman splitting van elektronen- en gatenniveaus, etc.). Deze neveneffecten verbergen het verwachte spin injectie signaal en kunnen volledig verantwoordelijk zijn voor de gemeten waarden.

Tijdens het werk, voorgesteld in deze thesis, werd er een nieuwe aanpak ontwikkeld om optische toegang te hebben tot elektrische spin injectie in een halfgeleider vanuit een ferromagnetisch metaal dat een magnetische anisotropie heeft loodrecht op de observatierichting. Dit is gebaseerd op spinmanipulatie in de halfgeleider (schuin ('Oblique') Hanle effect) eens de spingepolariseerde ladingsdragers geïnjecteerd zijn. In het algemeen kunnen dergelijke metingen uitgevoerd worden in een relatief zwak, uitwendig, schuin magnetisch veld, dat de magnetizatierichting van de ferromagnetische film niet beduidend beïnvloedt. Overigens verschilt de spinmanipulatie in de halfgeleider dat door zo'n magnetisch veld veroorzaakt wordt, significant van het geval van spinmanipulatie in de ferromagnetische film (verandering van magnetizatierichting). Het biedt een unieke weerspiegeling van spininjectie omdat de experimentele afhankelijkheid van het magnetisch veld een Lorentziaanse vorm heeft, terwijl de neveneffecten lineair of bijna lineair zijn met het uitwendige magnetische veld. Daarenboven onthult de aanpak met het schuine Hanle effect tegelijkertijd belangrijke informatie over spinkinetica in de halfgeleider. Dit is het geval omdat de verandering van het uit-het-vlak gemiddelde elektron spin deel, veroorzaakt door spinprecessie, bepaald wordt door het langste proces, dat gedomineerd wordt door de spinprecessie gedurende de elektronlevensduur beneden in de conductieband (na thermalisatie) van de halfgeleider. De tijdsschaal van elektron injectie en van thermalisatie is een paar grootteordes kleiner. Het schuin Hanle effect is, gecombineerd met de volledig optische karakterisatie van het spindetectie deel van de component (de halfgeleider heterostructuur), een krachtig instrument voor de kwantitatieve evaluatie van spininjectie.

Tenslotte, de overwegingen hierboven gemaakt volgend, werd er een set van verschillende MIS spin-LED’s gefabriceerd op IMEC (zie Hoofdstuk 6). De optische studie van elektrische spin injectie in deze componenten heeft een experimentele demonstratie mogelijk gemaakt van zeer efficiënte elektrische spin injectie in een halfgeleider vanuit een ferromagnetisch metaal in het directe elektrische contact, tot op kamertemperatuur, iets wat voordien volledig onmogelijk werd verondersteld. Daarenboven tonen de voorgestelde resultaten de manier om de efficiëntie van elektrische spin injectie in zulke componenten te verhogen van 2% op 80K (Typisch waargenomen door het merendeel van de groepen rondom de wereld) tot meer dan 60% op lage en kamer-temperatuur (bereikt in deze thesis). Bovendien is het belang van elektronthermalisatie-effecten en de impact van het doperingsniveau van de halfgeleider aangetoond voor een praktische studie van elektrische spininjectie op een optische manier. Verder hebben in spin-LED’s die specifiek ontworpen werden met het oog op een hoge elektronlokalisatie in het bulk-type actief gebied van de component, metingen op lage temperatuur het bestaan van een effectief magnetisch veld van nucleaire oorsprong onthuld. Deze nucleaire spinpolarisatie blijkt afkomstig van de hyperfijn interactie van nucleaire spins met de spins van elektrisch geïnjecteerde elektronen. Het toont zich in het experiment als een bijkomend magnetisch veld, toegevoegd aan het uitwendig magnetisch veld dat werkt op de spins van elektrisch geïnjecteerde elektronen.

# Besluiten en Vooruitzicht

De voorgestelde resultaten tonen aan dat de aanpak met het schuin Hanle effect een nuttig instrument is om optische toegang te krijgen tot elektrische spininjectie in halfgeleiders. Het scheidt spininjectie van neveneffecten, bijvoorbeeld van magneto-optische effecten en door Zeemansplitting geïnduceerde spinpolarisatie. Daarenboven biedt het zeer waardevolle informatie aangaande spinkinetica in de halfgeleider. Gecombineerd met de volledig optische karakterisatie van het spindetectie deel van de component, is het een krachtig instrument voor de kwantitatieve evaluatie van spininjectie.

Om snelle feedback te krijgen over de kwaliteit van spininjectoren gebruikt voor spintronica toepassingen, is het zeer belangrijk een onafhankelijk karakterisatie instrument te hebben. In het geval van MIS heterostructuren bieden tunneljuncties gefabriceerd in hetzelfde sputtersysteem essentiële feedback over de kwaliteit van de ferromagnetische metaal/ tunnelbarrière spininjectoren. Verder kan dergelijke TMR data gebruikt worden om de spininjectie-efficiëntie gemeten in de MIS-type heterostructuren te schatten. De polarisatie van geïnjecteerde elektronen gemeten in de gemaakte spin-LED's ligt reeds tamelijk dicht bij deze waarden.

Het kan lijken alsof GaAs niet eens het beste materiaal is voor de optische studie van elektrische spininjectie in de schuine Hanle effect geometrie, aangezien de elektron g-factor zeer laag is ($g^* = -0.44$ [198]). Voor andere materialen met hogere g-factoren [199], bijvoorbeeld GaSb ($g^* = -9.3$ [200]), zal dezelfde waarde van spinverstrooiingstijd $T_S$ een veel smallere (in het geval van GaSb- ~ 20 keer smallere) Hanle curve geven. In dit geval kunnen de Hanle metingen (zelfs op kamertemperatuur) uitgevoerd worden in een zeer laag magnetisch veld, waarin de invloed van het uitwendig magnetische veld op de magnetisatie van ferromagnetische film volledig genegeerd kan worden.

Tegenwoordig lijkt het dat de spininjectie in de hybride ferromagneet/ oxide/ halfgeleider componenten verhoogd kan worden op dezelfde manier als bij de TMR juncties, namelijk door de kwaliteit van de oxide barrière en zijn grensvlakken te verbeteren, en door ferromagnetische materialen met een hogere spin polarisatie te gebruiken.

De voorgestelde resultaten duiden aan dat het gebruik van een tunnelbarrière-injector inderdaad een interessante aanpak is om spins in een halfgeleider te injecteren. Het inbrengen van een oxidelaag laat toe stabielere en robuustere injectoren te maken. Er kan een grote variëteit aan ferromagnetische materialen op de oxidelaag gedeponeerd worden, om zo een universele spinbron te bekomen.

Alhoewel deze studie enkel spininjectie experimenteel aantoont in een enkel direct ferromagneet/halfgeleider contact, lijkt elektrische spininjectie met een tweede ferromagnetisch metaal in een dergelijk systeem geen probleem. Immers zijn dezelfde overwegingen geldig voor spin detectie op zo’n grensvlak, des te meer daar het reeds werd aangetoond in gelijkaardige systemen [201-204].

Deze resultaten lijken erg veelbelovend voor toekomstige spintronica componenten, die op kamertemperatuur werken, en dit gebruik makend van stabiele tunnelbarrière injectoren zoals $Al_2O_3$ of AlN op III-V (e.g. GaAs, GaN) of state-of-the-art $SiO_2$ voor Si/SiGe componenten.

De mogelijkheid om dynamische nucleaire spinpolarisatie door elektrisch geïnjecteerde spingepolariseerde elektronen te benutten, opent een nieuwe weg voor de praktische realisatie van grootschaalse integratie van vaste stof quantum computing, gebruik makend van reeds voorgestelde principes [192, 205] of volledig nieuwe ideeën ‘diegenen waaraan we nog niet eens denken’ [10].

# Scientific Contributions

## Publications

V.F.Motsnyi, P.Van Dorpe, W.Van Roy, E.Goovaerts, V.I.Safarov, G.Borghs, J.De Boeck, '*Dynamic nuclear spin polarization by electrical spin injection into a semiconductor heterostructure*', (2003, to be submited).

P.Van Dorpe, Z.Liu, W.Van Roy, V.F. Motsnyi, M.Sawicki, G.Borghs, J.De Boeck , '*Very high electron spin polarization in GaAs by injection from a (Ga,Mn)As Zener diode*', (2003 to be submited).

V.F.Motsnyi, P.Van Dorpe, W.Van Roy, E.Goovaerts, V.I.Safarov, G.Borghs, J.De Boeck, '*Optical Investigation of Electrical Spin Injection Into Semiconductors*', Phys.Rev.B (2003, submitted).

V.F.Motsnyi, J.De Boeck, J.Das, W.Van Roy, G.Borghs, E.Goovaerts, V.I.Safarov, '*Electrical spin injection in a ferromagnet'tunnel barrier / semiconductor heterostructure*', Appl.Phys.Lett. 81, 265 (2002).

P.Van Dorpe, V.F.Motsnyi, M.Nijboer, E.Goovaerts, V.I.Safarov, J.Das, W.Van Roy, G.Borghs, J. De Boeck, '*Highly efficient room-temperature spin injection in a metal-insulator-semiconductor light emitting diode*', Jpn.J.Appl.Phys. 42, L502 (2003).

V.F.Motsnyi, V.I.Safarov, P.van Dorpe, J.Das, W.Van Roy, E.Goovaerts, G.Borghs, J.De Boeck, '*Electricals spin injection in a ferromagnetic Metal/ Insulator/ Semiconductor tunnel heterostructure*', J.Superconductivity 16, 671 (2003) [Proceedings of the NATO ARW on Frontiers of Spintronics & Optics in Semiconductors ( Symposium in Honor of E.I.Rashba), 20-22 June 2002, Cambridge, MA USA].

J.De Boeck, W.Van Roy, J.Das, V.Motsnyi, Z.Liu, L.Lagae, H.Boeve, K.Dessein and G.Borghs, '*Technology and Materials Issues in Semiconductor-based Magnetoelectronics*', Semiconductor Science and Technology 17 (N4), 342 (2002), Special Issue: Semiconductor Spintronics with Guest Editor Hideo Ohno, *Invited*.

J.De Boeck, W.Van Roy, V.Motsnyi, Z.Liu, K.Dessein, G.Borghs, '*Hybrid epitaxial structures for spintronics*', Thin Solid Films 412, 3 (2002), *Invited*.

Jo De Boeck, Vasyl Motsnyi, Liu Zhiyu, Jo Das, Liesbet Lagae, Willem Van Roy, Viacheslav Safarov, Etienne Goovaerts, Gustaaf Borghs, '*Magnetoelectronics: Effective use of the electron spin in magnetic / semiconductor hybrid components*', in *Frontiers of Multifunctional Nanosystems* edited by E.Buzaneva & P.Scharff (Kluwer Academic Publishers, Dordrecht / Boston / London, ICBN 1-4020-0560-1 (HB), ICBN 1-4020-0561-X (PB) ), p.453: Proceedings of NATO ARW, 9-12 September 2001, Kyiv, Ukraine, *Invited*.

## Conference Contributions

V.F.Motsnyi, P.Van Dorpe, W.Van Roy, E.Goovaerts, V.I.Safarov, G.Borghs, J.De Boeck, '*Dynamic nuclear spin polarization by electrical spin injection into a semiconductor heterostructure*', Abstracts of Spintech II (International Conference and School on Semiconductor Spintronics and Quantum Information Technology), 4-8 August 2003, Brugge, Belgium, *Oral Contribution.*

P.Van Dorpe, Z.Liu, W.Van Roy, V.F. Motsnyi, M.Sawicki, G.Borghs, J.De Boeck , '*50% electron spin polarization in GaAs by injection from a (Ga,Mn)As Zener diode*', Abstracts of Spintech II (International Conference and School on Semiconductor Spintronics and Quantum Information Technology), 4-8 August 2003, Brugge, Belgium, *Oral Contribution.*

V.I.Safarov, V.F.Motsnyi, J.De Boeck, P.Van Dorpe, W.Van Roy, E.Goovaerts, and G.Borghs, '*Highly Efficient Spin Injection in Ferromagnetic Metal/Insulator/ Semiconductor Tunnel Structures*', Abstracts of MRS Fall meeting, 2-6 December 2002, Boston, USA, *Invited Oral Contribution*.

W.Van Roy, P.Van Dorpe, V.F.Motsnyi, J.De Boeck, '*Spin injection from epitaxial NiMnSb into GaAs through a Schottky tunnel barrie*', Abstracts of Spintech II (International Conference and School on Semiconductor Spintronics and Quantum Information Technology), 4-8 August 2003, Bruges, Belgium.

P.Van Dorpe, V.Motsnyi, M.Nijboer, W.Van Roy, J.Das, E.Goovaerts, V.Safarov, G.Borghs, J.De Boeck, "*The influence of the tunnel barrier properties on electrical spin injection in a MIS-spin LED*", Abstracts of 2$^{nd}$ International Conference on Physics and Application of Spin Related Phenomena in Semiconductors (PASPS), 23-26 July 2002, Wurzburg, Germany, *Invited Oral Contribution.*

J.De Boeck, W.Van Roy, V.F.Motsnyi, Z.Liu, P.Van Dorpe, M.Nijboer, J.Das, E.Goovaerts, V.I.Safarov, G.Borghs, '*Magnetic / Semiconductor Heterostructures for Spintronic Devices*', Abstracts of Intermag 2002, 28 April- 2 May 2002, Amsterdam, *Invited Oral Contribution*.

Vasyl Motsnyi, Viatcheslav Safarov, Jo De Boeck, Jo Das, Wim Van Roy, Etienne Goovaerts, Staf Borghs, ‘*Oblique Hanle Effect for Reliable Assessment of Electrical Spin Injection*’, Abstracts of the Intermag 2002, 28 April- 2 May 2002, Amsterdam, *Oral Contribution*.

W.Van Roy, P.Van Dorpe, V.F. Motsnyi, G. Borghs, J.De Boeck, ‘*Electrical spin injection from NiMnSb into GaAs*’, Abstracts of the AVS 49th International Symposium, 4-8 November 2002, Denver, Colorado, USA.

V.Motsnyi, P.van Dorpe, M.Nijboer, W.van Roy, J.Das, E.Goovaerts, J.De Boeck, V.Safarov, G.Borghs, ‘*Electrical spin injection in a semiconductor in the MIS heterostructure. Influence of the tunnel Barrier’*, Abstracts of the 26-th International Conference on the Physics of Semiconductors (ICPS), 29 July - 2 August 2002, Edinburgh, UK.

J.De Boeck, W.Van Roy, V.Motsnyi, Z.Liu, K.Dessein, G.Borghs, ‘*Hybrid epitaxial structures for spintronics*’, Abstracts of the 4th Workshop on MBE and VPE Growth, 24-28 September 2001, Warsaw, Poland, *Invited Oral Contribution*.

Willem Van Roy, Vasyl F.Motsnyi, Zhiyu Liu, Marek Wójcik, Ewa Jędryka, Stefan Nadolski, Viatcheslav I.Safarov, Gustaaf Borghs, and Jo De Boeck, *"Sources for Spin Injection into Semiconductors: Half-Metallic NiMnSb and Others,"* Abstracts of the 2001 JRCAT International Symposium on Atom Technology, 11-12 December 2001, Tokyo, Japan.

Jo De Boeck, Vasyl Motsnyi, Liu Zhiyu, Jo Das, Liesbet Lagae, Willem Van Roy, Viacheslav Safarov, Etienne Goovaerts, Gustaaf Borghs, ‘*Magnetoelectronics: effective use of the electron spin in magnetic / semiconductor hybrid components*’, Abstarcts of the NATO ARW ‘Frontiers of Multifunctional Nanosystems’, 9-12 September 2001, Kyiv, Ukraine, *Invited Oral Contribution*.

Vasyl Motsnyi, Willem Van Roy, Jo Das, Etienne Goovaerts, Gustaaf Borghs, Jo De Boeck, ‘*Ferromagnetic metal/tunnel barrier/ semiconductor devices for optical detection of spin-polarized current injection into a semiconductor*’, Abstracts of the 1st Joint European Magnetic Symposia, 28th August-1st September 2001, Grenoble, France.

Vasyl Motsnyi, Wilem Van Roy, Jo Das, Hans Boeve, Barun Dutta, Etienne Goovaerts, Staf Borghs, Jo De Boeck, ‘*Ferromagnetic metal/tunnel-barrier/semiconductor devices for optical detection of spin-polarized current injection into a semiconductor*’, Abstracts of the NEVAC/NNV Symposium "Applications of Magnetic Nanostructures", 17 November 2000, Eindhoven, Netherlands.

# Curriculum Vitae

**November 21, 1975,** Born in Bila Tzerkva (District of Kyiv), Ukraine.

**1992-1997** Student at National Taras Shevchenko University of Kyiv, Faculty of Radiophysics. Received the M.Sc. Degree in physics, option solid-state electronics, in June 1997.

**1998-2003** Ph.D. research at University of Antwerp, Faculty of Physics, and performed at IMEC, Interuniversity Micro-Electronics Centre.

**September 2003** Ph.D. Defense: ‘Optical investigation of electrical spin injection into semiconductors’.

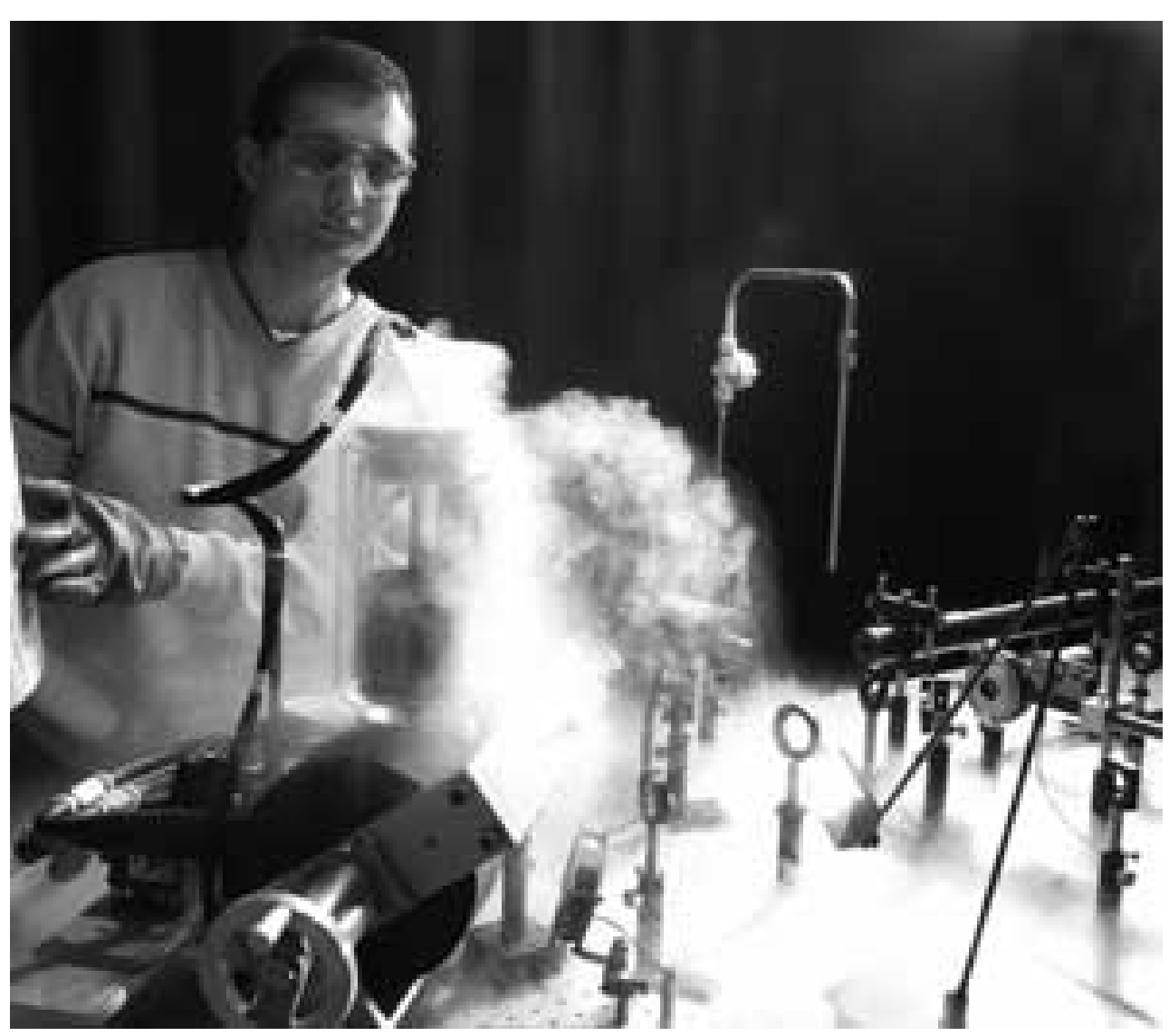

# Appendix A: Measurements of Electron Lifetime and Spin Scattering Time in p-GaAs Samples with Different Doping Concentration

In this section the measurements of all characteristic electron lifetimes in the p-GaAs samples having different doping concentration are presented. The samples were grown by MBE on similar p-GaAs substrates at different occasion. The doping concentrations and corresponding sample numbers are presented in Table A.1. The thickness of grown GaAs layers is $\sim 1.5 \ldots 2.5\ \mu m$ .

**Table A.1. GaAs samples and corresponding doping levels.**

| *Sample* | Doping level, [$cm^{-3}$] |
|---|---|
| G2358<br>G2458 | $p=5 \cdot 10^{16}$ |
| G2319<br>G2508 | $p=5 \cdot 10^{17}$ |
| G2247 | $p=6 \cdot 10^{17}$ |
| G2444 | $p=1.5 \cdot 10^{18}$ |

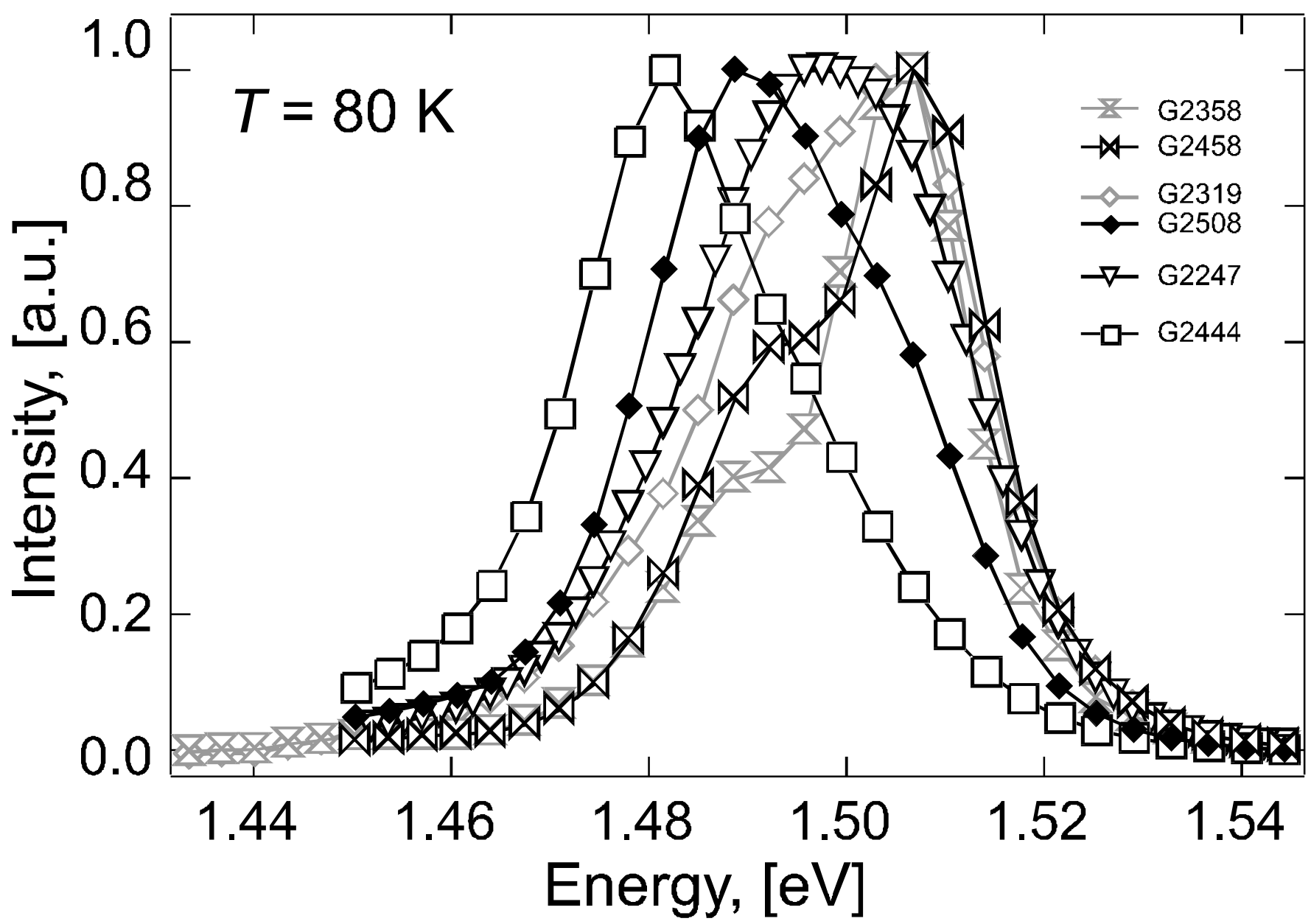

Fig.A.1. Experimental photoluminescence spectra of the GaAs samples having different doping level (Table A.1). Short pulses of femtosecond Ti/ Sapphire laser $h \cdot \upsilon = 1.63$ eV were used for optical excitation.

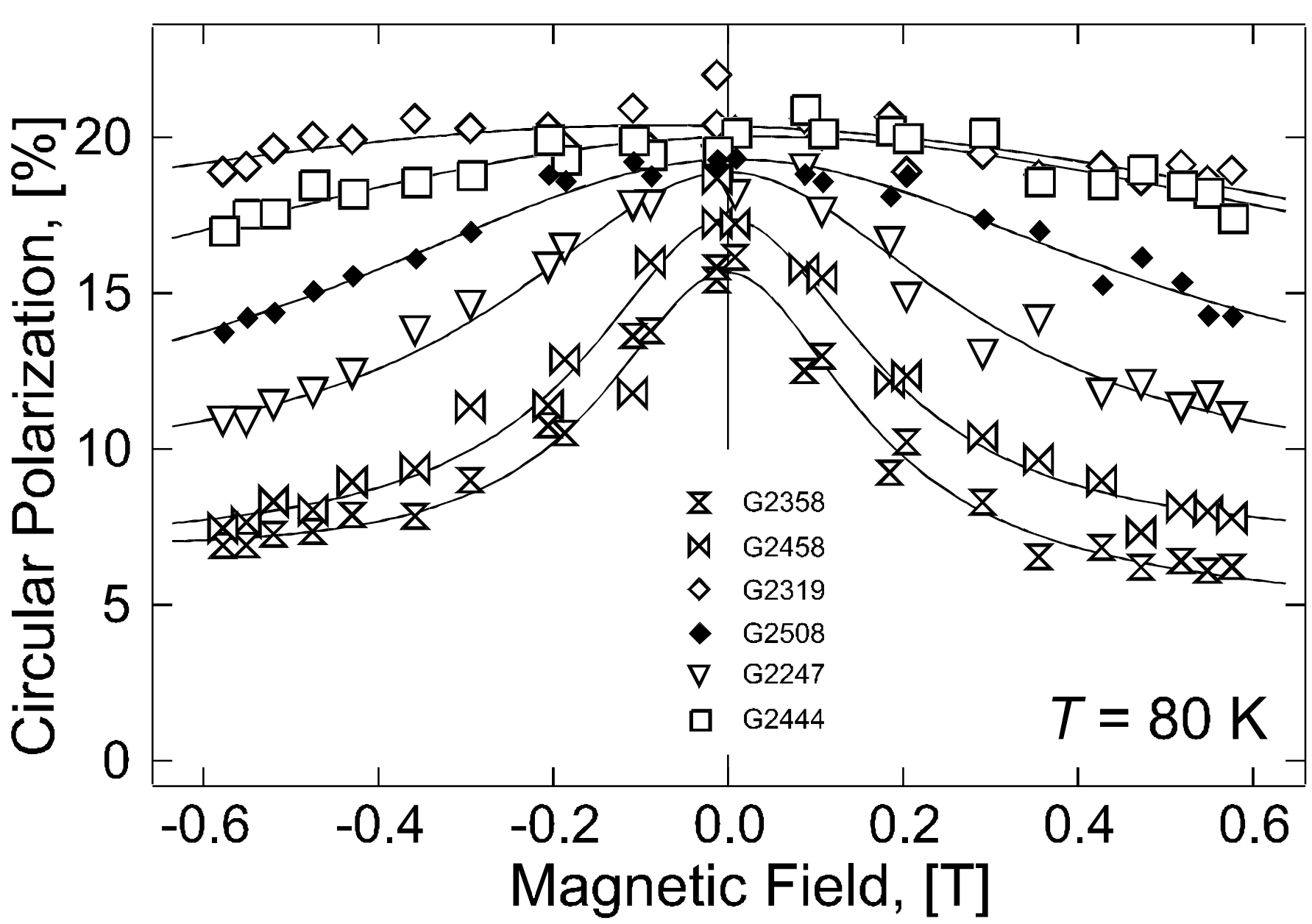

Fig.A.2. Depolarization of photoluminescence in the oblique Hanle effect geometry ($\varphi = \pi/4$) for the GaAs samples having different doping concentration (Table A.1). For optical excitation the 100% circularly polarized short pulses of femtosecond Ti/Sapphire laser ($h \cdot \upsilon = 1.63$ eV ) were used.

The electron lifetime and spin relaxation time measurements were performed in the all-optical experiment under optical spin injection and detection in the oblique Hanle effect geometry ( $\varphi = \pi/4$ , see Sections 4.2). For optical excitation the 100% circularly polarized short pulses of femtosecond Ti/Sapphire laser ( $h \cdot \upsilon = 1.63$ eV ) were used. Fig.A.1 shows the experimental photoluminescence spectra observed on these samples.

The experimental Hanle curves were measured at the maxima of the photoluminescence spectra. Fig.A.2 shows the measured depolarization of photoluminescence in the oblique Hanle effect geometry. The electron lifetime $\tau$ , spin relaxation time $\tau_S$ and spin lifetime $T_S$ determined from these measurements are presented in Fig.A.3. The same parameters for the sample *Type C* measured in an independent experiment under optical excitation with continuous wave semiconductor laser ( $h \cdot \upsilon = 1.58$ eV ) are shown for comparison (see Section 6.3.1). The filled circles correspond to the electron spin relaxation time for p-GaAs samples with different doping concentration reported early [206, 18] (see also Fig.3.8).

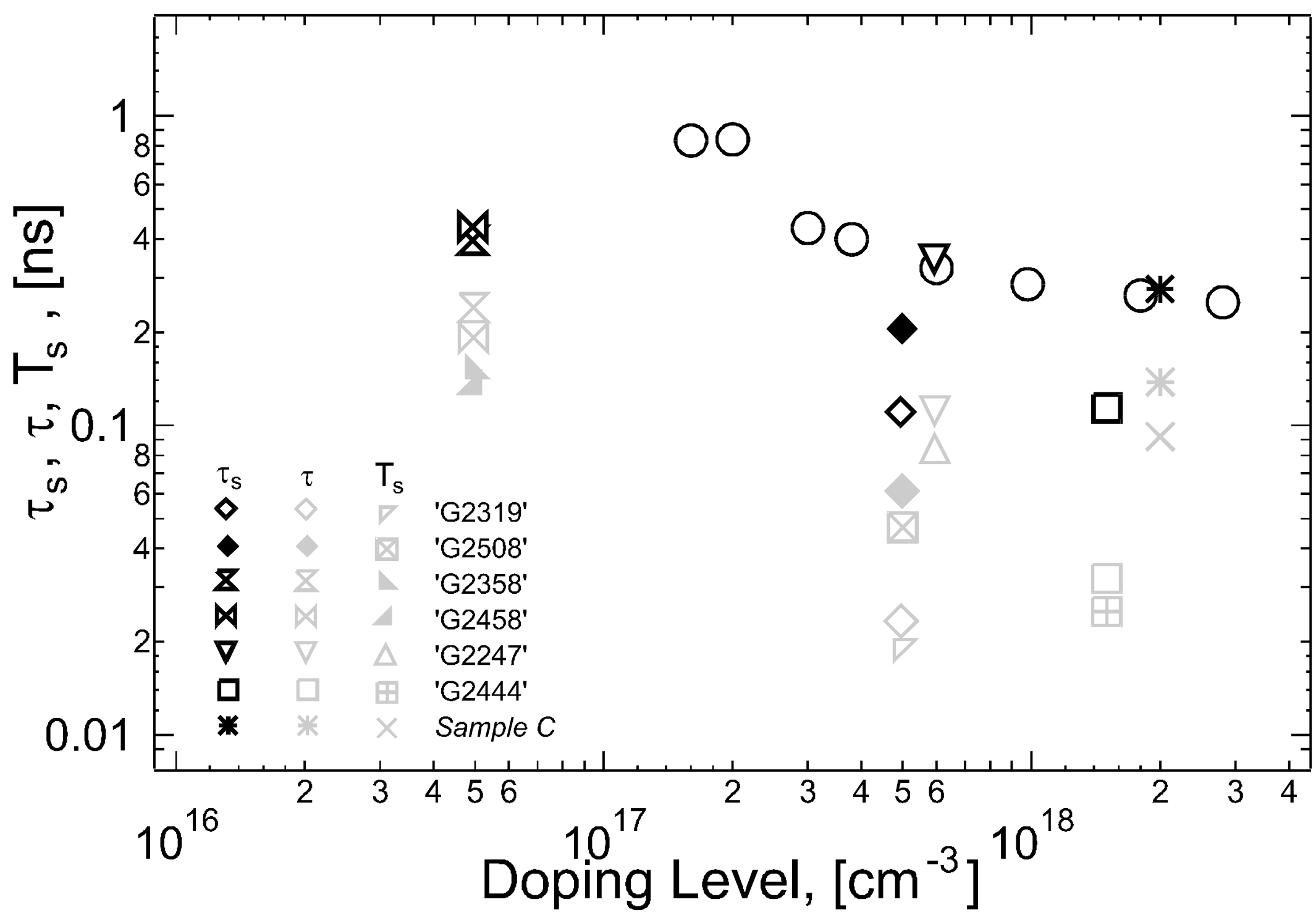

Fig.A.3. The electron lifetime $\tau$ , spin relaxation time $\tau_S$ and spin lifetime $T_S$ determined from the measurements presented in Fig.A.2. The same parameters for the sample *Type C* (see Section 6.3.1 for details) and the spin relaxation times for p-GaAs samples with different doping levels reported in Ref.[ 206, 18] (see also Fig.3.8).

The small variation of the spin relaxation time in these samples comparing to Ref.[206, 18] can be explained by different factors. First, for the measurements of all characteristic electron lifetimes the optical excitation was performed by short pulses, hence, the steady state approach (see Section 4.2) is not strictly valid. Second, after growth the samples were kept in air for some time. So the contamination by other elements can influence the measured parameters. The third one is that due to higher concentration of the structural defects the samples grown by MBE have worse optical efficiency as compared to the samples grown by MOSCVD, for example. As consequence, these defects may influence the measured quantities. And the last one is different nature of the dopant itself, which may influence the spin relaxation time in the semiconductor.